\documentclass[prd,aps,showpacs,twocolumn,superscriptaddress,floatfix,nofootinbib]{revtex4-1}
\usepackage[utf8]{inputenc}
\usepackage{bm}
\usepackage{bbold}
\usepackage{mathtools}
\usepackage[toc,page]{appendix}
\usepackage{dsfont}
\usepackage{combelow}
\usepackage[colorlinks=true,linkcolor=blue,citecolor=blue, urlcolor=blue]{hyperref} 
\usepackage[a4paper,top=2.4cm,bottom=2.4cm,right=1.0cm,bindingoffset=-1.2cm]{geometry}
\usepackage{tikz}
\usepackage{graphicx}
\usepackage{subfigure}
\usepackage{scalerel,amssymb}
\DeclareMathAlphabet{\pazocal}{OMS}{zplm}{m}{n}

\renewcommand{\d}{\mathrm{d}}
\newcommand{\epn}{\frac{\d E_\perp^{(0)}}{\d \eta}}

\newcommand{\F}{\pazocal{F}}
\newcommand{\xT}{\mathbf{x_\perp}}
\newcommand{\txT}{\tilde{\mathbf{x}}_\perp}
\newcommand{\nT}{\mathbf{n_\perp}}
\newcommand{\hxT}{\hat{\mathbf{x}}_\perp}
\newcommand{\pT}{\mathbf{p_\perp}}
\newcommand{\vT}{\mathbf{v_\perp}}

\newcommand{\NT}{\mathbf{\nabla_\perp}}

\newcommand{\ebar}{\bar{\epsilon}}

\begin{document}
\title{Development of transverse flow at small and large opacities in conformal kinetic theory}
\author{Victor E. Ambru\cb{s}}
\affiliation{Institut f\"ur Theoretische Physik, Johann Wolfgang Goethe-Universit\"at, Max-von-Laue-Strasse 1, D-60438 Frankfurt am Main, Germany}
\affiliation{Department of Physics, West University of Timi\cb{s}oara, \\
Bd.~Vasile P\^arvan 4, Timi\cb{s}oara 300223, Romania}
\author{S.~Schlichting}
\affiliation{Fakultät für Physik, Universität Bielefeld, D-33615 Bielefeld, Germany}
\author{C.~Werthmann}
\email{cwerthmann@physik.uni-bielefeld.de}
\affiliation{Fakultät für Physik, Universität Bielefeld, D-33615 Bielefeld, Germany}
\date{\today}

\begin{abstract}
We employ an effective kinetic description, based on the Boltzmann equation in the relaxation time approximation, to study the space-time dynamics and development of transverse flow of small and large collision systems. By combining analytical insights in the small opacity limit with numerical simulations at larger opacities, we are able to describe the development of transverse flow from very small to very large opacities. Suprisingly, we find that deviations between kinetic theory and hydrodynamics persist even in the limit of very large opacities, which can be attributed to the presence of the early pre-equilibrium phase.
\end{abstract}

\maketitle
\section{Introduction}\label{sec:intro}

 Over the past decades, the collective flow of soft hadrons produced in high-energy heavy-ion collisions at the Relativistic Heavy-Ion Collider (RHIC) and the Large Hadron Collider (LHC), has become one of the cornerstones to establish the existence of deconfined Quark Gluon Plasma (QGP), and to characterize the properties of strong-interaction matter under extreme conditions. 
 The space-time dynamics of relativistic heavy-ion collisions is commonly described in terms of relativistic viscous hydrodynamics~\cite{Teaney:2009qa,Gale:2013da,Luzum:2013yya,Heinz:2013th,Jeon:2015dfa},
 which provides an accurate description of experimental measurements of soft hadron production and collective flow at RHIC and LHC. 

 Based on the tremendous success in quantfying properties of the QGP produced in heavy-ion collisions~\cite{JETSCAPE:2020mzn,Nijs:2020roc,Gardim:2019xjs,Schenke:2020mbo}, different groups have performed
 hydrodynamic calculations for small systems~\cite{Bozek:2011if,Bozek:2012gr,Bozek:2013df,Bozek:2013uha,Bozek:2013ska,Bzdak:2013zma,Qin:2013bha,Werner:2013ipa,Kozlov:2014fqa,Schenke:2014zha,Romatschke:2015gxa,Shen:2016zpp,Weller:2017tsr,Mantysaari:2017cni,Schenke:2019pmk}, which also provide a reasonable description of the
 experimentally observed collective flow in proton-nucleus and proton-proton collisions~\cite{Dusling:2015gta,Loizides:2016tew,Nagle:2018nvi}. However, in constrast to nucleus-nucleus collisions, such calculations  
 are subject to much larger uncertainties, where in addition to the poorly constrained initial state geometry~\cite{Schenke:2014zha,Schenke:2021mxx,Demirci:2021kya}, one may question the theoretical justification 
 for employing a hydrodynamic description for a system, which features a very short lifetime and consists of very few degrees of freedom.
 
 Despite significant progress in understanding the onset of hydrodynamic behavior in QCD plasmas~(see e.g. \cite{Schlichting:2019abc,Berges:2020fwq} for recent reviews), calculations performed in this regard are typically
 subject to simplifying assumptions, e.g. modeling the early stages of heavy-ion collisions in terms of a transversely homogenous Bjorken flow, and are therefore
 not (yet) able to capture the competing effects of longitudinal and transverse expansion in small collision systems. 
 
 Beyond studies based on effective macroscopic descriptions of QCD, there have also been attempts to explain the onset of collective behavior in small systems by invoking a microscopic origin of the correlations. 
 Examples include calculations within the Color Glass Condensate (CGC) effective field theory of high-energy QCD~\cite{Schenke:2015aqa,McLerran:2015sva,Schenke:2016lrs,Dusling:2017dqg,Dusling:2017aot,Greif:2017bnr,Mace:2018vwq,Mace:2018yvl,Kovner:2018fxj,Greif:2020rhi,Agostini:2021xca}, as well as more conventional approaches extending 
 general purpose event generators such as PYTHIA or HERWIG to include space-time dependent final state interactions~\cite{Abramovsky:1988zh,OrtizVelasquez:2013ofg,Bierlich:2017vhg,Bierlich:2020naj}.

 Clearly, the development of a unified description, that encompasses  both paradigmes in the respective limit is an outstanding challenge~\cite{Wiedemann:2021bwz}. Beyond microscopic calculations that are rooted in the underlying theory of QCD, it is therefore an important achievement that -- at least within simpler microscopic descriptions --
 it is possible to include a non-trivial transverse expansion~\cite{Heiselberg:1998es,Borghini:2010hy,Romatschke:2018wgi,Kersting:2018qvi,Kurkela:2019kip,Kurkela:2018ygx,Borghini:2018xum,Kurkela:2021ctp}, and in some cases even detailed event-by-event geometries~\cite{He:2015hfa,Greif:2017bnr,Kurkela:2020wwb,Roch:2020zdl} to describe the onset of collective flow. In this spirit, the central objective of this paper is to scrutinize the development of transverse flow and investigate possible changes in the space-time dynamics and flow response of small and large systems.

We follow previous works~\cite{Kurkela:2018ygx,Kurkela:2019kip,Kurkela:2020wwb} and employ a simple kinetic description of the system described in Sec.~\ref{sec:theory},
where the system size and energy dependence is characterized by a single opacity parameter.  
Within this framework, we perform (semi-)analytic calculations to leading order in opacity in Sec.~\ref{sec:linear} and subsequently in Sec.~\ref{NumericalProcedure} develop two different 
numerical schemes that allow us to study the evolution of the system all the way from very low to very high opacity. Numerical results are presented in Sec.~\ref{sec:results}, where
we analyze the longitudinal cooling and flow response in kinetic theory as a function of opacity and compare it to analogous calculations in ideal and viscous hydrodynamics. 
We conclude with Sec.~\ref{sec:conclusions}. Several appendices contain additional details and explicit expressions for the (semi-) analytic calculations.
\newpage

\section{Effective kinetic description of anisotropic flow}\label{sec:theory}
\subsection{Setup}\label{sec:theory:setup}

We will describe the system via an averaged phase space distribution $f(x,p)$ of (massless) quasi-particles, for which we assume boost invariance along the longitudinal (beam) direction. Effectively, this reduces the dimensionality of the problem to (2+1)+3 dimensions, which can be efficiently described using Milne coordinates $x^\mu=(\tau,\xT,\eta)$ and $p^\mu=(p^\tau,\pT,p^\eta)$, where
\begin{align}
    \tau=&\sqrt{(x^0)^2-(x^3)^2}, & \eta=&\mathrm{artanh}(x^3/x^0),
    \label{eq:Milne}
\end{align}
such that $\tau$ is invariant and $\eta$ behaves additively under boosts in the longitudinal direction. Defining similarly
\begin{align}
    y=\mathrm{artanh}(p^z/p^t),
\end{align}
it follows from boost invariance that the phase-space distribution $f$ can only depend on $\eta$ and $y$ via their difference. We denote transverse positions in terms of regular cartesian coordinates $\xT=(x^1,x^2)$, such that the metric of the coordinate system $(\tau,\xT,\eta)$ is given by $g_{\mu\nu}=\mathrm{diag}(1,-1,-1,-\tau^2)$. The corresponding momenta $(p^\tau,\pT,p^\eta)$ are the cartesian transverse momentum $\pT$ and
\begin{align}
    p^\tau=p_\perp\,\mathrm{cosh}(y-\eta) \ \ , \quad p^\eta=\frac{p_\perp}{\tau}\,\mathrm{sinh}(y-\eta) \ \ ,
\end{align}
where $p_\perp=|\pT|$. Based on a kinetic description of the non-equilibrium dynamics, the evolution of the phase-space distribution is governed by the Boltzmann equation\footnote{We note that in Eq.~(\ref{eq:BoltzmannRTA}), coordinate derivatives on the lhs are to be evaluated at constant $p^{\mu}$ in Minkowksi space. Throughout this manuscript, we will employ different parametrizations of the spatial and momentum coordinates, which give rise to additional terms on the lhs. Specifically for a boost-invariant system in Milne coordinates, one finds~\cite{Mueller:1999pi} 
\begin{equation*}
    p^\tau\partial_\tau f + \bm{p}_\perp \cdot \partial_{\bm{x}_\perp} f 
    - \frac{p^\tau p^\eta}{\tau} \frac{\partial f}{\partial p^\eta}
    = C[f].
\end{equation*}
}
\begin{equation}
    p^\mu\partial_\mu f =C[f],\label{eq:BoltzmannRTA}
\end{equation}
For the collision kernel, we employ the relaxation time approximation (RTA)
\begin{equation}
C_{RTA}[f]=-\frac{p_\mu u^\mu}{\tau_R}\left[f-f_{eq}\left(\frac{p_\mu u^\mu}{T}\right)\right]\label{eq:BoltzmannRTA_C},
\end{equation}
where we choose a temperature dependent relaxation time 
\begin{align}
    \tau_R=\frac{5\eta/s}{T} \ \ ,
\end{align}
to describe a conformal system with constant shear-viscosity to entropy density ratio $\eta/s$. Energy-momentum conservation requires that the local equilibrium temperature $T$ and rest-frame velocity $u^\mu$ are determined via the Landau matching condition~\cite{Anderson:1974,Anderson:1974b,Cercignani:2002,Rezzolla:2013,Rocha:2021zcw}
\begin{align}
    u_\nu T^{\mu\nu}=\epsilon u^\mu 
    \label{eq:Landau}
\end{align}
with timelike four-velocity eigenvector $u_\mu u^\mu=+1$ and eigenvalue $\epsilon$, representing the energy in the local rest-frame. The temperature $T$ can be computed from the energy density $\epsilon$ via the equation of state
\begin{align}
    \epsilon=\frac{\pi^2}{30}\nu_{\rm eff} T^4 \ \ ,
\end{align}
which introduces a proportionality factor $\frac{\pi^2}{30}\nu_{\rm eff}$, which can be absorbed into redefintions of the related quantities (c.f. Sec.~\ref{sec:theory:scaling_properties}).
The stress-energy tensor $T^{\mu\nu}$ is obtained from the distribution function $f$ via
\begin{equation}
 T^{\mu\nu}(x) = \nu_{\rm eff} \sqrt{-g} \int \frac{\d^3p}{(2\pi)^3\,p^\tau} p^\mu p^\nu f(x,p)\;, \label{eq:Tmunu}
\end{equation}
where $\sqrt{-g}=\tau$ denotes the metric determinant, $\d^3p = \d^2 \pT \d p^\eta$ is the integration measure in Milne coordinates and $\nu_{\rm eff}$ represents the degeneracy factor.

Since the computation of observables will involve weighted integration of $f$ over momentum space, we define a shorthand notation $\langle\, \cdot\, \rangle$ as
\begin{equation}
    \langle \mathcal{O} \rangle = \nu_{\rm eff} \sqrt{-g} \int \frac{\d^3p}{(2\pi)^3\, p^\tau} \,\mathcal{O}(\pT, p^\eta)\, f \ \ ,\label{eq::defMomentumInt}
\end{equation}
which allows for example to write the energy momentum tensor as $T^{\mu\nu} = \left\langle {p^\mu p^\nu }\right\rangle$.

We consider initial conditions motivated by the Color-Glass condensate (CGC) effective field theory of high energy QCD~\cite{Gelis:2010nm}, where the initial state very shortly after the collision $(\tau \sim 1/Q_{s})$, can be viewed as a highly anisotropic collection of gluons with typical transverse momenta $\sim Q_{s}$ and vanishing longitudinal momenta in the local rest frame~\cite{Greif:2017bnr,Greif:2020rhi}. Specifically, we will consider initial conditions of the form 
\begin{equation}
    f(\tau_0,\xT,\pT,y-\eta)
    = \frac{(2\pi)^3}{\nu_{\rm eff}}\frac{\delta(y-\eta)}{\tau_0 p_\perp}\frac{\d N_{0}}{ \d^2\xT\d^2 \pT\d y},
    \label{eq:fini}
\end{equation}
which has vanishing longitudinal pressure ($T^{\eta}_{~\eta}=0$). Strikingly, it can be shown (c.f. Sec.~\ref{sec:theory:scaling_properties}) that  -- due to the particularly simple nature of RTA -- certain energy weighted observables do not depend on a particular form of the (isotropic) momentum distribution $\frac{\d N_{0}}{ \d^2\xT\d^2 \pT\d y}$~\cite{Kurkela:2019kip}, and we will therefore not specify it further. Instead the dynamics is entirely described by the initial energy density distribution, which for the initial conditions in Eq.~(\ref{eq:fini}) is determined by\footnote{Strictly speaking, the integration in Eq. (\ref{eq:epsilon-f-relation}) yields a density in $\d^2\xT \d y$. However, in the boost-invariant case, the phase-space distribution $f$ only depends on $y-\eta$, meaning that integration over $y$ and $\eta$ is interchangeable and densities w.r.t.+ $\d y$ and $\d \eta$ are identical.}

\begin{align}
 \epsilon(\tau_0, \xT)
 =& \frac{1}{\tau_0} \int d^2\pT~p_\perp~\frac{\d N_{0}}{ \d^2\xT \d^2 \pT\d y} \ \ . \label{eq:epsilon-f-relation}
\end{align}
We take the initial energy density $\epsilon(\tau_0, \xT)$  as a superposition of an isotropic background $\ebar(\tau_0,x_\perp)$ depending only on $x_\perp=|\xT|$ and an anisotropic component $\delta\epsilon(\tau_0,\xT)$, i.e.
\begin{align}
    \epsilon(\tau_0,\xT)&=\ebar(\tau_0,x_\perp)+\delta\epsilon(\tau_0,\xT) \ . \label{eq:initialEnergyDecomp}
\end{align}
We follow previous works
\cite{Kersting:2018qvi,Kurkela:2020wwb}  and consider the background to be of a rotationally symmetric Gaussian shape
\begin{align}
    \ebar(\tau_0,x_\perp) &= \frac{1}{\pi R^2 \tau_0} \epn \mathrm{exp}\left(-\frac{x_\perp^2}{R^2}\right), \label{eq:epsilon_isotropic_def}
\end{align}
where $R$ denotes the transverse system size and $\epn$ denotes the initial energy per unit rapidity. 
Similarly, the anisotropic perturbations are taken as\footnote{Note that the anisotropic perturbations contains a variance modification factor $\alpha$; a similar factor in the isotropic Gaussian could always be absorbed into the definition of $R$.}
\begin{multline}
    \delta\epsilon(\tau_0,\xT) =\ebar(\tau_0,x_\perp)\,\delta_n\,\mathrm{exp}\left(-\alpha\frac{x_\perp^2}{R^2}\right)\left(\frac{x_\perp}{R}\right)^n\\
    \times \mathrm{cos}(n\phi^{(n)}_{\xT\nT}) \ \ .
    \label{eq:deltaeps_def}
\end{multline}
such that upon integrating over the transverse coordinates $\xT$ the perturbations do not contribute to the total energy. 
By $\phi_{\xT\nT}^{(n)}$, we denote the angle 
\begin{align}
    \phi_{\xT\nT}^{(n)}=\phi_{x}-\Psi_{n} \ \ ,
    \label{eq:phin_def}
\end{align}
where $\phi_{x}$ is the position space azimuthal angle, i.e. $\phi_{x}=\arctan(x^2/x^1)$, and $\Psi_n$ is the symmetry plane angle of the $n$-th order angular harmonic mode. To compactify the notation, in the following we will drop the superscript $(n)$ and write $\phi_{\xT\nT}$. We note that in accordance with Eq.~\eqref{eq:deltaeps_def}, we will restrict ourselves to including only one anisotropic mode at a time, which means that we need not specify $\Psi_n$ (or rather the relative angles between different symmetry planes). We leave the parameter $\alpha$ unspecified for analytical calculations, and if not stated otherwise employ $\alpha=1/2$ when presenting numerical results.

By varying the amplitude $\delta_n$ of the anistropic perturbations, we can adjust the eccentricities $\epsilon_{n}$ of the initial state energy distribution. Following standard procedure~\cite{Teaney:2010vd,Bhalerao:2011yg}, the initial state eccentricities $\epsilon_{n}$ are determined as
\begin{align}
    \epsilon_{n}=-\frac{\int_\xT x_\perp^n \epsilon(\xT)\,\mathrm{cos}\left[n(\phi_x-\Psi_n)\right]}{\int_\xT x_\perp^n \epsilon(\xT)}\label{eccentricity definition} \ \ ,
\end{align}
which can be computed analytically for our form of the initial condition. Defining $\bar{\alpha}=1+\alpha$, the results are
\begin{align}
    \epsilon_{n}=-\delta_n \frac{n!}{2\Gamma(\frac{n}{2}+1)}\bar{\alpha}^{-n-1} \ \ .\label{eq:EpsilonPerDelta}
\end{align}
Values of the ratio $\epsilon_n/\delta_n$ for $n=2,\cdots,6$ in the case $\alpha=\frac{1}{2}$ are summarized in Table~\ref{tab:eps}, along with the maximally allowed values $\epsilon_{n}^{{\rm max}}$ for which a positive energy density is retained throughout the entire transverse plane. 
\begin{table}[h]
\begin{center}\begin{tabular}{c||c|c|c|c|c}
   $n$  & 2&3&4&5&6 \\
   \hline
   $\epsilon_n/\delta_n$& $-\frac{8}{27}$&$-\frac{64}{81\sqrt{\pi}}$&$-\frac{64}{81}$&$-\frac{2048}{729\sqrt{\pi}}$&$-\frac{2560}{729}$\\
   \hline
   $\epsilon_{n}^{{\rm max}}$&$0.4027$&$0.3845$&$0.3649$&$0.3454$&$0.3265$
\end{tabular}
\end{center}
\caption{Eccentricities $\epsilon_n/\delta_{n}$ and maximum allowed eccentricity $\epsilon_{n}^{{\rm max}}$ for which positivity of the initial energy density is satisfied.}
\label{tab:eps}
\end{table}

\subsection{Scaling properties}\label{sec:theory:scaling_properties}

Based on the above setup, the development of anisotropic flow in small systems constitutes a complicated multi-scale problem, which in general is sensitive to the typical energy of quasi-particles $Q_{s}$, the total energy per unit rapidity $dE_{\bot}^{(0)}/d\eta$, the system size $R$, as well as the dimensionless coupling strength $\eta/s$. However, due to the particular simplicity of the conformal relaxation time approximation in Eq.~(\ref{eq:BoltzmannRTA_C}), the entire dependence on these quantities can be expressed in terms of a single dimensionless opacity parameter $\hat{\gamma}$~\cite{Kurkela:2019kip}, as we will now demonstrate.

The starting point is the Boltzmann equation~\eqref{eq:BoltzmannRTA}, in which we assume that the phase-space distribution $f(x,p)$ can be expressed as an explicit function of the curvilinear coordinates $\tau$ and $\bm{x}_\perp$, as well as of the momentum space coordinates $p^\tau$, $v_z$ and $\phi_p$, defined via 
\begin{equation}
\begin{pmatrix}
 p^\tau \\ p^\eta
\end{pmatrix} = p^\tau 
\begin{pmatrix}
 1 \\ \tau^{-1} v_z
\end{pmatrix}, \quad 
\pT = p^\tau \sqrt{1 - v_z^2} 
\begin{pmatrix}
 \cos\phi_p \\ \sin \phi_p
\end{pmatrix}.\label{eq:pdofs}
\end{equation}
In this case, the Boltzmann equation \eqref{eq:BoltzmannRTA} reduces to \cite{Kurkela:2019kip}
\begin{multline}
    \left(\partial_\tau+\vT \cdot \NT-\frac{v_z(1-v_z^2)}{\tau}\partial_{v_z}-\frac{v_z^2p^\tau}{\tau}\partial_{p^\tau}\right)f\\
    = -\left(5 \frac{\eta}{s}\right)^{-1}T[f]v^\mu u_\mu[f]  (f-f_{eq}[f]),
    \label{eq:BoltzmannRTA_aux}
\end{multline}
where we denote $v^\mu = p^\mu / p^\tau$, while $T$ and $u^\mu$ are determined from the phase-space distribution $f$ via Landau matching, as described in the previous section. 

Now the general strategy to establish the scaling properties of the system is to first integrate out the momentum dependence and subsequently express all quantities in terms of dimensionless variables. Since the Landau matching condition in Eq.~(\ref{eq:Landau}) only requires the knowledge of energy weighted moments of the phase-space distribution, we reformulate the problem in terms of the reduced distribution \footnote{ Note that our definition for $\F$ differs from the one in Ref.~\cite{Kurkela:2019kip} by a factor of $\tau$, which is introduced in order to absorb trivial effects of the longitudinal expansion.}
\begin{multline}
    \F(\tau,\xT;\phi_p,v_z)
    =\frac{\nu_{\rm eff}~\pi R^2~\tau}{(2\pi)^3}\left(\epn\right)^{-1} \\ \times 
    \int_0^\infty \d p^\tau (p^\tau)^3
    f(\tau,\xT;p^\tau,\phi_p,v_z).
    \label{eq:Fdef}
\end{multline}
where the constant prefactor is simply chosen to cancel explicit dependencies on $\nu_{\rm eff}$ and $\epn$ in the resulting equations. Since Eq.~(\ref{eq:Fdef}) takes into account the correct energy $(p^{\tau})$ weigthing, the energy momentum tensor $T^{\mu\nu}$ can simply be expressed in terms of the reduced distribution as
\begin{equation}
 T^{\mu\nu}(\tau,\xT) = \frac{1}{\tau R^2} \epn \int \d\Omega_p v^\mu v^\nu \F(\tau,\xT;\phi_p,v_z),\label{eq:Tmunu_F}
\end{equation}
where $\d\Omega_p = \d v_z \d\phi_p$ denotes the solid angle element in momentum space. By multiplying Eq.~\eqref{eq:BoltzmannRTA_aux} with the appropriate pre-factors and performing the integration in Eq.~\eqref{eq:Fdef}, we then arrive at
\begin{multline}
 \left(\tilde{\partial}_\tau+\Vec{v}_\perp \cdot \tilde{\partial}_{\Vec{x}_\perp}-\frac{v_z(1-v_z^2)}{\tilde{\tau}}\partial_{v_z}+\frac{4v_z^2-1}{\tilde{\tau}}\right)\F \\
 =-\hat{\gamma} v^\mu u_\mu[\F]\,\tilde{\tau}^{-1/4} \tilde{T}[\F]\left(\F-\F_{eq}[\F]\right)\label{eq:boltz_F},
\end{multline}
where all quantities denoted with a tilde $\tilde{\phantom{\tau}}$ are explicitly dimensionless and defined as follows. Dimensionless coordinates are expressed with respect to the system size $R$ as
\begin{equation}
 \tilde{\tau}=\tau/R, \qquad \txT=\xT/R,
\end{equation}
while the dimensionless energy density $\tilde{\epsilon}$ and temperature $\tilde{T}$ of the system are defined according to
\begin{align}
 \tilde{\epsilon}&=\frac{\tau \pi R^2 }{dE_{\perp}^{(0)} / d\eta}\,\epsilon\ \ ,& 
 \tilde{T}&=\left(\frac{\tau \pi R^2 \frac{ \pi^2}{30}\nu_{\rm eff}}{dE_{\perp}^{(0)} / d\eta} \right)^{1/4} T\ \;.
 \label{eq:adim}
\end{align}
 Defining the stress-energy tensor with respect to the same non-dimensionalization employed for the energy density, we have
\begin{align}
 \widetilde{T}^{\mu\nu} = \frac{\tau \pi R^2}{dE_{\perp}^{(0)} / d\eta} \,T^{\mu\nu} 
 = \int \d\Omega_p v^\mu v^\nu \F
\end{align}
such that the Landau macthing contion in Eq.~(\ref{eq:Landau}) reduces to
\begin{equation}
 u_\nu\widetilde{T}^{\mu\nu}  = \tilde{\epsilon} u^\mu \ \;,
\end{equation}
and the equation of state takes the particularly simple form $\tilde{\epsilon} = \tilde{T}^4$ in terms of the dimensionless variables. By considering the fact that the local equilibrium distribution is determined as $f_{eq}(x,p)=f_{eq} [p^\tau (v \cdot u) / T]$ , the corresponding distribution $\F_{eq}$ can be expressed as
\begin{equation}
 \F_{eq} = \frac{\nu_{\rm eff}~\pi R^2~\tau}{(2\pi)^3} \left(\epn\right)^{-1}~ \frac{T^4}{(u \cdot v)^4}~\int_0^\infty dx\, x^3\, f_{eq}(x)\;,
\end{equation}
where the last integral can be computed in terms of the energy density $\epsilon$ as 
 $\int_0^\infty dx\, x^3\, f_{eq}(x) = \frac{(2\pi)^3 \epsilon}{4\pi~\nu_{\rm eff}~T^4}$
such that $\F_{eq}$ takes the simple form
\begin{align}
 \F_{eq} 
 =& \frac{\tilde{\epsilon}}{4 \pi~(u \cdot v)^4}.
\end{align}
Similarly, the initial condition for $\F$ can also be obtained by integrating Eq.~\eqref{eq:fini}, where assuming an azimuthally isotropic momentum distribution, one can express $\F$ in terms of the initial energy density as
\begin{equation}
 \F(\tilde{\tau}_0, {\txT}, \phi_p, v_z) = \frac{\delta(v_z)}{2\pi} \tilde{\epsilon}(\tilde{\tau}_0, \txT)\;,
\end{equation}
such that the pre-factors in the definition of $\tilde{\epsilon}(\tilde{\tau}_0, \txT)$ in Eq.~(\ref{eq:adim}) cancel with the ones in Eq.~(\ref{eq:Fdef}).

By performing the above transformations, all dependencies on the system size $R$, initial energy $\epn$ and number of degrees of freedom $\nu_{\rm eff}$ have thus been subsumed into a single dimensionless opactiy parameter
\begin{equation}
    \hat{\gamma}=\left(5\frac{\eta}{s}\right)^{-1}~R^{3/4}~\left(\frac{1}{\pi R^2~\frac{\pi^2}{30}\nu_{\rm eff}} \epn \right)^{1/4} \ \ .
    \label{eq:ghat_def}
\end{equation}
which appears on the right hand side of Eq.~(\ref{eq:boltz_F}) and controls the relaxation towards equilibrium. In order to get an idea of the typical magnitude of $\hat{\gamma}$, we can estimate its value as
\begin{multline}
   \hat{\gamma} \approx 0.88\,\left(\frac{{\eta}/{s}}{0.16}\right)^{-1}\left(\frac{R}{0.4\,\mathrm{fm}}\right)^{1/4}\\\times
   \left(\frac{\mathrm{d}E^{(0)}_\perp/\mathrm{d}\eta}{5\,\mathrm{GeV}}\right)^{1/4} 
\left(\frac{\nu_{\rm eff}}{40}\right)^{-1/4} \ \ .\label{eq:ghat_pp}
\end{multline}
which indicates that in small systems realized in $p+p$ and $p+Pb$, one should typically expect $\hat{\gamma}$ of the order unity. Conversely, in large systems, the opacity can be significantly larger, such that e.g. in central $Pb+Pb$ collisions at LHC energies one obtains
\begin{multline}
 \hat{\gamma} \approx 9.2\,\left(\frac{{\eta}/{s}}{0.16}\right)^{-1}\left(\frac{R}{6\,\mathrm{fm}}\right)^{1/4}\\\times 
 \left(\frac{\mathrm{d}E^{(0)}_\perp/\mathrm{d}\eta}{4000\,\mathrm{GeV}}\right)^{1/4} 
 \left(\frac{\nu_{\rm eff}}{40}\right)^{-1/4} \ \ .
 \label{eq:ghat_PbPb}
\end{multline}
Based on a combination of (semi-)analytic and numerical studies, we will therefore explore the full range of opacities $\hat{\gamma} \ll1$, $\hat{\gamma} \sim 1$ and $\hat{\gamma} \gg 1$ in order to investigate possible changes in the reaction dynamics for small and large systems.

\subsection{Observables}\label{sec:theory:observables}
Before we discuss the details of the solution of the previously stated problem, it is instructive to introduce the observables which we will use to quantify the time evolution of the system and the development of transverse flow. Below, we will define all observables in terms of the original phase space density $f$ and additionally express them in terms of the shorthand notation $\langle\,\cdot\,\rangle$ that was previously introduced in (\ref{eq::defMomentumInt}). Based on the above discussion, we will restrict ourselves to energy-weighted observables, which can be formulated in terms of moments of the reduced distribution $\F$ as
\begin{equation}
    \langle (p^\tau)^2 \mathcal{O}(v_z,\phi_{p}) \rangle = \frac{1}{\pi R^2} \epn \int \d \Omega_p\, \mathcal{O}(v_z,\phi_p)\, \F.
\end{equation}
One of the basic observables to look at is the transverse energy per unit rapidity $dE_{\bot}/d\eta$, computed via
\begin{equation}
    \frac{\d E_\perp}{\d \eta}= \nu_{\rm eff} \int_{\xT} \int\frac{d^3p}{(2\pi)^3}~p_\perp~f =\int_{\xT} \left\langle (p^\tau)^2\sqrt{1-v_z^2}\right\rangle,
\end{equation}
whose decrease in time is a measure of the work performed against the longitudinal expansion of the system. Since we are interested in azimuthal momentum anisotropies, the most important observables are the flow harmonics $v_n$, given as the normalized Fourier modes of the particle distribution in the azimuthal momentum angle~\cite{Voloshin:1994mz,Borghini:2000sa}. We note that, in accordance with the above discussion, we also weight the $v_{n}$'s with the transverse momentum $p_\perp$ to acquire an energy-weighted version of these flow harmonics, i.e. we will study the moments
\begin{equation}
    v_n^{E}=\frac{\int_{\xT} \int \frac{d^3p}{(2\pi)^3} ~p_\perp~e^{in\phi_{p}}~f}{\int_{\xT} \int \frac{d^3p}{(2\pi)^3} ~p_\perp~f} 
    =\frac{\int_\xT \left\langle (p^\tau)^2\sqrt{1-v_z^2}e^{in\phi_p}\right\rangle}{\int_{\xT} \left\langle (p^\tau)^2\sqrt{1-v_z^2}\right\rangle}\;, 
    \label{eq:vn_def}
\end{equation}
Beyond the $v_n^{E}$s which describe azimuthal anisotropies of the momentum distribution, another energy weighted elliptic momentum anisotropy can also be defined on the level of $T^{\mu\nu}$ without the need of full knowledge of $f$. Explicitly, this elliptic anisotropy of the energy flow $\epsilon_p$ is defined as
\cite{Ollitrault:1992bk,Song:2007ux,Karpenko:2013wva}
\begin{equation}
    \epsilon_p    =\frac{\int_\xT\, \left(T^{11}-T^{22}+2iT^{12}\right)}{\int_\xT\, \left(T^{11}+T^{22}\right)}
    =\frac{\int_\xT \left\langle (p^\tau)^2 \left(1-v_z^2\right) e^{2i\phi_p} \right\rangle}{\int_\xT \left\langle (p^\tau)^2 \left(1-v_z^2\right) \right\rangle}
    \label{eq:ep_def}
\end{equation}
and we will employ this measure in Sec.~\ref{sec:hydroLimit} to compare the kinetic evolution to relativistic viscous hydrodynamics, in order to avoid possible ambiguities of the freeze-out prescription.

\section{Solution to linear order in opacity $\hat{\gamma}$ and eccentricity $\epsilon_n$}\label{sec:linear}

While the Boltzmann equation \eqref{eq:BoltzmannRTA} as an integro-differential equation is in general too complicated to solve analytically, important conclusions can be obtained in the weakly interacting regime close to free-streaming, which corresponds to the limit $\hat{\gamma} \to 0$. We are primarily interested in the development of anisotropic flow, i.e. the final state momentum space anisotropy quantified by the harmonic coefficients $v_n$, in response to the initial state coordinate space eccentricity quantified by the amplitudes $\epsilon_n$ (or equivalently $\delta_n$) of the harmonic perturbations introduced in Eq.~\eqref{eq:deltaeps_def}. Starting from the free-streaming regime, where there is no production of $v_n$, we seek to follow previous works~\cite{Heiselberg:1998es,Borghini:2010hy,Romatschke:2018wgi,Kurkela:2018ygx,Borghini:2018xum,Kurkela:2021ctp} in deriving analytical expressions for $v_n(\tilde{\tau})$ which are accurate to linear order for small $\hat{\gamma}$ and small $\epsilon_n$.\\

Since in the free-streaming system, the momenta of the particles remain unchanged, the free-streaming dynamics is effectively 2+1 dimensional, and we will continue to work in spatial Milne coordinates, where in contrast to other sections, we use $y$ for longitudinal momentum parametrization instead of $p^\eta$ or $v_z$. Another feature is that the analytical setup will quite straightforwardly also allow to treat the problem more generally without restricting it to energy-weighted degrees of freedom. However, this requires to specify the initial condition in (\ref{eq:fini}) in more detail -- in particular with regards to the initial momentum distribution $\d N_0/\d^2 \xT \d^2 \pT \d y$, which will introduce additional scales that non-energy-weighted degrees of freedom will depend on. We will assume that this distribution is (initially) isotropic in transverse momentum and depends only on some non-specific but fixed function $F$ of the ratio of $p_\perp$ to the momentum scale $Q_s(\xT)$, i.e.
\begin{align}
    \frac{\d N_{0}}{ \d^2\xT\d^2 \pT\d y} = F\left(\frac{Q_s(\xT)}{p_\perp}\right) \ \ \label{eq:defFana}.
\end{align}
where the characteristic energy scale $Q_s(\xT)$ is related to the local energy density $\epsilon(\tau_0,\xT)$ via Eq.~(\ref{eq:epsilon-f-relation}). 

Below, we outline the calculation of observables to leading order in an expansion in opacity $\hat{\gamma}$ and eccentricity $\epsilon_{n}$ and quote the results for the flow harmonics $v_{n}$ and the longitudinal cooling of $dE_{\bot}/d\eta$. Details of the analytic calculation are compiled in Appendices~\ref{sec:free-streaming anisotropies}-\ref{details of linearized calculation}.

\subsection{Expansion scheme}\label{sec:linear:scheme}

To linearize the solution in opacity, we expand around the free-streaming limit corresponding to zero opacity,  denoted as $f^{(0)}$, which satisfies
\begin{align}
    p^\mu\partial_\mu f^{(0)} = 0\label{eq:Boltzmannfreestreaming}
\end{align}
The first order correction  $f^{(1)}$ is obtained by computing the effect of the first scattering of each particle, with the scattering rates determined by the zeroth order result
\begin{align}
    p^\mu\partial_\mu f^{(1)} = C[f^{(0)}]\label{eq:Boltzmannfirstorder}
\end{align}
This type of expansion was conceptionalized in~\cite{Heiselberg:1998es,Borghini:2010hy} and has recently also been used in other works examining weakly interacting systems~\cite{Romatschke:2018wgi,Kurkela:2018ygx,Borghini:2018xum,Kurkela:2021ctp}. As reasoned in the previous section, we can factor out from $C[f]$ the opacity parameter $\hat{\gamma}$ as a proportionality constant containing all parametric dependencies. Therefore $\hat{\gamma}$ can be identified as the expansion parameter of this expansion scheme.
In the following, we will denote observables $X$ computed in the free-streaming limit as $X^{(0)}$ and their first order corrections in opacity by $X^{(1)}$.

Similarly, for the expansion in eccentricity, we recall from Section~\ref{sec:theory:setup} that the initial energy density is of the form
\begin{align}
    \epsilon(\tau_0,\xT)=\ebar(\tau_0,x_\perp)+\delta\epsilon(\tau_0,\xT)
\end{align}
with isotropic $\ebar$ and purely anisotropic $\delta\epsilon$, which introduces a finite eccentricity $\epsilon_n$. Evidently, in free-streaming, the isotropic and anisotropic components of the phase-space distribution $f$ evolve independently of each other and the anisotropic perturbation can be computed exactly. However, when computing the induced changes of the phase-space distribution $f^{(1)}$, one is required to perform the Landau matching at the level of the full energy-momentum tensor emerging from $f^{(0)}$, which introduces a non-trivial coupling of the isotropic and anisotropic components. Hence, for simplicity, we will solve the corresponding eigenvalue equation only to linear order in the anisotropic perturbations, which formally corresponds to a leading order expansion in $\epsilon_n$. In the following, we will denote the linearized corrections to observables $X$ due to the anisotropic perturbation as $\delta X$.

\subsection{Observables}\label{sec:analytical:observables}

Since we want to examine momentum anisotropies, all observables of interest will be derived from the momentum distribution $\frac{\d N}{\d^2 p_\perp \d y}$, which can be obtained from the phase space density $f$ by integrating over coordinate space variables. Specifically in Milne coordinates, the four-volume transformation entails an extra functional determinant for the 3d hypersurface integration at fixed proper time $\tau$, such that
\begin{align}\label{cooper-frye formula}
    \frac{\d N}{\d ^2 \pT \d y} &= \frac{\nu_{\rm eff}}{(2\pi)^3} \int_\xT \int \d \eta\ p_\perp \tau \,\mathrm{cosh}(y-\eta)\, f \ \ .
\end{align}
Based on the momentum distribution $\frac{\d N}{\d^2 p_\perp \d y}$ we will extract the following moments \footnote{Note that, in contrast to the different treatments described in the other sections, the analytical treatment allows to describe more than just the energy-weighted version of the flow harmonics. Nevertheless, there are two important reasons for extracting moments of the distributions, rater than differential observables such as $\frac{\d N}{\d^2 p_\perp \d y}$ or $v_n(p_\perp)$. The first is that the integral over $p_\perp$ will be crucial in facilitating further analytical integrations later on. But perhaps the more convincing reason is the aforementioned simplification of the problem when restricting it to the case of $m=1$ for energy weighted observable.}
\begin{align}
    V_{mn} &=\int_\pT \, e^{in\phi_\pT} p_\perp^m \frac{\d N}{\d^2 \pT \d y}\label{definition of Vmn}
    &=\int_\xT \left\langle p^\tau p_\perp^m e^{in\phi_p}\right\rangle\ \ .
\end{align}
that can be directly related to the observables that are of interest to us. Specifically, one has $\frac{\d E_\perp}{\d \eta}=V_{10}$ and $v^{E}_n=\frac{V_{1n}}{V_{10}}$.

\subsection{Free streaming solution}

The free-streaming solution of (\ref{eq:Boltzmannfreestreaming}) can be computed e.g. via the method of characteristics to be
\begin{align}
    f^{(0)}(\tau,\xT,\pT,y&-\eta)
   =
   f^{(0)}\bigg(\tau_0,\xT-\vT t(\tau,\tau_0,y-\eta),\label{free streaming solution} \nonumber\\
     &\pT,\mathrm{arcsinh}\left(\frac{\tau}{\tau_0}\mathrm{sinh}(y-\eta)\right)\bigg) \ \ ,
\end{align}
where $\vT=\pT/|\pT|$ and
\begin{align}
    t(\tau,\tau_0,y-\eta)=\tau\mathrm{cosh}(y-\eta)-\sqrt{\tau_0^2+\tau^2\mathrm{sinh}^2(y-\eta)} \ \ .
\end{align}
We note that the free-streaming result simplifies significantly for $f^{(0)}(\tau_0,\xT,\pT,y-\eta) \propto \delta(y-\eta)$, as in this case
\begin{align}
    &\delta\left(\mathrm{arcsinh}\left(\frac{\tau}{\tau_0}\mathrm{sinh}(y-\eta)\right)\right)=\frac{\tau_0}{\tau} \delta (y-\eta)
\end{align}
and
\begin{align}
    t(\tau,\tau_0,0)=\tau-\tau_0 =\Delta\tau \ \ .
    \label{eq:FS_t0}
\end{align}
By applying this simplification to our initial condition in Eq.~(\ref{eq:fini}), we obtain
\begin{align}
   &f^{(0)}(\tau,\xT,\pT,y-\eta)\nonumber\\ &\quad =\frac{(2\pi)^3}{\nu_{\rm eff}}\frac{\delta(y-\eta)}{\tau{p}_\perp}F\left(\frac{Q_s(\xT-\vT\Delta\tau)}{p_\perp}\right) \ \ \label{eq:fIniAnalytical}.
\end{align}
Evidently, the free-streaming evolution will not change the momentum distribution $\frac{\d N}{\d^2 p_\perp \d y}$ since there are no scatterings and therefore also the moments $V^{(0)}_{mn}$ will remain constant
\begin{align}
 V_{m,n=0}^{(0)}(\tau) =& V_{m,n=0}^{(0)}(\tau_0), &
 V_{m,n>0}^{(0)}(\tau) =&0,
\end{align}
where the last equality follows by noting that
the initial condition is isotropic in momentum space.
Subsequently, all $v_{n}^{E}$s with $n\neq 0$ vanish identically at all times $\tau$, while the energy per unit rapidity remains constant
\begin{align}
 v_{n}^{E,(0)} =& 0, & 
 \frac{dE^{(0)}_\perp}{d\eta}(\tau) =& \frac{dE_\perp}{d\eta}(\tau_0).
\end{align}

% Since the initial condition is isotropic in momentum space, all $v_{n}^{E}$s with $n\neq 0$ vanish identically at all times $\tau$, while the energy per unit rapidity remains constant, i.e. $dE_{\bot}/d\eta=dE_{\bot}^{(0)}/d\eta$ to zeroth order in opacity. 

\subsection{Landau matching}\label{Landau matching}

Next, the free-streaming result can be used to compute the energy momentum tensor of the isotropic background $T^{(0)\mu\nu}$ and its anisotropic perturbations $\delta T^{(0)\mu\nu}$, which will be needed to obtain the local thermodynamic variables that enter the collision integral $C[f]$. Defining
\begin{align}
    v^\mu_\perp=p^\mu/p_\perp|_{y-\eta=0}=(1,\vT,0)
\end{align}
the isotropic part $T^{(0)\mu\nu}$ of the energy momentum tensor is given by
\begin{align}
    T^{(0)\mu\nu}=\frac{\tau_0}{\tau}\int \frac{\d \phi_p}{2\pi}\,v_\perp^\mu v_\perp^\nu\, \ebar(\tau_0,\xT-\Delta\tau\vT) \ \ .\label{eq:isotropicTmunu}
\end{align}
Due to isotropy, it has only four independent entries and can be written as
\begin{align}
    T^{(0)\mu\nu}=\left(\begin{array}{ccc}
        T^{(0)\tau\tau} & T^{(0)\tau \perp} \hxT^t & 0\\
        T^{(0)\tau \perp} \hxT & T^{(0)\mathbb{1}}\mathbb{1}+T^{(0)\perp\perp} \hxT\hxT^t & 0\\
        0&0&0
    \end{array}\right) \ \ .\label{eq:isotropicTmunuComponents}
\end{align}
where we denote $\hxT=\xT/|\xT|$. Based on the symmetries of $T^{(0)\mu\nu}$, the corresponding eigenvectors satisfying the relations
\begin{align}
    u_\mu T^{(0)\mu\nu} &= \epsilon u^\nu \ \ ,\\
    t_\mu T^{(0)\mu\nu} &= p_t t^\nu \ \ ,\\
    s_\mu T^{(0)\mu\nu} &= p_s s^\nu
\end{align}
can be parametrized as
\begin{align}
    u^\mu&=\gamma(1,\beta \hxT,0) \ \ ,\\
    t^\mu&=\gamma(\beta,\hxT,0) \ \ ,\\
    s^\mu&=(0,i\sigma_2 \hxT,0) \ \ ,
\end{align}
where 
\begin{eqnarray}
    \beta&=\frac{3T^{(0)\tau\tau}+T^{(0)\perp\perp}}{4T^{(0)\tau \perp}}-\sqrt{\left(\frac{3T^{(0)\tau\tau}+T^{(0)\perp\perp}}{4T^{(0)\tau \perp}}\right)^2-1}
\end{eqnarray}
is the local rest frame velocity, $\gamma=(1-\beta^2)^{-1/2}$ and $i\sigma_{2}=\begin{pmatrix}
0 & 1\\
-1 & 0
\end{pmatrix}$; the corresponding eigenvalues are given by
\begin{align}
    \epsilon&=T^{(0)\tau\tau}-\beta T^{(0)\tau \perp} \ \ ,\label{eq:epsilonfromTmunu}\\
    p_t&=\beta T^{(0)\tau \perp} - T^{(0)\perp\perp} - T^{(0)\mathbb{1}} \ \ ,\\
    p_s&=-T^{(0)\mathbb{1}} \ \ \label{eq:psfromTmunu},
\end{align}

Now, similarly to the isotropic background in Eq.~(\ref{eq:isotropicTmunu}), the anisotropic part of the energy-momentum tensor can be computed as
\begin{align}
    \delta T^{(0)\mu\nu}=\frac{\tau_0}{\tau}\int \frac{\d \phi_p}{2\pi}\,v_\perp^\mu v_\perp^\nu\, \delta\epsilon(\tau_0,\xT-\Delta\tau\vT),
\end{align}
which -- due to the absence of isotropy -- features six independent entries and its eigenvalues and eigenvectors will be a complicated function of all of them. Obtaining their exact result would be cumbersome and ultimately pointless, as it would be too complex to perform further calculations with them, so instead we will linearize the Landau matching condition in the perturbation, meaning that the corrections $\delta \epsilon$ and $\delta u^\mu$ are computed from

\begin{align}
    \delta u_\mu T^{(0)\mu\nu}+u_\mu \delta T^{(0)\mu\nu}&=\delta \epsilon u^\mu + \epsilon \delta u^\mu \label{linearizedLandaumatching}\ \ ,\\
    u_\mu \delta u^\mu &= 0 \ \ .
\end{align}
The second condition ensures that the perturbation preserves the correct normalization of $u^\mu$ to linear order. In order to solve this system of equations, we can expand
\begin{align}
    \delta u^\mu = \delta u_t t^\mu + \delta u_s s^\mu
\end{align}
and use the orthogonality of the eigenbasis of $T^{(0)\mu\nu}$ to obtain via contraction with the eigenvectors from (\ref{linearizedLandaumatching}) the following results:
\begin{align}
    \delta \epsilon = u_\mu \delta T^{(0)\mu\nu} u_\nu \ \ \label{eq:deltaepsilonfromTmunu},\\
    \delta u_t = \frac{u_\mu \delta T^{(0)\mu\nu} t_\nu}{p_t - \epsilon} \ \ ,\\
    \delta u_s = \frac{u_\mu \delta T^{(0)\mu\nu} s_\nu}{p_s - \epsilon} \ \ \label{eq:deltausfromTmunu}.
\end{align}

While Eqns.~(\ref{eq:epsilonfromTmunu}-\ref{eq:psfromTmunu},\ref{eq:deltaepsilonfromTmunu}-\ref{eq:deltausfromTmunu}) provide schematic expressions for the $\epsilon,\delta \epsilon,u^{\mu}$ and $\delta u^{\mu}$, the exact forms of $T^{(0)\mu\nu}$ and $\delta T^{(0)\mu\nu}$ that determine these quantities consist of lengthy expressions which are provided in Appendix~\ref{IntegralsLandauMatching}.

\vspace{20pt}
\subsection{First order corrections in \texorpdfstring{$\hat{\gamma}$}{Gammahat}}\label{first order corrections}

Now that we have obtained the local energy densities and flow velocities, computing the corrections $\delta f^{(1)}$ due to the first scatterings according to
\begin{align}
    \frac{p^\mu }{p^\tau}\partial_\mu f^{(1)} = \frac{C[f^{(0)}]}{p^\tau}
\end{align}
is comparetively straightforward after realizing that this is just the inhomogeneous case of the PDE we already solved for free streaming. The solution (\ref{free streaming solution}) allows to read off the Green's function for propagation in time and compute $f^{(1)}$ via

\begin{widetext}

\begin{align}
     f^{(1)}(\tau,\xT,\pT,y-\eta)
    =\int_{\tau_0}^\tau \mathrm{d}\tau ' \ \frac{C[f^{(0)}]}{p^\tau}\left(\tau ',\xT-\vT t(\tau,\tau ',y-\eta),\pT,\mathrm{arcsinh}\left(\frac{\tau}{\tau  '}\mathrm{sinh}(y-\eta)\right)\right)
\end{align}

\end{widetext}

Since we will integrate over space to compute $\frac{\d N^{(1)}}{\d^2 p_\perp \d y}$ according to (\ref{cooper-frye formula}), we can simplify this expression by performing the following substitutions
\begin{align}
    \xT '&= \xT-\vT t(\tau,\tau ',y-\eta), \nonumber\\
    \eta ' &= y-\mathrm{arcsinh}\left(\frac{\tau}{\tau '}\mathrm{sinh}(y-\eta)\right),
\end{align}
such that
\begin{align}
    \d^2 \xT ' &= \d^2 \xT, & 
    \d \eta '&=\frac{\tau}{\tau '}\frac{\mathrm{cosh}(y-\eta)}{\mathrm{cosh}(y-\eta ')} \d \eta,
\end{align}
yielding the following result for the changes in the momentum space distribution $\frac{\d N^{(1)}}{\d^2 p_\perp \d y}$
    \begin{align}
     \frac{\d N^{(1)}}{\d ^2 p_\perp \d y}(\tau,\pT)
    &=\int_{\tau_0}^\tau \mathrm{d}\tau '\int _{\xT '} \int \d \eta ' \ \tau ' \frac{\nu_{\rm eff}}{(2\pi)^3}\nonumber\\
    &\quad \times C[f^{(0)}]\left(\tau ',\xT ',\pT,y-\eta '\right) \ \;,
\end{align}
where in the following, we will drop the primes on all integration variables except for $\tau'$. Since, as stated in section~\ref{sec:analytical:observables}, the final observables we want to compute correspond to $\d^2p_\perp$-integrated moments of $\frac{\d N}{\d^2 p_\perp \d y}$, one is then left with the calculation of the following six dimensional integral
\begin{align}
    V^{(1)}_{mk}(\tau) &=\int_\pT \ e^{in\phi_p} p_\perp^m \int_{\tau_0}^\tau \mathrm{d}\tau ' \int_\xT  \int \d \eta  \ \tau '  \frac{\nu_{\rm eff}}{(2\pi)^3}\nonumber\\
    &\quad \times C[f^{(0)}]\left(\tau ' ,\xT ,\pT,y-\eta\right) \ \ . \label{V1}
\end{align}
We find that four of these integrals can be carried out analytically, while the remaining two integrals over $\d\tau’$ and $\d x_\perp$ require numerical methods. Below we provide a brief 
outline of the four analytical integrations and explain how different terms can be categorized. Explicit expressions and further details of the analytic calculation can be found in Appendix~\ref{details of linearized calculation}.

The integration over $p_\perp$ is performed first to obtain moments of $F$ and $f_{eq}$, which will facilitate the other integrations. Since the integrand depends mostly on $u\cdot v$, we substitute integration over the position space azimuthal angle $\phi_{x}$ for integration over $\phi_{\xT\pT}=\phi_x-\phi_p$. The integral over $\eta$ is straightforward for the term containing a Dirac-Delta but for other terms it is of similar complexity to the integral over $\phi_{\xT\pT}$ and both are performed together. After these integrations, the only remaining dependence on the azimuthal momentum angle takes the form $e^{ik\phi_p}\mathrm{cos}(n\phi_{\pT\nT})$ ($\phi_{\pT\nT}=\phi_p-\Psi_n$) and the last integral becomes a trigonometric orthogonality relation, which signifies that eccentricities do not mix, as is to be expected due to the linearization.

Due to the fact that we consider an isotropic background with a purely anisotropic perturbation, the leading order expansion of the flow harmonics is given by
\begin{align}
    v_n^{(m)}=\frac{V_{mn}}{V_{m0}}=\frac{\delta V_{mn}^{(1)}}{V_{m0}^{(0)}}+\mathrm{nonlinear~terms}
    \label{eq:vnm_linear}
\end{align}
Due to this symmetry, one also finds that the leading order opacity contributions to the observables $V_{mn}$ conveniently separate into the first order isotropic corrections $V_{m0}^{(1)}\propto \hat{\gamma}$ for $n=0$ on one hand and the first order anisotropic corrections 
$\delta V_{mn}^{(1)}\propto \hat{\gamma}\delta_{n}$ to only moments with $n\neq 0$ on the other hand. While the former ($V_{m0}^{(1)}$) represent opacity corrections to the evolution of the isotropic background, the latter ($\delta V_{mn}^{(1)}$) describe the systems response to the anisotropic energy perturbations.

Besides, another important mathematical as well as physical distinction is that between terms $V^{(1,0)}_{mn}$ coming from the decay of $f^{(0)}$ and terms $V^{(1,eq)}_{mn}$ coming from the buildup of $f_{eq}$. In our calculation we treat these terms separately and then sum them to find the total observable. However, it is important to point out that in many cases the two terms turn out to have different parametric dependencies. By construction of the expansion scheme, all of them are proportional to $\hat{\gamma}$. However due to the different functional forms of $f^{(0)}$ and $f_{eq}$ we obtain that the isotropic $(n=0)$ corrections are given by
\begin{align}
    V^{(1,0)}_{m0} &=  -\hat{\gamma} V^{(0)}_{m0}~\mathcal{P}_{m}(\tilde{\tau}) \ \ ,\\
    V^{(1,eq)}_{m0}&= +\hat{\gamma} \nu_{\rm eff} R^{-m}\left(\nu_{\rm eff}^{-1} \epn R\right)^{\frac{m+3}{4}} \mathcal{Q}_{m}(\tilde{\tau}) \ \ ,
\end{align}
and similarly for the anistropic corrections $(n\neq 0)$
\begin{align}
    V^{(1,0)}_{mn} &=  -\hat{\gamma} \delta_{n} V^{(0)}_{m0}~\mathcal{P}_{mn}(\tilde{\tau}) \ \ ,\\
    V^{(1,eq)}_{mn}&= +\hat{\gamma} \delta_{n} \nu_{\rm eff} R^{-m}\left(\nu_{\rm eff}^{-1} \epn R\right)^{\frac{m+3}{4}}\mathcal{Q}_{mn}(\tilde{\tau}) \ \ .
\end{align}
Detailed expressions of the functions $\mathcal{P}_{m}(\tilde{\tau}),\mathcal{Q}_{m}(\tilde{\tau}),\mathcal{P}_{mn}(\tilde{\tau}),\mathcal{Q}_{mn}(\tilde{\tau})$ are given in Eqns.~(\ref{eq:Pmtau},\ref{eq:Qmtau},\ref{eq:Pmntau},\ref{eq:Qmntau}).
% (see Eqns.~(\ref{eq:Pmtau},\ref{eq:Qmtau},\ref{eq:Pmntau},\ref{eq:Qmntau}) for detailed expressions of the functions $\mathcal{P}_{m}(\tilde{\tau}),\mathcal{Q}_{m}(\tilde{\tau}),\mathcal{P}_{mn}(\tilde{\tau}),\mathcal{Q}_{mn}(\tilde{\tau})$)
Of course, the appearance of a different parametric behavior is not too surprising, as $f^{0}$ depends on the entire momentum distribution,  whereas $f_{eq}$ only depends on the local energy density. Generally, to fix the relative size of decay and buildup for $V^{(1)}_{mn}$, we need an input for $\epn$ and $V^{(0)}_{m0}$, which means specifying the related moments of the initial momentum distribution $F$ in Eq.~(\ref{eq:defFana}). Clearly, the only exception to this rule is the case $m=1$ of energy weighted observables, where $V^{(0)}_{10}=\epn$ and the calculation of $V^{(1)}_{1n}/V^{(0)}_{10}$ does not require any
further specification of the initial momentum distribution $F$.\\

By restricting our attention to energy weighted observables, we can then perform the residual integrals numerically, to obtain the leading order changes in the initial energy per unit rapidity
\begin{align}
    \left.\frac{\d E_\perp^{(1)}}{d \eta}(\tau \rightarrow \infty) \right/ \left(\hat{\gamma} \epn\right)&=-0.210\label{eq:dEdyanalyticalasymptote}
\end{align}
and the flow response
\begin{align}
     \left. v_2^{E}(\tau \rightarrow \infty)\right/(\hat{\gamma}\epsilon_2)&=0.212\label{eq:v2analyticalasymptote}\\
     \left. v_3^{E}(\tau \rightarrow \infty)\right/(\hat{\gamma}\epsilon_3)&=0.0665\\
     \left. v_4^{E}(\tau \rightarrow \infty)\right/(\hat{\gamma}\epsilon_4)&=-0.00914\;,\label{eq:v4analyticalasymptote}
\end{align}
which we will compare to full numerical solutions of the RTA Boltzmann equation in the following. Beyond the results in Eqns.~(\ref{eq:dEdyanalyticalasymptote}-\ref{eq:v4analyticalasymptote}), which provide the asymptotic ($\tau \to \infty$) values of the transverse energy and flow coefficients, it is clear that Eq.~(\ref{V1}) also gives access to the time evolution of these quantities which we will further investigate in Sec.~\ref{sec:results}.

We note that the above result are obtained for the initial condition in Eqs.~(\ref{eq:initialEnergyDecomp}-\ref{eq:deltaeps_def}) with $\alpha=1/2$, which is different than the case $\alpha=0$ considered in~\cite{Kurkela:2018ygx}.  If we choose $\alpha=0$ instead, we find 
\begin{align}
     \left. v_2^{E}(\tau \rightarrow \infty)\right/(\hat{\gamma}\epsilon_2)&=0.213\label{eq:v2analyticalasymptote_alpha0}\\
     \left. v_3^{E}(\tau \rightarrow \infty)\right/(\hat{\gamma}\epsilon_3)&=0.0621\\
     \left. v_4^{E}(\tau \rightarrow \infty)\right/(\hat{\gamma}\epsilon_4)&=-0.00483\;,\label{eq:v4analyticalasymptote_alpha0}
\end{align}
in agreement with~\cite{Kurkela:2018ygx}.\footnote{Note that for comparison with ~\cite{Kurkela:2018ygx}, one also need to account for the factor of $\epsilon_n/\delta_n$ in Eq.~(\ref{eq:EpsilonPerDelta}).} By comparing the results for different $v_{n}$s in Eq.~(\ref{eq:v2analyticalasymptote}-\ref{eq:v4analyticalasymptote},\ref{eq:v2analyticalasymptote_alpha0}-\ref{eq:v4analyticalasymptote_alpha0}), one finds that $v_{2}$ appears to be rather insensitive to $\alpha$, whereas the higher order $v_{n}$s 
are more sensitive to $\alpha$, as we will further discuss in Sec.~\ref{sec:results_vn}. Especially $v_4$ changes by approximately a factor of two between the two cases, and can even turn out to have different signs for different values of $\alpha$, indicating a strong dependence on the initial profile in the low opacity regime.

\section{Numerical procedure for non-linear solution}
\label{NumericalProcedure}
We will now discuss two different schemes to obtain numerical solutions of the RTA Boltzmann equation, which are based on a momentum moment expansion discussed in subsec.~\ref{sec:MomentumMomentExpansion} and the relativistic Lattice Boltzmann method discussed in subsec.~\ref{NumericalProcedure:RLB}.

\subsection{Expansion in spherical harmonic moments} 
\label{sec:MomentumMomentExpansion}
Within our first approach, we follow previous works~\cite{Kamata:2020mka}, where instead of describing the evolution of the phase space density $f$, the numerical algorithm solves time evolution equations only for some energy weighted momentum moments on a two-dimensional lattice in transverse space. Specifically, we consider the following energy weighted moments $C_l^m$ of the phase-space distribution
\begin{align}
    C_l^m&:=\tau\int \frac{\d^3 p}{(2\pi)^3}\, Y_l^m(\theta_p,\phi_p) \, p^\tau\, f \label{eq:ClmDefinition}\\
    &=\int\frac{\d^2p_\perp}{(2\pi)^2}\int\frac{\d p_\eta}{2\pi}\,Y_l^m(\theta_p,\phi_p)\,\sqrt{p_\perp^2+\frac{p_\eta^2}{\tau^2}}\,f \ \ .\nonumber
\end{align}
where $Y_{l}^{m}$ denote the spherical harmonics, which are given in terms of the associated Legendre polynomials $P_l^m$ as
\begin{align}
    Y_l^m(\theta,\phi)=y_l^m \, P_l^m(\mathrm{cos}\,\theta)\,e^{im\phi}
\end{align}
with normalization
\begin{align}
    y_l^m=\sqrt{\frac{2l+1}{4\pi}\frac{(l-m)!}{(l+m)!}} \ \,
\end{align}
and the momentum space angles in Eq.~(\ref{eq:ClmDefinition}) are parametrized by the polar and azimuthal angles $\theta_p$ and $\phi_p$ defined as
\begin{align}
    \mathrm{cos}\,\theta_p=\frac{p_\eta/\tau}{p^\tau} \,  \ \ , \quad
    \mathrm{tan}\, \phi_{p}= \frac{p^{2}}{p^{1}}\;.
\end{align}
Since only a finite number of moments can be described numerically, the algorithm only keeps track of the moments with $l<l_{max}$ for an adjustable large enough value of $l_{max}$ to achieve apparent convergence. 

\subsubsection{Initial conditions \& Evolution equation for moments}
 By taking the corresponding moments of the initial conditions in Eq.~(\ref{eq:fini}), one obtains the initial conditions for the spherical harmonic moments as
\begin{equation}
C_l^m(\tau_0)=\tau_0 \,\epsilon(\tau_0,\xT)\,y_l^0\,P_l^0(0)\,\delta^{m0}\;.
\end{equation}
This expression contains as a factor the Legendre polynomial evaluation $P_l^0(0)$, which vanishes for odd $l$ and is otherwise given by
\begin{align}
    P_l^0(0)=(-1)^{l/2}\pi^{-1/2}\frac{\Gamma\left(\frac{l+1}{2}\right)}{\Gamma\left(\frac{l}{2}+1\right)} \ \ .
\end{align}\\

Simlarly, denoting the local rest-frame velocity as $u^\mu=\gamma\, (1,\beta_1,\beta_2,0)$ and taking the corresponding moments of the Boltzmann equation~(\ref{eq:BoltzmannRTA},\ref{eq:BoltzmannRTA_C}) then yields the following equation of motion for the spherical harmonic moments

\begin{widetext}
\begin{align}
    \partial_\tau C_l^m&=\frac{1}{\tau}(b_{l,+2}^mC_{l+2}^m+b_{l,0}^mC_l^m+b_{l,-2}^mC_{l-2}^m)
    +\frac{1}{2}\left(\frac{\gamma \beta_1}{\tau_R}-\partial_1\right) [u_{l,+}^mC_{l+1}^{m+1}+u_{l,-}^mC_{l-1}^{m+1}+d_{l,+}^mC_{l+1}^{m-1}+d_{l,-}^mC_{l-1}^{m-1}]\nonumber\\
    &\quad +\frac{1}{2i}\left(\frac{\gamma \beta_2}{\tau_R}-\partial_2\right) [u_{l,+}^mC_{l+1}^{m+1}+u_{l,-}^mC_{l-1}^{m+1}-d_{l,+}^mC_{l+1}^{m-1}-d_{l,-}^mC_{l-1}^{m-1}] + \frac{1}{\tau_R} E_l^m(u^\mu,T)-\frac{\gamma}{\tau_R} C_l^m \ \ ,\label{eq:Clmeom}
\end{align}
where the terms with spatial derivatives proportional to the coefficients $u,d$ describe free-streaming while the terms proportional to the coefficients $b$ are related to the longitudinal expansion. The corresponding coefficients are given by~\cite{Kamata:2020mka}\footnote{Note that here $b_{l,0}^m$ is smaller by $\frac{1}{3}$ compared to~\cite{Kamata:2020mka} because the $C_l^m$ are weighted with $\tau$ instead of $\tau^{4/3}$.}
\begin{align}
    u_{l,-}^m&=\sqrt{\frac{(l-m)(l-m-1)}{4l^2-1}}  \ \ , &\,u_{l,+}^m&=-\sqrt{\frac{(l+m+1)(l+m+2)}{4l(l+2)+3}} \ \ ,\\
    d_{l,-}^m&=-\sqrt{\frac{(l+m)(l+m-1)}{(4l^2-1)}} \ \ , &\,d_{l,+}^m&=\sqrt{\frac{(l-m+1)(l-m+2)}{4l(l+2)+3}} \ \ ,\\
    b_{l,-2}^m&=-\frac{l+2}{2l-1}\sqrt{\frac{(l-m)(l-m-1)(l+m)(l+m-1)}{(2l-3)(2l+1)}} \ \ ,
    &\,b_{l,0}^m&=\frac{1-3(l^2-l)+5m^2}{4l(l+1)-3}\ \ ,\nonumber\\
    b_{l,+2}^m&=\frac{l-1}{2l+3}\sqrt{\frac{(l-m+1)(l-m+2)(l+m+2)(l+m+1)}{(2l+5)(2l+1)}} \ \ ,
    \end{align}
Finally all terms that are inversely proportionals to the relaxation time $\tau_{R}$ result from the collision kernel and describe the relaxation of the system towards local equilibrium, with the equilibrium moments $E^{l}_{m}$ given by
    \begin{align}
    E_l^m&=\int\frac{\d^2p_\perp}{(2\pi)^2}\int\frac{\d p_\eta}{2\pi}\,Y_l^m(\theta_p,\phi_p)\,p^\mu u_\mu \, f_{eq}=2^{-l-2}\pi^{1/2}\tau\epsilon\, Y_l^m\left(\frac{\pi}{2},\phi_u\right)\gamma^{-3}\beta^l\frac{\Gamma (l+3)}{\Gamma\left(l+\frac{3}{2}\right)}\ _2F_1\left(\frac{l+3}{2},\frac{l+4}{2};l+\frac{3}{2};\beta^2\right)
\end{align}
\end{widetext}
where we denote $\mathrm{tan}\,\phi_u=\beta_1/\beta_2$ and $\beta^2=\beta_1^2+\beta_2^2$ and refer to Appendix~\ref{appendix equilibrium moments} for further details of the calculation. We also note for later convenience, that by separately keeping track of the free-streaming, long. expansion and collisional contributions in Eq.~(\ref{eq:Clmeom}), we can compute the respective contributions to the rate of change of any observables.

Evaluation of Eq.~(\ref{eq:Clmeom}) also requires an input for the local energy density $\epsilon$ and flow velocity $u^{\mu}$, which have to be computed via Landau matching. In practice, we first compute the various components of the energy momentum tensor from the following linear combintations of spherical harmonic moments
\begin{align}
    \tau T^{\tau\tau}&=\sqrt{4\pi}C_0^0\\
    \tau T^{\tau 1}&=\sqrt{\frac{2\pi}{3}}(C_1^{-1}-C_1^1)\\
    \tau T^{\tau 2}&=\sqrt{\frac{2\pi}{3}}i(C_1^{-1}+C_1^1)\\
    \tau T^{11}&=\sqrt{\frac{4\pi}{9}}\left(C_0^0-\sqrt{\frac{1}{5}}C_2^0\right)+\sqrt{\frac{2\pi}{15}}(C_2^2+C_2^{-2})\\
    \tau T^{22}&=\sqrt{\frac{4\pi}{9}}\left(C_0^0-\sqrt{\frac{1}{5}}C_2^0\right)-\sqrt{\frac{2\pi}{15}}(C_2^2+C_2^{-2})\\
    \tau T^{12}&=\sqrt{\frac{2\pi}{15}}i(C_2^{-2}-C_2^{2})
\end{align}
and subsequently perform a numerical diagonalization of $T^{\mu}_{~\nu}$ using the Eigen C++ library~\cite{eigenweb}, to obtain the rest-frame velocity $u^{\mu}$ and local energy density $\epsilon$ as the timelike eigenvector and eigenvalue.

With regards to the numerical implementation of Eq.~(\ref{eq:Clmeom}), we also note that the terms containing spatial derivatives can be efficiently computed in Fourier space, and we employ an $\mathcal{O}(a_S^2)$ improved five-point stencil derivative. Concerning the discretization of the time step, we employ a fourth order Runge-Kutta scheme with adaptive time step of typically $\delta \tau=0.01\, \text{min}(\tau,R/10)$.

\subsubsection{Observables}

Since the numerical setup is restricted to energy weighted moments of the phase-space distribution, all observables that can be computed are necessarily weighted with energy as well. Evidently, to compute an observable, it has to be expressed as a linear combination of the moments $C_l^m$, meaning that it has to be expanded in spherical harmonics. By making use of their orthogonality relation, one can then express observables of the form
\begin{align}
    \tau \,\langle (p^\tau)^2\, \mathcal{O} \rangle &= \int \frac{\d^2p_\perp}{(2\pi)^2}\int\frac{\d p_\eta}{2\pi}\,p^\tau \,\mathcal{O}(\theta_p,\phi_p)\, f
\end{align}
as a linear combination of the coefficients
\begin{align}
    \tau \,\langle (p^\tau)^2 \,\mathcal{O} \rangle &= \sum_{(l,m)} \alpha_{m,l}^\mathcal{O} \, C_l^m \ \ ,
\end{align}
where the expansion coefficients $\alpha_{m,l}^\mathcal{O}$ are determined as
\begin{align}
    \alpha_{m,l}^\mathcal{O}=\int \d\Omega_p \, \mathcal{O}(\theta_p,\phi_p)\,Y_l^{m*}(\theta_p,\phi_p) \ \ .
\end{align}
Specifically, for calculating the observables  $dE_{\perp}/dy$ and $v_{n}^{E}$ that are of interest to us, we need to consider integrals of expressions of the form \begin{align}
    \mathcal{O}_n(\theta_p,\phi_p)=e^{in\phi_p}\,\mathrm{sin}(\theta_p) \ \ .
\end{align}
Since the $\phi_p$-dependence of $Y_l^{m*}$ is simply given by $e^{-im\phi_p}$, it is already obvious that $\alpha_{m,l}^{\mathcal{O}_n}$ vanishes for all $m\neq n$. Additionally, for increasing $l$ the spherical harmonics $Y_l^{n*}$ alternate between being symmetric and antisymmetric w.r.t. $\theta_p$, such that $\alpha_{n,l}^{\mathcal{O}_n}=0$ for $l-n$ odd, while for $l-n$ even, the coefficients can be computed as
\begin{align}
    \alpha_{n,l}^{\mathcal{O}_n}&=2\pi 2^{-l} y_l^n \sum_{k=0}^{\frac{l-n}{2}}(-1)^{k+n}{l\choose n}{2l-2k\choose l} \\
    &\quad \times \frac{(l-2k)!}{(l-2k-n)!}\frac{\Gamma \left(\frac{n+3}{2}\right)\Gamma \left(\frac{l-2k-n+1}{2}\right)}{\Gamma \left(\frac{l-2k+4}{2}\right)}\nonumber \ \ .
\end{align}
In the special case $n=1$, only one coefficient is nonvanishing, but otherwise there are infinitely many. However, their values are decreasing with $l$ quickly enough so that cutting off the expansion at $l_{max}$ yields a reasonable approximation.

\subsection{Relativistic Lattice Boltzmann solver}\label{NumericalProcedure:RLB}

Within our second approach, we employ a relativistic lattice Boltzmann solver inspired by the finite difference Relativistic Lattice Boltzmann (RLB) algorithm discussed in Refs.~\cite{Romatschke:2011hm,Ambrus:2018kug,Succi:2018,Gabbana:2019ydb,Gabbana:2020oka}. The strategy for devising the numerical method is split into three main parts, that are described in this subsection. The structure of the kinetic equation is presented in Sec.~\ref{NumericalProcedure:RLB:eq} in two forms. The first form is based on the standard Bjorken coordinates $(\tau, \xT, \eta)$, while the second relies on a set of hybrid free-streaming coordinates, inspired by the approach in Ref.~\cite{Kurkela:2019kip}. The momentum space discretization is discussed in Sec.~\ref{NumericalProcedure:RLB:discretization}. The spatial and temporal discretization, as well as the numerical schemes employed for the advection and time stepping, are briefly summarized in Sec.~\ref{NumericalProcedure:RLB:numsch}.

\subsubsection{Boltzmann equation for the RLB approach}\label{NumericalProcedure:RLB:eq}

In the RLB method, we employ the factorization of the momentum space with respect to the coordinates ($p$, $v_z$, $\phi_p$) introduced in Eq.~\eqref{eq:pdofs}. Starting from Eq.~\eqref{eq:BoltzmannRTA_aux}, 
we apply the LB algorithm at the level of the function $\F_{\rm RLB} = \frac{\tau_0}{\tau} \F$, where $\F$ is introduced in Eq.~\eqref{eq:Fdef}. Specifically, 
\begin{equation}
 \F_{\rm RLB} = \frac{\pi \nu_{\rm eff} R^2 \tau_0}{(2\pi)^3}\left(\epn\right)^{-1} 
 \int_0^\infty \d p^\tau \, (p^\tau)^3 f.\label{eq:F_RLB}
\end{equation}
The non-dimensionalization of the coordinates $\tau$ and $\xT$ is performed with respect to $R^{3/4} \tau_0^{1/4}$ \cite{Kurkela:2018qeb}, i.e.
\begin{align}
 {\bar \tau} =& \frac{\tau}{\tau_0^{1/4} R^{3/4}} = \left(\frac{R}{\tau_0}\right)^{1/4} \tilde{\tau}, \nonumber\\
 \bar{\mathbf{x}}_{\perp} =& \frac{\mathbf{x}_{\perp}}{\tau_0^{1/4} R^{3/4}} = \left(\frac{R}{\tau_0}\right)^{1/4} \tilde{\xT},
\end{align}
while the energy density and temperature are non-dimensionalized with respect to constant quantities:
\begin{align}
 \bar{\epsilon} =& \frac{\tau_0 \pi R^2 \epsilon}{dE_{\perp}^{(0)} / d\eta} = \frac{\tau_0}{\tau} \tilde{\epsilon}, \nonumber\\
 \overline{T} =& \left(\frac{\tau_0 \pi R^2 \frac{\pi^2}{30} \nu_{\rm eff}}{dE_{\perp}^{(0)} / d\eta} \right)^{1/4} T = \left(\frac{\tau_0}{\tau}\right)^{1/4} \tilde{T},
\end{align}
such that ${\bar \epsilon} = \overline{T}^4$. In this section, we use an overhead bar $\bar{\phantom{\epsilon}}$ to denote dimensionless quantities obtained using the above convention, in contrast to the overhead tilde $\tilde{\phantom{\epsilon}}$ employed in Sec.~\ref{sec:theory} (note that $\overline{T}$ and $\bar{\epsilon}$ are related to $T$ and $\epsilon$ through constant factors). The Boltzmann equation \eqref{eq:BoltzmannRTA_aux} written for $\F_{\rm RLB}$ introduced above reads \cite{Kurkela:2018qeb}:
\begin{multline}
 \hspace{-5pt} \left(\frac{\partial}{\partial {\bar \tau}}
 + \vT \cdot \overline{\nabla}
 + \frac{1 + v_z^2}{\tau}\right) \F_{\rm RLB}
 - \frac{1}{\bar \tau} \frac{\partial[v_z(1 - v_z^2) \F_{\rm RLB}]}{\partial v_z} \\= -\hat{\gamma} (v^\mu u_\mu) \overline{T} 
 (\F_{\rm RLB} - \F^{eq}_{\rm RLB}).
 \label{eq:BoltzmannRTA_RLB}
\end{multline}
The components of the stress-energy tensor can be non-dimensionalized in the same way as the energy density, being related to $\F_{\rm RLB}$ through
\begin{equation}
 \overline{T}^{\mu\nu} = \int d\Omega_p \, v^\mu v^\nu \F_{\rm RLB}.
\end{equation}
The energy-weighted flow harmonics \eqref{eq:vn_def} can be obtained via
\begin{equation}
 v_n^{E}=\frac{\int_{\xT} \int d\Omega_p \sqrt{1 - v_z^2} e^{in \phi_p} \F_{\rm RLB}}
 {\int_{\xT} \int d\Omega_p \sqrt{1 - v_z^2} \F_{\rm RLB}}.
 \label{eq:RLB_vn}
\end{equation}

The Boltzmann equation in the form given in Eq.~\eqref{eq:BoltzmannRTA_RLB} serves as the basis of the algorithm employed for large values of the opacity $\hat{\gamma}$. At small values of the opacity, we find the form in Eq.~\eqref{eq:BoltzmannRTA_RLB} unsuitable and instead employ free-streaming coordinates in momentum space. This approach is inspired from Ref.~\cite{Kurkela:2018qeb} but differs from the aforementioned approach because the spatial coordinates are left unchanged. Defining
\begin{align}
 p_{\rm fs}^\tau =& p^\tau \Delta, &
 v^{\rm fs}_z =& \frac{\tau v_z}{\tau_0 \Delta},\nonumber\\
 p^\tau =& p^\tau_{\rm fs} \Delta_{\rm fs}, &
 v_z =& \frac{\tau_0 v^{\rm fs}_z}{\tau \Delta_{\rm fs}},
\end{align}
where 
\begin{align}
 \Delta =& \sqrt{1 + \left(\frac{\tau^2}{\tau_0^2} - 1\right) v_z^2}, \nonumber\\
 \Delta_{\rm fs} = \frac{1}{\Delta} =& \sqrt{1 - \left(1 - \frac{\tau_0^2}{\tau^2}\right) v_{z; {\rm fs}}^2}.
\end{align}
the Boltzmann equation \eqref{eq:BoltzmannRTA_aux} becomes
\begin{equation}
 \frac{\partial f}{\partial \bar{\tau}} + \frac{1}{\Delta_{\rm fs}} \mathbf{v}_{\perp;{\rm fs}} \cdot \overline{\nabla}_{\perp} f 
 = -\hat{\gamma} (v^\mu u_\mu) \overline{T}
 (f - f_{eq}),
 \label{eq:BoltzmannRTA_fs_aux}
\end{equation}
where $\mathbf{v}_{\perp;{\rm fs}} = \sqrt{1 - v_{z; {\rm fs}}^2} (\cos\phi_p, \sin\phi_p)$, while $v_\mu u^\mu = u^\tau - \frac{1}{\Delta_{\rm fs}} \mathbf{v}_{\perp; {\rm fs}} \cdot \mathbf{u}_{\perp}$. We now introduce the function $\F_{\rm fs}$, defined in analogy to Eq.~\eqref{eq:F_RLB} using integration with respect to $p_{\rm fs}^\tau$:
\begin{equation}
 \F_{\rm fs} = \frac{\pi \nu_{\rm eff} R^2 \tau_0}{(2\pi)^3}\left(\epn\right)^{-1} 
 \int_0^\infty \d p^\tau_{\rm fs}\, (p^\tau_{\rm fs})^3 f,\label{eq:F_fs}
\end{equation}
such that Eq.~\eqref{eq:BoltzmannRTA_fs_aux} becomes 
\begin{multline}
 \frac{\partial \F_{\rm fs}}{\partial \bar{\tau}} + \frac{1}{\Delta_{\rm fs}} \bm{v}_{\perp; {\rm fs}} \cdot \overline{\nabla}_{\perp} \F_{\rm fs}\\
 = -\hat{\gamma} (v^\mu u_\mu) \overline{T}
 (\F_{\rm fs} - \F^{eq}_{\rm fs}).
 \label{eq:BoltzmannRTA_fs}
\end{multline}
Due to the changes to the momentum space degrees of freedom, the computation of the components of the stress-energy tensor becomes more involved. Taking into account the transformation of the measure $dp^\tau\,(p^\tau)^2 d\Omega_p = (\tau_0 / \tau) dp^\tau_{\rm fs} (p^\tau_{\rm fs})^2 d\Omega_{p;{\rm fs}}$, the dimensionless components $\overline{T}^{\mu\nu}$ can be computed as
\begin{align}
 \overline{T}^{\tau\tau} =& \frac{\tau_0}{\tau} \int d\Omega_{p; \rm fs}
 \,\Delta_{\rm fs} \, \F_{\rm fs},\nonumber\\
 \overline{T}^{\tau i}
 =& \frac{\tau_0}{\tau} \int d\Omega_{p; \rm fs}
 v_{\perp;{\rm fs}}^i \F_{\rm fs},\nonumber\\
 \overline{T}^{ij}
 =& \frac{\tau_0}{\tau} \int d\Omega_{p; \rm fs}
 \frac{v_{\perp; {\rm fs}}^i v_{\perp; {\rm fs}}^j}{\Delta_{\rm fs}} \F_{\rm fs},\nonumber\\
 \tau^2 \overline{T}^{\eta\eta}  =& \frac{\tau^3_0}{\tau^3} \int d\Omega_{p; \rm fs} \frac{v_{z; {\rm fs}}^2}{\Delta_{\rm fs}} \, \F_{\rm fs},
 \label{eq:FS_SET}
\end{align}
where $v_{\perp;{\rm fs}}^i = v_{\perp; {\rm fs}} (\cos\phi_p, \sin\phi_p)$ and $v_{\perp; {\rm fs}} = \sqrt{1 - v_{z;{\rm fs}}^2}$. Based on the the equilibrium Bose-Einstein distribution at vanishing chemical potential
\begin{equation}
 f_{\rm eq} = \left(e^{p_\mu u^\mu / T} - 1 \right)^{-1}.
\end{equation}
the functions $\F^{eq}_{\rm RLB}$ and $\F^{eq}_{\rm fs}$ in Eq.~\eqref{eq:F_RLB}, are readily obtained as
\begin{align}
 \F_{\rm RLB}^{eq} =& \frac{1}{4\pi} \frac{\overline{\epsilon}}{(u^0 - \vT \cdot \bm{u}_{\perp})^4},\nonumber\\
 \mathcal{F}^{eq}_{\rm fs} =& \frac{1}{4\pi}
 \frac{\overline{\epsilon}}{(u^0 \Delta_{\rm fs} - \bm{v}_{\perp;{\rm fs}} \cdot \bm{u}_{\perp})^4},
\end{align}
where $\overline{\epsilon} = \overline{T}^{\,4}$.

The system is initialized using the Romatschke-Strickland distribution \cite{Romatschke:2003ms} for Bose-Einstein statistics \cite{Molnar:2016gwq}
\begin{equation}
 f_{\rm RS} = \left\{\exp \left[\frac{1}{\Lambda} \sqrt{(p \cdot u)^2 + \xi_0 (p \cdot \hat{\eta})^2}\right] - 1\right\}^{-1},
 \label{eq:RS_gen}
\end{equation}
where $\hat{\eta}^\mu$ is the unit-vector along the rapidity coordinate. Simplifying to the initial state considered in this paper, Eq.~\eqref{eq:RS_gen} reduces to
\begin{equation}
 f_{\rm RS} = \left[\exp \left(\frac{p^\tau}{\Lambda} \sqrt{1 + \xi_0 v_z^2}\right) - 1\right]^{-1}.\label{eq:RS}
\end{equation}
The function $\Lambda \equiv \Lambda(\xT)$ is determined from 
\begin{equation}
 \Lambda^4(\xT) = 2T^4(\tau_0, \xT) \left(\frac{\arctan \sqrt{\xi_0}}{\sqrt{\xi_0}} + \frac{1}{1 + \xi_0}\right)^{-1},
 \label{eq:RS_aux}
\end{equation}
where $T(\tau_0, \xT)$ is obtained from the initial energy density $\epsilon(\tau_0, \xT)$ via the equation $\epsilon = a T^4$, where $a = \frac{\pi^2 \nu_{\rm eff}}{30}$ for Bose-Einstein statistics.
The anisotropy parameter $\xi_0$ can be used to set the ratio of longitudinal and transverse pressures $\mathcal{P}_{L;0} / \mathcal{P}_{T;0}$ via 
\begin{align}
 \frac{\mathcal{P}_{L;0}}{\mathcal{P}_{T;0}} = \frac{2}{1 + \xi_0} \frac{(1 + \xi_0) \frac{\arctan\sqrt{\xi_0}}{\sqrt{\xi_0}} - 1}
 {1 + (\xi_0 - 1)  \frac{\arctan\sqrt{\xi_0}}{\sqrt{\xi_0}}}.
\end{align}
The case $\mathcal{P}_{L;0} / \mathcal{P}_{T;0} = 0$ implied by the initial conditions in Eq.~\eqref{eq:fini} can be reached only as the limit $\xi_0 \rightarrow \infty$. In this paper, we consider finite (large) values of $
\xi_0$ and, for simplicity, we employ the same value of $\xi_0$ throughout the whole transverse plane.
Since at initial time $\tau = \tau_0$ $p^\tau_{\rm fs} = p^\tau$ and $v_{z; {\rm fs}} = v_z$ , it can be seen that the initial conditions
$\F^{\rm RS}_{\rm RLB}$ and $\F^{\rm RS}_{\rm fs}$ are equal and given by
\begin{equation}
 \mathcal{F}^{\rm RS}_{\rm RLB} = \mathcal{F}^{\rm RS}_{\rm fs} = \frac{\overline{\epsilon} / 2\pi}{(1 + \xi_0 v_z^2)^2} \left(\frac{\arctan \sqrt{\xi_0}}{\sqrt{\xi_0}} + \frac{1}{1 + \xi_0}\right)^{-1}.
\end{equation}

\subsubsection{Momentum space discretization}\label{NumericalProcedure:RLB:discretization}

In this paper, we employ the discretization of the momentum space discussed in Ref.~\cite{Ambrus:2018kug}. In this scheme, we employ $Q_{\phi_p} \times Q_z$ discrete values for $\phi_p$ and $v_z$ ($v_{z; {\rm fs}}$ in the case of the free-streaming variables), such that $(\phi_p, v_z)$ or $(\phi_p, v_{z; {\rm fs}})$ are replaced by $(\phi_{p;i}, v_{z;j})$ and $(\phi_{p;i}, v^{\rm fs}_{z; j})$, respectively. The discrete set of distributions $\mathcal{F}^*_{ij}$ (with $* \in \{{\rm RLB}, {\rm fs}\}$) are related to the original distribution function $\mathcal{F}_*$,  via \cite{Ambrus:2018kug}
\begin{equation}
 \begin{pmatrix}
  \mathcal{F}^{\rm RLB}_{ij} \\
  \mathcal{F}^{\rm fs}_{ij}
 \end{pmatrix}
  = \frac{2\pi}{Q_{\phi_p}} w_j 
 \begin{pmatrix}
  \mathcal{F}_{\rm RLB}(\phi_{p;i}, v_{z;j}) \\
  \mathcal{F}_{\rm fs}(\phi_{p;i}, v^{\rm fs}_{z;j}).
 \end{pmatrix}.
\end{equation}
The weight $2\pi / Q_{\phi_p}$ is computed in both RLB and fs cases following the prescription of the Mysovskikh (trigonometric) \cite{Mysovskikh:1987} quadrature, by which the trigonometric circle is discretized equidistantly, $\phi_{p;j} = \phi_0 + \frac{2\pi}{Q_{\phi_p}} \left(j - \frac{1}{2}\right)$, with $1 \le j \le Q_{\phi_p}$. For definiteness, we set the arbitrary offset $\phi_0$ to $0$. For the discretization of $v_z$ (RLB) and $v_{z; {\rm fs}}$ (FS), we employ two different strategies as discussed below.

{\bf RLB case}. In the case of large values of $\hat{\gamma}$, when Eq.~\eqref{eq:BoltzmannRTA_RLB} is considered, we employ the Gauss-Legendre quadrature, such that $w_j$ represent the Gauss-Legendre weights and $v_{z;j}$ are the roots of the Legendre polynomial of order $Q_z$, i.e. $P_{Q_z}(v_{z;j}) = 0$. Their values up to quadrature orders $Q_z = 1000$ can be found in the supplementary material of Ref.~\cite{Ambrus:2018kug}. In this approach, the term $\partial [v_z(1 - v_z^2) \F_{\rm RLB}] / \partial v_z$ is computed by projection onto the space of Legendre polynomials,
\begin{equation}
 \left[\frac{\partial[v_z(1 - v_z^2) \F_{\rm RLB}]}{\partial v_z}\right]_{ji} = \sum_{j' = 1}^{Q_z} \mathcal{K}^P_{j,j'} \F^{\rm RLB}_{j'i}.
\end{equation}
The matrix elements $\mathcal{K}^P_{j,j'}$, given in Eq.~(3.54) of Ref.~\cite{Ambrus:2018kug}, are
\begin{multline}
 \mathcal{K}^P_{j,j'} = w_j \sum_{m = 1}^{Q_z - 3} 
 \frac{m(m+1)(m+2)}{2(2m+3)} P_m(v_{z;j}) P_{m+2}(v_{z;j'}) \\
 - w_j \sum_{m = 1}^{Q_z - 1} \frac{m(m+1)}{2} P_m(v_{z;j}) \Bigg[\frac{(2m + 1) P_m(v_{z;j'})}{(2m-1)(2m+3)}\\
 + \frac{m-1}{2m-1} P_{m-2}(v_{z;j'})\Bigg].
\end{multline}
The components of the stress-energy tensor are obtained by replacing the integration with respect to $\d\Omega_p$ with a double sum over $i$ and $j$:
\begin{equation}
 \overline{T}^{\mu\nu} = \sum_{i = 1}^{Q_{\phi_p}} \sum_{j = 1}^{Q_z} \F^{\rm RLB}_{ij} v^\mu_{ij} v^{\nu}_{ij},\label{eq:RLB_Tmunu}
\end{equation}
where $v^{\tau}_{ij} = 1$, $(v^1_{ij}, v^2_{ij}) = \sqrt{1 - v_{z;j}^2} (\cos\phi_{p;i}, \sin\phi_{p;j})$ and $v^\eta_{ij} = \tau^{-1} v_{z;j}$. A similar prescription is employed for the computation of the $\d\Omega_p$ integral in the energy-weighted flow harmonics $v_n^E$ \eqref{eq:RLB_vn}.

{\bf FS case.}
For small values of $\hat{\gamma}$, the free-streaming coordinate $v_{z;{\rm fs}}$ is discretized in a logarithmic scale. Inspired from Eq. (A61) of Ref.~\cite{Kurkela:2019kip}, we first perform the change of coordinate to
\begin{equation}
 v_{z;{\rm fs}} = \frac{1}{A} \tanh \chi,
\end{equation}
where $0 < A < 1$ and $\chi$ takes values between $\pm {\rm artanh}\, A$. In order to increase the accuracy of the momentum space integration, we consider the rectangle method and take the discrete values $\chi_j$ at the center of the $Q_z$ equidistant intervals, such that
\begin{equation}
 \chi_j = \left(\frac{2j - 1}{Q_z} - 1\right) {\rm artanh} A, \qquad 
 v^{\rm fs}_{z;j} = \frac{1}{A} \tanh \chi_j.
\end{equation}
The quadrature weights $w_j$ are then computed based on the Jacobian due to the change of integration variable from $v_{z; {\rm fs}}$ to $\chi$, 
\begin{equation}
 w_j = \frac{2 {\rm artanh} A}{A Q_z \cosh^2 \chi_j}.
\end{equation}
Since the discretization of $v_{z;{\rm fs}}$ presented above is no longer given by a Gauss quadrature prescription, we note that the FS approach gives rise to a numerical scheme which is more similar to the Discrete Velocity Method (DVM) approach \cite{Mieussens:2000,Weih:2020qyh}. 
As before, the components of the stress-energy tensor can be obtained by replacing the integral with respect to $\d \Omega_p$ in Eq.~\eqref{eq:FS_SET} with quadrature sums:
\begin{align}
 \overline{T}^{\tau\tau} =& \frac{\tau_0}{\tau} \sum_{i,j}
 \Delta^{\rm fs}_j \F^{\rm fs}_{ij},\nonumber\\
 \begin{pmatrix}
  \overline{T}^{\tau 1} \\ \overline{T}^{\tau 2}
 \end{pmatrix}
 =& \frac{\tau_0}{\tau} \sum_{i,j}
 v^{\rm fs}_{\perp;j}
 \begin{pmatrix}
  \cos\phi_{p;i} \\ \sin\phi_{p;i}
 \end{pmatrix} \F^{\rm fs}_{ij},\nonumber\\
 \begin{pmatrix}
  \overline{T}^{11} \\ \overline{T}^{12} \\ \overline{T}^{22}
 \end{pmatrix}
 =& \frac{\tau_0}{\tau} \sum_{i,j}
 \frac{(v^{\rm fs}_{\perp; j})^2}{\Delta^{\rm fs}_j} 
 \begin{pmatrix}
  \cos^2\phi_{p,i} \\ \sin\phi_{p,i} \cos\phi_{p,i} \\ \sin^2 \phi_{p,i}
 \end{pmatrix} \F^{\rm fs}_{ij},\nonumber\\
 \tau^2 \overline{T}^{\eta\eta}  =& \frac{\tau^3_0}{\tau^3} \sum_{i,j} \frac{(v^{\rm fs}_{z; j})^2}{\Delta^{\rm fs}_j} \, \F^{\rm fs}_{ij},
\end{align}
where $\Delta^{\rm fs}_j = [1 - (1 - \tau_0^2 /\tau^2) (v^{\rm fs}_{z;j})^2]^{1/2}$ and $v^{\rm fs}_{\perp;j} = \sqrt{1 - (v^{\rm fs}_{z;j})^2}$. A similar procedure is employed for the computation of $v_n^E$ \eqref{eq:RLB_vn}.

\subsubsection{Finite difference schemes}\label{NumericalProcedure:RLB:numsch}

In order to obtain the numerical solution of Eqs.~\eqref{eq:BoltzmannRTA_RLB} and \eqref{eq:BoltzmannRTA_fs}, we consider an equidistant discretization of the temporal and spatial variables. Setting the time step as $\delta \tau$, the time coordinate is discretized according to $\tau_n = \tau_0 + n \delta \tau$. Writing the Boltzmann equation as 
\begin{equation}
 \frac{\partial \mathcal{F}}{\partial \tau} = L[\mathcal{F}],
\end{equation}
where $L[\mathcal{F}]$ can be found from Eqs.~\eqref{eq:BoltzmannRTA_RLB} or \eqref{eq:BoltzmannRTA_fs}, we employ the third-order total variation diminishing (TVD) Runge-Kutta method proposed in Ref.~\cite{Shu:1988,Gottlieb:1998}. This algorithm allows the values $\mathcal{F}_{n + 1}$ of the distribution functions at the new time step to be obtained from the old ones using two intermediate stages.

The advection along the transverse directions
is performed using the flux-based upwind-biased fifth order weighted essentially non-oscillatory (WENO-5) scheme \cite{Jiang:1996,Rezzolla:2013}. Considering that the spatial domain of extent $L_1 \times L_2$ is discretized using $N_1 \times N_2$ equidistant nodes, the coordinates of the discrete points are
\begin{align}
 x_{1,s} =& x_{1,\rm left} + \frac{L_1}{N_1}\left(s - \frac{1}{2}\right), \nonumber\\
 x_{2,r} =& x_{2,\rm bot} + \frac{L_2}{N_2}\left(r - \frac{1}{2}\right),
\end{align}
with $1 \le s \le N_1$ and $1 \le r \le N_2$.
Focusing without loss of generality on the derivative with respect to $x_1$, the algorithm entails
\begin{equation}
 c_1 \left(\frac{\partial \mathcal{F}}{\partial x_1}\right)_{s,r} = 
 \frac{\mathbb{F}_{s+\frac{1}{2}, r} - \mathbb{F}_{s-\frac{1}{2}, r}}{\delta x_1},
\end{equation}
where $\delta x_1 = L_1 / N_1$. The velocity $c_1$ is given in the case when $\hat{\gamma}$ is large, when Eq.~\eqref{eq:BoltzmannRTA_RLB} is solved, by $c_1 = \sqrt{1 - v_z^2} \cos\phi_p$, being independent of position and space. In the case of small values of $\hat{\gamma}$, Eq.~\eqref{eq:BoltzmannRTA_fs} shows that the advection velocity $c_1 = \frac{1}{\Delta} \sqrt{1 - \tilde{v}_z^2} \cos\phi_p$ depends on the Bjorken time $\tau$, however it remains constant throughout the entire transverse plane. The stencils required to compute the fluxes $\mathbb{F}_{s \pm \frac{1}{2}, r}$ are chosen in an upwind-biased manner based on the sign of $c_1$. Since the algorithm to compute these stencils is rather lengthy, we do not repeat it here and instead refer the interested reader to Refs.~\cite{Jiang:1996,Rezzolla:2013,Ambrus:2018kug,Busuioc:2019} for details.

\section{Results}
\label{sec:results}
We will now analyze the space-time evolution of the system and the development of transverse flow as a function of the opacity parameter $\hat{\gamma}$ (c.f. Eq.~\eqref{eq:ghat_def}). We focus on the range of opacities $0.01 \le \hat{\gamma} \le 400$ and consider different initial eccentricities $\epsilon_n \in \{0.01, 0.05, 0.1, 0.2, 0.32, 0.36\}$ (c.f. Eqns.~(\ref{eq:epsilon_isotropic_def},\ref{eq:deltaeps_def})). 

If not stated otherwise, open symbols/dashed lines correspond to results obtained using the expansion in spherical harmonic moments in Sec.~\ref{sec:MomentumMomentExpansion}, where we typically use $l_{\rm max}=32$,$N_{S}=256$, $a_S=0.0625~R$.\footnote{ We note that results for $\hat{\gamma}\leq 1$ require a larger accuracy, and we use $l_{\rm max}=48$,$N_{S}=320$, $a_S=0.05~R$. Similarly, for accurate calculations of $\d E_\perp/\d y$ we need a larger value of $l_{\rm max}$ and we employ $l_{\rm max}=96$,$N_{S}=160$, $a_S=0.06~R$ in this case.} Conversely, results obtained with the relativistic lattice Boltzmann (RLB) method are represented by solid symbols/solid lines. The RLB simulations are divided in two batches. The first batch includes systems with $\hat{\gamma} \ge 2$. For these simulations, we used the RLB algorithm for large $\hat{\gamma}$ described in Sec.~\ref{NumericalProcedure:RLB} with $Q_z = 40$ and $Q_{\phi_p} = 80$, while the number of nodes on each semiaxis is taken to be $X = 100$ for $\epsilon_n \ge 0.05$ and $X = 200$ for $\epsilon_n < 0.05$. The anisotropy parameter in the initial state is set to $\xi_0 = 20$, corresponding to an initial ratio $\mathcal{P}_L / \mathcal{P}_T \simeq 0.08$. The second batch comprises the systems with $\hat{\gamma} \le 2$ for which we employ the hybrid free-streaming algorithm described in Sec.~\ref{NumericalProcedure:RLB} with $Q_z = 500$ and $Q_{\phi_p} = 80$. In this case, the anisotropy parameter is set to $\xi_0 = 100$, corresponding to $\mathcal{P}_L / \mathcal{P}_T \simeq 0.02$ and the spatial resolution is $X = 100$ nodes per semiaxis.

\subsection{Cooling due to longitudinal expansion ($dE_{\perp} / d\eta$)}
\label{sec:results_cooling}

\begin{figure}
\centering
\begin{tabular}{c}
 \includegraphics[width=\linewidth]{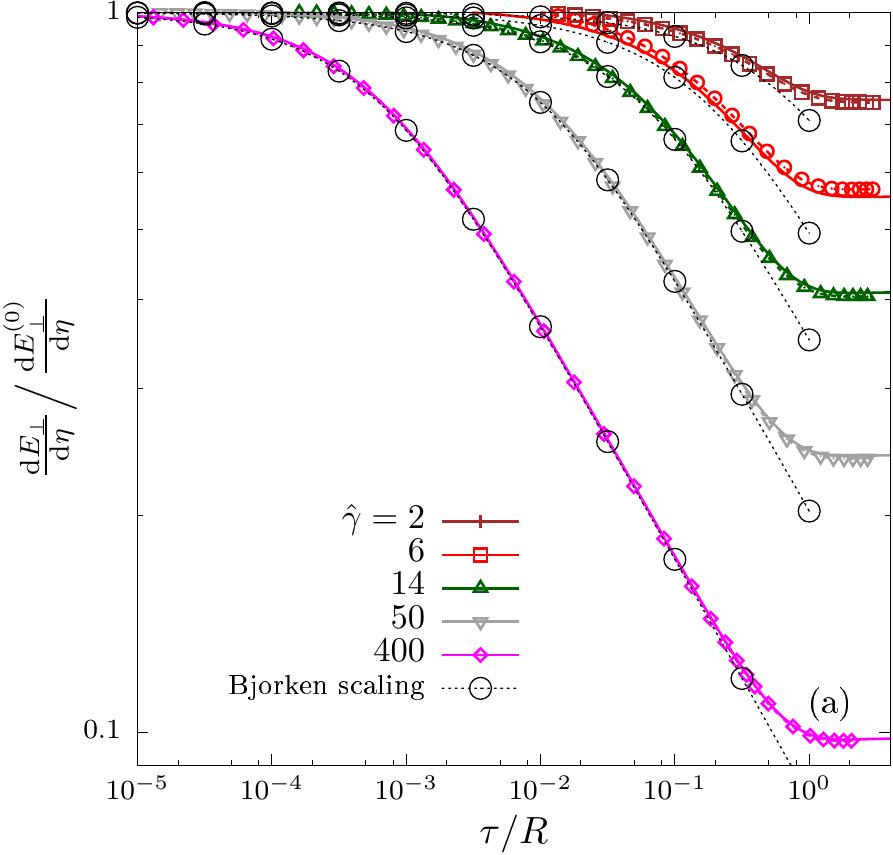} \\
 \includegraphics[width=\linewidth]{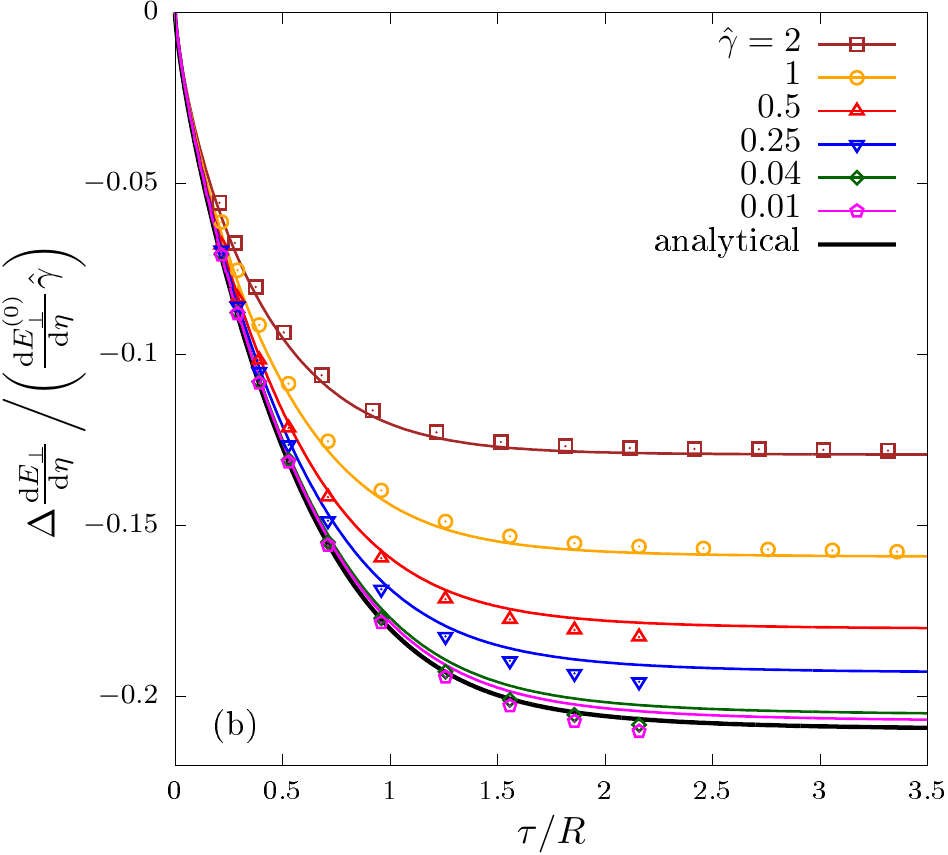}
\end{tabular}
\caption{Evolution of $dE_{\perp} / d\eta$ (top) and  $\Delta dE_{\perp} / d\eta = dE_{\perp} / d\eta - dE_{\perp}^{(0)} / d\eta$ (bottom) normalized with respect to the initial value $dE^{(0)}_{\perp} / d\eta$ for various large (top) and small opacities (bottom). Colored solid lines were obtained with the RLB method, while open symbols denote results from the moment method. The solid black line shows the first order result in opacity expansion. Dashed black lines with black circles corresponds to the Bjorken flow prediction derived in Eq.~\eqref{eq:dEdyBjorkenApprox}, and the curves in the top are presented on double logarithmic scale. All results are obtained for initial eccentricity $\epsilon_2 = 0.05$.}
\label{fig:dEdy}
\end{figure}

Before we discuss the development of transverse flow, we first investigate the cooling of the system due to work performed against the longitudinal expansion, which is quantified by the decrease of the transverse energy per rapidity $\d E_\perp/\d \eta$. We first note that for a free-streaming system $\d E_\perp/\d \eta$ is constant. Increasing the opacity will initially only have a small effect, which can be quantified in terms of the linear decrease in $\hat{\gamma}$ calculated in Sec.~\ref{first order corrections}. However, for large opacities $\hat{\gamma} \gg 1$, the system has sufficient time to undergo pressure isotropization at early times, leading to an extended phase of longitudinal cooling,  which results in a significant decrease of $\d E_\perp/\d \eta$. Hence, when presenting our results for $\d E_\perp/\d y (\tau)$ in Fig.~\ref{fig:dEdy}, we have grouped  them into two plots for large opacities in the upper panel and small opacities in the lower panel. While for large opacities, the curves are normalized by the initial value $\d E_\perp^{(0)}/\d \eta$ and plotted on a doubly logarithmic scale to visualize the power law decay of $\d E_\perp/\d \eta$ at intermediate times, for small opacities we show the difference of $\d E_\perp/\d \eta - \d E_\perp^{(0)}/\d \eta$, normalized by the initial value and $\hat{\gamma}$ to account for the linear behaviour in opacity. We also show a comparison with the analytical result from Section~\ref{first order corrections}, which provides a good description of the curves for $\hat{\gamma}\lesssim 1$.

Qualitatively, all curves exhibit a similar behavior starting out from the early time fixed point of kinetic theory, where longitudinal pressure vanishes and energy per rapidity stays almost constant. Subsequently, as longitudinal pressure develops due to interactions work is being performed, which starts to happen earlier and earlier the larger the opacity. Eventually,  at late times $\tau/R\gtrsim 1$, the transverse expansion becomes dominant and the system rapidly cools down, resulting in a late time plateau of the $\d E_\perp/\d \eta (\tau)$-curves.

We find that for large opacities $\hat{\gamma}\gtrsim 10$, the pressure isotropization at early times and the onset of the transverse expansion at later times are sufficiently well separated to observes an intermediate $\tau^{-1/3}$-scaling of $\d E/\d \eta$, which -- as we will see shortly -- can be related to the usual $\epsilon \sim \tau^{-4/3}$ decrease of the energy density in Bjorken flow. It stands to reason that, at early times, the transverse gradients in the system are negligible compared to the longitudinal expansion, and the system will locally behave like a one-dimensional Bjorken system. Based on the following considerations, this behavior can be quantified further, and cast into a parameter-free prediction for the evolution of $dE_{\bot}/d\eta$ in Eq.~\eqref{eq:dEdyBjorkenApprox}, which is indicated by black circles in the upper panel of Fig.~\ref{fig:dEdy} and agrees remarkably well with numerical results for the large opacities up to times $\tau/R \lesssim 0.1$.

Defining the conformal scaling variable $\tilde{w}(\tau,\xT)=\frac{T(\tau,\xT)\tau}{4\pi\eta/s}$, the Bjorken flow exhibits a universal attractor curve~\cite{Giacalone:2019ldn} for
\begin{eqnarray}
\label{eq:BjorkenAttractor}
\epsilon(\tau)\tau^{4/3}=(4\pi \eta/s)^{4/9}
a^{1/9}
(\epsilon\tau)_{0}^{8/9}~C_{\infty}~\mathcal{E}(\tilde{w})\;, \nonumber \\
\tau^{1/3} \frac{\d E_\perp}{d^2\xT \d \eta} = (4\pi \eta/s)^{4/9}
a^{1/9}
(\epsilon\tau)_{0}^{8/9}~C_{\infty}~f_{E_{\bot}}(\tilde{w}) \nonumber \\
\end{eqnarray}
where the aysmptotic limits of $\mathcal{E}(\tilde{w})$ are known~\cite{Giacalone:2019ldn} and given by
\begin{align}
\mathcal{E}(\tilde{w}\gg1)=& 1-\frac{2}{3\pi \tilde{w}},\nonumber\\
\mathcal{E}(\tilde{w}\ll1) =& C_{\infty}^{-1} \tilde{w}^{4/9}. \label{eq:universal_E_lim}
\end{align}
Similarly one finds for $f_{E_{\bot}}(\tilde{w})$ that
\begin{align}
f_{E_{\bot}}(\tilde{w} \gg 1)&=\frac{\pi}{4}, \nonumber\\
f_{E_{\bot}}(\tilde{w}\ll1)&=C_{\infty}^{-1} \tilde{w}^{4/9},
\label{eq:universal_fEp_lim}
\end{align}
where for the RTA Boltzmann equation $C_{\infty} \approx 0.9$~\cite{Giacalone:2019ldn,Kamata:2020mka}, and the leading constant $\pi/4$ can be deduced from an integral of the thermal equilibrium distribution. By use of the equation of state $\epsilon = a T^4$, Eq.~(\ref{eq:BjorkenAttractor}) can be re-cast as a self-consistency conditions for $\tilde{w}$, which takes the form
\begin{multline}
\label{eq:wTildeSC}
\tilde{w}(\tau,\xT) = (4\pi \eta/s)^{-8/9} 
a^{-2/9}
(\epsilon\tau)_{0}^{2/9}(\xT) \\
\times \tau^{2/3}~[C_{\infty} \mathcal{E}(\tilde{w}(\tau,\xT))]^{1/4}\;,
\end{multline}
and can be used to relate the differential with respect to $\tilde{w}$ at fixed $\tau$ and that with respect to the transverse plane coordinates $\xT$ as
\begin{eqnarray}
\frac{d\tilde{w}}{\tilde{w}} \left[1 - \frac{1}{4} \frac{\tilde{w}\mathcal{E}'(\tilde{w})}{\mathcal{E}(\tilde{w})} \right]= \frac{2}{9} \frac{\frac{\partial (\epsilon\tau)_{0}(\xT)}{\partial |\xT|}}{(\epsilon\tau)_{0}(\xT)}~d|\xT|\;,
\end{eqnarray}
Specifically, for the initial Gaussian profile in Eq.~(\ref{eq:epsilon_isotropic_def}), one finds
\begin{eqnarray}
|\xT| d|\xT| = - \frac{9 R^2}{4} \frac{d\tilde{w}}{\tilde{w}} \left[1 - \frac{1}{4} \frac{\tilde{w}\mathcal{E}'(\tilde{w})}{\mathcal{E}(\tilde{w})} \right]
\end{eqnarray}
which can be used to calculate the resulting change of the energy per unit rapidity as follows. Starting from the definition in Eq.~(\ref{eq:BjorkenAttractor}), one can express the energy per unit rapdity in terms of the scaling function
\begin{eqnarray}
\frac{dE_{\bot}}{d\eta}&=& \tau^{-1/3} (4\pi \eta/s)^{4/9}
a^{1/9}
C_{\infty} \\&& \times (2\pi) \int d|\xT|~|\xT|~f_{E_{\bot}}(\tilde{w}(\tau,\xT))~(\epsilon\tau)_{0}^{8/9}(\xT)\;. \nonumber
\end{eqnarray}
Expressing the remaining factor of the energy density $(\epsilon\tau)_{0}^{8/9}$ in terms of the conformal scaling variable $\tilde{w}$ according to
\begin{equation}
(\epsilon\tau)_{0}^{8/9}(\xT)=(4\pi\eta/s)^{32/9} 
a^{8/9}
\tau^{-8/3}~\frac{\tilde{w}^{4}}{C_\infty \mathcal{E}(\tilde{w})},
\end{equation}
which follows from Eq.~(\ref{eq:wTildeSC}), one can then express
\begin{multline}
\frac{dE_{\bot}}{d\eta}= \frac{9\pi a}{2 R} \left(\frac{R}{\tau}\right)^{3} (4\pi \eta/s)^{4}
\\\times 
\int_{0}^{\tilde{w}(\tau,\xT=0)} 
\frac{\tilde{w}^3 d\tilde{w}}{\mathcal{E}(\tilde{w})} \left[1 - \frac{1}{4} \frac{\tilde{w} \mathcal{E}'(\tilde{w})}{\mathcal{E}(\tilde{w})} \right]~f_{E_{\bot}}(\tilde{w}). 
\end{multline}
By considering the ratio $\frac{dE_{\bot}}{d\eta} / \frac{dE^{0}_{\bot}}{d\eta}$ and identifying 
\begin{eqnarray}
\frac{(4\pi \eta/s)^{4}
a}{dE^{0}_{\bot}/d\eta~R}= \frac{1}{\pi}\left(\frac{4\pi}{5 \hat{\gamma}}\right)^{4}
\end{eqnarray}
one then obtains the final result
\begin{multline}
\label{eq:dEdyBjorkenApprox}
\frac{dE_{\bot}/d\eta}{dE^{0}_{\bot}/d\eta}=\frac{9}{2}~\left(\frac{4\pi}{5\hat{\gamma}}\right)^{4}~\left(\frac{R}{\tau}\right)^{3} \\ 
\times \int_{0}^{\tilde{w}(\tau,\xT=0)} \frac{\tilde{w}^3 d\tilde{w}}{\mathcal{E}(\tilde{w})} \left[1 - \frac{\tilde{w}}{4} \frac{\mathcal{E}'(\tilde{w})}{\mathcal{E}(\tilde{w})} \right]~f_{E_{\bot}}(\tilde{w}),
\end{multline}
where $\tilde{w}(\tau,\xT=0)$ in the center of the collision can be expressed in terms of $\hat{\gamma}$ via
\begin{equation}
\tilde{w}(\tau,\xT=0)=\left(\frac{5 \hat{\gamma}}{4\pi} \right)^{8/9} \left(\frac{\tau}{R}\right)^{2/3} [C_{\infty}\mathcal{E}(\tilde{w})]^{1/4}.
\label{eq:w_center}
\end{equation}

\begin{figure}
    \centering
    \includegraphics[width=.9\linewidth]{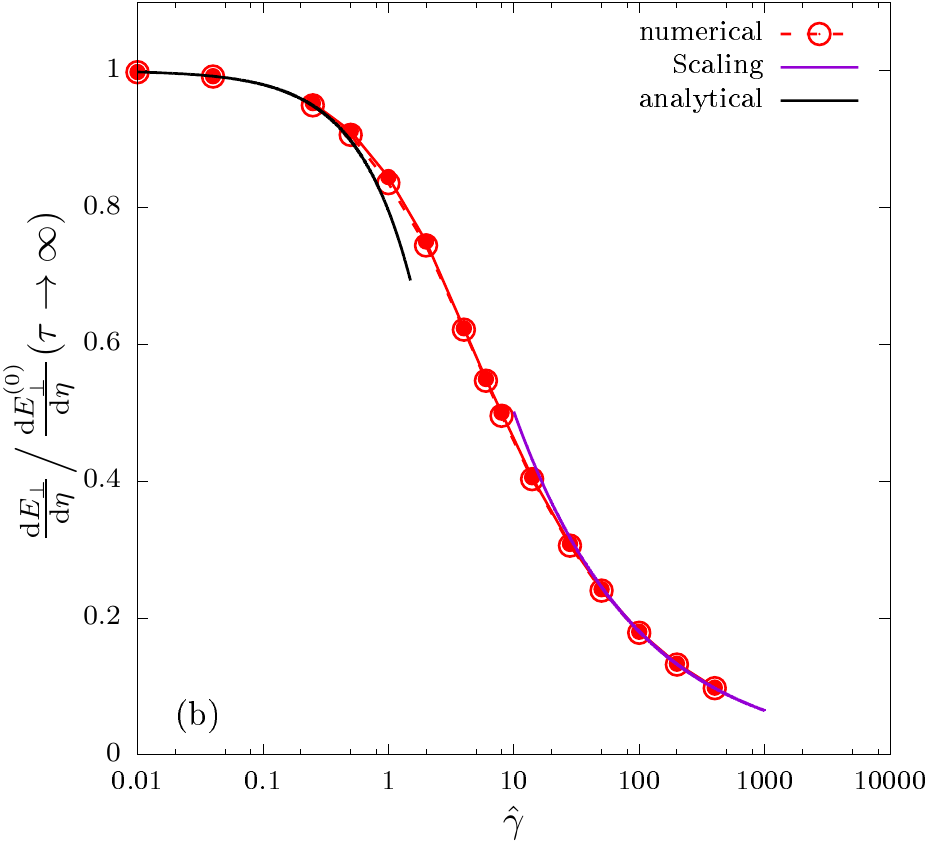}
    \caption{Opacity ($\hat{\gamma}$) dependence of the ratio of final to initial (transverse) energy per-unit rapidity $\frac{\mathrm{d}E_\perp}{\mathrm{d}\eta}\Big/\frac{\mathrm{d}E_\perp^{(0)}}{\mathrm{d}\eta}$. The red solid line with filled circles denotes results from the RLB method, while the red dotted line with open circles was obtained in the moment method. Numerical results are compared to analytical results obtained in leading order opacity expansion (black solid curve), and a power-law scaling fit $\frac{\mathrm{d}E_\perp}{\mathrm{d}\eta}\Big/\frac{\mathrm{d}E_\perp^{(0)}}{\mathrm{d}\eta} \approx 1.4\, \hat{\gamma}^{-4/9}$ at large opacities (purple solid line)}.
    \label{fig:dEdy_gamma_dependence}
\end{figure}

The asymptotic behavior of Eq.~\eqref{eq:dEdyBjorkenApprox} can be understood as follows. In the limit $\hat{\gamma} \left(\frac{\tau}{R}\right)^{3/4} \ll 1$, we have $\tilde{w}\approx \frac{5}{4\pi} \hat{\gamma} \left(\frac{\tau}{R}\right)^{3/4}  \ll1$ and we can approximate $\mathcal{E}(\tilde{w})=f_{E_{\bot}}(\tilde{w})=C_{\infty}^{-1} \tilde{w}^{4/9}$ to obtain
\begin{equation}
\frac{dE_{\bot}/d\eta}{dE^{0}_{\bot}/d\eta} =1,
\end{equation}
as expected. Conversely, in the limit $\hat{\gamma}^{3/4} \left(\frac{\tau}{R}\right) \gg 1$, we have $\tilde{w}\approx \left(\frac{5}{4\pi} \right)^{8/9} \hat{\gamma}^{8/9}~C_{\infty}^{1/4}~\left(\frac{\tau}{R}\right)^{2/3} \gg 1$, such that we can approximate $\mathcal{E}(\tilde{w}) \approx 1$ and evaluate Eq.~\eqref{eq:dEdyBjorkenApprox} as
\begin{align}
 \frac{dE_{\bot}/d\eta}{dE^{0}_{\bot}/d\eta} = \frac{9}{8} \left(\frac{4\pi}{5 \hat{\gamma}}\right)^{4/9} \left(\frac{R}{\tau}\right)^{1/3} C_\infty f_{E_{\perp}}(\infty).
 \label{eq:dEdeta_late}
\end{align}
which predicts that $dE_{\bot}/d\eta$ decreases as $\tau^{-1/3}$ at intermediate times, before the transverse expansion becomes dominant. By comparing the results in Fig.~\ref{fig:dEdy}, one finds that for sufficiently large opacities the power law behavior in Eq.~\eqref{eq:dEdeta_late} is indeed realized at intermediate times, and discontinues once $\tau/R \simeq 1$, when the transverse expansion becomes dominant, such that the estimate \eqref{eq:dEdeta_late} is no longer applicable and $\frac{dE_{\bot}/d\eta}{dE^{0}_{\bot}/d\eta}$ attains a constant asymptotic value. 
The details regarding the computation of the integral in Eq.~\eqref{eq:dEdyBjorkenApprox} are presented in Appendix~\ref{app:attractor}.

We note that our estimate in Eq.~(\ref{eq:dEdeta_late}) also shows that for sufficiently large opacities, where longitudinal cooling occurs predominantly before the transverse expansion sets in, the final value of $\frac{dE_{\bot}/d\eta}{dE^{0}_{\bot}/d\eta}$ is propotional to $\hat{\gamma}^{-4/9}$, as previously argued in~\cite{Kurkela:2019kip}. Numerical results for the asymptotic values of $\frac{\d E_{\bot}/d\eta}{\d E^{0}_{\bot}/d\eta}$, extracted by performing extrapolations of the curves of the form $a + b \tilde{\tau}^{-c}$, where $a$, $b$ and $c$ are fitting parameters, are shown in Fig.~\ref{fig:dEdy_gamma_dependence} as a function of the opacity parameter $\hat{\gamma}$. We find that at low $\hat{\gamma}$, the analytical result to leading order in opacity (c.f. Eq.~\eqref{eq:dEdyanalyticalasymptote}), represented with a solid black line, provide an accurate description up to $\hat{\gamma} \lesssim 1$.  Conversely, for large opacities $\hat{\gamma}$, the decrease of the energy per unit rapidity $\frac{\d E_{\bot}/d\eta}{\d E^{0}_{\bot}/d\eta}$ exhibits the expected scaling behavior, with $\frac{\d E_{\bot}/d\eta}{\d E^{0}_{\bot}/d\eta} \approx 1.4 ~\hat{\gamma}^{-4/9}$ for $\hat{\gamma} \gtrsim 10$, as indicated by the purple line. By comparing the numerical coefficient with Eq.~\eqref{eq:dEdeta_late}, this result seems to indicate that cooling stops at a time $\tau_{\rm stop} \simeq 0.6 R$, which is consistent with the trend seen for the high $\hat{\gamma}$ curves in Fig.~\ref{fig:dEdy}(a).

\subsection{Development \& opacity dependence of transverse flow harmonics $(v_n)$}\label{sec:results_vn}

\def\lw{.32\linewidth}
\begin{figure*}
\centering 
\begin{tabular}{ccc}
 \includegraphics[width=\lw]{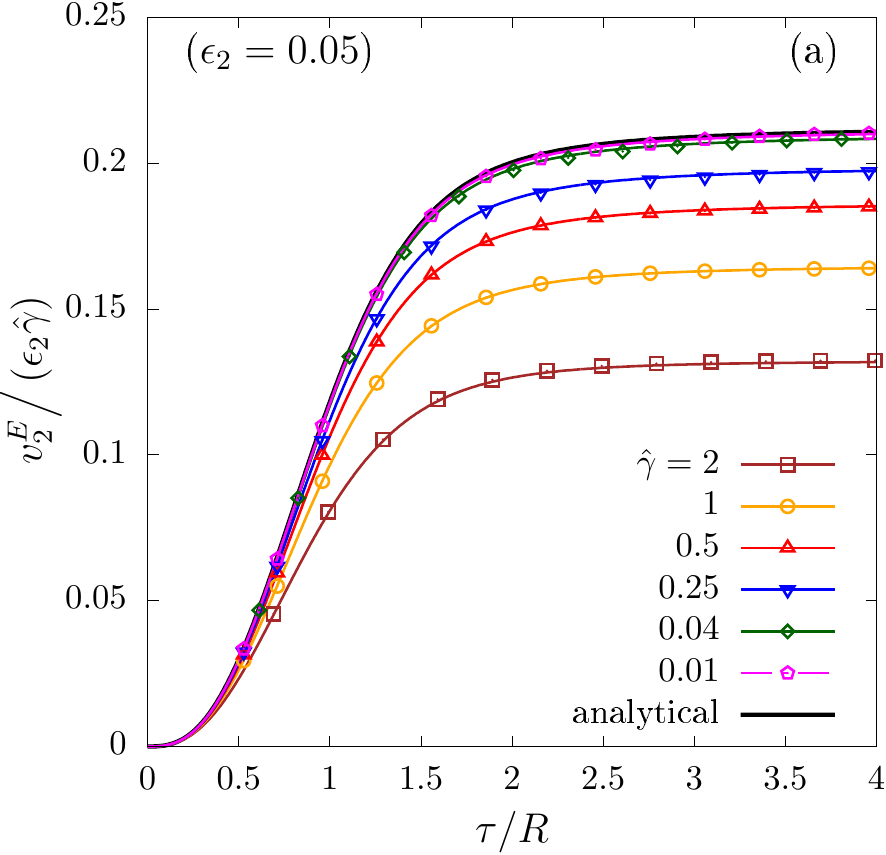} &
 \includegraphics[width=\lw]{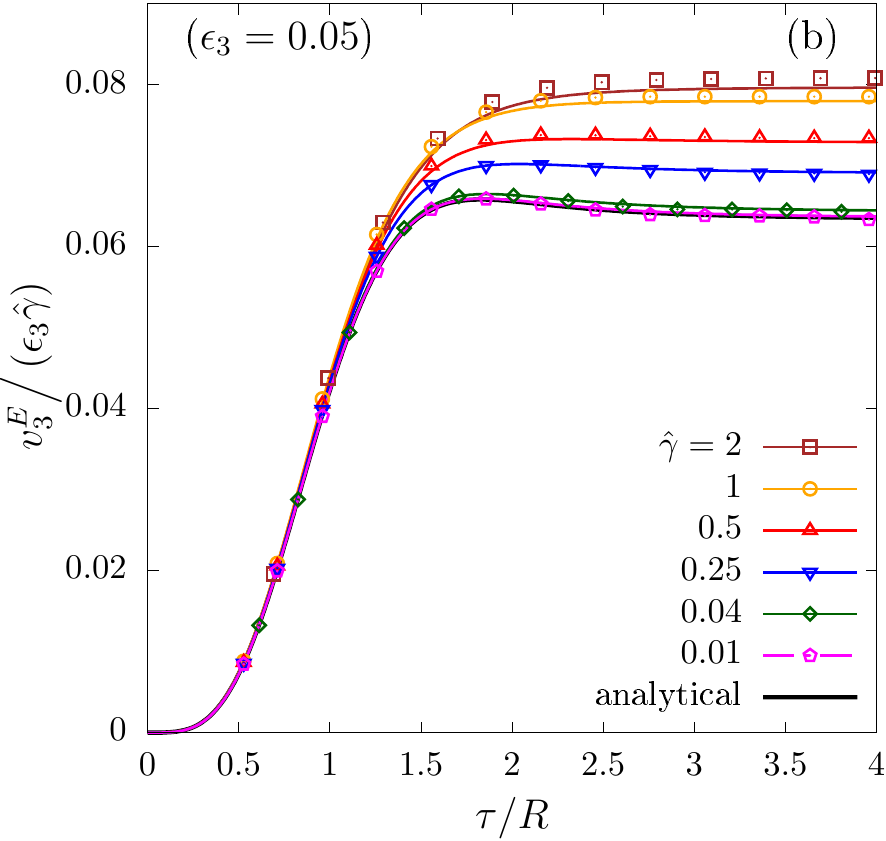} &
 \includegraphics[width=\lw]{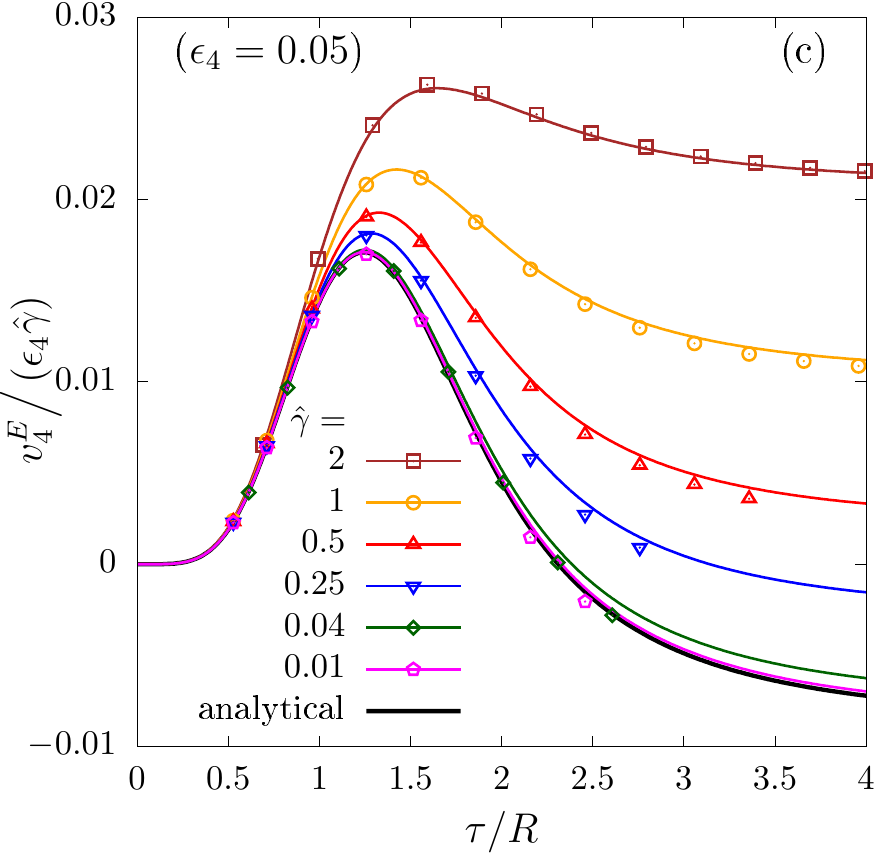} \\
 \includegraphics[width=\lw]{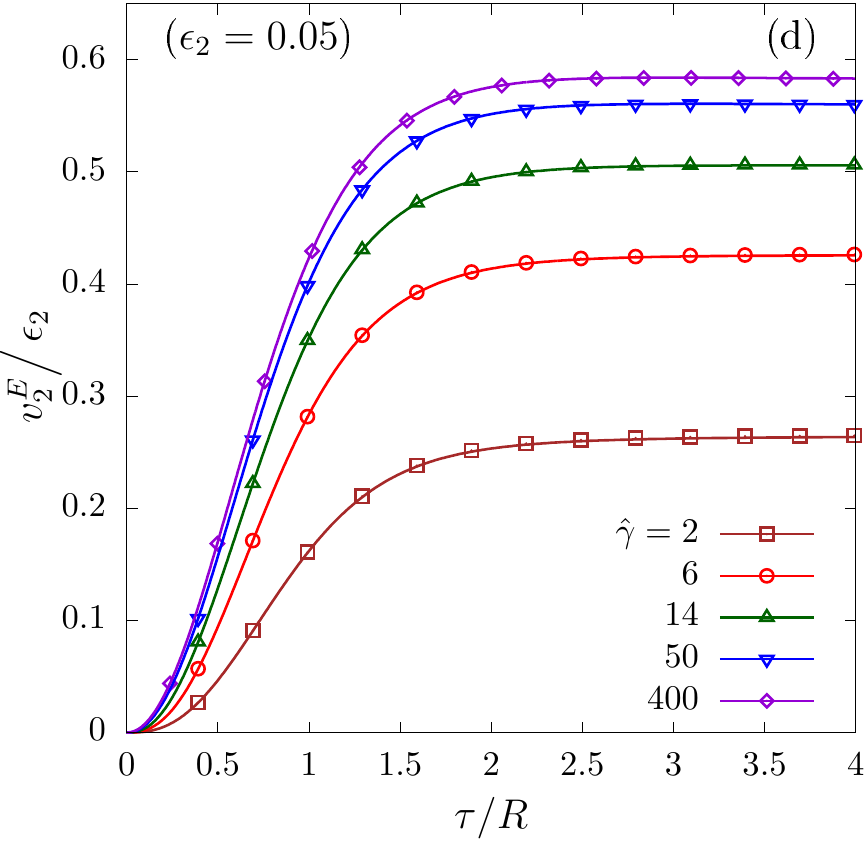} &
 \includegraphics[width=\lw]{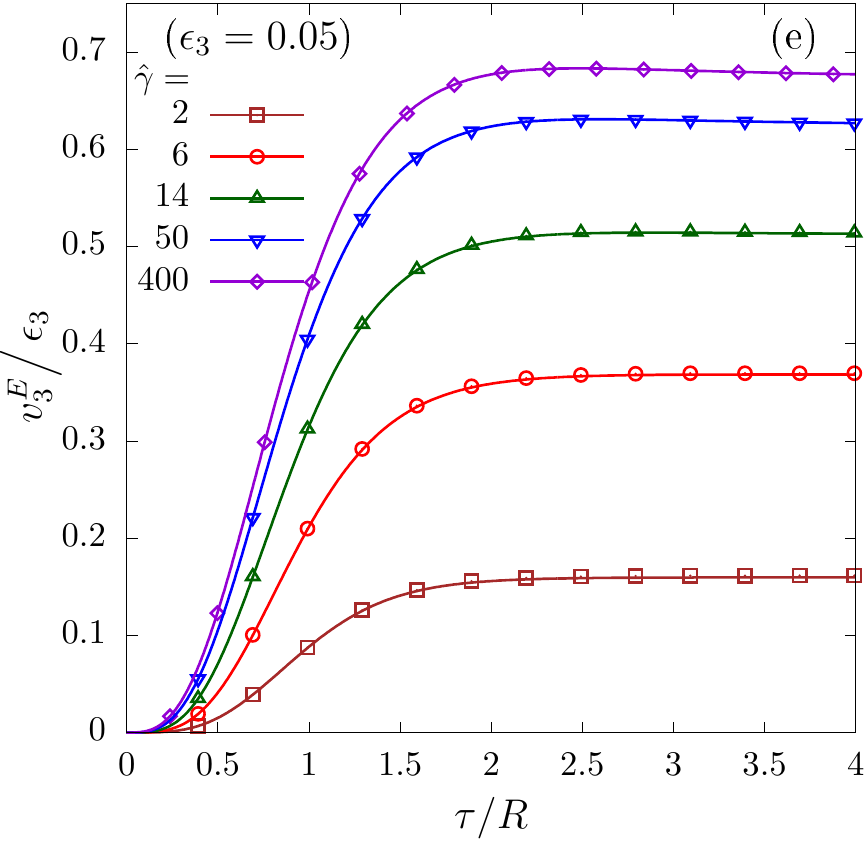} &
 \includegraphics[width=\lw]{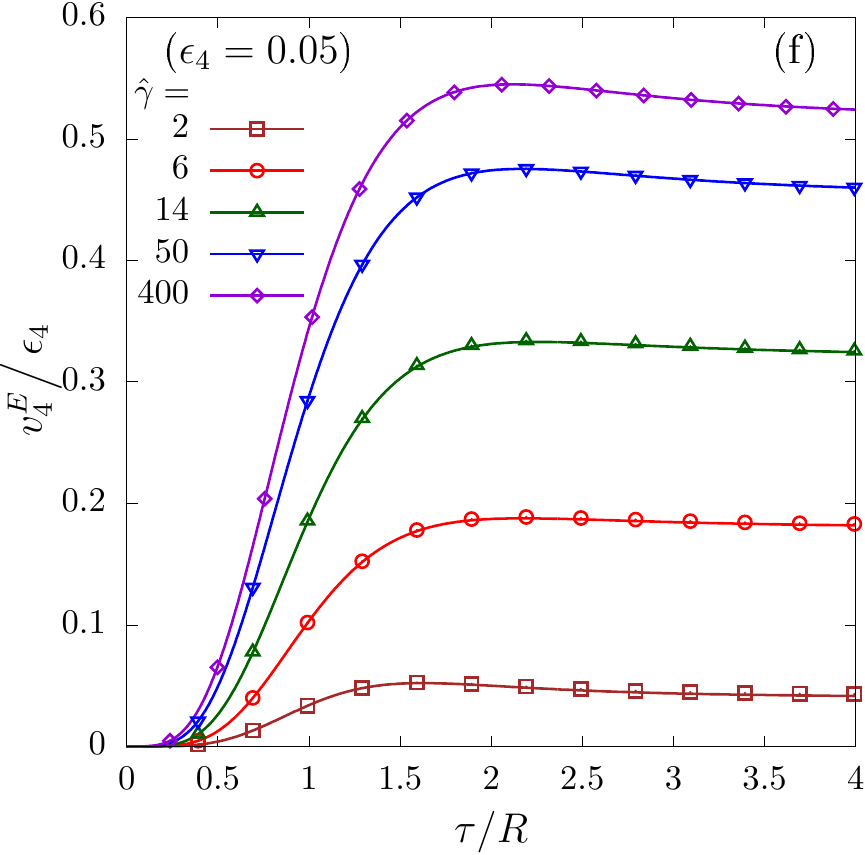}
\end{tabular}
\caption{Evolution of the linear flow response
$v_n^E / \epsilon_n \hat{\gamma}$ at small opacities 
${\hat \gamma}$ (top) and 
$v_n^E / \epsilon_n$ at large opacities ${\hat \gamma}$ (bottom). Different columns correspond to elliptic flow
$n = 2$ (left), triangular flow $n = 3$ (middle) and quadrangular flow
$n = 4$ (right). Colored solid lines were obtained in the RLB method, while open symbols denote results from the moments method. Analytical results are plotted as solid black lines. All results were obtained for an initial eccentricity
$\epsilon_n = 0.05$.}
\label{fig:vn_vs_tau_e0p05}
\end{figure*}

\def\lw{.32\linewidth}
\begin{figure*}
\centering 
\begin{tabular}{ccc}
 \includegraphics[width=\lw]{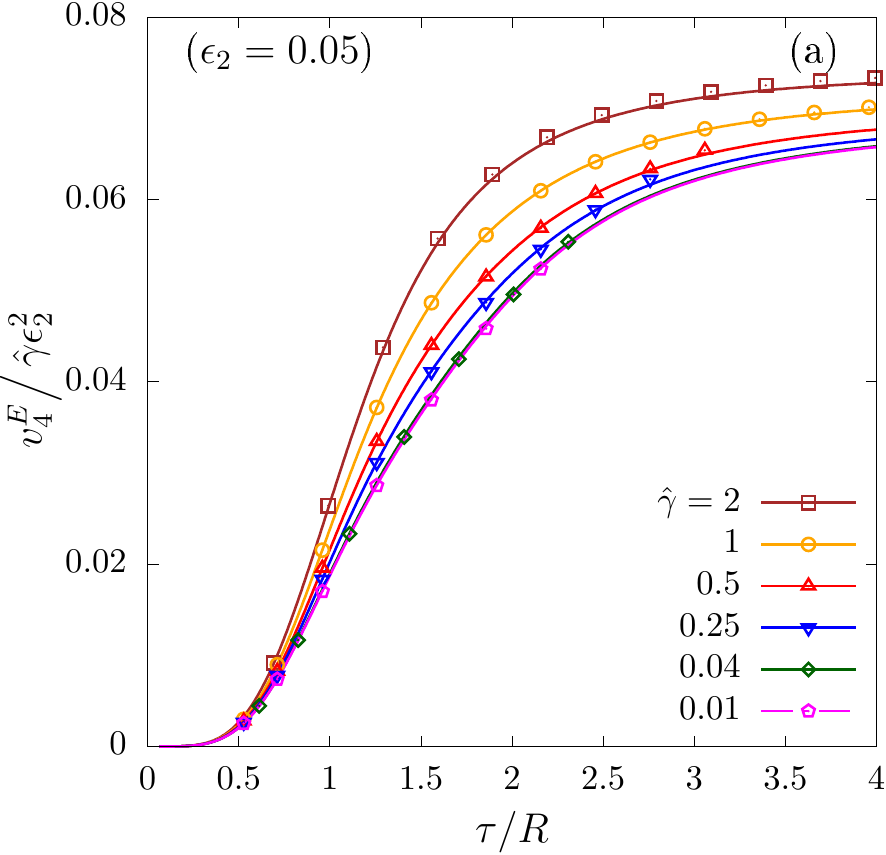} &
 \includegraphics[width=\lw]{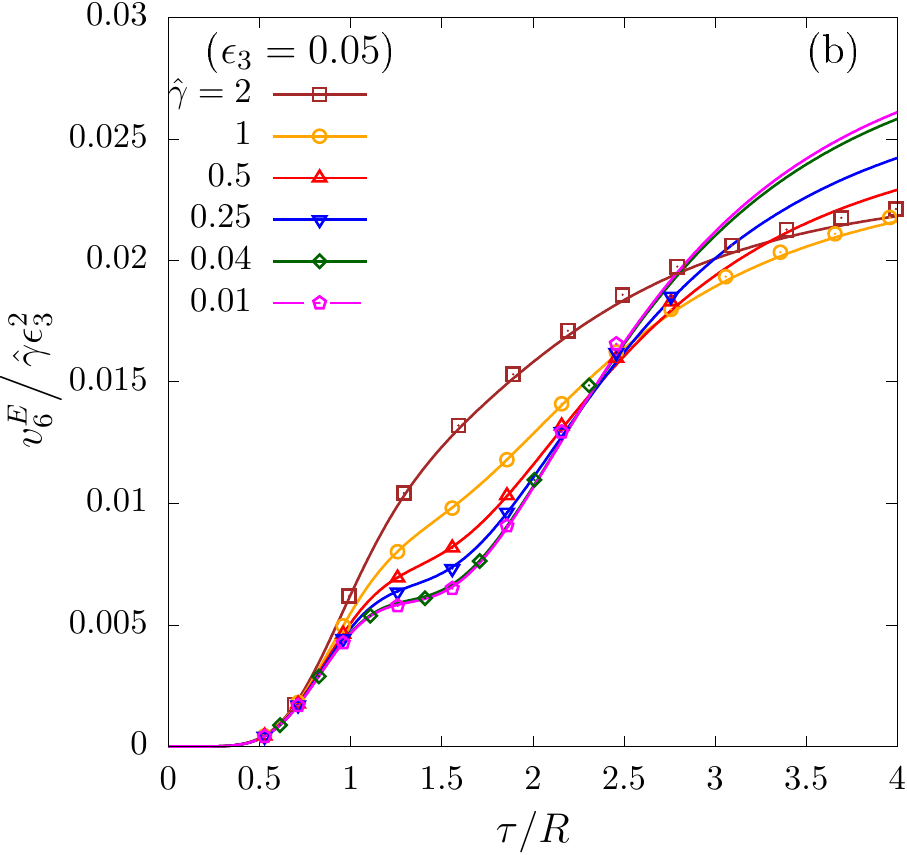} &
  \includegraphics[width=\lw]{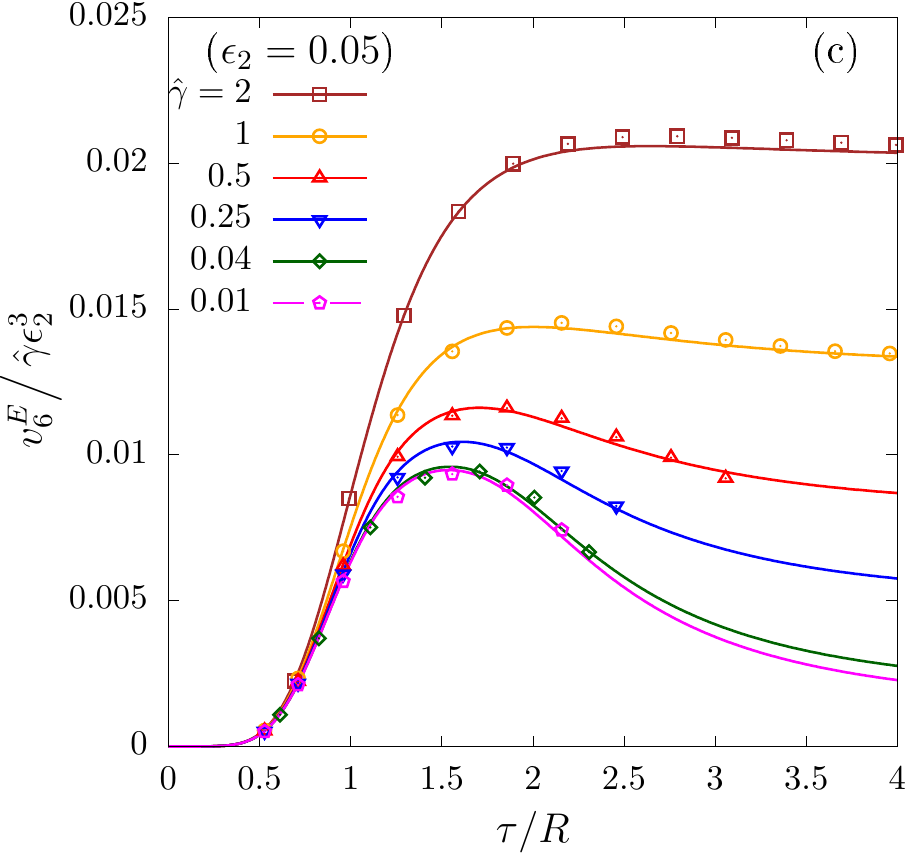} \\
  \includegraphics[width=\lw]{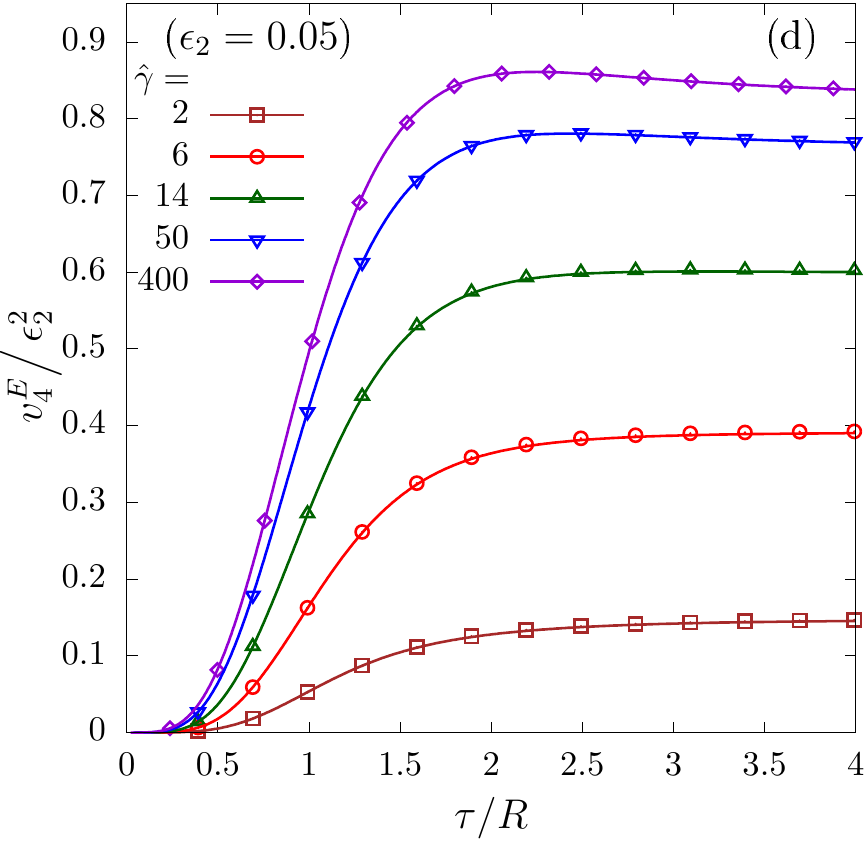} &
   \includegraphics[width=\lw]{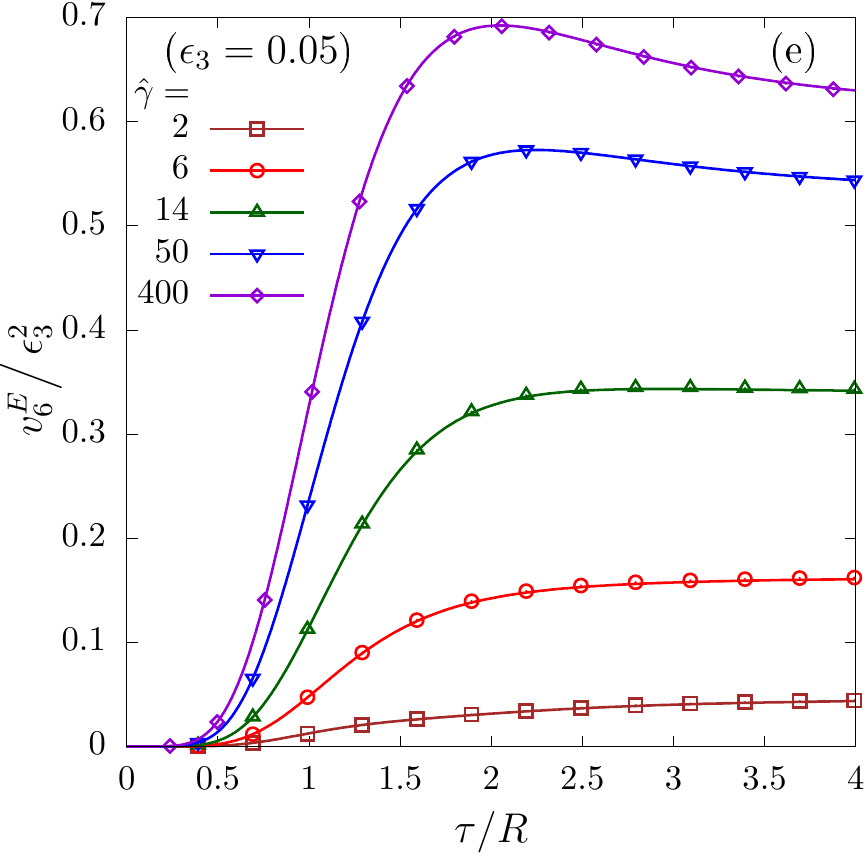} &
   \includegraphics[width=\lw]{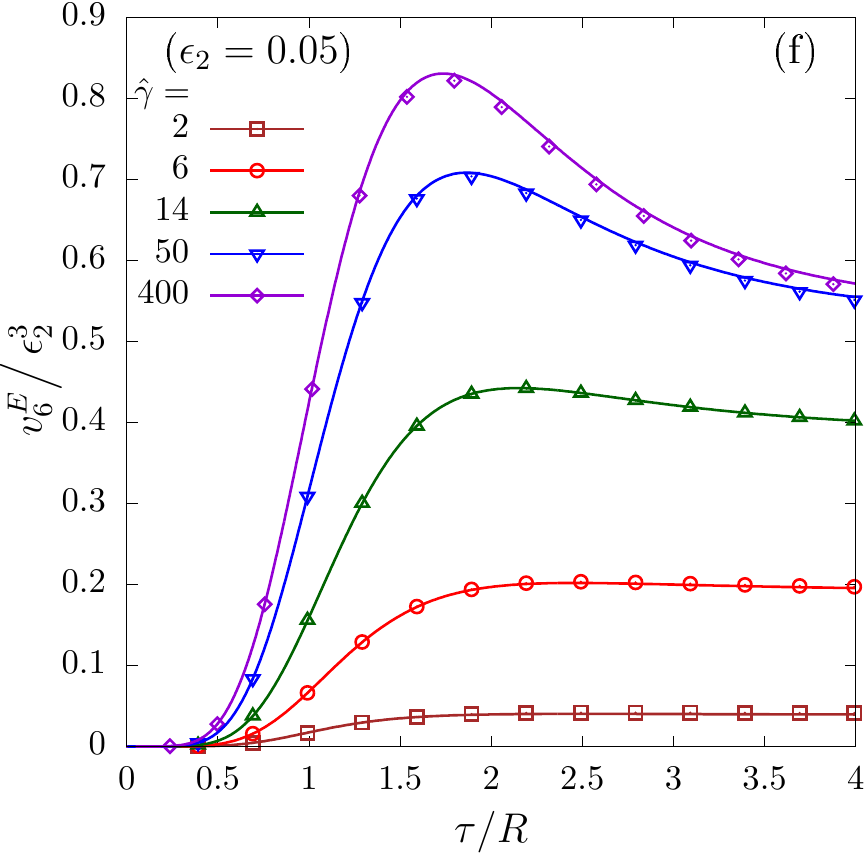} \\
\end{tabular}
\caption{(top line) Evolution at low opacities of the non-linear flow response coefficients
$v_4^E / \epsilon_2^2 {\hat \gamma}$ (left), $v_6^E / \epsilon_3^2 {\hat \gamma}$ (middle), and $v_6^E / \epsilon_2^3 {\hat \gamma}$ (right). (bottom line) Evolution at large opacities of $v_4^E / \epsilon_2^2$, $v_6^E / \epsilon_3^2$ (middle), and $v_6^E / \epsilon_2^3$ (right). Colored solid lines were obtained in the RLB method, while open symbols denote results from the moments method. All results were obtained for an initial eccentricity
$\epsilon_n = 0.05$.}
\label{fig:vn_vs_tau_e0p05_nl}
\end{figure*}

Next, we will analyze the development of anisotropic flow in terms of the time dependence of
the harmonic transverse flow coefficients $v_n^{E}$ for different opacities. We recall, that the initial anisotropies are modeled using a single harmonic ($n$) perturbation and first look at the time dependence of $v_{2}^{E},v_{3}^{E}$ and $v_{4}^{E}$ for different opacities, where in each case the eccentricities are fixed to $\epsilon_n=0.05$, which serves as a good approximation to the small eccentricity limit. We measure the magnitude 
of the linear response ratio $v_n /\epsilon_n$ for each harmonic; in addition we also extract the non-linear 
response of the fourth and sixth order harmonics via the ratios
$v_4 / \epsilon_2^2$, $v_6 / \epsilon_3^2$ and $v_6 / \epsilon_2^3$.

Since the qualitative behaviour of $v_n$
is somewhat different in the regimes of small and large opacities $\hat{\gamma}$, we again divide our results into two categories corresponding to $\hat{\gamma} \ge 2$
and $\hat{\gamma} \le 2$, as in
Fig.~\ref{fig:dEdy}. Since the analytical calculation in
Sec.~\ref{sec:linear} indicates that at small $\hat{\gamma}$, all 
response coefficients increase linearly with $\hat{\gamma}$, 
we will further normalize our low opacity results $(\hat{\gamma} \le 2)$ by division with 
respect to $\hat{\gamma}$.

Our results are compactly summarized in Figs.~\ref{fig:vn_vs_tau_e0p05} and ~\ref{fig:vn_vs_tau_e0p05_nl}, where we present numerical results for the linear 
($v_2/ \epsilon_2,v_3/ \epsilon_3,v_4/ \epsilon_4$) and non-linear 
($v_4/ \epsilon_2^2,v_6/ \epsilon_3^2,v_6/ \epsilon_2^3$) response coefficients obtained
for small (top panels) and large (bottom panels)
values of $\hat{\gamma}$.
We find that for small values of $\hat{\gamma} \lesssim 0.04$, the leading order
linear dependence of $v_n/\epsilon_n$ on $\hat{\gamma}$ computed 
in Eq.~(\ref{eq:v2analyticalasymptote}-\ref{eq:v4analyticalasymptote}) is nicely confirmed by the asymptotic approach of
our numerical results to the analytical results, represented by 
a solid black line. Similarly, a linear dependence with respect to
$\hat{\gamma}$ is also found for the non-linear response coefficients
$v_4/ \epsilon_2^2,v_6/ \epsilon_3^2,v_6/ \epsilon_2^3$, which for $v_4 / \epsilon_2^2$ is in line with the result obtained in 
Ref.~\cite{Kurkela:2018ygx} for a slightly different initial setup.
While for $\hat{\gamma} \gtrsim 0.25$, all linear flow coefficients
exhibit a positive response with respect to the initial
eccentricities, the quadrangular flow $v_4 /\epsilon_4 \hat{\gamma}$ in Fig.~\ref{fig:vn_vs_tau_e0p05}
shows a negative response for $\hat{\gamma} \lesssim 0.25$.

Based on a closer inspection, one finds that the curves of $v_4/(\epsilon_4 \hat{\gamma})$ exhibit
an early time increase similar to the behavior seen for the other harmonic flow coefficients, however in contrast to $v_{2}$, $v_{3}$ the initial rise of $v_{4}$ peaks around $\tau \simeq 1.5 R$, followed by a decrease due to negative contributions received at large times. When increasing 
the opacity, non-linear effects
cause the elliptic flow response $v_2/(\epsilon_2 \hat{\gamma})$ to decrease,
while $v_3 / (\epsilon_3 \hat{\gamma})$, 
$v_4 / (\epsilon_4 \hat{\gamma})$ as well as the non-linear $v_4 / (\epsilon_2^2 \hat{\gamma})$ and $v_2 / (\epsilon_2^3 \hat{\gamma})$ exhibit an increasing trend; due to the rather complicated time dependence, the behavior of $v_6 / (\epsilon_3^2 \hat{\gamma})$ appears non-monotonic. Clearly, the largest effect is seen in the case of the $v_4/\epsilon_{4}$-response which changes sign as the late time contributions become less and less prominent. 

When considering large opacities $\hat{\gamma} \gtrsim 2$ shown in
the bottom panels of Figs.~\ref{fig:vn_vs_tau_e0p05} and ~\ref{fig:vn_vs_tau_e0p05_nl}, 
the curves for linear 
($v_2/ \epsilon_2,v_3/ \epsilon_3,v_4/ \epsilon_4$) and non-linear 
($v_4/ \epsilon_2^2,v_6/ \epsilon_3^2,v_6/ \epsilon_2^3$) response coefficients retain the same qualitative time dependence and monotonically increase as a function of $\hat{\gamma}$, seemingly approaching a finite large opacity limit, which we will further examine in the following. Generally, we find that the linear anisotropic flow response develops pre-dominantly in the regime 
$0.5 \lesssim \tau/R \lesssim 2$
and then stay almost constant, with the exception of the aforementioned late time decrease of linear $v_4$. In the case of the non-linear coefficients $v_4/ \epsilon_2^2,v_6/ \epsilon_3^2$ and $v_6/ \epsilon_2^3$
the response takes a little longer to  develop,
but nevertheless the asymptotic late time value is reached on similar timescales $0.5 \lesssim \tau/R \lesssim 4$.

\begin{figure*}
\centering
\begin{tabular}{cccc}
\includegraphics[width=.278\linewidth]{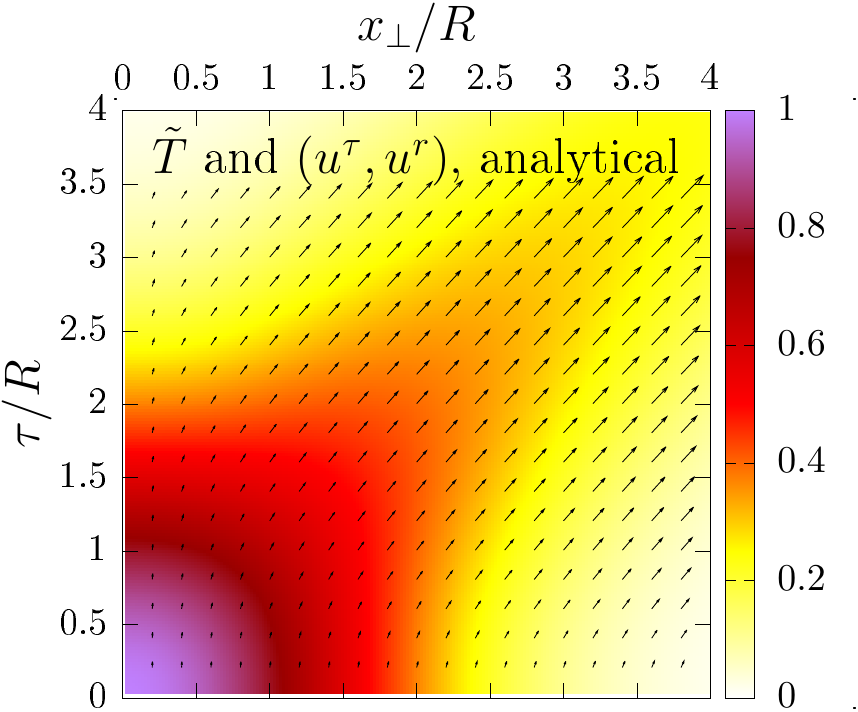} &
 \includegraphics[width=.24\linewidth]{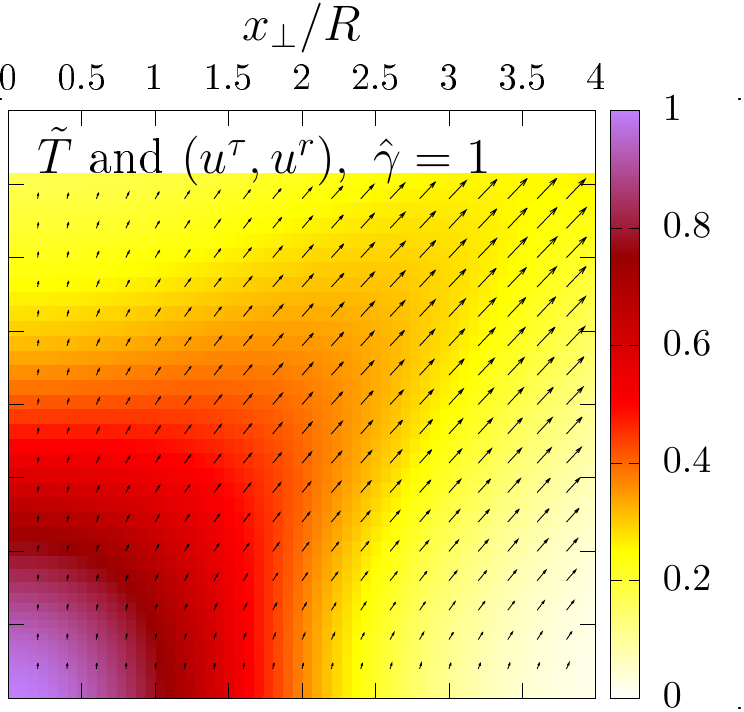} &
 \includegraphics[width=.24\linewidth]{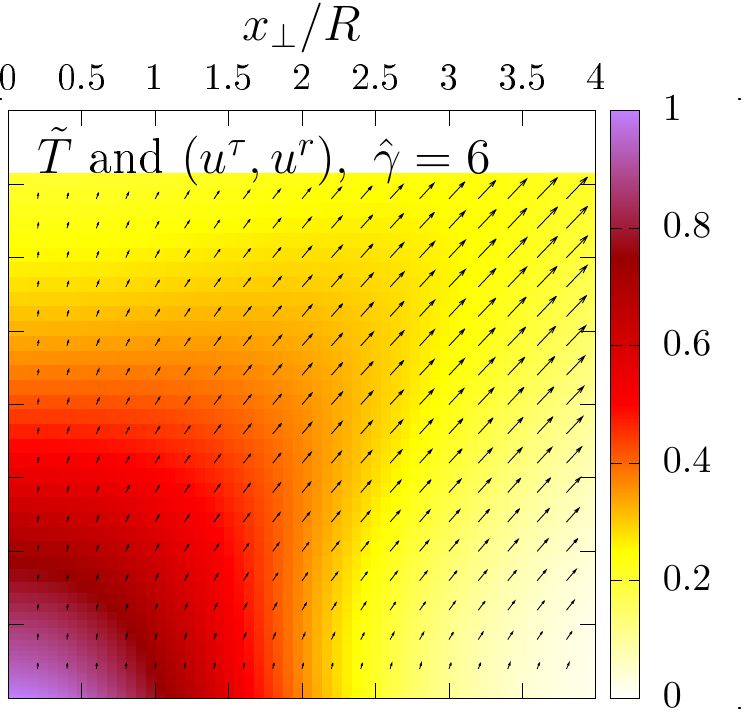} &
 \includegraphics[width=.24\linewidth]{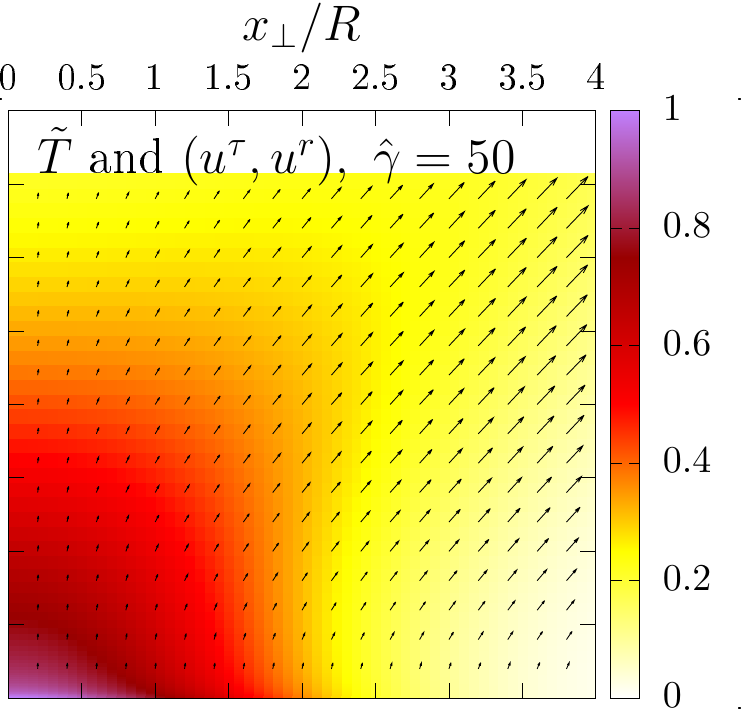}\\
 \includegraphics[width=.278\linewidth]{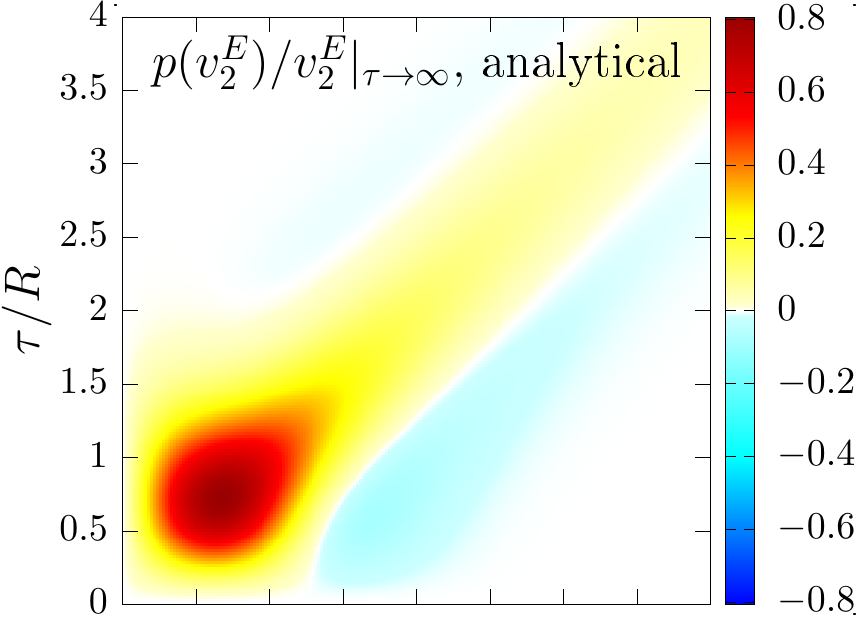} &
 \includegraphics[width=.24\linewidth]{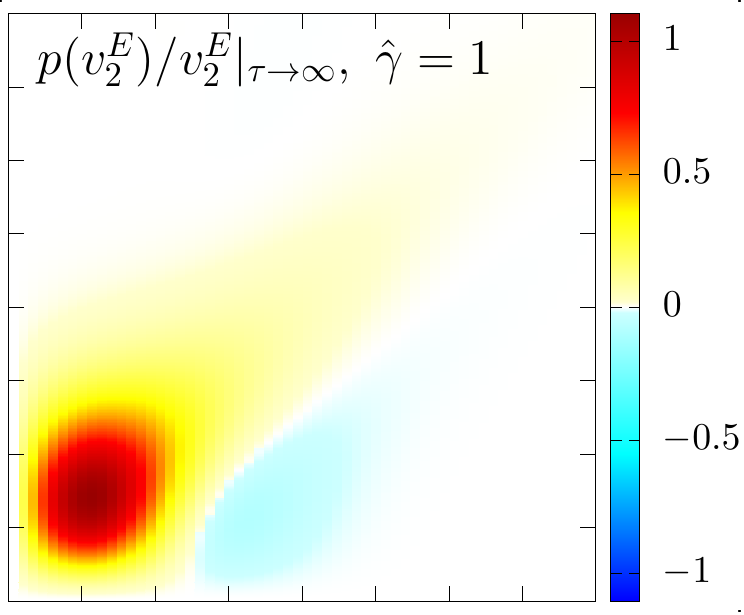} &
 \includegraphics[width=.24\linewidth]{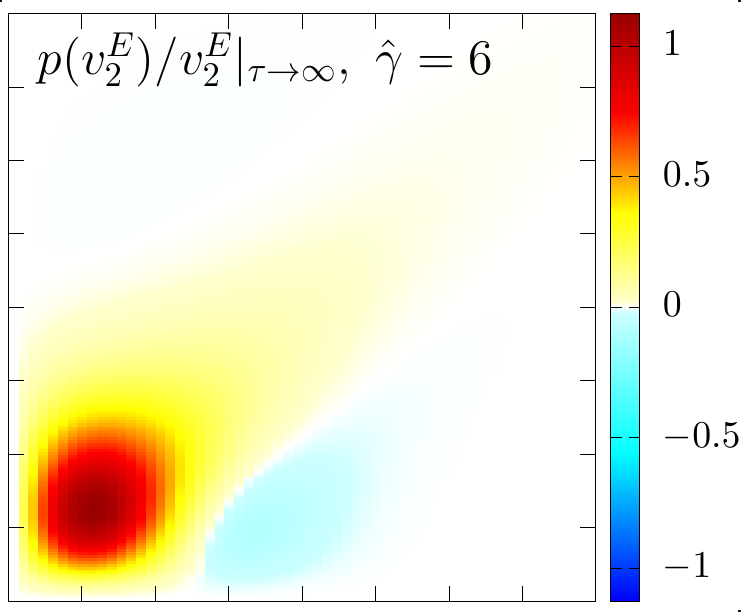} &
 \includegraphics[width=.24\linewidth]{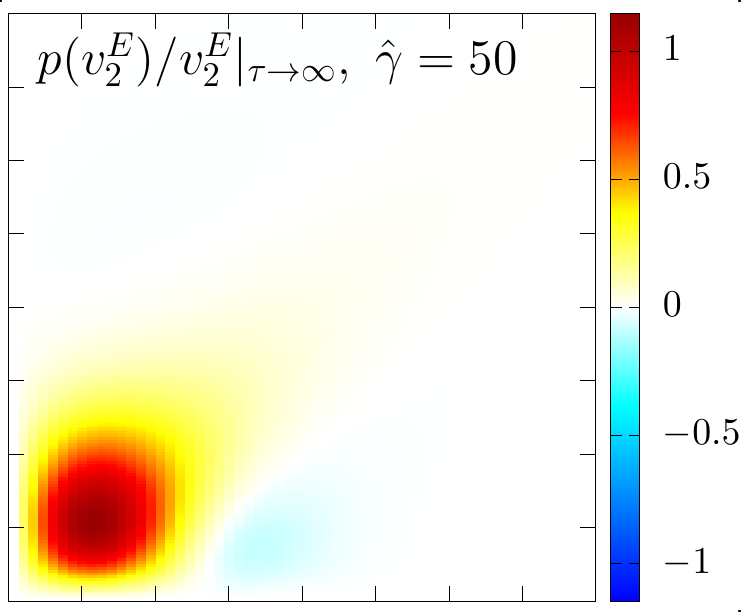} \\
 \includegraphics[width=.278\linewidth]{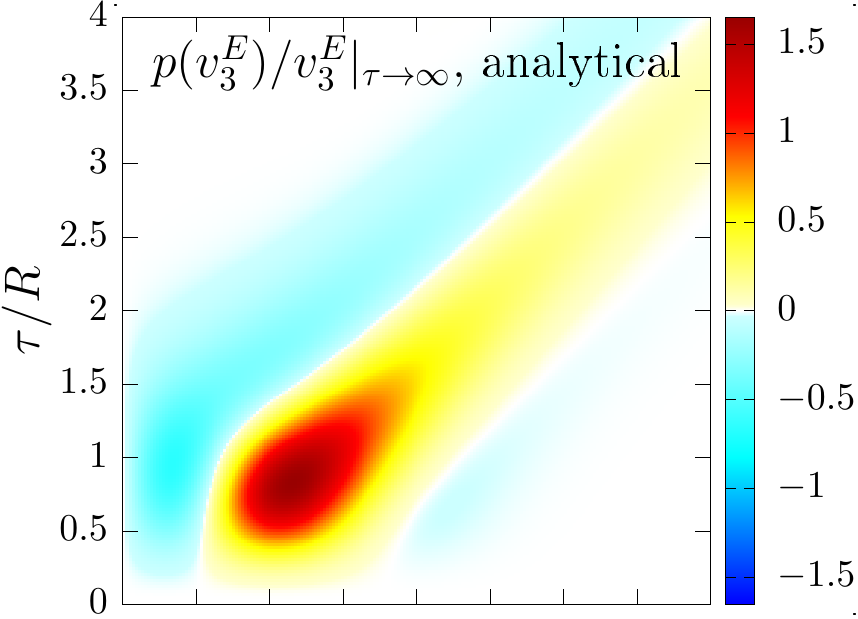} &
 \includegraphics[width=.24\linewidth]{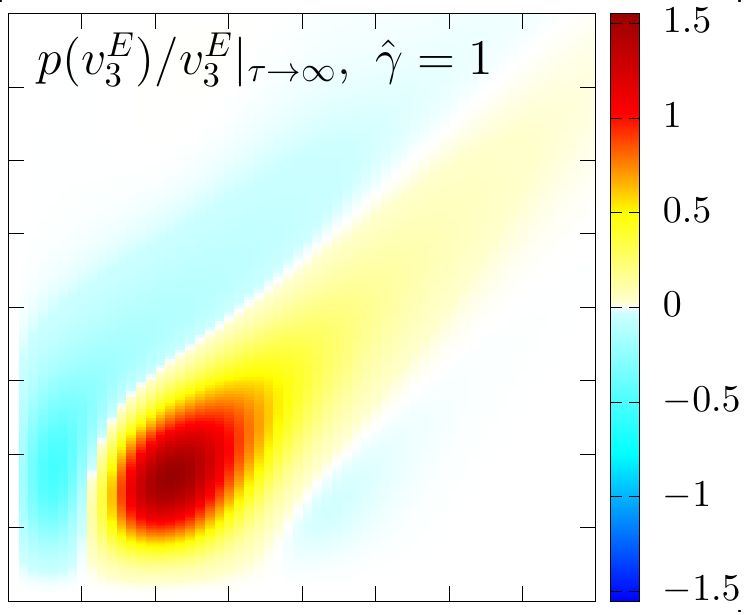} &
 \includegraphics[width=.24\linewidth]{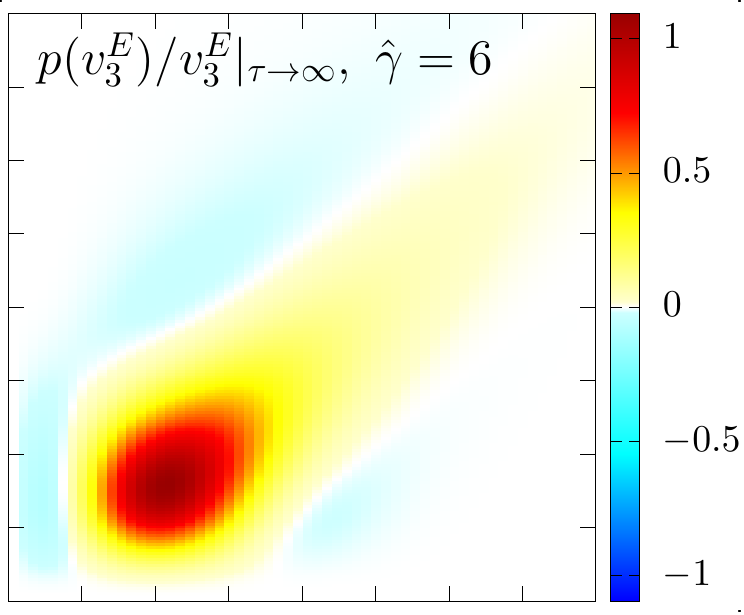} &
 \includegraphics[width=.24\linewidth]{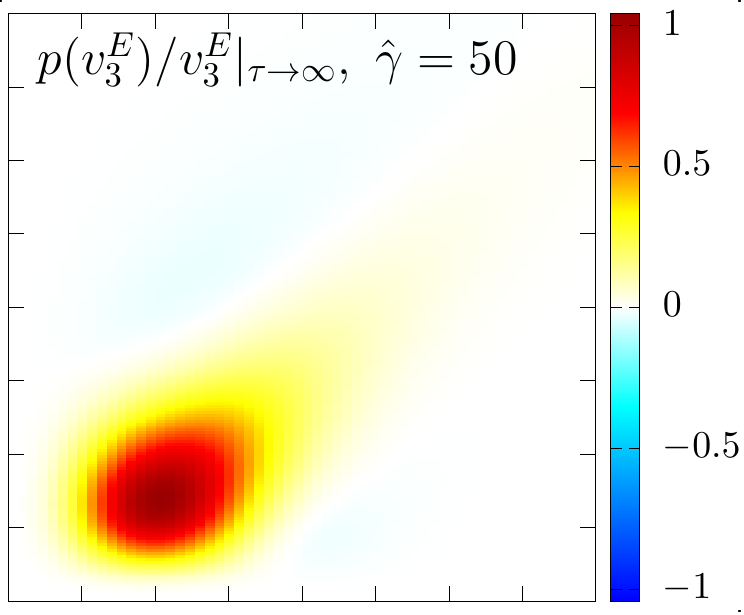} \\
 \includegraphics[width=.278\linewidth]{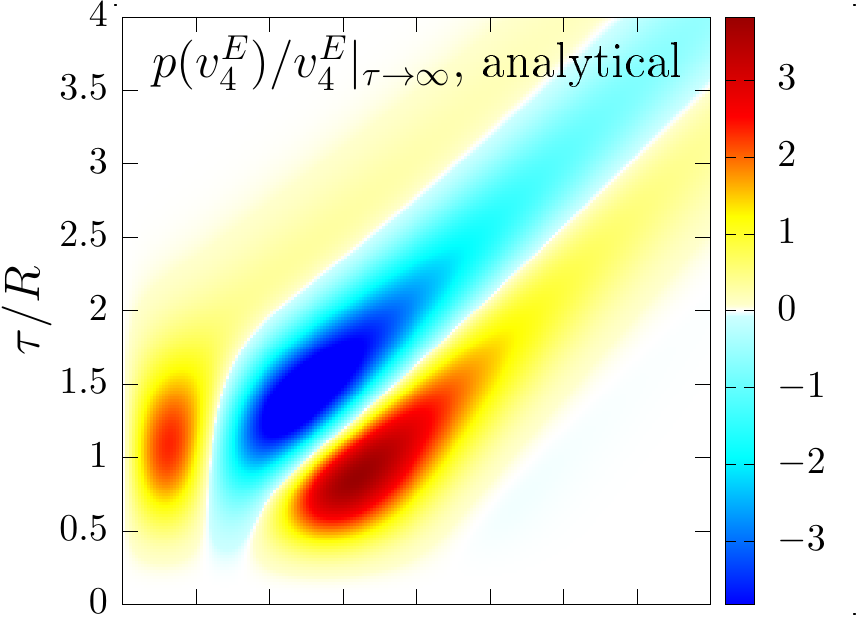} &
 \includegraphics[width=.24\linewidth]{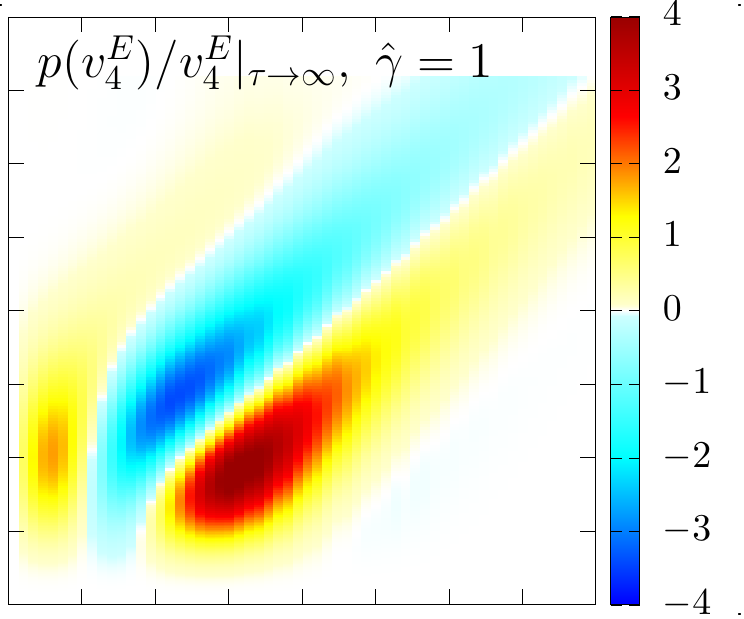} &
 \includegraphics[width=.24\linewidth]{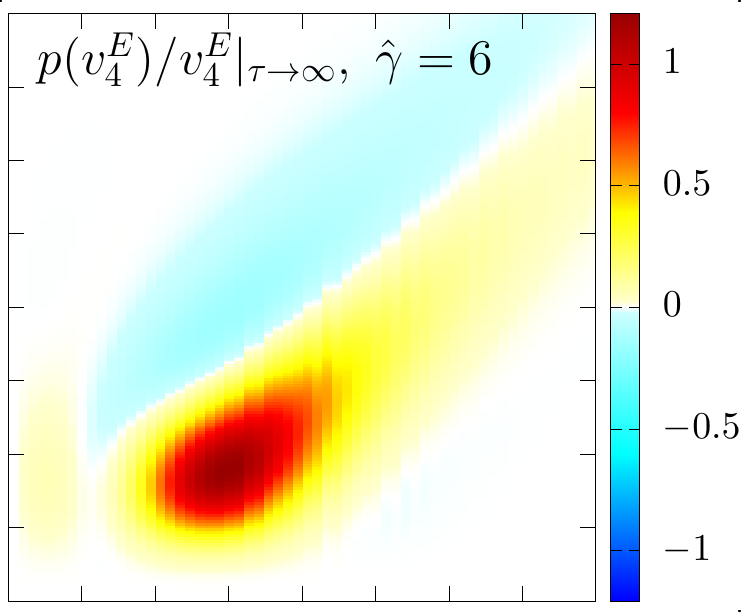} &
 \includegraphics[width=.24\linewidth]{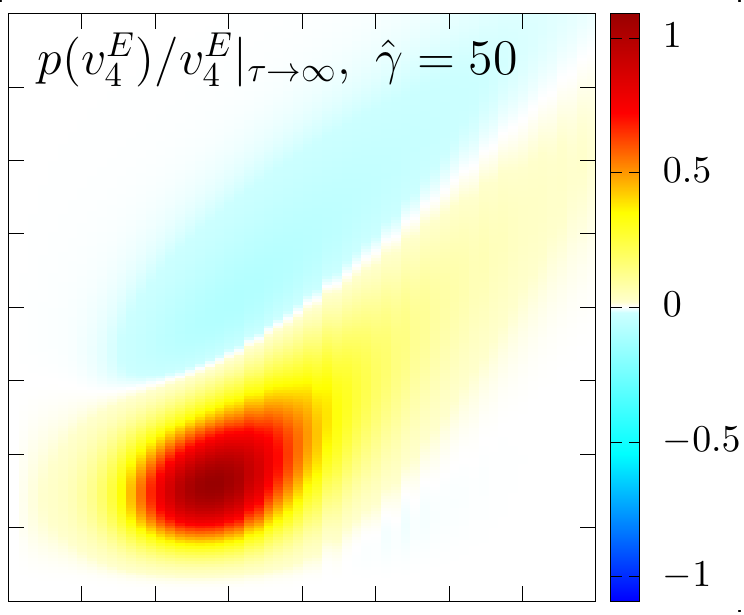}
 \\
 \multicolumn{2}{r}{\includegraphics[width=.278\linewidth]{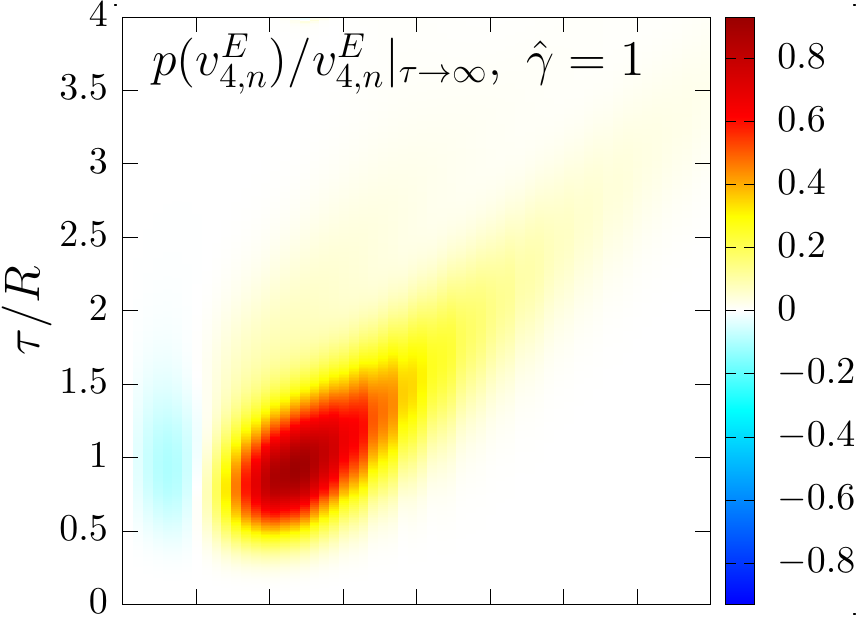}}
  &
 \includegraphics[width=.24\linewidth]{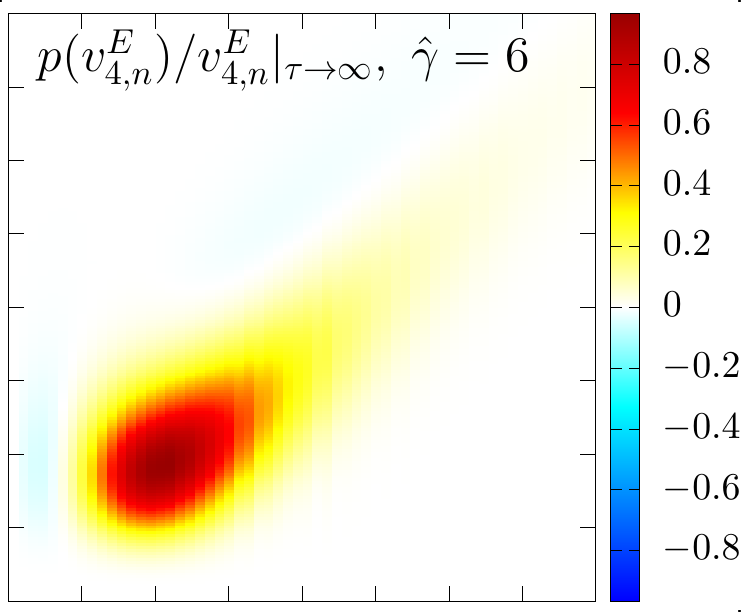} &
 \includegraphics[width=.24\linewidth]{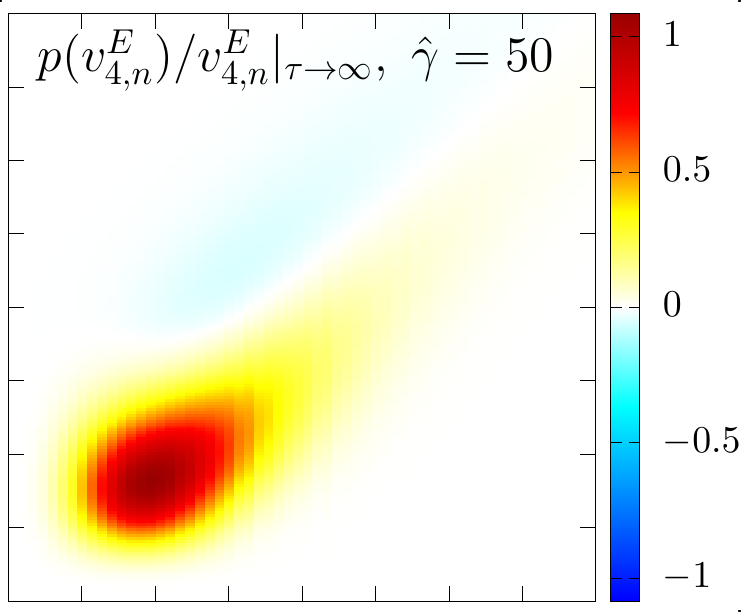}
 
\end{tabular}
    \caption{(top row) Space-time profiles of  the effective temperature $\tilde{T}$ along with the temporal and radial components of the vector field $u^\mu$, presented in the $x_\perp$-$\tau$-plane for $\epsilon_n=0.05$. (bottom rows)  Space-time profiles of the production rates of linear $v_2^E,v_3^E$ and $v_4^E$ response (second to fourth row) as well as nonlinear $v_4^E$ response (fifth row) . Numerical results were obtained in the moment method.}
    \label{fig:hmap}
\end{figure*}

Beyond the time evolution of the different flow harmonics, additional insights into the development of anisotropic flow can be gained
from their production rates $p(v_{n})$, which correspond the local rate of change of these quantities. Since free-streaming and longitudinal expansion do not change the (transverse) momentum distribution of particles, the build up of anisotropic flow is solely due to interactions. We can thus determine the production rate $p(O)$ of a flow observable $O=\int d^2 x_\perp \int \frac{\d^3 p}{(2\pi)^3}\,
\mathcal{O} \,f$ as
\begin{align}
    p(O)=\left. \frac{\d O}{\d x_\perp \d \tau } \right|_{\rm coll}
    =\int \d \phi_{x_\perp}\, x_\perp \int \frac{\d^3p}{(2\pi)^3} \, \mathcal{O} \,  \left.\frac{\d}{\d \tau} f \right|_{\rm coll} \ \ ,
\end{align}
where the rate of change of the phase-space distribution $f$ due to collisions is given by
\begin{align}
    \left.\frac{\d}{\d \tau} f \right|_{\rm coll} = \frac{v^\mu u_\mu}{\tau_R}(f_{eq}-f)\;.
\end{align}
Specifically, the observables $v_n^{E}$ are defined according to Eq.~(\ref{eq:vn_def}) as quotients of two such terms, such that the production rate $p(O)$ receives two contributions coming from the numerator and denominator according to the quotient rule for differentiation.

Fig.~\ref{fig:hmap} features heat maps in the $x_\perp$-$\tau$-plane for $p(v_2)$, $p(v_3)$ and $p(v_4)$ as well as $p(v_{4,n})$ referring to the nonlinear response, normalized by the respective late time asymptotic values of $v_n$ for several different opacities ranging from the analytical results for small opactites $\hat{\gamma} \ll1$ all the way to $\hat{\gamma}=50$. Besides the production rates of different $v_{n}$, the top panel of Fig.~\ref{fig:hmap}, also shows a heat-map of the dimensionless temperature $\tilde{T}$ and the flow components $u^\tau$ and $u^r$ to allow for a comparison with the spatial distribution and expansion of the system. The latter showcase how with increasing opacity, the system cools more rapidly in the center and the transverse expansion proceeds much slower, resulting in a longer lifetime of the central fireball. Strong correlations of the temperature profile in $\tau$ and $x_\perp$ only develop at much later times when compared to the free streaming limit, which exhibits a prominent diagonal line in the $\tilde{T}$-heatmap. Inspection of the $p(v_n)$-heatmaps reveals that different regions in the $x_\perp$-$\tau$-plane contribute  with different signs to the development of anisotropic flow $v_{n}$. By comparing the results for $v_2,v_3$ and $v_4$, one also observes that for larger $n$ the relevant regions extend more towards larger $x_\perp$, while at the same time more of these regions appear, causing large cancellations between the different contributions. Specifically for small opacitites, the structure of the heatmaps of the $v_{n}$ production rates can be related to the weight
\begin{align}
     &|\xT-\vT \Delta \tau| ^{n} \cos(n\phi_{\xT-\vT \Delta \tau,\nT}) = \nonumber \\ 
     &\sum_{j=0}^{n} (-1)^{j} \binom{n}{j}  x_{\bot}^{n} \left(\frac{\Delta \tau}{x_{T}}\right)^{j} \Big[\cos(n \phi_{\xT\nT}) \cos(j \phi_{\xT\pT})  \nonumber \\
     & \qquad - \sin(n \phi_{\xT\nT}) \sin(j\phi_{\xT\pT})\Big]
\end{align}
with which the anisotropic perturbations of the initial phase-space distribution propagate in free streaming. Since the evolution of the perturbation is expressed as a sum of $n+1$ terms containing different powers of $\Delta\tau/x_\perp$ with alternating signs, it will divide the $x_\perp$-$\tau$-plane into $n+1$ regions of alternating signs depending on which one of these terms dominates. In addition, the production of the anisotropic flow $v_{n}$s will be weighted with the local effective temperature $\tilde{T}$ of the system, such that for small opacities most of the contributions originate from the $\tau \sim x_\perp$ diagonal, so only $\Delta\tau/x_\perp$-terms that dominate close to that region will have a significant impact on the total $v_n(\tau)$. Specifically, for $n=2$, there is only one dominant term, which explains the monotonic increase of $v_{2}$ as a function of time seen in Fig.~\ref{fig:vn_vs_tau_e0p05}. Conversely, for $n=3$, one positive and one negative contribution are competing, with the positive one being slightly larger than the negative one, which is why for small opacities $v_3/e_{3}$ is significantly smaller than $v_2/e_{2}$ and features a slight negative trend at late times. Finally, in the case of $n=4$, there are three relevant terms. At early times, the two positive contribution from the inner and outer border of the system win and $v_4$ increases, but the one negative contribution surrounded by them in the $x_\perp$-$\tau$-plane is closest to the diagonal and dominates at late times, resulting in a sign change for $v_4$ observed for the smallest opacities in Fig.~\ref{fig:vn_vs_tau_e0p05}.

With increasing opacity one observes a clear change in the shapes of the regions, resulting in a shift of $v_{n}$ production towards earlier $\tau$ and smaller $x_\perp$ in Fig.~\ref{fig:hmap}. However, more strikingly the increase of opacity also leads to a change of the relative weights of different regions, developing towards a scenario with only one dominant positive contribution for all the $v_{n}$s at large opacity.

We finally note that the weighting with the effective temperature $\tilde{T}$ plays an important role in this mechanism, such that a different initial condition could result in different relative weights of the regions with different sign of the production rates, which can have notable effects on the buildup of the different flow harmonics. Clearly, one should expect that the higher order flow harmonics, where more cancellations appear are more sensitive to changes of the initial conditions, and indeed we find  that varying the parameter $\alpha$ that controls the radial profile (c.f. Sec.~\ref{sec:theory:setup}) will have a notable influence on the $v_3(\tau)$ and $v_4(\tau)$-curves at small opacities.

\begin{figure}[t!]
\centering 
 \includegraphics[width=0.8\linewidth]{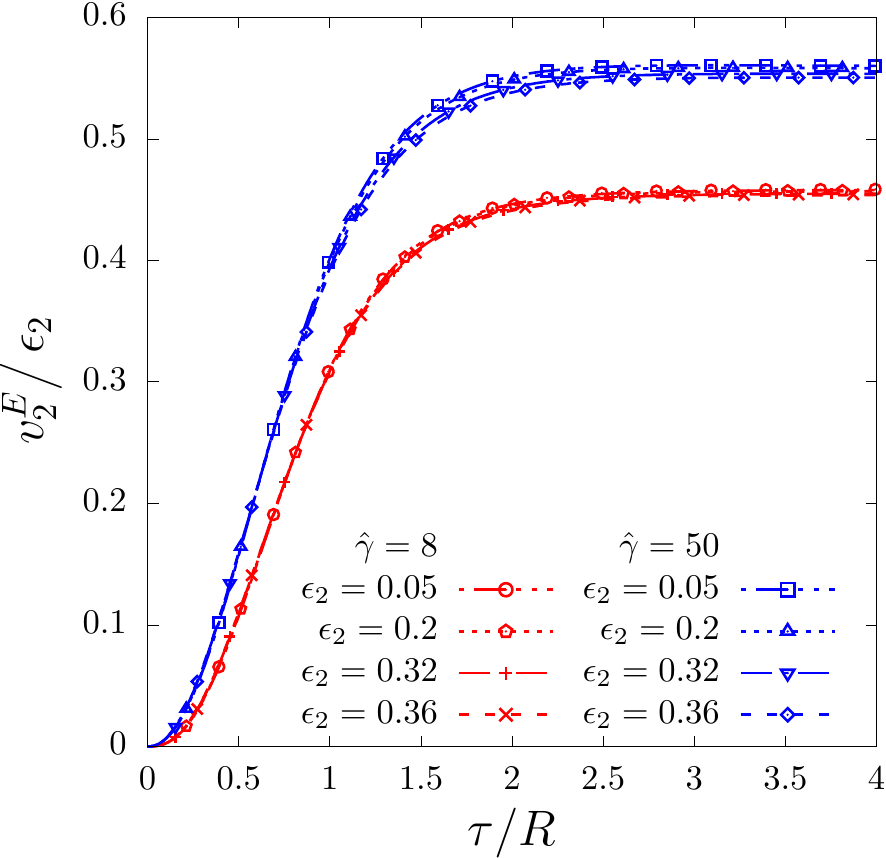} 
 \includegraphics[width=0.8\linewidth]{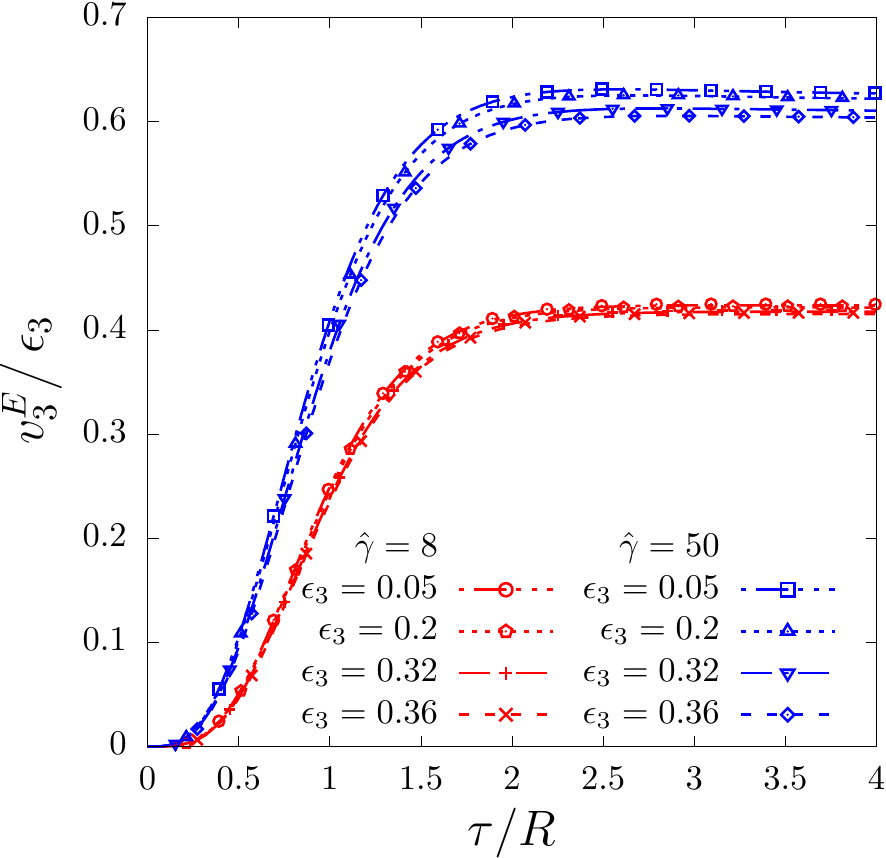}
 \includegraphics[width=0.8\linewidth]{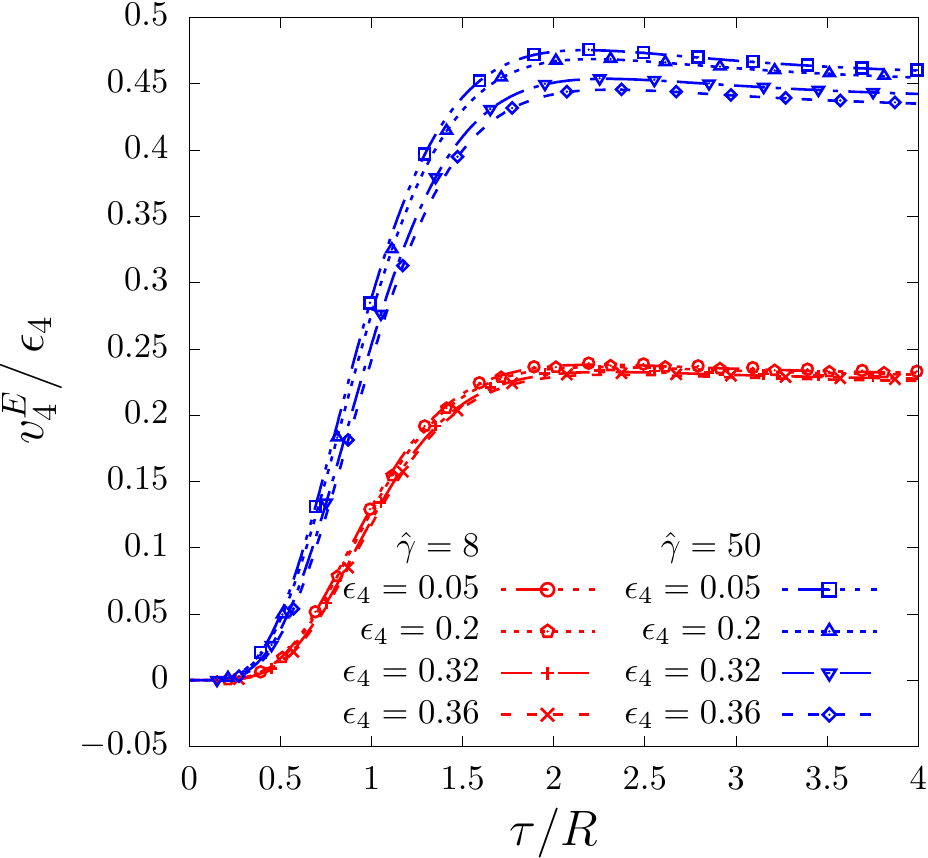} 
\caption{Linear response coefficients $v_n^E / \epsilon_n$ as a function of $\tau/R$ for different opacities ${\hat \gamma =8}$ and ${\hat \gamma =50}$ and various different eccentricities $\epsilon_{n}$. Lines denote results from the RLB method and symbols show results from the moment method.} 
\label{fig:vn_vs_tau_g50p0}
\end{figure}

\begin{figure}[t!]
\centering 
 \includegraphics[width=0.8\linewidth]{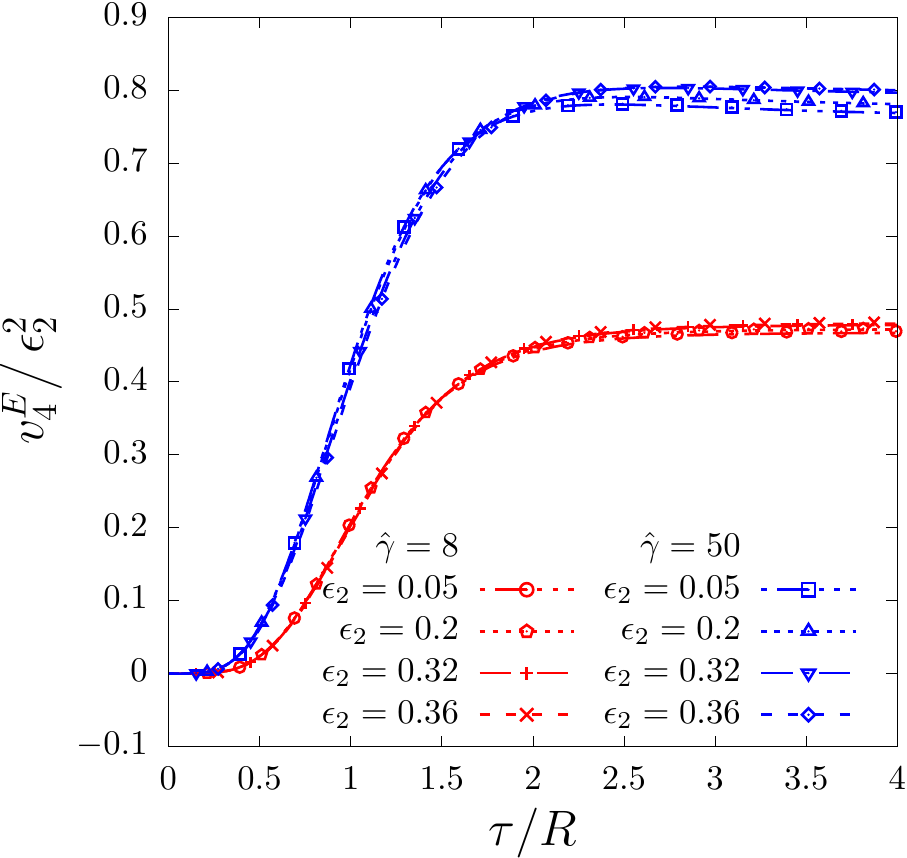}\\
 \includegraphics[width=0.8\linewidth]{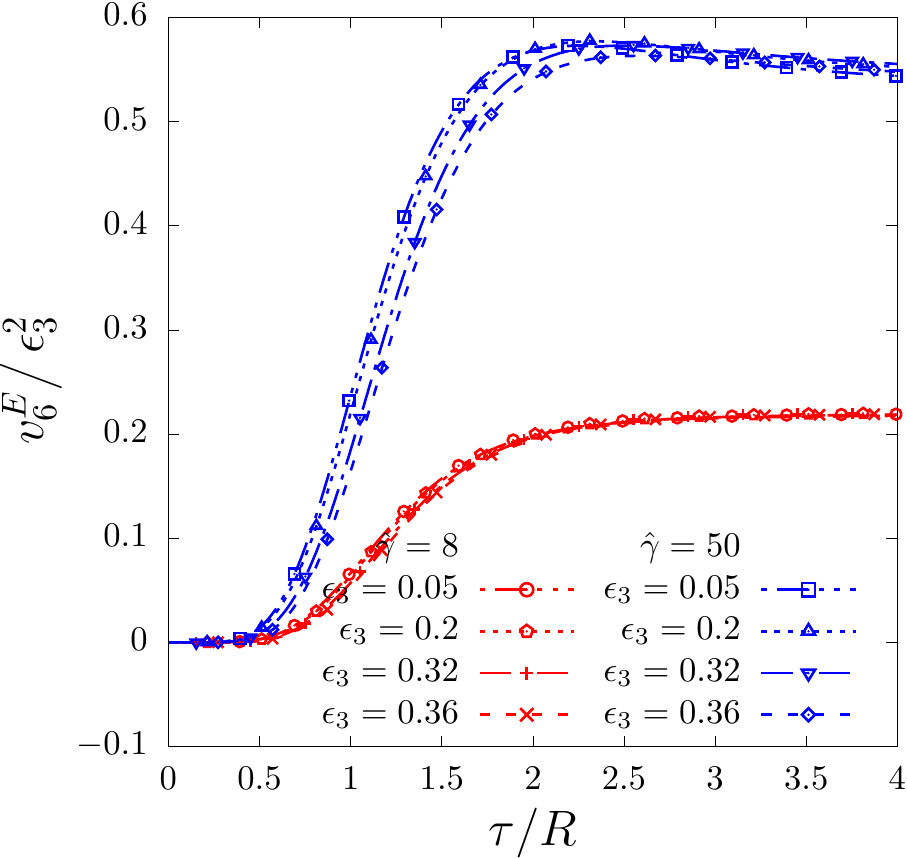}\\
 \includegraphics[width=0.8\linewidth]{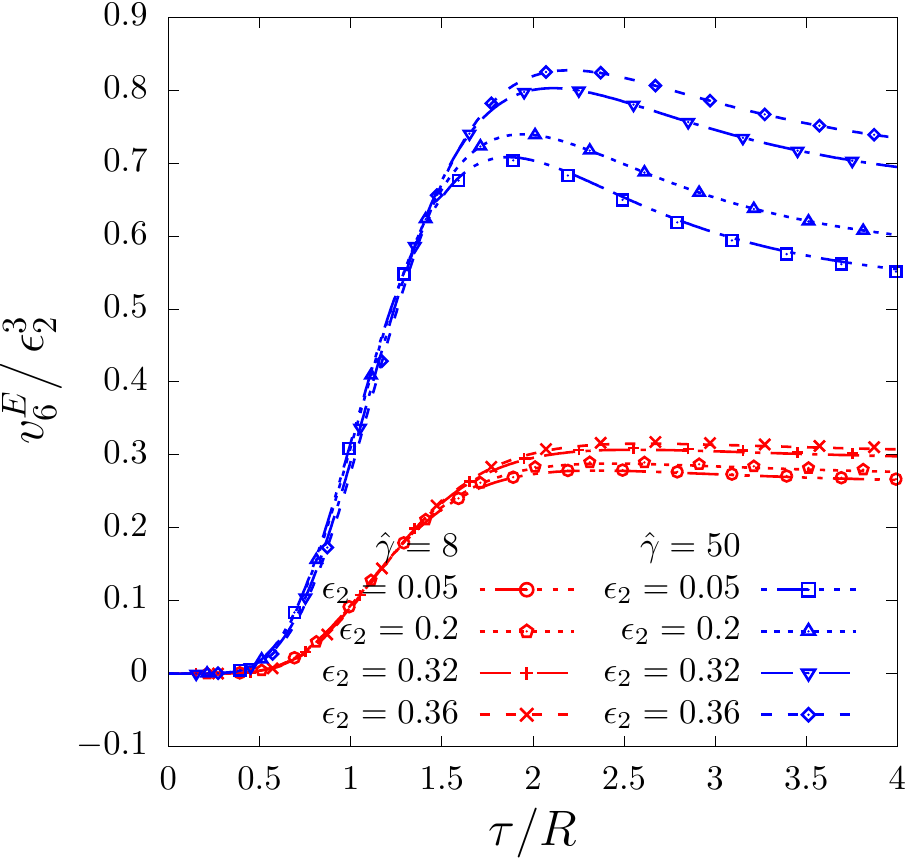}
\caption{Non-linear response coefficients $v_4^E / \epsilon_2^2$ (top),  $v_6^E / \epsilon_3^2$ (middle) and $v_6^E / \epsilon_2^3$ (bottom) as a function of $\tau/R$ for different opacities ${\hat \gamma =8}$ and ${\hat \gamma =50}$ and various different eccentricities $\epsilon_{n}$. Lines denote results from the RLB method and symbols show results from the moment method.} 
\label{fig:vn_vs_tau_g50p0_nl}
\end{figure}

Beyond the opacity dependence, one may also examine how the development of anisotropic flow $v_n(\tau)$ changes with the amplitude $\epsilon_{n}$ of the respective initial eccentricity. Fig.~\ref{fig:vn_vs_tau_g50p0} and~\ref{fig:vn_vs_tau_g50p0_nl} showcase how the curves of normalized flow spread with eccentricity for two representative fixed values of $\hat{\gamma}$. Somewhat surprisingly, we find that the curves exhibit only very small deviations from an entirely linear (quadratic) dependence on eccentricity in the linear $v_2$, $v_3$ and $v_4$ (quadratic $v_4$ and $v_6$) flow response, even for rather large eccentricities. The only response featuring a significant dependence on eccentricity is the cubic $v_6$ response to $\epsilon_2$. While this holds true not only for the final values but also for the entire build up and evolution as a function of $\tau/R$, we remark however, that these findings are probably specific to the particularly simple geometry considered in our setup, and it will therefore be important to extend such systematic studies of the opacity dependence of the flow response towards more realistic profiles of the transverse geometry.

\begin{figure}[!t]
    \centering
     \includegraphics[width=.8\linewidth]{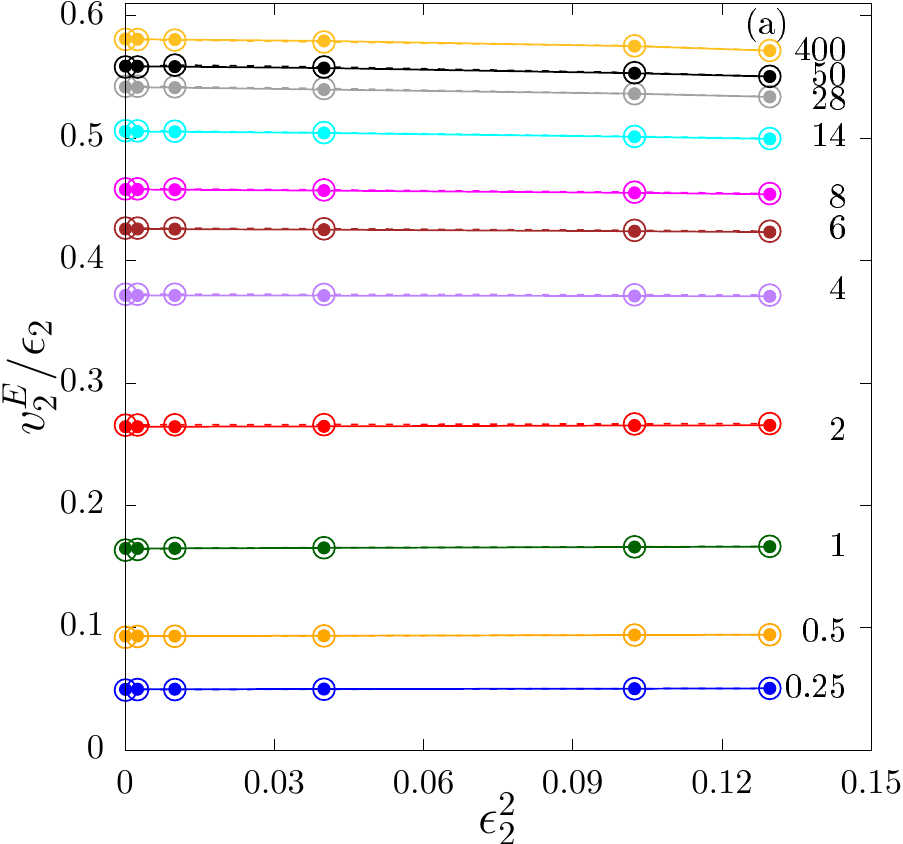} \\
     \includegraphics[width=.8\linewidth]{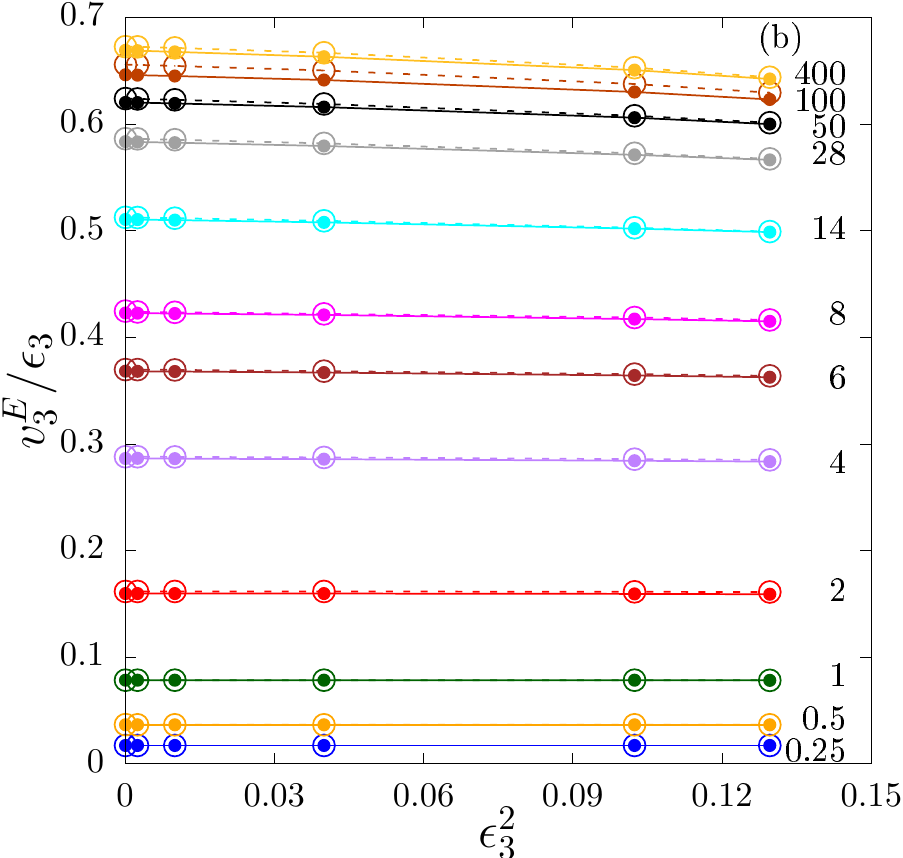} \\
     \includegraphics[width=.8\linewidth]{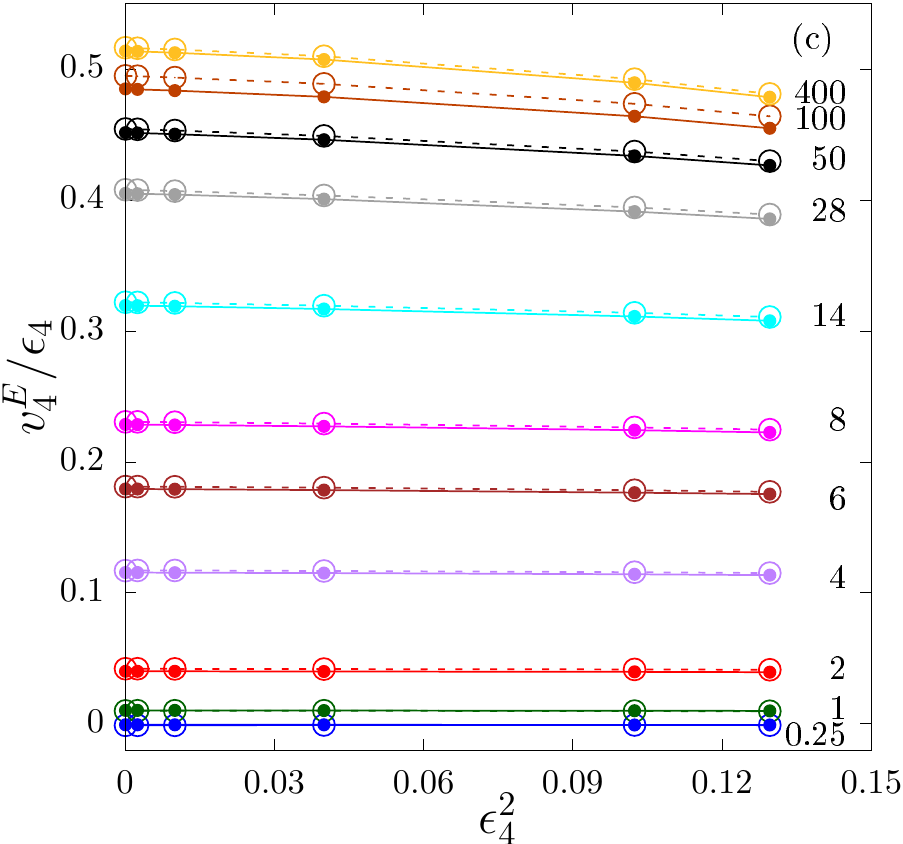} 
    \caption{Eccentricity $\epsilon_n$-dependence of the linear response coefficients
    $\kappa_{n,n}=v_n^E(\tau\rightarrow\infty)/\epsilon_n$ for elliptic flow $n=2$ (top), triangular flow $n=3$ (middle) and quadrangular flow $n=4$ (bottom). Solid lines with filled circles denote results from the RLB method, while dotted lines with open circles were obtained in the moment method.}
    \label{fig:vn_epsilon_dependence}
\end{figure}

\begin{figure}[!t]
    \centering
     \includegraphics[width=.8\linewidth]{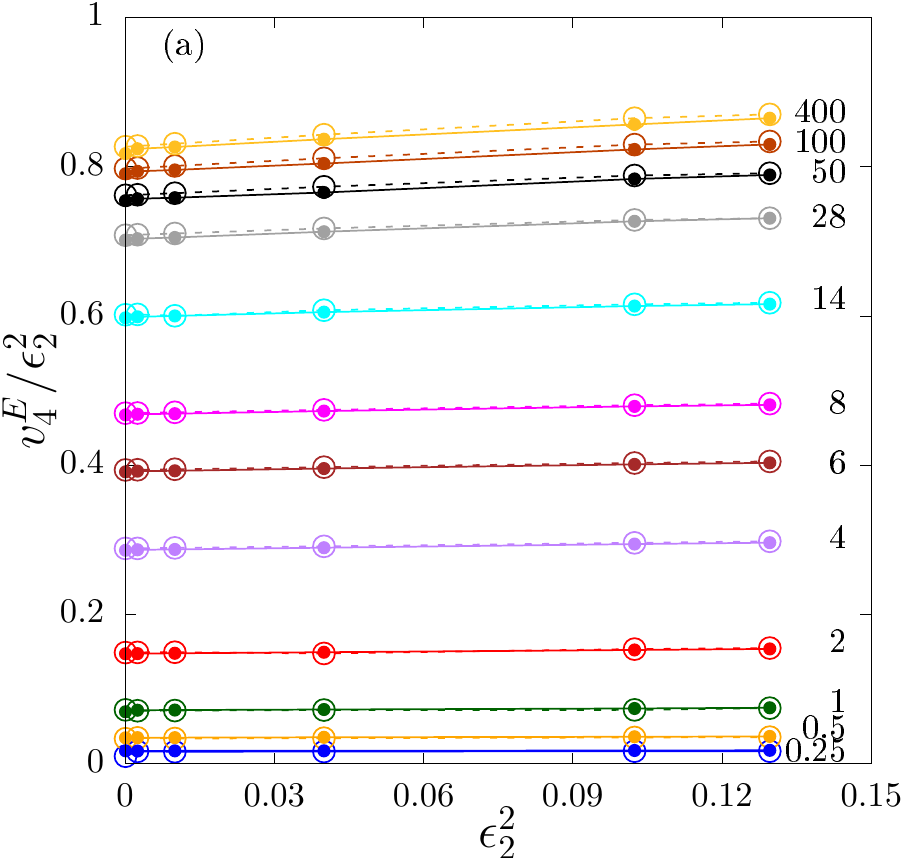}\\
     \includegraphics[width=.8\linewidth]{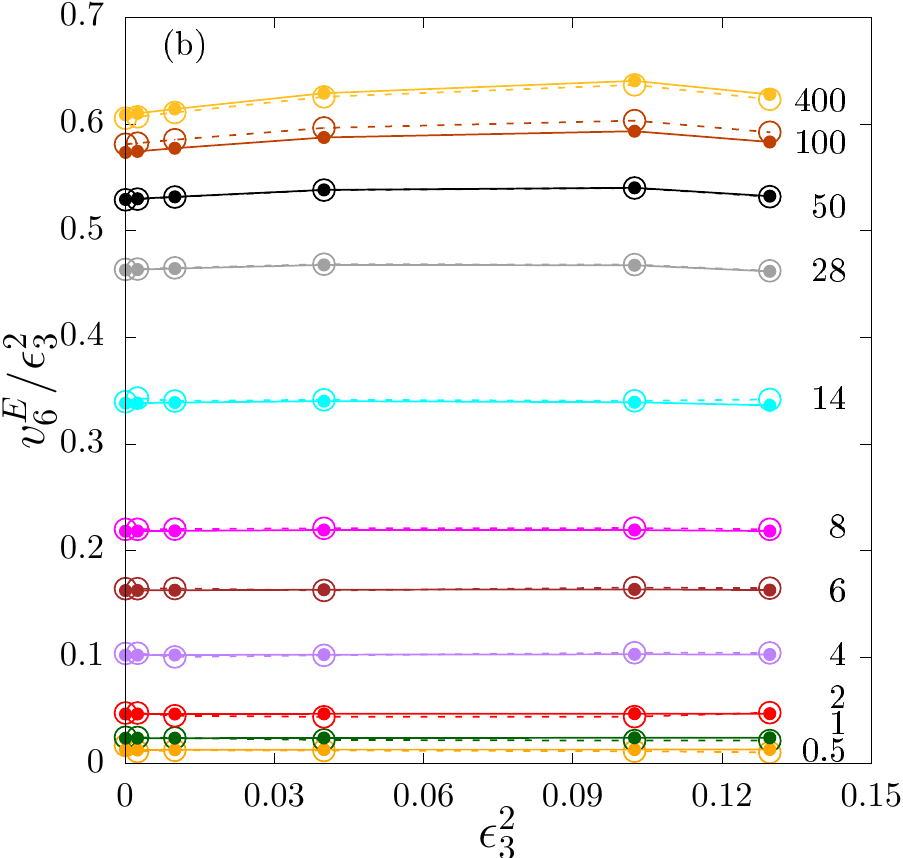}\\
     \includegraphics[width=.8\linewidth]{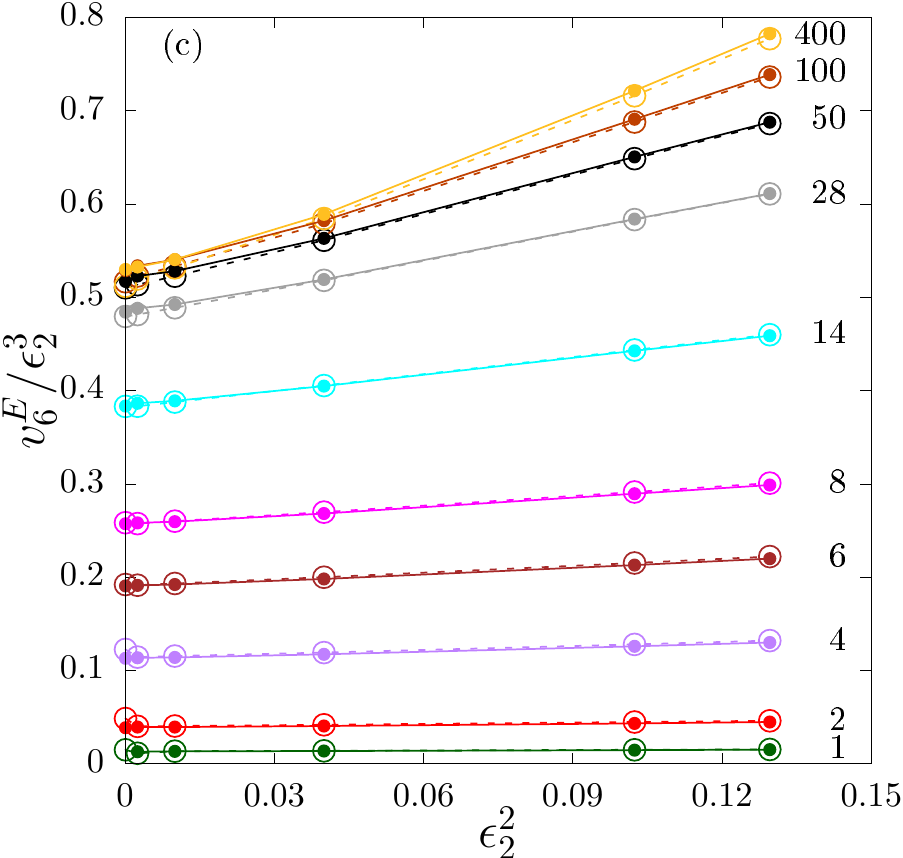}
    \caption{Eccentricity ($\epsilon_n$) dependence of the non-linear response coefficients $\kappa_{4,22} = v_4^E(\tau \rightarrow \infty) / \epsilon_2^2$ (top), $\kappa_{6,33} = v_6^E(\tau \rightarrow \infty) / \epsilon_3^2$ (middle) and $\kappa_{6,222} = v_6^E(\tau \rightarrow \infty) / \epsilon_2^3$ (bottom). Solid lines with filled circles denote results from the RLB method, while dotted lines with open circles were obtained in the moment method.}
    \label{fig:vn_epsilon_dependence_nl}
\end{figure}

Next, in order to further scrutinize the eccentricity dependence, we extract the extrapolated final values of $v_n/\epsilon_n$ resp. nonlinear $v_4/\epsilon_2^2$, $v_6/\epsilon_3^2$ and $v_6/\epsilon_2^3$ at late times and plot them as a function of the square of the relevant eccentricity for several different opacities. Our results shown in Fig.~\ref{fig:vn_epsilon_dependence} and~\ref{fig:vn_epsilon_dependence_nl} again confirm the surprisingly small deviations from perfect linear (quadratic) scaling of the flow response, with only very slight negative (positive) trends at large opacity and eccentricity. Our results in Fig.~\ref{fig:vn_epsilon_dependence} appear to be in conflict with results previously obtained by Kurkela et al.~\cite{Kurkela:2020wwb} in the same setup.
We note once again, that although the absence of significant non-linearity in the eccentricities may seem in conflict with conventional knowledge~(see e.g.~\cite{Noronha-Hostler:2015dbi,Niemi:2015qia,Roch:2020zdl}), we attribute this to the specific initial conditions considered within our setup, and we have explicitly checked that hydrodynamic simulations of the same initial conditions also lead to similar results for $v_{2}/\epsilon_2$. Vice versa, the absence non-linearities within our setup also indicates that the significant non-linearity observed for more realistic initial state models should be attributed to other features of the initial states considered in hydrodynamic simulations of heavy-ion collisions, which are not solely characterized in terms of the usual eccentricties.

\begin{figure*}[!ht]
\begin{center}
\begin{tabular}{cc}
    \includegraphics[width=.48\linewidth]{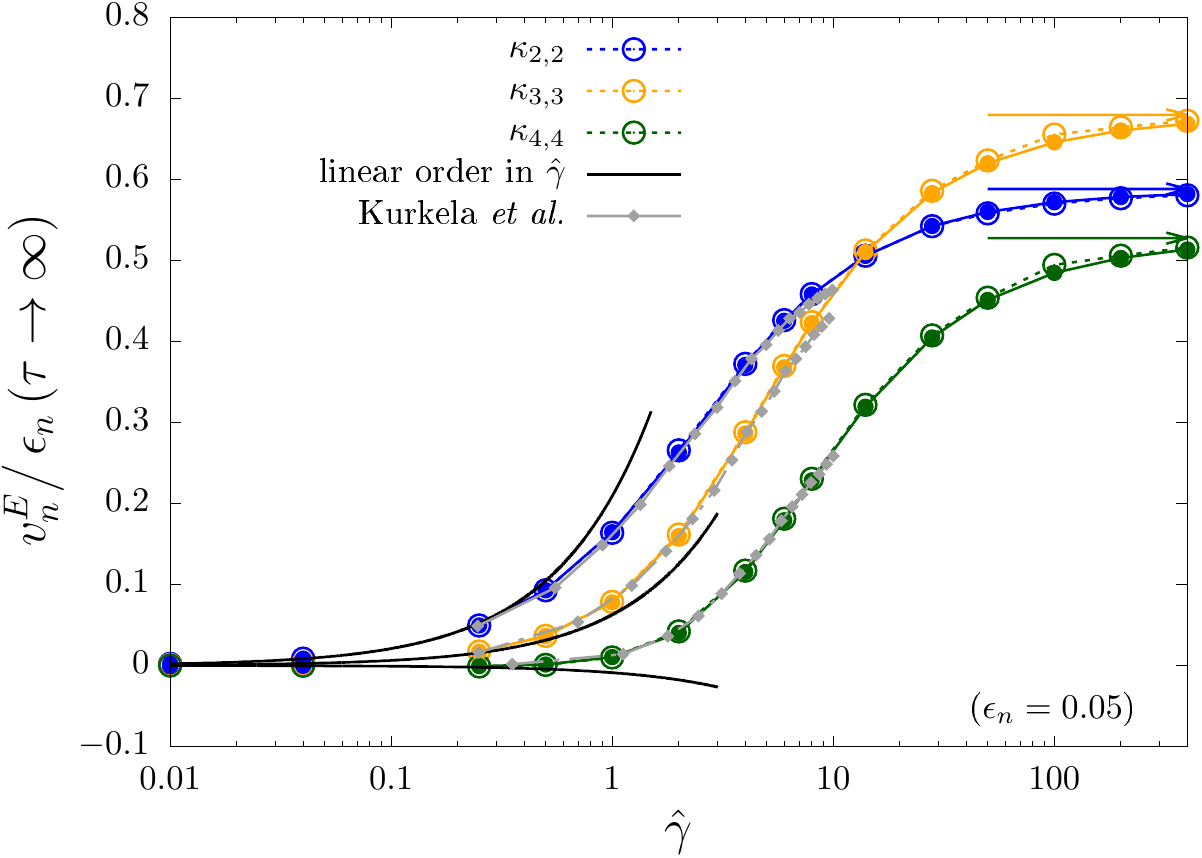} & 
    \includegraphics[width=.48\linewidth]{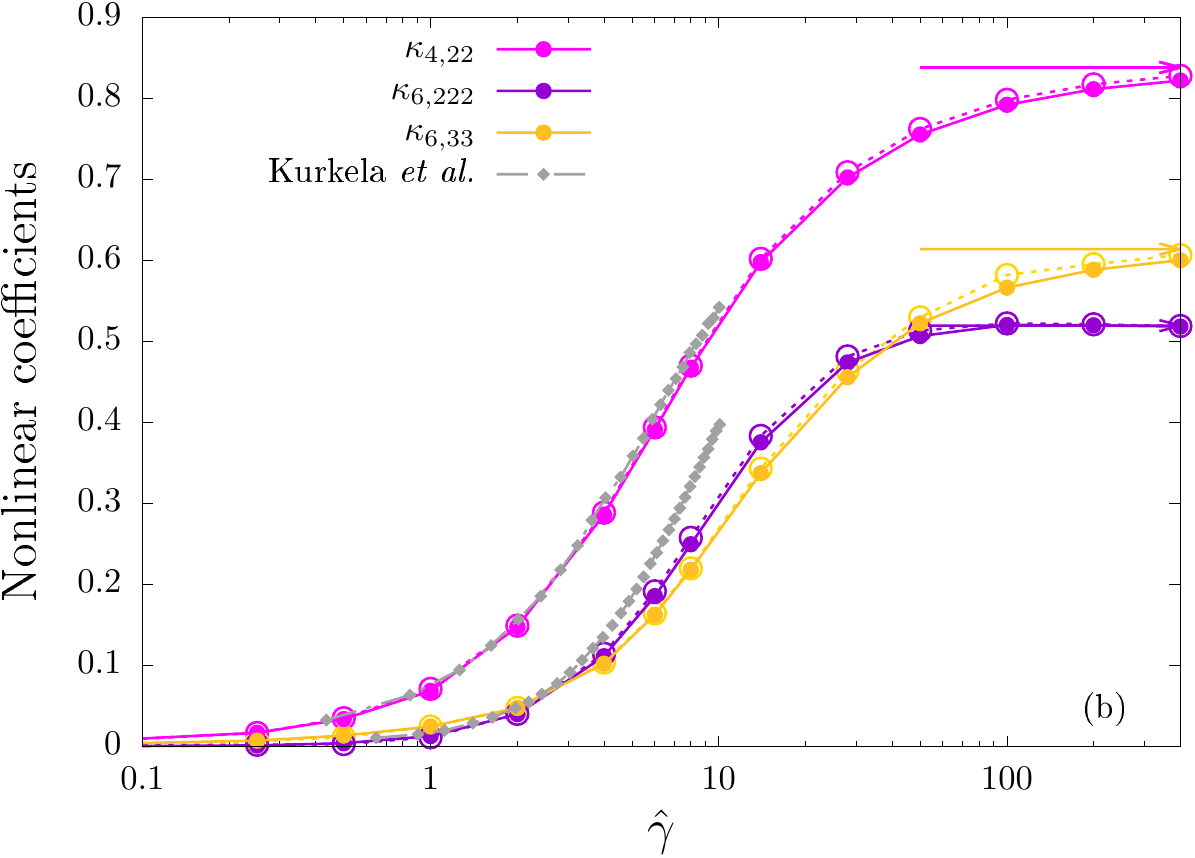}
\end{tabular}
\end{center}
\caption{ Opacity ($\hat{\gamma}$) dependence of (left) the linear $\kappa_{n,n}=\lim_{\epsilon_n\to 0}v_n^E(\tau\rightarrow\infty)/\epsilon_n$ and (right) the non-linear $\kappa_{n,mm}=\lim_{\epsilon_m\to 0}v_n^E(\tau\rightarrow\infty)/\epsilon_m^2$, $\kappa_{n,mmm}=\lim_{\epsilon_m\to 0}v_n^E(\tau\rightarrow\infty)/\epsilon_m^3$  response coefficients. Colored solid lines with filled circles denote results from the RLB method, while colored dotted lines with open circles were obtained in the moment method. The black solid lines show the results obtained to leading order in opacity expansion for the linear coefficients (left). Gray lines represent the results of Kurkela et al. in~\cite{Kurkela:2020wwb} (no such results are available for $\kappa_{6,33}$ in panel (b)). Horizontal arrows indicate asymptotic values extracted from a fit to the numerical data at large opacities (see text). \label{fig:vn_gamma_dependence}
}
\end{figure*}

Since the flow response to the initial eccentricity is essentially linear within our setup, our findings for the development of transverse flow can be compactly summarized in Fig.~\ref{fig:vn_gamma_dependence}, where we present results for the $\hat{\gamma}$-dependence of the response coefficients $\kappa_{n,n}=\lim_{\epsilon_n \to 0} v_n/\epsilon_n$ as well as $\kappa_{4,22}=\lim_{\epsilon_2 \to 0} v_4/\epsilon_2^2$, $\kappa_{6,33}=\lim_{\epsilon_3 \to 0} v_6/\epsilon_3^2$ and $\kappa_{6,222}=\lim_{\epsilon_2 \to 0} v_6/\epsilon_2^3$ estimated from our data at $\epsilon=0.05$. Besides the numerical results, we also indicate the linearized analytical approximation in Eqns.~(\ref{eq:v2analyticalasymptote}-\ref{eq:v4analyticalasymptote}) and the numerical results of Kurkela et al.~\cite{Kurkela:2020wwb}. Despite the discrepancy in the results for the eccentricity dependence, we generally find good agreement with Kurkela et al. in the linear response at low opacities ($\hat{\gamma}\lesssim 10$), which only starts to deviate slightly at larger opacities. %The only coefficient for which we find significant deviation for opacities $\hat\gamma \gtrsim 8$ is the cubic response $\kappa_{6,222}$. 

Concerning the opacity dependence, one finds that at low opacities up to $\hat{\gamma}\lesssim 1$, the linear response coefficients are reasonably well described by the leading order opacity expansion $\kappa_{n,n}\sim \hat{\gamma}$ in Eqns.~(\ref{eq:v2analyticalasymptote}-\ref{eq:v4analyticalasymptote}). However, one should note that, due to the intricate space-time structure of $v_{n}$ production, the higher harmonic coefficents are increasingly sensitive to changes in the underlying dynamics, such that e.g.  $\kappa_{4,4}$, starts to deviate from the leading order opacity expansion already at smaller values of $\hat{\gamma}$. When increasing the opacity further, one observes a sizeable change in the linear and non-linear flow response coefficients for $1\lesssim\hat{\gamma}\lesssim 100$, which is no longer captured by the leading order opacity expansion. Eventually, for very large opacities $\hat{\gamma} \gtrsim 100$, the opacity dependence of the linear and nonlinear response coefficients becomes weaker and weaker, indicating a saturation towards a finite large opacity limit. Empirically, we find that in this regime, the opacity dependence of the response coefficients can be well approximated by a constant asymtptotic value and a power law correction, with the asymptotic values $\kappa(\hat{\gamma}\to \infty)$ indicated by horizontal arrows in Fig.~\ref{fig:vn_gamma_dependence}.

\subsection{Energy flow \& hydrodynamic limit}
\label{sec:hydroLimit}

\begin{figure}
\begin{center}
 \includegraphics[width=.95\linewidth]{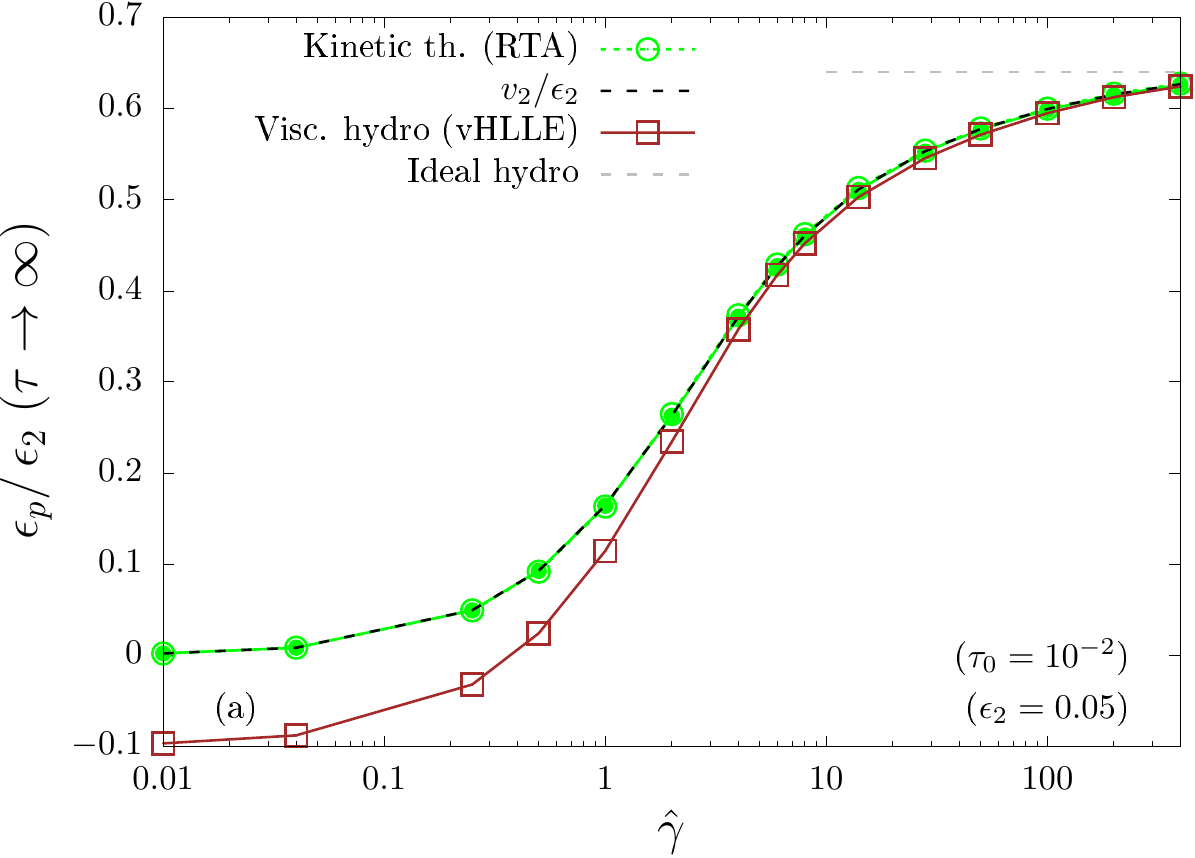} \\
  \includegraphics[width=.95\linewidth]{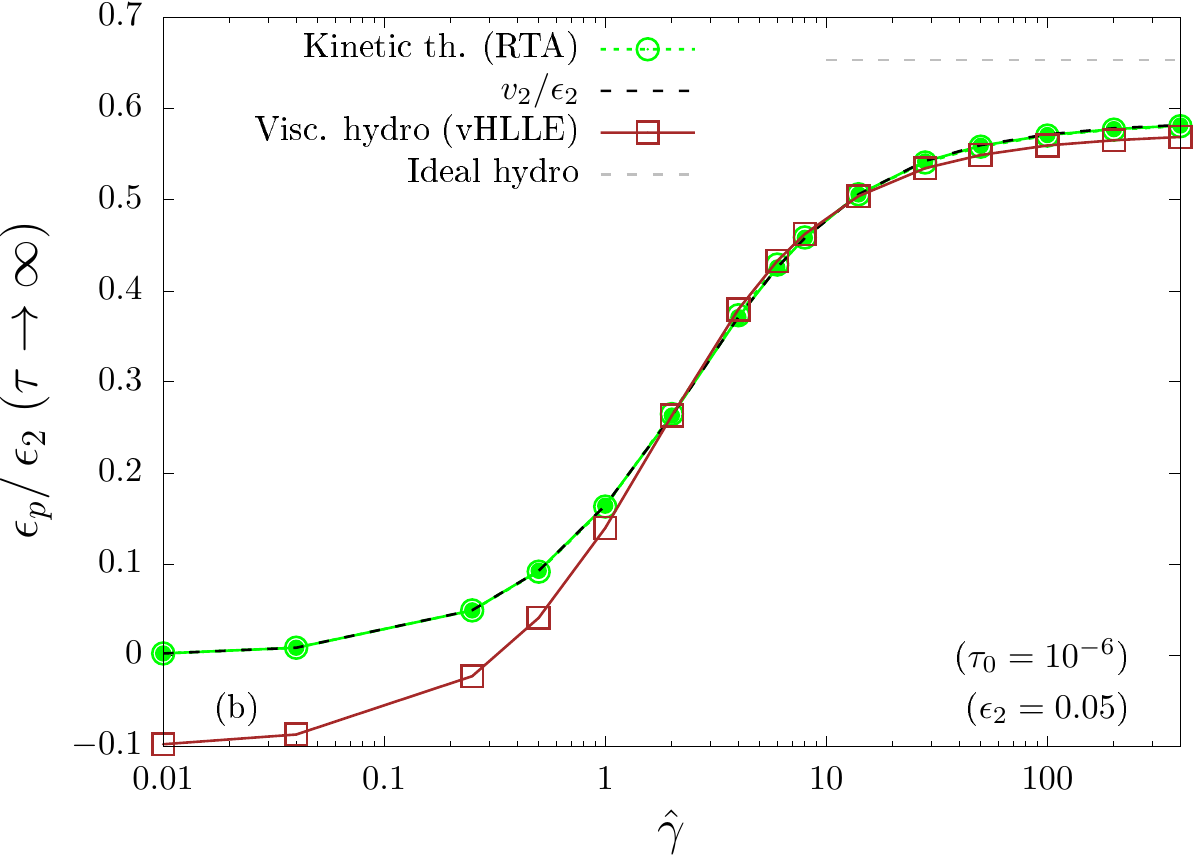} \\
\end{center}
\caption{Opacity ($\hat{\gamma}$) dependence of the energy-flow response $\epsilon_p/\epsilon_2$ for two different initialization times $\tau_0/R=10^{-2}$ (top) and $\tau_0/R=10^{-6}$ (bottom). Two results are plotted for kinetic theory: those from the RLB method are plotted as a green solid line with filled circles and those from the moments method are plotted as a green dashed line with open circles. All results are for $\epsilon_2=0.05$. 
\label{fig:epsp_gamma_dependence_RTA}
}
\end{figure}

So far we have employed an effective kinetic description to study longitudinal cooling and the development of transverse flow as a function of the opacity parameter $\hat{\gamma}$. While at small opacities $\hat{\gamma}\ll 1$ the results from numerical simulations are well described by the first interaction correction to free-streaming, one generally expects that in the opposite limit of large opacities $\hat{\gamma} \gg 1$, the effective kinetic description approaches the limit of dissipative and eventually ideal hydrodynamics. Hence in order to investigate, to what extent this expectation holds true, we will now compare our results from kinetic theory with numerical simulations in Mueller-Israel-Stewart type second order relativistic viscous hydrodynamics.

We employ the publicly available vHLLE code originally introduced in Ref.~\cite{Karpenko:2013wva}, and extend the latest GitHub branch \footnote{
Commit number to be inserted upon publication %\texttt{7283047c5c93589bb256065aac1b2ae3ca75af9b} on 23.04.2021
} to include the initial conditions considered in this paper.  Apart from the conservation equation for the stress-energy tensor, $\nabla_\nu T^{\mu\nu} = 0$, the code implements the M\"uller-Israel-Stewart equations for the evolution of the pressure deviator $\pi^{\mu\nu}$, which for the case of a conformal fluid reduce to \cite{Bernhard:2016tnd}
\begin{multline}
 \dot{\pi}^{\langle \mu\nu\rangle} = 
 \frac{2\eta \sigma^{\mu\nu} - \pi^{\mu\nu}}{\tau_\pi} - \frac{\delta_{\pi\pi}}{\tau_\pi} \pi^{\mu\nu} \theta + 
 \frac{\phi_7}{\tau_\pi} \pi^{\langle\mu}_\alpha \pi^{\nu \rangle\alpha} \\
 - \frac{\tau_{\pi\pi}}{\tau_\pi} \pi^{\langle\mu}_{\alpha} \sigma^{\nu\rangle \alpha},
 \label{eq:vHLLE_pi}
\end{multline}
where $\sigma_{\mu\nu} = 2 \nabla_{\langle \mu} u_{\nu \rangle}$ is the shear tensor, $\theta = \nabla_\mu u^\mu$ is the expansion scalar, while the transport coefficients appearing above satisfy \cite{Denicol:2014vaa}
\begin{align}
 \tau_\pi =& \frac{5\eta}{s T}, &
 \frac{\delta_{\pi\pi}}{\tau_\pi} =& \frac{4}{3}, &
 \phi_7 =& \frac{9}{70p}, &
 \frac{\tau_{\pi\pi}}{\tau_\pi} =& \frac{10}{7}.
\end{align}

We note already at this stage, that the early time behavior in ideal and viscous hydrodynamics does not agree with the early time free-streaming limit of kinetic theory, which as pointed in ~\cite{Kurkela:2020wwb,Kurkela:2019set} leads to an unphysical behavior of $dE_{\bot}/d\eta$ at early times, that makes the scaling variable $\hat{\gamma}$ ill-defined in the limit $\tau_0\to 0$. While in~\cite{Kurkela:2020wwb}, this problem was addressed by modifying the initial conditions and matching the energy per unit rapidity at a later time $\tau/R=1$ of the evolution, we follow the more common procedure, and choose a finite initial time $\tau_0$, where we initialize the energy density as in Eqns.~(\ref{eq:epsilon_isotropic_def}) and (\ref{eq:deltaeps_def}), 
and set the components of the shear stress tensor, $\pi^{\mu\nu}$, to\footnote{We employ a conformal equation of state $e=3p$.}
\begin{equation}
\tau_0^{-2} \pi^{\eta\eta} = -2\pi^{xx} = -2\pi^{yy} = -p,
\end{equation}
which ensures vanishing longitudinal pressure, to comply with the initial conditions for kinetic theory in Eq.~\eqref{eq:fini}. \footnote{Since at very early times, the evolution in viscous hydrodynamics and kinetic theory does not agree, another conceivable option is to initialize the hydrodnamic simulation on the hydrodynamic attractor for Bjorken flow~\cite{Kurkela:2019set,Kurkela:2020wwb}. We have also performed such simulations, and find no significant differences regarding the development of transverse flow.
}
Similarly, we fix the value of the shear viscosity to entropy density ratio $\eta / s$ for a given value of $\hat{\gamma}$ in the same way as for RTA, via Eq.~\eqref{eq:ghat_def}, evaluated at initial time $\tau_0$. By comparing kinetic theory and hydrodynamic simulations with the same finite $\tau_0$, we can then achieve a direct comparison and in addition investigate the dependence on the initialization time $\tau_0$ in the two different theories.

Evaluating the energy-weighted flow
 harmonics $v_n^E$ considered in this paper, 
a Cooper-Frye-like mechanism should be considered
to reconstruct the phase-space distribution function from the 
hydrodynamic fields $e$, $u^\mu$ and $\pi^{\mu\nu}$.
We circumvent this ambiguity by instead referring to 
the stress-energy anisotropy $\epsilon_p$, which according to Eq.~\eqref{eq:ep_def}, can be defined directly in terms of the components of the energy-momentum tensor.
Since $T^{\mu\nu}$ is fundamentally accessible in both 
kinetic theory and hydrodynamics, a comparison between the 
two theories can be made unambiguously at the 
level of $\epsilon_p$. While the quantity $\epsilon_{p}$ measures the second harmonic modulation of the energy flow, and in our kinetic theory simulations exhibits almost identical behavior to $v_{2}^{E}$, we are not aware of generalizations of $\epsilon_{p}$ to higher order flow harmonics, and will therefore restrict our attention to $n =2$ perturbations, with initial eccentricity $\epsilon_2 = 0.05$.\footnote{We have checked that, similar to the kinetic theory results in Fig.~\ref{fig:vn_epsilon_dependence}, non-linear contributions $\epsilon_{p}\sim e_{2}^{3}$ are sufficiently small to be neglected for the linear response analysis of $\epsilon_{p}/e_{2}$.}

Our results for the elliptic energy-flow response are compactly summarized in Fig.~\ref{fig:epsp_gamma_dependence_RTA}, where we compare the opacity dependence of $\epsilon_{p}/\epsilon_{2}$ in kinetic theory (RTA) and hydrodynamics (vHLLE) for two different initialization times $\tau_0/R=10^{-2},10^{-6}$ in the top and bottom panels. When considering the larger initialization time $\tau_0/R=10^{-2}$, one finds that viscous hydrodynamics provides a reasonable description of kinetic theory for $\hat{\gamma} \gtrsim 5$, with both curves smoothly approaching the ideal hydrodynamic limit for large opacities, as indicated by the gray dashed line. When considering a much smaller initialization time, $\tau_0/R=10^{-6}$,  we find small deviations between kinetic theory and hydrodynamics in the same opacity range. While these deviations might not be very sizeable, they notably do not steadily decrease with increasing opacity, as one would naively expect. Moreover, a perhaps more evident observation is that neither of the two curves appears to approach the ideal hydrodynamics result, such that even when extrapolated to infinite opacity the RTA value ($\simeq 0.59$) slightly differs from the vHLLE value ($\simeq 0.57$) and both fall about 10\% short of the ideal hydrodynamic limit ($\simeq 0.64$).

Even though this behavior may appear counterintuitive at first sight, it can ultimately be traced back to the non-commutativity of the limits $\tau_0 \to 0$, where the system is subject to a rapid longitudinal expansion, and $\hat{\gamma} \to \infty$, where hydrodynamics emerges from kinetic theory as the system undergoes rapid equilibration. Starting from kinetic theory, it is clear that for any finite opacity $\hat{\gamma}$ the system is initially far-from equilibrium and behaves as approximately free-streaming, until on time scales $\tau_{\rm eq}/R \sim \hat{\gamma}^{-4/3}$ the system undergoes equilibration, and the subsequent evolution can be approximately described by viscous or even ideal fluid dynamics. While in the limit $\hat{\gamma} \to \infty$, the equilibration time $\tau_{\rm eq}/R \to 0$ and fluid dynamics becomes applicable at earlier and earlier times, the early time free-streaming and initial approach towards equilibrium is never correctly described by fluid dynamics. The results in Fig.~\ref{fig:epsp_gamma_dependence_RTA}, thus provide a clear illustration of the fact that at very early times, the system is necessarily out-of-equilibrium and the two limits $\hat{\gamma} \to \infty$ and $\tau_0 \to 0$ are in general not commutative.

Even though at large opacities the mismatch between kinetic theory and hydrodynamics occurs only at very early times, this affects e.g. the longitudinal cooling and can still have a notable effect on the development of anisotropic flow at later times, which is seen in Fig.~\ref{fig:epsp_gamma_dependence_RTA}.  We are thus lead to conclude that a non-equilibrium description of the early time dynamics is inevitable to accurately describe the development of anisotropic flow, even at relatively large opacities.

\begin{figure}[!t]
\begin{center}
\includegraphics[width=0.96\linewidth]{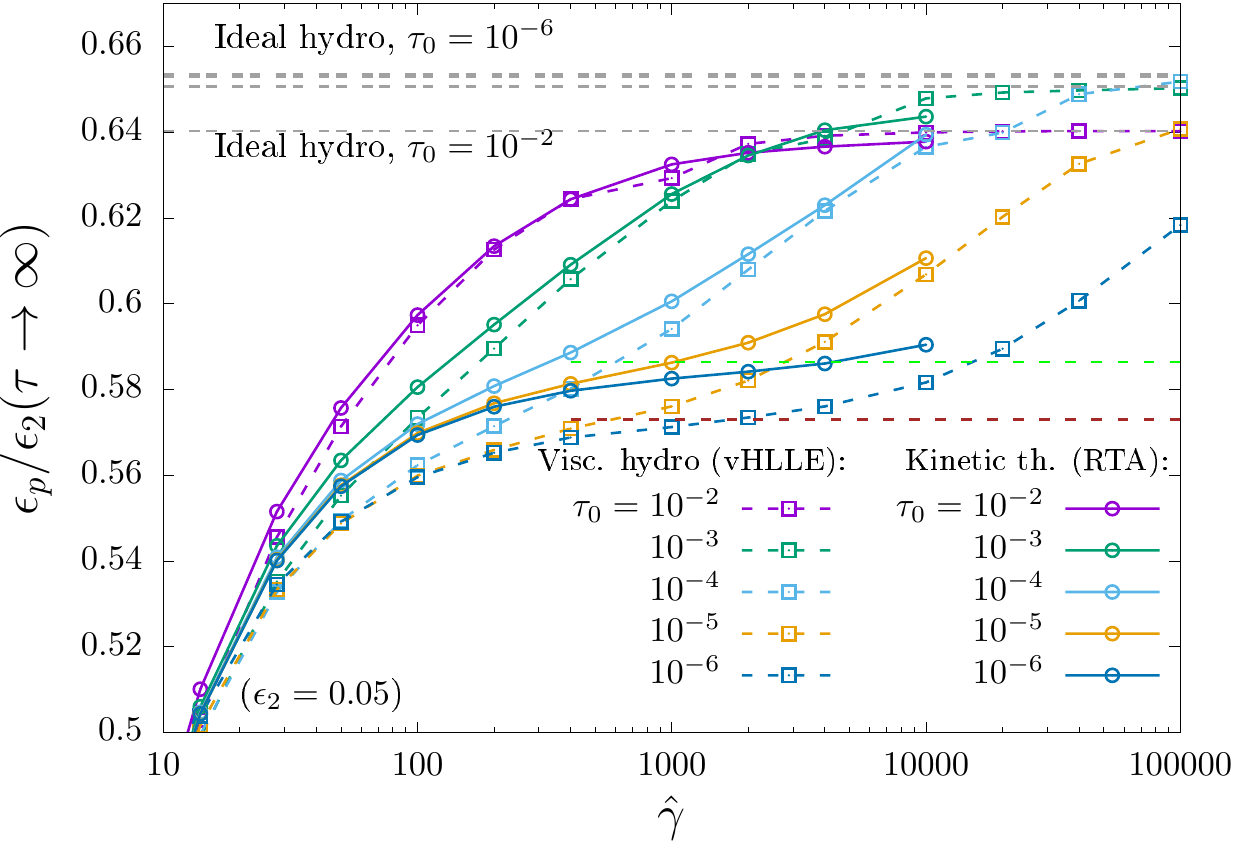}
\end{center}
\caption{Opacity ($\hat{\gamma}$) dependence of the response
coefficient $\epsilon_p/\epsilon_2$ in kinetic theory (RTA, obtained using the RLB method), viscous (vHLLE) and ideal hydrodynamics for different initialization times $\tau_0/R=10^{-2} - 10^{-6}$. Convergence towards ideal hydrodynamics is only observed when the initialization time becomes smaller than the equilibration time of the system.
\label{fig:hydro}}
\end{figure}

 As a final remark to the comparison of opacity dependencies in the different descriptions, we note that for any finite $\tau_0$ kinetic theory and viscous hydrodynamics will approach ideal hydrodynamics for sufficiently large opacities where the equilibration time $\tau_{\rm eq}$ becomes smaller than the initialization time $\tau_0$. While the results shown in Fig.~\ref{fig:hydro} provide an explicit illustration of this behavior, the convergence towards ideal hydrodynamics at large opacities corresponds to the incorrect order of limits, as physically one needs to account for the entire space-time evolution of the system, i.e. the limit $\tau_0 \to0$ has to be taken before $\hat{\gamma} \to \infty$.

One may wonder, how the increasingly short period of non-equilibrium evolution at early times can have such a significant impact on the transverse flow, which only develops on much later times scales $\tau/R \gtrsim 0.1$. While it is true that at very early times, the system does not develop a significant amount of transverse expansion and can locally be described by Bjorken flow as discussed in Section~\ref{sec:results_cooling}, it is equally important to realize that the early-time dynamics is nevertheless inhomogeneous in the transverse plane. Due to the fact that the  initial energy density locally sets the scale for the Bjorken evolution, some regions will experience a faster cooling relative to others, thereby changing the shape of the energy density distribution in transverse space.
 Due to this phenomenon of inhomogenous longitudinal cooling, the geometric eccentricities will be modified even before the transverse expansion 
 sets in. Since the anisotropic flow is built up solely due to transverse expansion, its magnitude is determined by the value of the eccentricity at the onset of transverse expansion. 
 We therefore conclude that differences in the longitudinal cooling at early times are ultimately responsible for the observed differences in the transverse flow.

\begin{figure}
    \centering
    \includegraphics[width=\linewidth]{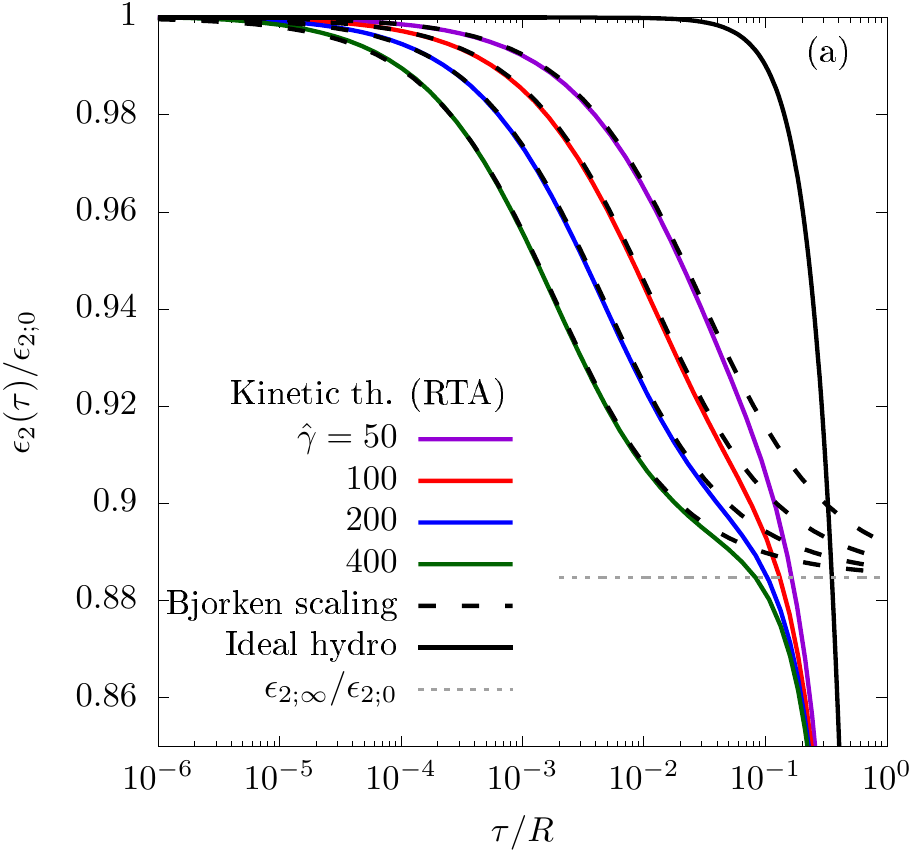} \\
    \includegraphics[width=\linewidth]{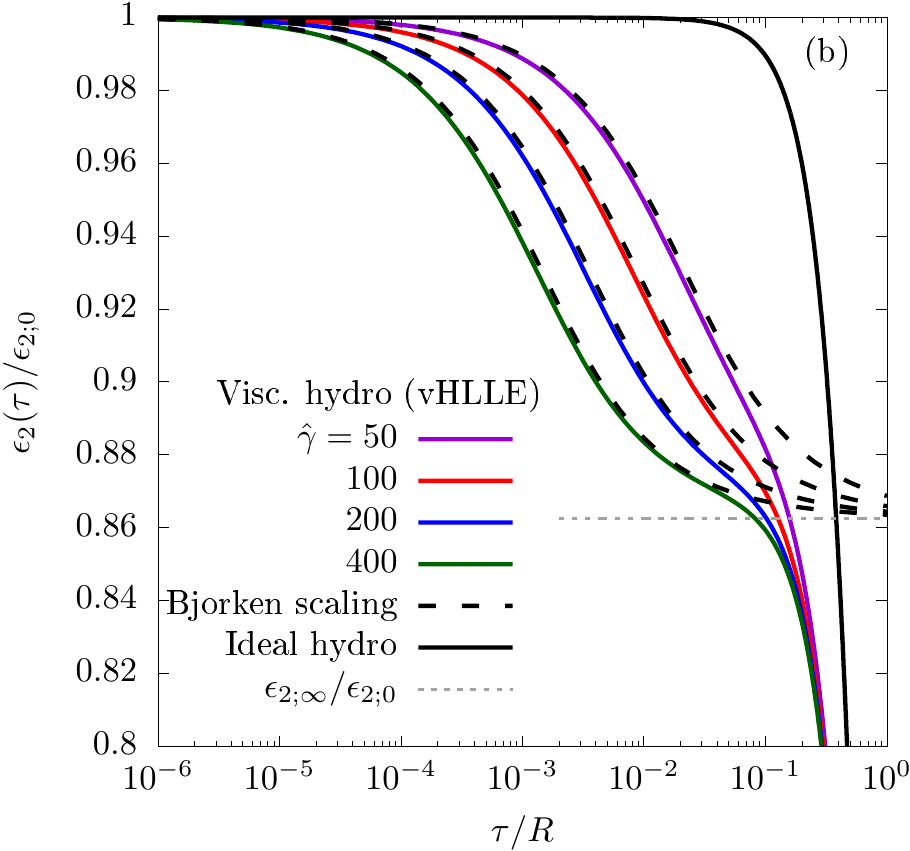} \\
    \caption{Comparison of the evolution of $\epsilon_2$ normalized to its initial value $\epsilon_{2;0}$ on a logarithmic timescale for kinetic theory obtained using the moments method (top) and viscous hydrodynamics (bottom). Also shown are the corresponding results in Bjorken flow scaling approximation (dashed black lines) and ideal hydrodynamics (solid black lines). Gray dashed lines show the limit in Eq.~\eqref{eq:eps_running} in the absence of transverse expansion.}
    \label{fig:eccentricity_decay}
\end{figure}

We illustrate this behavior in Fig.~\ref{fig:eccentricity_decay}, where we present the evolution of the coordinate space eccentricity $\epsilon_2$ as a function of time $\tau/R$. Different colored curves in the top panel show the evolution of $\epsilon_2$ in kinetic theory for different opacities. Similarly, the bottom panel shows the  corresponding results obtained in viscous hydrodynamics (vHLLE). 
The ideal hydrodynamics result is shown for comparison as a solid black line in both panels. Starting around $\tau\sim 0.1 R$ all curves exhibit a significant drop  due to the onset of transverse expansion. However, in kinetic theory and viscous hydrodynamics, the eccentricity decreases even before that due to the previously discussed phenomenon of inhomogeneous longitudinal cooling. Strikingly, this effect can also be described (semi-) analytically by
approximating the dynamics as a collection of local Bjorken flows in a similar way to what was discussed in Section~\ref{sec:results_cooling}, which yields results for the decrease of $\epsilon_2$ that we plotted as dashed black lines. We note that the limiting behavior for this decrease can be obtained as
\begin{equation}
 \lim_{\tau \rightarrow \infty} \frac{\epsilon_{2}(\tau)}{\epsilon_{2;0}} = \frac{(1 - q / 4)^3}{(1 - q / 6)^3},\label{eq:eps_running}
\end{equation}
where $q$ is related to the behaviour of the universal function $\mathcal{E}(\tilde{w}) \sim \tilde{w}^q$ at small $\tilde{w}$, such that in kinetic theory, $q = 4/9$ as indicated in Eq.~\eqref{eq:universal_E_lim}, whereas for the hydrodynamic theory in Eq.~\eqref{eq:vHLLE_pi}, one has $q = (\sqrt{505} - 13)/18 \simeq 0.526$. Evaluating Eq.~\eqref{eq:eps_running} for the above values of $q$, one obtains a $\sim 11.5\%$ (RTA) and $13.7\%$ (VHLLE) decrease of $\epsilon_2$ solely due to the longitudinal expansion, as indicated by the gray dashed lines in Fig.~\ref{fig:eccentricity_decay}. Hence,  this effect indeed takes on the correct magnitude to be able to describe the difference of $\sim 10\%$ in the large opacity limits of kinetic theory and viscous hydro compared to ideal hydrodynamics.

\section{Conclusions \& Outlook}\label{sec:conclusions}

We employed the Boltzmann equation in the (conformal) relaxation time approximation as a simple model 
to study the space-time dynamics of small and large systems created in high-energy hadronic collisions. Within the simple effective kinetic description described in Sec.~\ref{sec:theory}, the evolution of the system depends
on a single dimensionless opacity parameter $\hat{\gamma}$ that combines the system size 
and energy dependences, and we estimate $\hat{\gamma}$ to range from values $\lesssim 1$ in p+Pb collisions 
to $\approx 10$ in Pb+Pb collisions at LHC energies (c.f. Eq.~(\ref{eq:ghat_pp}) and (\ref{eq:ghat_PbPb})).

We performed (semi-)analytic calculations at leading order in opacity $\hat{\gamma}$ (c.f. Sec.~\ref{sec:linear}) and developed first principles numerical simulations (c.f. Sec.~\ref{NumericalProcedure})
to investigate the longitudinal cooling of the transverse energy per unit rapidity, $\frac{\d E_{\bot}/d\eta}{\d E_{\bot}^{(0)}/d\eta}$, and the development of transverse flow quantified by the (energy weighted) flow harmonics
$v_n^{E}$ for a large range of opacities.

We find that with increasing opacities, pressure isotropization takes place at earlier and earlier times, 
such that for large opacities $\hat{\gamma} \gtrsim 1$ the onset of longitudinal cooling of the system 
is well described by one dimensional Bjorken dynamics, until at later times $\tau/R \gtrsim 0.1$ the effects of the transverse expansion 
can no longer be ignored.

By studying the response to anisotropic perturbations of the initial energy density, we investigated the development of
transverse flow from low to high opacities. While for small opacities, $\hat{\gamma} \lesssim 1$, 
the development of transvserse flow is reasonably well described by the leading order opacity corrections to free-streaming,
we find that for $1 \lesssim \hat{\gamma} \lesssim 100$ the linear and non-linear flow response exhibits a strong opacity dependence,
and eventually saturates for large opacities $\hat{\gamma} \gtrsim 100$.

Even though one naively expects the results for large opacities $\hat{\gamma} \gg 1$ to approach the hydrodynamic limit,
it turns out that subtleties of the limits $\hat{\gamma} \to \infty$ and $\tau_0 \to0$ provide a restriction on the accuracy of  hydrodynamic descriptions.
Since the early time pre-equilibrium dynamics of the system cannot be accurately described by ordinary viscous or ideal hydrodynamics, deviations between all approaches persists even at very large opacities. With respect to RTA results, we found discrepancies of the viscous and ideal hydro results of the order of $\sim 2.5\%$ and $\sim 12\%$, respectively. However, as these discrepancies can be mostly attributed to the phenomenon of inhomogeneous longitudinal cooling, we believe that the inclusion of a more appropriate pre-equilibrium description as in K{\o}MP{\o}ST~\cite{Kurkela:2018wud,Kurkela:2018vqr} may significantly improve the agreement between microscopic and macroscopic descriptions~(see also~\cite{Kurkela:2018qeb}). Similarly, it is also conceivable that resummed hydrodynamic approaches such as anisotropic hydroynamics (aHydro) \cite{Martinez:2010sc,Florkowski:2010cf,Florkowski:2013lya,Martinez:2012tu,McNelis:2021zji} can accurately describe the inhomogeneous longitudinal cooling and it will be interesting to further investigate this in the future.

While our current study provides a detailed assessment of the development of transverse flow from very small to very large opacities, some of the shortcomings should be addressed
prior to inferring phenomenological conclusions for proton-proton, proton-nucleus and nucleus-nucleus collisions. Evidently, it would be important to perform event-by-event studies with a 
more realistic transverse collision geometry, which is conceptually straightforward but will require significantly larger computation time. Beyond such straightforward extensions, 
it would also be interesting to consider more realistic collision kernels~\cite{Kurkela:2021ctp} and investigate the effects of a non-conformal equation of state, which however will require additional theoretical developments.

\textit{Acknowledgements:} We thank N.~Borghini, H.~Elfner, N.~Kersting, A.~Mazeliauskas, B.~Schenke, U.~A.~Wiedemann, H.~Roch, M.~Borrell~Martinez, A.~Shark and P. Aasha for valuable discussions. This work is supported by the Deutsche Forschungsgemeinschaft (DFG, German Research Foundation)
through the CRC-TR 211 ’Strong-interaction matter under extreme conditions’– project
number 315477589 – TRR 211. V.E.A. gratefully acknowledges the support of the Alexander von Humboldt Foundation through a Research
Fellowship for postdoctoral researchers.
Numerical calculations presented in this work were performed at Paderborn Center for Parallel Computing (PC2) and the Center for Scientific Computing (CSC) at the Goethe-University of Frankfurt and we gratefully acknowledge their support.

\begin{appendix}

\section{Free-streaming the anisotropies}\label{sec:free-streaming anisotropies}

In linear order of the opacity expansion, the dynamics of the anisotropies is naturally dominated by the free-streaming limit, as will also be more evident from the calculations in the following sections. It is therefore pivotal to examine how the anisotropic factors of $x_\perp^n\,\cos(n\phi_{\xT\nT})$
that are part of the initial condition in Eqs.~(\ref{eq:epsilon_isotropic_def},\ref{eq:deltaeps_def}) 
behave under free-streaming according to the propagation as given in Eq. (\ref{eq:fIniAnalytical}). 
 The notation $\phi_{\xT\nT} = \phi_x - \Psi_n$ was introduced in Eq.~\eqref{eq:phin_def}.
% For this purpose, it is convenient to rewrite the cosine in terms of a Chebyshev polynomials of the first kind. Similarly, one can also rewrite $\sin(n\phi_{\xT\nT})$ in terms of Chebyshev polynomials of the second kind.
% \begin{align}
%   \cos(n\phi)&=T_n(\cos(\phi))= \frac{n}{2}\sum_{k=0}^{\left\lfloor \frac{n}{2}\right\rfloor}(-1)^k\frac{(n-k-1)!}{(n-2k)!k!}2^{n-2k}\,\cos^{n-2k}(\phi)\ \ , \label{splitcosine} \\
%   \sin(n\phi)&=\sin(\phi)\,U_{n-1}(\cos(\phi))= \sin(\phi)\sum_{k=0}^{\left\lfloor \frac{n-1}{2}\right\rfloor}(-1)^k\binom{n-k-1}{k}2^{n-2k-1}\,\cos^{n-2k-1}(\phi)\ \ . \label{splitsine}
% \end{align}
For this purpose, it is convenient to rewrite the $\cos(n\phi)$ and $\sin(n\phi)$ in terms of the Chebyshev polynomials of the first and second kind \cite[Ch.~18]{NIST:DLMF}:
\begin{align}
   \cos(n\phi)&=T_n(\cos(\phi)), & 
   \sin(n\phi)&=\sin(\phi)\,U_{n-1}(\cos(\phi)). 
   \label{eq:split_cos_sin}
\end{align}
The explicit form of the Chebyshev polynomials, 
\begin{align}
   T_n(z)=& \frac{n}{2}\sum_{k=0}^{\left\lfloor \frac{n}{2}\right\rfloor}(-1)^k\frac{(n-k-1)!}{(n-2k)!k!} (2z)^{n-2k},\nonumber\\
   U_n(z)=& \sum_{k=0}^{\left\lfloor \frac{n}{2}\right\rfloor}(-1)^k\binom{n-k}{k} (2z)^{n-2k}, \label{eq:chebyshev_exp}
\end{align}
can be used to express $\cos(n\phi)$ and $\sin(n\phi)$ in terms of powers of $\cos\phi$.
Under free-streaming \eqref{free streaming solution}, the factor $x_\perp^n \cos(n \phi^{(n)}_{\xT\nT})$ evolves to 
\begin{multline}
    |\xT-\vT \Delta \tau| ^{n} \cos(n\phi_{\xT-\vT \Delta \tau,\nT}) \\
    =|\xT-\vT \Delta \tau| ^{n} T_{n}\left( \frac{(\xT-\vT \Delta \tau) \cdot \nT}{|\xT-\vT \Delta \tau|}  \right),\label{eq:fscosine}
    %= \sum_{j=0}^{n}\binom{n}{j}x_\perp^{n-j}(-\Delta\tau)^j\,\cos(n\phi_{\xT\nT}+j\phi_{\xT\pT})\ \ . 
\end{multline}
where Eq.~\eqref{eq:split_cos_sin} was employed on the right hand side. In the above, the time lapse $t(\tau, \tau_0, y-\eta)$ was replaced by $\Delta \tau = \tau - \tau_0$ by virtue of Eq.~\eqref{eq:FS_t0}.

The Chebyshev polynomials obey the identity
\begin{align}
    z^n =|z|^n T_n\left(\frac{a}{|z|}\right)+ib|z|^{n-1}U_{n-1}\left(\frac{a}{|z|}\right),
\end{align}
where $a$ and $b$ are real numbers and $z = a + ib$ is a complex number. Denoting $\phi_{\pT \nT} = \phi_p - \Psi_n$, we set
\begin{align}
 a&\rightarrow x_\perp\,\cos(\phi_{\xT\nT})-\Delta\tau\,\cos(\phi_{\pT\nT}),\nonumber\\
 b&\rightarrow x_\perp\,\sin(\phi_{\xT\nT})-\Delta\tau\,\sin(\phi_{\pT\nT}),\nonumber\\
 z&\rightarrow x_\perp\,e^{i\phi_{\xT\nT}}-\Delta\tau\,e^{i\phi_{\pT\nT}},
\end{align}
such that $\mathrm{Re}(z^n)$ is just the expression on the left hand side of Eq.~\eqref{eq:fscosine}, an expression with a very simple dependence on $\phi_{\xT\pT} = \phi_x - \phi_p$ can be derived:
\begin{multline}
 |\xT-\vT \Delta \tau| ^{n} \cos(n\phi_{\xT-\vT \Delta \tau,\nT}) = \sum_{j=0}^{n} \binom{n}{j}
 \\\times x_{T}^{n-j} (-\Delta \tau)^{j} [\cos(n \phi_{\xT\nT}) \cos(j \phi_{\xT\pT}) \\
 - \sin(n \phi_{\xT\nT}) \sin(j\phi_{\xT\pT})].
\end{multline}

% This does not look insightful at first, but choosing specific values of $a$ and $b$, this expression can be identified as the real part of the following useful property of the Chebyshev polynomials:

% \begin{align}
%     z^n&=|z|^n T_n\left(\frac{a}{|z|}\right)+ib|z|^{n-1}U_{n-1}\left(\frac{a}{|z|}\right) \ \ ,\\
%     a&\rightarrow x_\perp\,\cos(\phi_{\xT\nT})-\Delta\tau\,\cos(\phi_{\pT\nT}) \ \ ,\\
%     b&\rightarrow x_\perp\,\sin(\phi_{\xT\nT})-\Delta\tau\,\sin(\phi_{\pT\nT})\\
%     \Rightarrow z&\rightarrow x_\perp\,e^{i\phi_{\xT\nT}}-\Delta\tau\,e^{i\phi_{\pT\nT}}
% \end{align}

% Therefore the free-streamed expression must be equal to $\mathrm{Re}(z^n)$ in this case. Straightforward calculations lead to an expression with a very simple $\phi_{\xT\pT}$-dependence.
% \begin{align}
%      |\xT-\vT \Delta \tau| ^{n} \cos(n\phi_{\xT-\vT \Delta \tau,\nT}) = \sum_{j=0}^{n} \binom{n}{j} x_{T}^{n-j} (-\Delta \tau)^{j} \left[\cos(n \phi_{\xT\nT}) \cos(j \phi_{\xT\pT}) - \sin(n \phi_{\xT\nT}) \sin(j\phi_{\xT\pT})\right]
% \end{align}

\begin{widetext}

\section{Solving integrals in Landau matching}\label{IntegralsLandauMatching}

In order to be able to perform the necessary integrals of the kernel according to Eq.~\ref{V1}, we will need specific expressions for $\epsilon$ and $u^\mu$, meaning that we need to compute the exact form of $T^{(0)\mu\nu}$ and $\delta T^{(0)\mu\nu}$ by solving the integrals given in Section~\ref{Landau matching}. 

As we will discuss below, we can break the integration down to solving integrals of an exponential of $\cos(\phi)$ multiplied with powers of trigonometric functions. These can be expressed in terms of modified Bessel functions of the first kind.

\begin{align}
    \int\d \phi\, e^{a\cos(\phi)}\,\cos(n\phi)&=2\pi I_n(a) \ \ ,\\
    \int\d \phi\, e^{a\cos(\phi)}\,\cos(n\phi)\,\cos^m(\phi)&=2\pi I^{(m)}_n(a) \ \ ,\\
    \int\d \phi\, e^{a\cos(\phi)}\,\sin(n\phi)&=0 \ \ ,\\
    \int\d \phi\, e^{a\cos(\phi)}\,\sin(n\phi)\,\sin(\phi)&=2\pi I'_n(a)-2\pi I_{n+1}(a) \ \ ,\\
    \int\d \phi\, e^{a\cos(\phi)}\,\sin(n\phi)\,\cos(\phi)\,\sin(\phi)&=2\pi I''_n(a)-2\pi I'_{n+1}(a) \ \ .\label{eq:integrals}
\end{align}

% It is straightforward to see that $T^{(0)\mu\nu}$ is indeed of this form as
% \begin{align}
%     T^{(0)\mu\nu}&=\frac{\tau_0}{\tau}\int \frac{\d \phi_p}{2\pi}\,v_\perp^\mu v_\perp^\nu\, \ebar(\tau_0,\xT-\Delta\tau\vT) \ \ , \\
%     \ebar(\tau_0,\xT) &= \frac{1}{\pi R^2 \tau_0} \epn \mathrm{exp}\left(-\frac{x_\perp^2}{ R^2}\right) \ \ , \\
%     v^\mu_\perp&=p^\mu/p_\perp=(\mathrm{cosh}(y-\eta),\vT,\mathrm{sinh}(y-\eta)) \ \ .
% \end{align}
% where the integral with respect to $p_\perp$ was performed according to Eq.~\eqref{eq:epsilon-f-relation}. The isotropic part of the initial energy density profile, introduced in Eq.~\eqref{eq:}  
It is straightforward to see that $T^{(0)\mu\nu}$ is indeed of this form as
\begin{equation}
 T^{(0)\mu\nu}=\frac{\tau_0}{\tau}\int \frac{\d \phi_p}{2\pi}\,v^\mu_\perp v^\nu_\perp\, \ebar(\tau_0,\xT-\Delta\tau\vT),
\end{equation}
where $v^\mu_\perp = (1, \cos\phi_p, \sin\phi_p, 0)$ has no longitudinal component under free-streaming due to the $\delta(y-\eta)$ function in Eq.~\eqref{eq:fIniAnalytical}. 
The integral with respect to $p_\perp$ was performed according to Eq.~\eqref{eq:epsilon-f-relation}. At zeroth order, we took into account only the isotropic part of the initial energy density profile $\overline{\epsilon}$, introduced in Eq.~\eqref{eq:epsilon_isotropic_def}, which is evaluated at shifted coordinates according to
\begin{equation}
 \ebar(\tau_0,\xT - \Delta \tau \vT) = \frac{1}{\pi R^2 \tau_0} \epn \mathrm{exp}\left(-\frac{x_\perp^2 + \Delta \tau^2 - 2 x_\perp \Delta \tau \, \cos(\phi_x - \phi_p)}{ R^2}\right).
\end{equation}

% Using the integral formulae given in Eq.~\eqref{eq:integrals}, the components of 
% \begin{align}
%     T^{(0)\mu\nu}=\left(\begin{array}{ccc}
%         T^{(0)\tau\tau} & T^{(0)\tau \perp} \hxT^t &0\\
%         T^{(0)\tau \perp} \hxT & T^{(0)\mathbb{1}}\mathbb{1}+T^{(0)\perp} \hxT\hxT^t&0\\
%         0&0&0
%     \end{array}\right)
% \end{align}
Using the integral formulae given in Eq.~\eqref{eq:integrals}, the components of the stress-energy tensor introduced in Eq.~\eqref{eq:isotropicTmunuComponents} 
can be computed to be
\begin{align}
    T^{(0)\tau\tau}&=\frac{1}{\tau}\frac{1}{\pi R^2} \epn \mathrm{exp}\left(-\frac{x_\perp^2+\Delta\tau^2}{R^2}\right)I_0(2b) \ \ ,\\
    T^{(0)\tau \perp}&=\frac{1}{\tau}\frac{1}{\pi R^2} \epn \mathrm{exp}\left(-\frac{x_\perp^2+\Delta\tau^2}{R^2}\right)I_1(2b) \ \ ,\\
    T^{(0) \perp\perp}&=\frac{1}{\tau}\frac{1}{\pi R^2} \epn \mathrm{exp}\left(-\frac{x_\perp^2+\Delta\tau^2}{R^2}\right)I_2(2b) \ \ ,\\
    T^{(0)\mathbb{1}}&=\frac{1}{\tau}\frac{1}{\pi R^2} \epn \mathrm{exp}\left(-\frac{x_\perp^2+\Delta\tau^2}{R^2}\right)[I_0(2b)-I_0''(2b)] \ \ ,
\end{align}
where we defined $b=\frac{x_\perp \Delta\tau}{R^2}$.

The anisotropic part $\delta T^{(0) \mu\nu}$,
\begin{equation}
 \delta T^{(0)\mu\nu} =\frac{\tau_0}{\tau}\int \frac{\d \phi_{\xT\pT}}{2\pi}\,v_\perp^\mu v_\perp^\nu\, \delta\epsilon(\tau_0,\xT-\Delta\tau\vT),
 \label{eq:FS_anisotropicTmunu_int}
\end{equation}
exhibits a dependence on the angle $\phi_{\xT\nT}$ due to the form \eqref{eq:deltaeps_def} of the anisotropic part of the initial energy profile, which is evaluated at shifted coordinates according to:
\begin{equation}
    \delta\epsilon(\tau_0,\xT - \Delta \tau \vT) =\frac{1}{\pi R^2 \tau_0} \epn\mathrm{exp}\left[-\bar{\alpha}\frac{|\xT -\Delta \tau \vT|^2}{R^2}\right]\,\delta_n\left(\frac{|\xT - \Delta \tau \vT|}{R}\right)^n \cos(n\phi_{\xT\nT}) \ \ .
\end{equation}
Solving the integrals in Eq.~\eqref{eq:FS_anisotropicTmunu_int} is a bit more difficult, requiring the computation of an angular integral of the free-streamed anisotropies. We have already seen in App.~\ref{sec:free-streaming anisotropies} how we can rewrite them into a term with a straightforward $\phi_{\xT\pT}$-dependence. An additional $\phi_{\xT\pT}$-dependence comes from the velocity vectors $v^\mu$. In the computation of $\delta\epsilon$, $\delta u_t$ and $\delta u_s$ they will be contracted with the eigenvectors $u^\mu$, $t^\mu$ and $s^\mu$ in the following ways:
\begin{align}
    u_\mu v^\mu v^\nu u_\nu &= \gamma^2[1-2\beta\cos(\phi_{\xT\pT})+\beta^2\cos^2(\phi_{\xT\pT})] \ \ ,\nonumber\\
    u_\mu v^\mu v^\nu t_\nu &= \gamma^2[\beta-\cos(\phi_{\xT\pT})-\beta^2\cos(\phi_{\xT\pT})+\beta\cos^2(\phi_{\xT\pT})] \ \ ,\nonumber\\
    u_\mu v^\mu v^\nu s_\nu &= -\gamma[\sin(\phi_{\xT\pT})-\beta\cos(\phi_{\xT\pT})\sin(\phi_{\xT\pT})] \ \ .
\end{align}
Taking into account all of the ingredients presented above, we indeed find that we can decompose all terms into integrals of the form in Eq.~\eqref{eq:integrals}. We can plug the results into Eqs.~(\ref{eq:deltaepsilonfromTmunu}-\ref{eq:deltausfromTmunu}) to obtain explicit expressions for the anisotropic corrections $\delta\epsilon$, $\delta u_t$ and $\delta u_s$:
\begin{align}
\delta \epsilon &= u_\mu\delta T^{(0)\mu\nu}u_\nu=\delta_n\frac{1}{\tau}\epn\frac{1}{2\pi R^{n+2}}e^{-\bar{\alpha}\frac{x_\perp^2+\Delta\tau^2}{R^2}} \int \frac{d\phi_{\xT\pT}}{2\pi} e^{2\bar{\alpha} b  \cos(\phi_{\xT\pT})}\gamma^2[1-2\beta\cos(\phi_{\xT\pT})+\beta^2\cos^2(\phi_{\xT\pT})]\nonumber\\ &\quad\times\sum_{j=0}^{n}\binom{n}{j}x_\perp^{n-j}(-\Delta\tau)^j\,\cos(n\phi_{\xT\nT}+j\phi_{\xT\pT})\\
&=\cos(n\phi_{\xT\nT})\delta_n\frac{1}{\tau}\epn\frac{1}{\pi R^{n+2}}e^{-\bar{\alpha}\frac{x_\perp^2+\Delta\tau^2}{R^2}} \gamma^2 \sum_{j=0}^{n}\binom{n}{j}x_\perp^{n-j}(-\Delta\tau)^j\,[I_j(2\bar{\alpha}b)-2\beta I'_j(2\bar{\alpha}b)+\beta^2I''_j(2\bar{\alpha}b)]\\
\delta u_t &= \frac{u_\mu\delta T^{(0)\mu\nu}t_\nu}{p_t-\epsilon}=\delta_n\frac{1}{\tau}\epn\frac{1}{\pi R^{n+2}}e^{-\bar{\alpha}\frac{x_\perp^2+\Delta\tau^2}{R^2}} \int \frac{d\phi_{\xT\pT}}{2\pi} e^{2\bar{\alpha} b  \cos(\phi_{\xT\pT})}\gamma^2[\beta-\cos(\phi_{\xT\pT})\label{eq:deltaepsilonresult}\\ 
&\quad \quad -\beta^2\cos(\phi_{\xT\pT})+\beta\cos^2(\phi_{\xT\pT})]\sum_{j=0}^{n}\binom{n}{j}x_\perp^{n-j}(-\Delta\tau)^j\,\cos(n\phi_{\xT\nT}+j\phi_{\xT\pT})\nonumber\\
&=\cos(n\phi_{\xT\nT})\delta_n\frac{1}{p_t-\epsilon}\frac{1}{\tau}\epn\frac{1}{\pi R^{n+2}}e^{-\bar{\alpha}\frac{x_\perp^2+\Delta\tau^2}{R^2}} \gamma^2 \sum_{j=0}^{n}\binom{n}{j}x_\perp^{n-j}(-\Delta\tau)^j\,[\beta I_j(2\bar{\alpha}b)-(1+\beta^2) I'_j(2\bar{\alpha}b)+\beta I''_j(2\bar{\alpha}b)]\\
\delta u_s &= \frac{u_\mu\delta T^{(0)\mu\nu}s_\nu}{p_s-\epsilon}=\delta_n\frac{1}{\tau}\epn\frac{1}{\pi R^{n+2}}e^{-\bar{\alpha}\frac{x_\perp^2+\Delta\tau^2}{R^2}} \int \frac{d\phi_{\xT\pT}}{2\pi} e^{2\bar{\alpha} b  \cos(\phi_{\xT\pT})}\gamma[\sin(\phi_{\xT\pT})-\beta\cos(\phi_{\xT\pT})\sin{\phi_{\xT\pT}}]\nonumber\\ &\quad\times\sum_{j=0}^{n}\binom{n}{j}x_\perp^{n-j}(-\Delta\tau)^j\,\cos(n\phi_{\xT\nT}+j\phi_{\xT\pT})\label{eq:deltausresult}\\
&=-\sin(n\phi_{\xT\nT})\delta_n\frac{1}{p_s-\epsilon}\frac{1}{\tau}\epn\frac{1}{\pi R^{n+2}}e^{-\bar{\alpha}\frac{x_\perp^2+\Delta\tau^2}{R^2}} \gamma \sum_{j=0}^{n}\binom{n}{j}x_\perp^{n-j}(-\Delta\tau)^j\,[ I'_j(2\bar{\alpha}b)- I_{j+1}(2\bar{\alpha}b)\\
&\quad \quad -\beta I''_j(2\bar{\alpha}b)+\beta I'_{j+1}(2\bar{\alpha}b)]\nonumber
\end{align}

\section{Details of linearized calculation}\label{details of linearized calculation}

As stated in Section~\ref{first order corrections}, the linear order corrections to the observables $V_{mn}$ can be computed as a six-dimensional integral of the kernel:
\begin{align}
    V^{(1)}_{mn}(\tau) &=\int \d^2 p_\perp \ e^{in\phi_p} p_\perp^m \int_{\tau_0}^\tau \mathrm{d}\tau ' \int \d ^2 x_\perp  \int \d \eta  \ \tau '  \frac{\nu_{\rm eff}}{(2\pi)^3} C[f^{(0)}]\left(\tau ' ,\xT ,\pT,y-\eta\right) \ \ , \label{Vmn1}\\
    C[f^{(0)}]&=-p^\mu u_\mu\left(5 \frac{\eta}{s}\right)^{-1}T \left(f^{(0)}-f_{eq}\right) \ \ .
\end{align}
We already outlined in that section how this problem can be split into different terms. The moments $V_{m0}$ depend only on the isotropic part, while the moments $V_{mn}$ with $n\neq0$ vanish in the isotropic case and have to be computed to linear order in the anisotropic perturbations.  Additionally, the linear order corrections to the moments split into buildup of equilibrium $V^{(1,eq)}_{mn}$ and decay of the initial condition $V^{(1,0)}_{mn}$ as computed from the corresponding parts of the kernel:
\begin{align}
    C_{eq}[f^{(0)}]&=p^\mu u_\mu\left(5 \frac{\eta}{s}\right)^{-1}T\ f_{eq} \ \ ,\label{eq:isoCeq}\\
    C_0[f^{(0)}]&=-p^\mu u_\mu\left(5 \frac{\eta}{s}\right)^{-1}T\ f^{(0)} \ \ .\label{eq:isoC0}
\end{align}
This section discusses how four of the integrals can be computed analytically for each of these terms. Many of the angular integrations will again take the forms of the integral formulae given in the beginning of Appendix~\ref{IntegralsLandauMatching}.
We will start with the moments $V_{m0}$ as they are independent of the anisotropic perturbation.

Exact expressions for the local theormodynamic quantities $T$, $u^\mu$ can be computed from the components of $T^{\mu\nu}$ that are discussed in Appendix~\ref{IntegralsLandauMatching} according to the formulae derived in Section~\ref{Landau matching}. In terms of $b=\frac{x_\perp\,\Delta\tau}{R^2}$, they read
\begin{align}
    T &=R^{-1}\left(\frac{1}{\pi}\left(\frac{\pi^2}{30}\nu_{\rm eff}\right)^{-1}\epn R\right)^{1/4} \left(\frac{R}{\tau}\right)^{1/4} \mathrm{exp}\left(-\frac{x_\perp^2+\Delta\tau^2}{4R^2}\right)\left[I_0(b)-\beta I_0'(b)\right]^{1/4} \ \ ,\\
    u^\mu&=\gamma(1,\beta\hxT,0) \ \ , \ \ \gamma=(1-\beta^2)^{-1/2} \ \ ,\\
    \beta&=\frac{I_0(b)}{I_1(b)}-\frac{1}{2b}- \sqrt{\left[\frac{I_0(b)}{I_1(b)}-\frac{1}{2b}\right]^2-1} \ \ .
\end{align}
Looking at the expression for $T$, it is immediately apparent that its dimensionless constant prefactor together with $\left(5 \frac{\eta}{s}\right)^{-1}$ constitutes a factor of $\hat{\gamma}$ in $C[f^{(0)}]$, as we have predicted in Section~\ref{sec:theory:scaling_properties}. We can immediately also compute
\begin{align}
    p^\mu u_\mu &=\gamma p_\perp\left[\mathrm{cosh}(y-\eta)-\beta\,\cos(\phi_{\xT\pT})\right] \ \ .
\end{align}
Reminding also of the form of $f^{(0)}$
\begin{align}
    f^{(0)}(\tau,\xT,\pT,y-\eta)=\frac{(2\pi)^3}{\nu_{\rm eff}}\frac{\delta(y-\eta)}{\tau{p}_\perp}F\left(\frac{Q_s(\xT-\vT\Delta\tau)}{p_\perp}\right) \ \ ,
\end{align}
where $Q_s$ is fixed by the isotropic energy density according to (\ref{eq:epsilon-f-relation},\ref{eq:epsilon_isotropic_def},\ref{eq:defFana}) to be of the form
\begin{align}
    Q_s(\xT)=Q_{s,0}\,\mathrm{exp}\left(-\frac{x_\perp^2}{3R^2}\right) \ \ ,
\end{align}
we can compute $V_{mn}^{(1,0)}$ by plugging the above expressions into the integral formula (\ref{Vmn1}) for the part of the kernel given in (\ref{eq:isoC0}). Due to the fact that in both cases we integrate $f^{(0)}$, the integral over $p_\perp$ is analogous to the computation of the zeroth-order moments $V^{(0)}_{m0}$, where
\begin{align}
    V_{m0}^{(0)}&=\int \d^2p_\perp p_\perp^m \frac{\d N^{(0)}}{\d^2 p_\perp \d y}\\
    &=2\pi\int \d^2 x_\perp\int \d \eta \int_0^\infty \d p_\perp p_\perp^{m+1} \delta(y-\eta) F\left(\frac{Q_s(\xT-\vT \Delta \tau)}{p_\perp}\right)\\
    &=2\pi\int \d^2 x_\perp Q_s^{m+2}(\xT) \int_0^\infty \d k\, k^{m+1} F\left(\frac{1}{k}\right)\\
    &=4\pi^2\frac{3R^2}{m+2}Q_{s,0}^{m+2}\int_0^\infty \d k\, k^{m+1} F\left(\frac{1}{k}\right)\ \ .
\end{align}

We can therefore express our result for $V_{m0}^{(1,0)}$ in terms of these zeroth-order moments and find
\begin{align}
    V_{m0}^{(1,0)}(\tau)
    &=- V_{m0}^{(0)}\left(5 \frac{\eta}{s}\right)^{-1} \frac{m+2}{3R^2} \int_{\tau_0}^{\tau} \mathrm{d}\tau' \int_0^\infty \d x_\perp \ x_\perp T\,\gamma\, \mathrm{exp}\left[-\frac{(m+2)(\Delta\tau'^2+x_\perp^2)}{3R^2}\right] \\
    &\quad\times\int_0^{2\pi} \frac{\d \phi_{\xT\pT}}{2\pi} \ \left[1-\beta\cos(\phi_{\xT\pT})\right] \mathrm{exp}\left(\frac{2(m+2)b\, \cos(\phi_{\xT\pT})}{3}\right)\nonumber\\
    &=-V_{m0}^{(0)}\,\hat{\gamma}\,\mathcal{P}_m(\tilde\tau) \ \ ,\\
    \mathcal{P}_m(\tilde\tau) &=\frac{(m+2)}{3} \int_{\tilde\tau_0}^{\tilde\tau} \mathrm{d}\tilde{\tau} ' \int_0^\infty \d \tilde{x}_\perp \ \tilde{x}_\perp \tilde{T}\,\gamma\, \mathrm{exp}\left[-\frac{(m+2)}{3}(\Delta\tilde{\tau}'^2+\tilde{x}_\perp^2)\right] \left[I_0\left(\frac{2m+4}{3}b\right)-\beta I_0 ' \left(\frac{2m+4}{3}b\right)\right] \ \ .\label{eq:Pmtau}
\end{align}
where one has to keep in mind that in the integrand $b$, $\beta$, $\gamma$ and $T$ are to be understood as functions of $\tau'$ instead of $\tau$. In the last step, the result was rewritten into the tilded coordinates introduced in Section~\ref{sec:theory:scaling_properties} to make the parametric dependences more apparent.

For $V_{mn}^{(1,eq)}$ given by Eq. (\ref{Vmn1}) with the partial kernel (\ref{eq:isoCeq}), computing the moments of $f_{eq}$ via the $p_\perp$-integration yields
\begin{align}
    V_{m0}^{(1,eq)}(\tau)
     &=\left(5 \frac{\eta}{s}\right)^{-1}\frac{\nu_{\rm eff}}{(2\pi)^2}\Gamma (m+3) \ \zeta (m+3)\int_{\tau_0}^{\tau} \mathrm{d}\tau ' \int_0^\infty \d x_\perp \ x_\perp \tau'\, T^{m+4} \\
     &\quad \times\int_0^{2\pi} \d\phi_{\xT\pT} \int \d \eta \  \lbrace\gamma\left[\mathrm{cosh}(y-\eta)-\beta\cos(\phi_{\xT\pT})\right]\rbrace^{-m-2} \nonumber \\
     &=\left(5 \frac{\eta}{s}\right)^{-1}\frac{\nu_{\rm eff}}{2\pi^{1/2}}\, \Gamma (m+3)\ \zeta (m+3)\ \frac{\Gamma\left(\frac{m+2}{2}\right)}{\Gamma\left(\frac{m+3}{2}\right)} \int_{\tau_0}^{\tau} \mathrm{d}\tau' \int_0^\infty \d x_\perp \ x_\perp \tau'\, T^{m+4} \gamma^{-m-2}\\
     &\quad \times  \ _2F_1\left(\frac{m+2}{2},\frac{m+2}{2};1;{\beta^2}\right) \nonumber\\
     &=\hat{\gamma}\nu_{\rm eff}R^{-m}\left(\frac{1}{\pi}\nu_{\rm eff}^{-1}\epn R\right)^{(m+3)/4}\,\mathcal{Q}_m(\tilde\tau) \ \ , \label{eq:parametricVm01eq}\\
     \,\mathcal{Q}_m(\tilde\tau)&=\left(\frac{\pi^2}{30}\right)^{-(m+3)/4}\frac{1}{2\pi^{1/2}}\, \Gamma (m+3)\ \zeta (m+3)\ \frac{\Gamma\left(\frac{m+2}{2}\right)}{\Gamma\left(\frac{m+3}{2}\right)} \int_{\tilde\tau_0}^{\tilde\tau} \mathrm{d}\tilde{\tau}' \int_0^\infty \d \tilde{x}_\perp \ \tilde{x}_\perp \tilde{\tau}'\, \tilde{T}^{m+4} \gamma^{-m-2}\\
     &\quad \times  \ _2F_1\left(\frac{m+2}{2},\frac{m+2}{2};1;{\beta^2}\right) \nonumber\ \ .\label{eq:Qmtau}
\end{align}

After absorbing one of the $T$-prefactors into $\hat{\gamma}$, the parametric dependence of this term given by (\ref{eq:parametricVm01eq}). As the basic structure of the integrands is the same, moments with $n\neq0$ will have the same parametric dependences except for the additional anisotropy parameter $\delta_n$.\\

Now to compute the anisotropic corrections $\delta V^{(1)}_{mn}$ for $n\neq0$, we first need to derive the change $\delta C[f^{(0)}]$ in the kernel
\begin{align}
    C[f^{(0)}]=-p^\mu u_\mu\left(5 \frac{\eta}{s}\right)^{-1}T \left(f_{eq}-f^{(0)}\right)
\end{align}
due to the anisotropies, so we can plug it into (\ref{Vmn1}). $C[f^{(0)}]$ depends on three quantities that receive anisotropic corrections: $f^{(0)}$, $T$ and $u^\mu$. Linearization in the corrections will yield three different contributions. Separating the terms proportional to $f^{(0)}$ from those proportional to $f_{eq}$, we can split the kernel into the following two parts:
\begin{align}
    \delta C_{0}[ f^{(0)}]&=-\left(5 \frac{\eta}{s}\right)^{-1}p_\mu \left(u^\mu T\, \delta f^{(0)}+u^\mu \delta Tf^{(0)}+\delta u^\mu Tf^{(0)}\right)\label{eq:deltaC0}\\
    \delta C_{eq}[ f^{(0)}]&=\left(5 \frac{\eta}{s}\right)^{-1}p_\mu \left[\left(u^\mu\delta T+\delta u^\mu T\right)f_{eq}\left(\frac{p_\nu u^\nu}{T}\right)
     +\left(-u^\mu\delta T+\delta u^\mu T\right)\frac{p_\rho u^\rho }{T}f_{eq}'\left(\frac{p_\nu u^\nu}{T}\right)\right]\label{eq::deltaCeq}
\end{align}
We can compute the anisotropic contributions to thermodynamic quantities that show up in the kernel from the results for $\delta\epsilon$, $\delta u_t$ and $\delta u_s$ given in Eq.s (\ref{eq:deltaepsilonresult}-\ref{eq:deltausresult}). The change in temperature $\delta T$ can be computed by linearizing the equation of state $T=\left(\frac{\pi^2}{30}\nu_{\rm eff}\right)^{-1/4}\epsilon^{-1/4}$ in $\delta \epsilon$, and the contraction $\delta u_\mu p^\mu$ can be expressed in terms of $\delta u_t$ and $\delta u_s$.
\begin{align}
\delta T &= \frac{1}{4} \left(\frac{\pi^2}{30}\nu_{\rm eff}\right)^{-1/4} \epsilon^{-3/4} \delta\epsilon=\frac{1}{4} \left(\frac{\pi^2}{30}\nu_{\rm eff}\right)^{-1} T^{-3} \delta \epsilon \ \ ,\\
    p^\mu \delta u_\mu &= p^\mu(\delta u_t t_\mu + \delta u_s s_\mu)
    =\delta u_t p_\perp \gamma [\beta\,\mathrm{cosh}(y-\eta)-\cos(\phi_{\xT\pT})]-\delta u_s p_\perp \sin(\phi_{\xT\pT}) \ \ .
\end{align}

The only anisotropic quantity that we still need to derive is the form of the perturbation $\delta f^{(0)}$ due to the energy density perturbation $\delta\epsilon$. Given that
\begin{align}
    f^{(0)}(\tau,\xT,\pT,y-\eta)=\frac{(2\pi)^3}{\nu_{\rm eff}}\frac{\delta(y-\eta)}{\tau p_\perp} F\left(\frac{Q_s(\xT-\Delta\tau)}{p_\perp}\right) \ \ ,
\end{align}

the change in $f$ is due to the change in $Q_s$ which is directly related to $\epsilon$. More specifically,

\begin{align}
    \delta f (\tau,\xT,\pT,y-\eta)=\frac{(2\pi)^3}{\nu_{\rm eff}} \frac{\delta(y-\eta)}{\tau p_\perp} \frac{\delta Q_s(\xT-\Delta\tau\vT)}{p_\perp}F'\left(\frac{Q_s(\xT-\Delta\tau\vT)}{p_\perp}\right)
\end{align}
where

\begin{align}
    \delta Q_s(\xT)&=\frac{1}{3}Q_s(\xT)\frac{\delta\epsilon(\tau_0,\xT)}{\epsilon(\tau_0,x_\perp)}=\frac{1}{3}Q_s(\xT) \,\delta_n\, \mathrm{exp}\left(-\alpha\frac{x_\perp^2}{R^2}\right)\left(\frac{x_\perp}{R}\right)^n\cos(n\phi_{\xT\nT}) \ \ .
    \end{align}

Evaluating $\delta Q_s$ at $\xT-\Delta\tau\vT$ will thus yield as a factor the free-streamed version of $x_\perp^n\cos(n\phi_{\xT\nT})$ that was computed in appendix~\ref{sec:free-streaming anisotropies}.

We now want to compute the moments $V_{mn}^{(1,0)}$ for $n\neq 0$ by computing the integrals in Eq. (\ref{Vmn1}) for the part of the kernel perturbation given in Eq. (\ref{eq:deltaC0}). As in the isotropic case, we can simplify the integral expression by identifying $V_{m0}^{(0)}$. This holds true also for the term containing $\delta f^{(0)}$ instead of $f^{(0)}$, since
\begin{align}
    \int \d p_\perp p_\perp^m \,Q_s(\xT-\vT \Delta \tau)\, F'\left(\frac{Q_s(\xT-\vT \Delta \tau)}{p_\perp}\right)=(m+2)\int \d p_\perp p_\perp^{m+1}\, F\left(\frac{Q_s(\xT-\vT \Delta \tau)}{p_\perp}\right) \ \ .
    \end{align}

The angular integrals are of the same type as the ones in Appendix~\ref{IntegralsLandauMatching}, however each of the three perturbations has a slightly different angular dependence, so we will discuss them one by one. The $\delta f$-term is proportional to
\begin{align}
    &\int \d \phi_{\pT\nT}\int \d \phi_{\xT\pT} e^{in\phi_{\pT\nT}}e^{2\left(\frac{m+2}{3}+\alpha\right)b\,\cos(\phi_{\xT\pT})}(1-\beta\cos(\phi_{\xT\pT}))\sum_{j=0}^{n}\binom{n}{j}x_\perp^{n-j}(-\Delta\tau)^j\,\cos(n\phi_{\xT\nT}+j\phi{\xT\pT})\\
    &=2\pi\int \d \phi_{\pT\nT} e^{in\phi_{\xT\pT}} \cos(n\phi_{\pT\nT})\sum_{j=0}^{n}\binom{n}{j}x_\perp^{n-j}(-\Delta\tau)^j\,\left[I_j\left(\left(\frac{m+2}{3}+\alpha\right)b\right)-\beta I'_j\left(\left(\frac{m+2}{3}+\alpha\right)b\right)\right]\\
    &=2\pi^2\sum_{j=0}^{n}\binom{n}{j}x_\perp^{n-j}(-\Delta\tau)^j\,\left[I_j\left(\left(\frac{m+2}{3}+\alpha\right)b\right)-\beta I'_j\left(\left(\frac{m+2}{3}+\alpha\right)b\right)\right] \ \ .
    \end{align}
The $\delta T$-perturbation contains via $\delta \epsilon$ a factor of $\cos(n\phi_{\xT\nT})=\cos(n\phi_{\pT\nT})\cos(n\phi_{\xT\pT})-\sin(n\phi_{\pT\nT})\sin(n\phi_{\xT\pT})$. The term that is odd in $\phi_{\xT\pT}$ will vanish, while the other integrates to

\begin{align}
    &\int \d \phi_{\pT\nT}\int \d \phi_{\xT\pT} e^{in\phi_{\pT\nT}}e^{2\frac{m+2}{3}b\,\cos(\phi_{\xT\pT})}[1-\beta\,\cos(\phi_{\xT\pT})]\,\cos(n\phi_{\xT\nT})\\
    &=2\pi\int \d \phi_{\pT\nT} e^{in\phi_{\xT\pT}} \cos(n\phi_{\pT\nT})\left[I_n\left(\frac{m+2}{3}b\right)-\beta I'_n\left(\frac{m+2}{3}b\right)\right]\\
    &=2\pi^2\,\left[I_n\left(\frac{m+2}{3}b\right)-\beta I'_n\left(\frac{m+2}{3}b\right)\right] \ \ .
\end{align}

Lastly, the $\delta u^\mu$-perturbation is of the form $\delta u_t \gamma [\beta-\cos(\phi_{\xT\pT})]-\delta u_s \sin(\phi_{\xT\pT})$. The term containing 
$\delta u_t\propto\cos(n\phi_{\xT\nT})$ behaves exactly like before:

\begin{align}
    &\int \d \phi_{\pT\nT} e^{in\phi_{\pT\nT}} \int \d \phi_{\xT\pT} e^{2\frac{m+2}{3}b\,\cos(\phi_{\xT\pT})}[\beta-\cos(\phi_{\xT\pT})]\cos(n\phi_{\xT\nT})\\
    &=2\pi^2\left[\beta I_n\left(\frac{m+2}{3}b\right)- I'_n\left(\frac{m+2}{3}b\right)\right] \ \ ,
\end{align}
while the other term contains a factor of $-\sin(\phi_{\xT\pT}) \, \delta u_s$ which has the total angular dependence $-\sin(\phi_{\xT\pT})\sin(n\phi_{\xT\nT})=-\sin(\phi_{\xT\pT})[\cos(n\phi_{\pT\nT})\sin(n\phi_{\xT\pT})+\sin(n\phi_{\pT\nT})\cos(n\phi_{\xT\pT})]$, so in angular integration, the $\phi_{\xT\pT}$-even part becomes
\begin{align}
    &\int \d \phi_{\pT\nT} e^{in\phi_{\pT\nT}} \int \d \phi_{\xT\pT} e^{2\frac{m+2}{3}\,b\,\cos(\phi_{\xT\pT})}(-\sin(\phi_{\xT\pT}))\sin(n\phi_{\xT\nT})\\
    &=-2\pi^2\left[\beta I'_n\left(\frac{m+2}{3}b\right)- I_{n+1}\left(\frac{m+2}{3}b\right)\right] \ \ .
\end{align}

Putting all of this together, we can find the 2d integral expression for $\delta V^{(1,0)}_{mn}$:

\begin{align}
    \delta V^{(1,0)}_{mn}&=-V^{(0)}_{m0}\,\delta_n\,\hat{\gamma}\,\mathcal{P}_{mn}(\tilde{\tau}) \ \ ,\\
    \mathcal{P}_{mn}(\tilde{\tau})&=\frac{m+3}{6}
    \int_{\tilde{\tau}_0}^{\tilde{\tau}}\d\tilde{\tau}'\int_0^{\infty}\d \tilde{x}_\perp \tilde{x}_\perp \gamma\, \tilde{T}\, 
    \mathrm{exp}\left[-\left(\frac{m+2}{3}+\alpha\right)(\tilde{x}_\perp^2+\Delta\tilde{\tau}'^2)\right]\label{eq:Pmntau}\\
    &\times\sum_{j=0}^{n}\binom{n}{j}\tilde{x}_\perp^{n-j}(-\Delta\tilde {\tau}')^j\left\lbrace\frac{m+2}{3}\left[I_j\left(\left(\frac{m+2}{3}+\alpha\right)b\right)-\beta I'_j\left(\left(\frac{m+2}{3}+\alpha\right)b\right)\right]\right.\nonumber\\
    &\quad +\frac{1}{4} \frac{1}{\tilde{\tau}'} \gamma^2 \tilde{T}^{-4}
    \left[I_n\left(\frac{m+2}{3}b\right)-\beta I'_n\left(\frac{m+2}{3}b\right)\right]
    [I_j(2\bar{\alpha}b)-2\beta I'_j(2\bar{\alpha}b)+\beta^2I''_j(2\bar{\alpha}b)]\nonumber\\
    &\quad +\left\lbrace\gamma^2\left[\left(2\beta+\frac{1}{2b}\right)I_1(2b)-2I_0(2b)\right]^{-1}
    [\beta I_j(2\bar{\alpha}b)-(1+\beta^2) I'_{j}(2\bar{\alpha}b)+\beta I''_j(2\bar{\alpha}b)]\right.\nonumber\\
    &\quad\times\left[\beta I_n\left(\frac{m+2}{3}b\right)- I'_n\left(\frac{m+2}{3}b\right)\right]-\left[\left(\beta-\frac{1}{2b}\right)I_1(2b)-I_0(2b)\right]^{-1}\nonumber\\
    &\quad \left.\left.\times[I'_j(2\bar{\alpha}b)- I_{j+1}(2\bar{\alpha}b)-\beta I''_j(2\bar{\alpha}b)+\beta I'_{j+1}(2\bar{\alpha}b)]\left[\beta I'_n\left(\frac{m+2}{3}b\right)- I_{n+1}\left(\frac{m+2}{3}b\right)\right]\right\rbrace\right\rbrace\nonumber
\end{align}

Next, we will compute $\delta V_{mn}^{(1,eq)}$ by plugging (\ref{eq::deltaCeq}) into (\ref{Vmn1}). Again, the most straightforward integration is the one over $p_\perp$, which equates to taking moments of $f_{eq}$. Terms containing $f_{eq}'$ can be cast into the same form as the others by partial integration, which yields 
\begin{align}
    \int \d p_\perp\,p_\perp^{m+2}\frac{p_\mu u^\mu}{T}\,f_{eq}'\left(\frac{p_\nu u^\nu}{T}\right)=-(m+3)\int \d p_\perp\,p_\perp^{m+2} \,f_{eq}\left(\frac{p_\nu u^\nu}{T}\right) \ \ .
\end{align}

To compute the angular integrals, as in the computation of $\delta V_{mn}^{(1,0)}$ we can rewrite the $\phi_{\xT\nT}$-dependence of $\delta \epsilon$, $\delta u_t$ and $\delta u_s$ into a dependence on $\phi_{\pT\nT}$ and $\phi_{\xT\pT}$, which makes the $\phi_{\pT\nT}$-integration trivial. However, the next step will be the trickiest one of this entire section, as the integrals over $\phi_{\xT\pT}$ and $\eta$ are highly non-trivial. The integrals that need to be computed for the different anisotropic correction terms are:
\begin{align}
   &\int \d \eta \int \d \phi_{\xT\pT}\,\left(\frac{p_\mu u^\mu}{p_\perp}\right)^{-m-2}\, \delta \epsilon \ \ ,\\
    &\int \d \eta \int \d \phi_{\xT\pT}\,\left(\frac{p_\mu u^\mu}{p_\perp}\right)^{-m-3}\, \delta u_t \, [\beta\mathrm{cosh}(y-\eta)-\cos(\phi_{\xT\pT})] \ \ ,\\
    &\int \d \eta \int \d \phi_{\xT\pT}\,\left(\frac{p_\mu u^\mu}{p_\perp}\right)^{-m-3}\, \delta u_s \,\sin(\phi_{\xT\pT}) \ \ .
\end{align}
Getting rid of all prefactors that do not depend on the integration variables, this amounts to computing the following integrals:

\begin{align}
     G_{\epsilon}(n,m,\beta)&=\int \d \eta \int \d \phi_{\xT\pT}\, [\mathrm{cosh}(y-\eta)-\beta\cos(\phi_{\xT\pT})]^{-m-2}\,\cos(n\phi_{\xT\pT}) \ \ ,\label{eeqphietaint}\\
     G_{u_t}(n,m,\beta)&=\int \d \eta \int \d \phi_{\xT\pT}\, [\mathrm{cosh}(y-\eta)-\beta\cos(\phi_{\xT\pT})]^{-m-3}\,[\beta\mathrm{cosh}(y-\eta)-\cos(\phi_{\xT\pT})]\,\cos(n\phi_{\xT\pT}) \ \ ,\\
     G_{u_s}(n,m,\beta)&=\int \d \eta \int \d \phi_{\xT\pT}, [\mathrm{cosh}(y-\eta)-\beta\cos(\phi_{\xT\pT})]^{-m-3}\,\sin(\phi_{\xT\pT})\,\sin(n\phi_{\xT\pT}) \ \ .\label{useqphietaint}
\end{align}
We have defined these integrals as the functions $G_{X}(n,m,\beta)$ to abbreviate the notation of our results. To compute them, we rewrite again $\sin(n\phi_{\xT\pT})$ and $\cos(n\phi_{\xT\pT})$ into Chebyshev polynomials as we did in Appendix~\ref{sec:free-streaming anisotropies}. Since the polynomial expression for $\sin(n\phi_{\xT\pT})$ also contains a factor of $\sin(\phi_{\xT\pT})$ which together with the in (\ref{useqphietaint}) already present sine combines to $1-\cos^2(\phi_{\xT\pT})$, now only different powers of $\cos(\phi_{\xT\pT})$ without any sines occur in the integrand, which can be integrated analytically as follows:

\begin{align}
    &\int \d \eta \int \d \phi_{\xT\pT} [\mathrm{cosh}(y-\eta)-\beta\cos(\phi_{\xT\pT})]^{-m-2}\,\cos^{l}(\phi_{\xT\pT})\\
    &=4\int_1^\infty \d x \int_{-1}^{1} \d y \frac{ [x-\beta y)]^{-m-2}y^{l}}{\sqrt{1+x^2}\sqrt{1-y^2}}\\
    &=\begin{cases}
    \frac{2\pi\Gamma\left(\frac{m+2}{2}\right)\Gamma\left(\frac{l+1}{2}\right)}{\Gamma\left(\frac{m+3}{2}\right)\Gamma\left(\frac{l+2}{2}\right)} \ _3F_2\left(\frac{m+2}{2},\frac{m+2}{2},\frac{l+1}{2};\frac{1}{2},\frac{l+2}{2};\beta^2\right)\ \ &, \ \mathrm{l\ even}\\
    \frac{4\pi\beta\Gamma\left(\frac{m+3}{2}\right)\Gamma\left(\frac{l+2}{2}\right)}{\Gamma\left(\frac{m+2}{2}\right)\Gamma\left(\frac{l+3}{2}\right)} \ _3F_2\left(\frac{m+3}{2},\frac{m+3}{2},\frac{l+2}{2};\frac{3}{2},\frac{l+3}{2};\beta^2\right)\ \ &, \ \mathrm{l\ odd}
    \end{cases} \ \ ,\\
     &\int \d \eta \int \d \phi_{\xT\pT} [\mathrm{cosh}(y-\eta)-\beta\cos(\phi_{\xT\pT})]^{-m-3}\,\cos^{l}(\phi_{\xT\pT})\,\mathrm{cosh}(y-\eta)\\
    &=4\int_1^\infty \d x \int_{-1}^{1} \d y \frac{ [x-\beta y)]^{-m-3}y^{l}x}{\sqrt{1+x^2}\sqrt{1-y^2}}\\
    &=\begin{cases}
    \frac{2\pi\Gamma\left(\frac{m+2}{2}\right)\Gamma\left(\frac{l+1}{2}\right)}{\Gamma\left(\frac{m+3}{2}\right)\Gamma\left(\frac{l+2}{2}\right)} \ _3F_2\left(\frac{m+2}{2},\frac{m+4}{2},\frac{l+1}{2};\frac{1}{2},\frac{l+2}{2};\beta^2\right)\ \ &, \ \mathrm{l\ even}\\
    \frac{4\pi\beta\Gamma\left(\frac{m+5}{2}\right)\Gamma\left(\frac{l+2}{2}\right)}{\Gamma\left(\frac{m+4}{2}\right)\Gamma\left(\frac{l+3}{2}\right)} \ _3F_2\left(\frac{m+3}{2},\frac{m+5}{2},\frac{l+2}{2};\frac{3}{2},\frac{l+3}{2};\beta^2\right)\ \ &, \ \mathrm{l\ odd}
    \end{cases}\ \ .
\end{align}
To simplify these expressions, we can make use of the following property of the $\Gamma$-function:
\begin{align}
    \frac{\Gamma\left(n+\frac{1}{2}\right)}{\Gamma\left(n+1\right)}=\frac{(2n)!}{4^n (n!)^2}\sqrt{\pi}
\end{align}
Then one finds for the integrals (\ref{eeqphietaint})-(\ref{useqphietaint}):
\begin{align}
    &G_{\epsilon}(n,m,\beta)\nonumber\\
    &=\int \d \eta \int \d \phi [\mathrm{cosh}(y-\eta)-\beta\cos(\phi)]^{-m-2}\frac{n}{2}\sum_{k=0}^{\left\lfloor \frac{n}{2}\right\rfloor}(-1)^k\frac{(n-k-1)!}{(n-2k)!k!}\,2^{n-2k}\,\cos^{n-2k}(\phi)\nonumber\\
    &=\pi^{3/2}n\begin{cases}
    \frac{\Gamma\left(\frac{m+2}{2}\right)}{\Gamma\left(\frac{m+3}{2}\right)}\sum_{k=0}^{\left\lfloor \frac{n}{2}\right\rfloor}(-1)^k \frac{(n-k-1)!}{\left(\frac{n-2k}{2}!\right)^2k!}  \ _3F_2\left(\frac{m+2}{2},\frac{m+2}{2},\frac{n-2k+1}{2};\frac{1}{2},\frac{n-2k+2}{2};\beta^2\right)\ \ &, \ \mathrm{n\ even}\\
    \beta\frac{\Gamma\left(\frac{m+2}{2}\right)}{\Gamma\left(\frac{m+3}{2}\right)}\sum_{k=0}^{\left\lfloor \frac{n}{2}\right\rfloor}(-1)^k \frac{(n-k-1)!}{\left(\frac{n-2k-1}{2}!\right)^2k!}  \ _3F_2\left(\frac{m+3}{2},\frac{m+3}{2},\frac{n-2k+2}{2};\frac{3}{2},\frac{n-2k+3}{2};\beta^2\right)\ \ &, \  \mathrm{n\ odd}
    \end{cases} \ \ ,\\
    &G_{u_t}(n,m,\beta)\nonumber\\
    &=\int \d \eta \int \d \phi [\mathrm{cosh}(y-\eta)-\beta\cos(\phi)]^{-m-3}\sum_{k=0}^{\left\lfloor \frac{n-1}{2}\right\rfloor}(-1)^k\binom{n-k-1}{k}\,2^{n-2k-1}\,\left[\cos^{n-2k-1}(\phi)-\cos^{n-2k+1}(\phi)\right]\nonumber\\
    &=2\pi^{3/2}\begin{cases}
    \frac{\Gamma\left(\frac{m+3}{2}\right)}{\Gamma\left(\frac{m+4}{2}\right)}\sum_{k=0}^{\left\lfloor \frac{n-1}{2}\right\rfloor}(-1)^k \frac{(n-k-1)!}{\left(\frac{n-2k-1}{2}!\right)^2k!} \left[\ _3F_2\left(\frac{m+3}{2},\frac{m+3}{2},\frac{n-2k}{2};\frac{1}{2},\frac{n-2k+1}{2};\beta^2\right) \right.\\
    \quad \quad \quad \quad \left. + \frac{n-2k}{n-2k+1} \ _3F_2\left(\frac{m+3}{2},\frac{m+3}{2},\frac{n-2k+2}{2};\frac{1}{2},\frac{n-2k+3}{2};\beta^2\right)\right]\ \ &, \ \mathrm{n\ odd}\\
    \beta\frac{\Gamma\left(\frac{m+4}{2}\right)}{\Gamma\left(\frac{m+3}{2}\right)}\sum_{k=0}^{\left\lfloor \frac{n-1}{2}\right\rfloor}(-1)^k \frac{(n-k-1)!}{\left(\frac{n-2k-2}{2}!\right)^2k!} \left[\ _3F_2\left(\frac{m+4}{2},\frac{m+4}{2},\frac{n-2k+1}{2};\frac{3}{2},\frac{n-2k+2}{2};\beta^2\right) \right.\\
    \quad \quad \quad \quad \left.+ \frac{n-2k+1}{n-2k+2} \ _3F_2\left(\frac{m+4}{2},\frac{m+4}{2},\frac{n-2k+3}{2};\frac{3}{2},\frac{n-2k+4}{2};\beta^2\right)\right]\ \ &, \  \mathrm{n\ even}
    \end{cases} \ \ ,\\
    &G_{u_s}(n,m,\beta)\nonumber\\
    &=\int \d \eta \int \d \phi [\mathrm{cosh}(y-\eta)-\beta\cos(\phi)]^{-m-3}\frac{n}{2}\sum_{k=0}^{\left\lfloor \frac{n}{2}\right\rfloor}(-1)^k\frac{(n-k-1)!}{(n-2k)!k!}\,2^{n-2k}\,\left[\beta\cos^{n-2k}(\phi)\mathrm{cosh}(y-\eta)-\cos^{n-2k+1}(\phi)\right]\nonumber\\
    &=\pi^{3/2}n\begin{cases}
    \sum_{k=0}^{\left\lfloor \frac{n}{2}\right\rfloor}(-1)^k \frac{(n-k-1)!}{\left(\frac{n-2k}{2}!\right)^2k!} \left[\beta \frac{\Gamma\left(\frac{m+2}{2}\right)}{\Gamma\left(\frac{m+3}{2}\right)} \ _3F_2\left(\frac{m+2}{2},\frac{m+4}{2},\frac{n-2k+1}{2};\frac{1}{2},\frac{n-2k+2}{2};\beta^2\right) \right.\\
    \quad \quad \quad \quad \left. -  2\beta \frac{\Gamma\left(\frac{m+4}{2}\right)}{\Gamma\left(\frac{m+3}{2}\right)}\frac{n-2k+1}{n-2k+2} \ _3F_2\left(\frac{m+4}{2},\frac{m+4}{2},\frac{n-2k+3}{2};\frac{3}{2},\frac{n-2k+4}{2};\beta^2\right)  \right]\ \ &, \ \mathrm{n\ even}\\
    \sum_{k=0}^{\left\lfloor \frac{n}{2}\right\rfloor}(-1)^k \frac{(n-k-1)!}{\left(\frac{n-2k}{2}!\right)^2k!} \left[\beta^2 \frac{\Gamma\left(\frac{m+5}{2}\right)}{\Gamma\left(\frac{m+4}{2}\right)} \ _3F_2\left(\frac{m+3}{2},\frac{m+5}{2},\frac{n-2k+2}{2};\frac{3}{2},\frac{n-2k+3}{2};\beta^2\right) \right.\\
    \quad \quad \quad \quad \left. -  \beta \frac{\Gamma\left(\frac{m+3}{2}\right)}{2\Gamma\left(\frac{m+4}{2}\right)} \ _3F_2\left(\frac{m+3}{2},\frac{m+3}{2},\frac{n-2k+2}{2};\frac{1}{2},\frac{n-2k+3}{2};\beta^2\right)  \right]\ \ &, \  \mathrm{n\ odd}
    \end{cases} \ \ .
\end{align}

The final step to computing the total expression for $\delta V_{mn}^{(1,eq)}$ is a bookkeeping task of combining all the above integration steps, at the end of which one acquires
\begin{align}
    \delta V_{mn}^{(1,eq)}&=\hat{\gamma}\delta_n\nu_{\rm eff}R^{-m}\left(\frac{1}{\pi}\nu_{\rm eff}^{-1}\epn R\right)^{(m+3)/4}\mathcal{Q}_{mn}(\tilde{\tau}) \ \ , \\
    \mathcal{Q}_{mn}(\tilde{\tau})&=\left(\frac{\pi^2}{30}\right)^{-(m+3)/4}\,\frac{1}{8\pi^2}\zeta(m+3)\int_{\tilde{\tau}_0}^{\tilde{\tau}} \d \tilde{\tau}' \int_0^\infty \d \tilde{x}_\perp \tilde{x}_\perp \gamma^{-m-2}\, \tilde{T}^m\,\mathrm{exp}\left[-\bar{\alpha}\left(\tilde{x}_\perp^2+\Delta\tilde{\tau}'^2\right)\right]\label{eq:Qmntau}\\ &\times\sum_{j=0}^{n}\binom{n}{j}\tilde{x}_\perp^{n-j}(-\Delta\tilde{\tau}')^j\,\left\lbrace\frac{1}{4}[\Gamma(m+3)+\Gamma(m+4)]\,\gamma^2\,[I_j(2\bar{\alpha}b)-2\beta I'_j(2\bar{\alpha}b)+\beta^2I''_j(2\bar{\alpha}b)]\, G_\epsilon(n,m,\beta)\right.\nonumber\\
    &\quad+[\Gamma(m+3)-\Gamma(m+4)]\,\tilde{\tau'}\,\tilde{T}^{4}\,\left\lbrace\gamma^2\left[\left(2\beta+\frac{1}{2b}\right)I_1(2b)-2I_0(2b)\right]^{-1}\right.\nonumber\\
    &\quad\quad\times[\beta I_j(2\bar{\alpha}b)- (1+\beta^2)I'_{j}(2\bar{\alpha}b)+\beta I''_j(2\bar{\alpha}b)]\,G_{u_t}(n,m,\beta)\nonumber\\
    &\quad \quad\left.\left.-\left[\left(\beta-\frac{1}{2b}\right)I_1(2b)-I_0(2b)\right]^{-1}[ I'_j(2\bar{\alpha}b)- I_{j+1}(2\bar{\alpha}b)-\beta I''_j(2\bar{\alpha}b)+\beta I'_{j+1}(2\bar{\alpha}b)]\,G_{u_s}(n,m,\beta)\right\rbrace\right\rbrace \ \ .\nonumber
\end{align}

\section{Equilibrium moments of the numerical setup}\label{appendix equilibrium moments}

In this appendix, the equilibrium moments $E_l^m$ emerging in the time evolution equations for the moments $C_l^m$ as derived in Section~\ref{sec:MomentumMomentExpansion} are computed. Since taking the integral $\int \d p^\tau (p^\tau)^3$ of the equilibrium distribution will yield the energy density, the expression simplifies in spherical coordinates.

\begin{align}
    E_l^m&=\int\frac{\d^2p_\perp}{(2\pi)^2}\int\frac{\d p_\eta}{2\pi}\,Y_l^m(\theta_p,\phi_p)\,p^\mu u_\mu \, f_{eq}\\
    &=\tau \int_0^\infty \d p^\tau (p^\tau)^3 \int_0^{2\pi} \frac{\d \phi_p}{2\pi} \int \frac{\d \cos\theta_p}{2} \, Y_l^m(\theta_p,\phi_p) \frac{1}{2\pi^2} v^\mu u_\mu f_{eq}\left(\frac{p^\mu u_\mu}{T}\right)\\
    &=\tau \epsilon \int_0^{2\pi} \frac{\d \phi_p}{2\pi} \int \frac{\d \cos\theta_p}{2}\,  Y_l^m(\theta_p,\phi_p)\, (v^\mu u_\mu)^{-3}
\end{align}

In this calculation, we have defined $v^\mu=p^\mu/p^\tau$. To compute the angular integral, we write 
\begin{align}
    v^\mu u_\mu=\gamma\,(1-\vec{\beta}\cdot\vec{v})=\gamma\,(1-\beta\cos\theta_{up})
\end{align}
and express the spherical harmonics in a rotated coordinate system, thus writing
\begin{align}
    Y_l^m(\theta_p,\phi_p)=\sum_{m'=-l}^{l} \left(D_{mm'}^{l}\right)^* Y_l^{m'}(\theta_{up},\phi_{up}) \ \ ,
\end{align}
where the Wigner D-matrix depends on the angles involved in the rotation from $(\theta_{up},\phi_{up})$ to $(\theta_p,\phi_p)$. In these coordinates, the $\phi_{up}$-integral becomes trivial, thus only an integral of the Legendre polynomials remains to be computed.

\begin{align}
     &\int_0^{2\pi} \frac{\d \phi_{up}}{2\pi} \int \frac{\d \cos\theta_{up}}{2}\,  Y_l^{m'}(\theta_{up},\phi_{up})\, \gamma^{-3}\,(1-\beta\,\cos\theta_{up})^{-3}=\delta^{m'0} y_l^0 \int \d x \frac{P_l(x)}{2\gamma^3(1-\beta x)^3}
\end{align}

For the case $m'=0$, the Wigner D-matrix simplifies to
\begin{align}
    \left(D_{m0}^{l}\right)^*=\sqrt{\frac{4\pi}{2l+1}}Y_l^m(\theta_{rot},\phi_{rot}) \ \ .
\end{align}
Since $\Vec{u}$ lies in the transverse plane, where its orientation is given by $\phi_u$, we can identify the rotation angles to be $\theta_{rot}=\frac{\pi}{2}$ and $\phi_{rot}=\phi_u$, which yields
\begin{align}
    E_l^m&=\tau\epsilon\, Y_l^m\left(\frac{\pi}{2},\phi_u\right)\int_{-1}^1 \d x\, \frac{P_l(x)}{2\gamma^3(1-\beta x)^{3}} \ \ .
\end{align}

Finally, the remaining integral can be solved analytically:

\begin{align}
    \int_{-1}^1 \d x \frac{P_l(x)}{2\gamma^3(1-\beta x)^3}=2^{-l-2} \pi^{1/2} \frac{\Gamma\left(l+3\right)}{\Gamma\left(l+\frac{3}{2}\right)}\gamma^{-3} \beta^l\ _2F_1\left(\frac{l+4}{2},\frac{l+3}{2};l+\frac{3}{2};\beta^2\right) \ \ .
\end{align}

\end{widetext}

\section{Early and intermediate time cooling based on $0+1$-D Bjorken attractor} \label{app:attractor}

\begin{figure}
\centering
\begin{tabular}{c}
 \includegraphics[width=\linewidth]{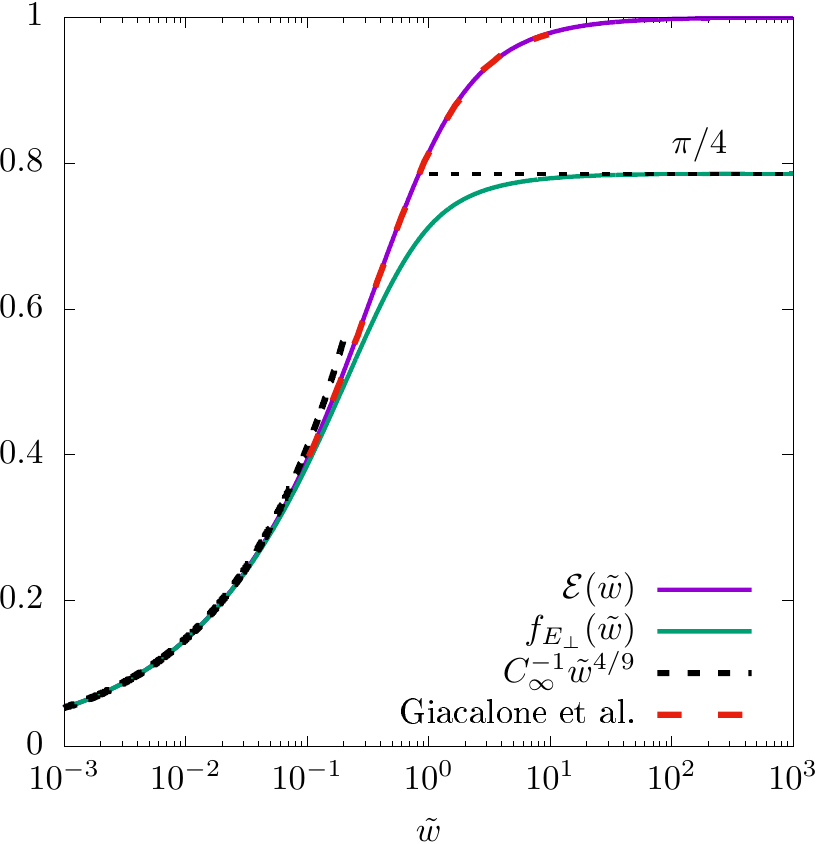} 
\end{tabular}
\caption{Universal functions $\mathcal{E}(\tilde{w})$ and 
$f_{E_\perp}(\tilde{w})$ obtained using the RLB method, 
represented with respect to the 
conformal parameter $\tilde{w} = \tau T / (4\pi \eta/s)$ 
for the $0+1$-D Bjorken flow.
 The red dashed curve shows the results for the 
RTA attractor reported in Fig.~1 of Ref.~\cite{Giacalone:2019ldn}.
}
\label{fig:universal}
\end{figure}
Below we describe the procedure
employed to perform the integration in Eq.~\eqref{eq:dEdyBjorkenApprox}, 
which provides the {\it Bjorken scaling} curve in Fig.~\ref{fig:dEdy}(a). 
The main ingredients that we require are the 
universal functions $\mathcal{E}(\tilde{w})$ and 
$f_{E_\perp}(\tilde{w})$. These are determined by 
performing a $0+1$-D simulation (i.e., for a system 
which is homogeneous with respect to the transverse 
plane) using the RLB method described in
Sec.~\ref{NumericalProcedure:RLB}. The initial time and
temperature were set to $\tau_0 = 10^{-4}\ {\rm fm}$ and 
$T_0 = 0.315\ {\rm GeV}$, while $4\pi \eta / s = 1$, 
giving $\tilde{w}_0 \simeq 1.6 \times 10^{-4}$. The 
initial distribution was taken to be of Romatschke-Strickland 
form \eqref{eq:RS} with anisotropy parameter 
$\xi_0 = 1000$, corresponding to an initial pressure 
ratio $\mathcal{P}_L / \mathcal{P}_T \simeq 0.002$. The
simulation was ran until $\tau/\tau_0 = 10^{10}$ 
or $\tilde{w} \simeq 1892$. During the 
simulation, the energy density and 
$dE_\perp / d^2\xT d\eta$ are computed and the 
universal functions $\mathcal{E}$ and $f_{E_\perp}$ are 
obtained using
\begin{align}
 C_\infty \mathcal{E}(\tilde{w}) =&
 \frac{\epsilon(\tilde{w})}{\epsilon_0}
 \left(\frac{\tau}{\tau_0}\right)^{4/3}
 \tilde{w}_0^{4/9},\nonumber\\
 C_\infty f_{E_\perp}(\tilde{w}) =&
 \frac{\tau_0}{\tau \epsilon_0} 
 \frac{dE}{d^2\xT d\eta} \left(\frac{\tau}{\tau_0}\right)^{4/3} \tilde{w}_0^{4/9}.
\end{align}
and the result are presented in Fig.~\ref{fig:universal}. 
 For completeness,
we provide a comparison with the results for $\mathcal{E}(\tilde{w})$ 
reported as ``Boltzmann RTA'' in Fig.~1 of Ref.~\cite{Giacalone:2019ldn}, 
which are shown using the red dashed line. The $C_\infty^{-1} \tilde{w}^{4/9}$
limit valid at small values of $\tilde{w}$ is shown as the black dotted line.

Next, in order to perform the integrals in Eq.~\eqref{eq:dEdyBjorkenApprox}, the top end of 
the integration $\tilde{w}(\tau, \xT = 0)$ must be 
found by numerically inverting Eq.~\eqref{eq:w_center}.
Considering the range $10^{-5} \le \tau/R \le 1$ and 
$2 \le \hat{\gamma} \le 400$, the minimum and 
maximum values of $\tilde{w}(\tau,\xT = 0)$ encountered 
are $1.4 \times 10^{-4}$ and $88$, corresponding to 
$(\tau / R, \hat{\gamma}) = (10^{-5}, 2)$ and $(1, 400)$,
respectively. In order to avoid ``boundary effects'' due to 
our choice of initial conditions, we considered 
the numerical data only for 
$\tilde{w} \gtrsim 3.4 \times 10^{-4}$, while for 
smaller values of $\tilde{w}$, we employed the 
analytical limits in Eqs.~\eqref{eq:universal_E_lim} 
and \eqref{eq:universal_fEp_lim}, namely 
$\mathcal{E}, f_{E_\perp} \simeq C_\infty^{-1} \tilde{w}^{4/9}$.

\end{appendix}
\bibliography{references.bib}

%merlin.mbs apsrev4-1.bst 2010-07-25 4.21a (PWD, AO, DPC) hacked
%Control: key (0)
%Control: author (8) initials jnrlst
%Control: editor formatted (1) identically to author
%Control: production of article title (-1) disabled
%Control: page (0) single
%Control: year (1) truncated
%Control: production of eprint (0) enabled
\begin{thebibliography}{103}%
\makeatletter
\providecommand \@ifxundefined [1]{%
 \@ifx{#1\undefined}
}%
\providecommand \@ifnum [1]{%
 \ifnum #1\expandafter \@firstoftwo
 \else \expandafter \@secondoftwo
 \fi
}%
\providecommand \@ifx [1]{%
 \ifx #1\expandafter \@firstoftwo
 \else \expandafter \@secondoftwo
 \fi
}%
\providecommand \natexlab [1]{#1}%
\providecommand \enquote  [1]{``#1''}%
\providecommand \bibnamefont  [1]{#1}%
\providecommand \bibfnamefont [1]{#1}%
\providecommand \citenamefont [1]{#1}%
\providecommand \href@noop [0]{\@secondoftwo}%
\providecommand \href [0]{\begingroup \@sanitize@url \@href}%
\providecommand \@href[1]{\@@startlink{#1}\@@href}%
\providecommand \@@href[1]{\endgroup#1\@@endlink}%
\providecommand \@sanitize@url [0]{\catcode `\\12\catcode `\$12\catcode
  `\&12\catcode `\#12\catcode `\^12\catcode `\_12\catcode `\%12\relax}%
\providecommand \@@startlink[1]{}%
\providecommand \@@endlink[0]{}%
\providecommand \url  [0]{\begingroup\@sanitize@url \@url }%
\providecommand \@url [1]{\endgroup\@href {#1}{\urlprefix }}%
\providecommand \urlprefix  [0]{URL }%
\providecommand \Eprint [0]{\href }%
\providecommand \doibase [0]{http://dx.doi.org/}%
\providecommand \selectlanguage [0]{\@gobble}%
\providecommand \bibinfo  [0]{\@secondoftwo}%
\providecommand \bibfield  [0]{\@secondoftwo}%
\providecommand \translation [1]{[#1]}%
\providecommand \BibitemOpen [0]{}%
\providecommand \bibitemStop [0]{}%
\providecommand \bibitemNoStop [0]{.\EOS\space}%
\providecommand \EOS [0]{\spacefactor3000\relax}%
\providecommand \BibitemShut  [1]{\csname bibitem#1\endcsname}%
\let\auto@bib@innerbib\@empty
%</preamble>
\bibitem [{\citenamefont {Teaney}(2010)}]{Teaney:2009qa}%
  \BibitemOpen
  \bibfield  {author} {\bibinfo {author} {\bibfnamefont {D.~A.}\ \bibnamefont
  {Teaney}},\ }\href {\doibase 10.1142/9789814293297_0004} {\enquote {\bibinfo
  {title} {{Viscous Hydrodynamics and the Quark Gluon Plasma}},}\ } (\bibinfo
  {year} {2010}),\ \Eprint {http://arxiv.org/abs/0905.2433} {arXiv:0905.2433
  [nucl-th]} \BibitemShut {NoStop}%
\bibitem [{\citenamefont {Gale}\ \emph {et~al.}(2013)\citenamefont {Gale},
  \citenamefont {Jeon},\ and\ \citenamefont {Schenke}}]{Gale:2013da}%
  \BibitemOpen
  \bibfield  {author} {\bibinfo {author} {\bibfnamefont {C.}~\bibnamefont
  {Gale}}, \bibinfo {author} {\bibfnamefont {S.}~\bibnamefont {Jeon}}, \ and\
  \bibinfo {author} {\bibfnamefont {B.}~\bibnamefont {Schenke}},\ }\href
  {\doibase 10.1142/S0217751X13400113} {\bibfield  {journal} {\bibinfo
  {journal} {Int. J. Mod. Phys. A}\ }\textbf {\bibinfo {volume} {28}},\
  \bibinfo {pages} {1340011} (\bibinfo {year} {2013})},\ \Eprint
  {http://arxiv.org/abs/1301.5893} {arXiv:1301.5893 [nucl-th]} \BibitemShut
  {NoStop}%
\bibitem [{\citenamefont {Luzum}\ and\ \citenamefont
  {Petersen}(2014)}]{Luzum:2013yya}%
  \BibitemOpen
  \bibfield  {author} {\bibinfo {author} {\bibfnamefont {M.}~\bibnamefont
  {Luzum}}\ and\ \bibinfo {author} {\bibfnamefont {H.}~\bibnamefont
  {Petersen}},\ }\href {\doibase 10.1088/0954-3899/41/6/063102} {\bibfield
  {journal} {\bibinfo  {journal} {J. Phys. G}\ }\textbf {\bibinfo {volume}
  {41}},\ \bibinfo {pages} {063102} (\bibinfo {year} {2014})},\ \Eprint
  {http://arxiv.org/abs/1312.5503} {arXiv:1312.5503 [nucl-th]} \BibitemShut
  {NoStop}%
\bibitem [{\citenamefont {Heinz}\ and\ \citenamefont
  {Snellings}(2013)}]{Heinz:2013th}%
  \BibitemOpen
  \bibfield  {author} {\bibinfo {author} {\bibfnamefont {U.}~\bibnamefont
  {Heinz}}\ and\ \bibinfo {author} {\bibfnamefont {R.}~\bibnamefont
  {Snellings}},\ }\href {\doibase 10.1146/annurev-nucl-102212-170540}
  {\bibfield  {journal} {\bibinfo  {journal} {Ann. Rev. Nucl. Part. Sci.}\
  }\textbf {\bibinfo {volume} {63}},\ \bibinfo {pages} {123} (\bibinfo {year}
  {2013})},\ \Eprint {http://arxiv.org/abs/1301.2826} {arXiv:1301.2826
  [nucl-th]} \BibitemShut {NoStop}%
\bibitem [{\citenamefont {Jeon}\ and\ \citenamefont
  {Heinz}(2015)}]{Jeon:2015dfa}%
  \BibitemOpen
  \bibfield  {author} {\bibinfo {author} {\bibfnamefont {S.}~\bibnamefont
  {Jeon}}\ and\ \bibinfo {author} {\bibfnamefont {U.}~\bibnamefont {Heinz}},\
  }\href {\doibase 10.1142/S0218301315300106} {\bibfield  {journal} {\bibinfo
  {journal} {Int. J. Mod. Phys. E}\ }\textbf {\bibinfo {volume} {24}},\
  \bibinfo {pages} {1530010} (\bibinfo {year} {2015})},\ \Eprint
  {http://arxiv.org/abs/1503.03931} {arXiv:1503.03931 [hep-ph]} \BibitemShut
  {NoStop}%
\bibitem [{\citenamefont {Everett}\ \emph {et~al.}(2021)\citenamefont {Everett}
  \emph {et~al.}}]{JETSCAPE:2020mzn}%
  \BibitemOpen
  \bibfield  {author} {\bibinfo {author} {\bibfnamefont {D.}~\bibnamefont
  {Everett}} \emph {et~al.} (\bibinfo {collaboration} {JETSCAPE}),\ }\href
  {\doibase 10.1103/PhysRevC.103.054904} {\bibfield  {journal} {\bibinfo
  {journal} {Phys. Rev. C}\ }\textbf {\bibinfo {volume} {103}},\ \bibinfo
  {pages} {054904} (\bibinfo {year} {2021})},\ \Eprint
  {http://arxiv.org/abs/2011.01430} {arXiv:2011.01430 [hep-ph]} \BibitemShut
  {NoStop}%
\bibitem [{\citenamefont {Nijs}\ \emph {et~al.}(2021)\citenamefont {Nijs},
  \citenamefont {van~der Schee}, \citenamefont {G\"ursoy},\ and\ \citenamefont
  {Snellings}}]{Nijs:2020roc}%
  \BibitemOpen
  \bibfield  {author} {\bibinfo {author} {\bibfnamefont {G.}~\bibnamefont
  {Nijs}}, \bibinfo {author} {\bibfnamefont {W.}~\bibnamefont {van~der Schee}},
  \bibinfo {author} {\bibfnamefont {U.}~\bibnamefont {G\"ursoy}}, \ and\
  \bibinfo {author} {\bibfnamefont {R.}~\bibnamefont {Snellings}},\ }\href
  {\doibase 10.1103/PhysRevC.103.054909} {\bibfield  {journal} {\bibinfo
  {journal} {Phys. Rev. C}\ }\textbf {\bibinfo {volume} {103}},\ \bibinfo
  {pages} {054909} (\bibinfo {year} {2021})},\ \Eprint
  {http://arxiv.org/abs/2010.15134} {arXiv:2010.15134 [nucl-th]} \BibitemShut
  {NoStop}%
\bibitem [{\citenamefont {Gardim}\ \emph {et~al.}(2020)\citenamefont {Gardim},
  \citenamefont {Giacalone}, \citenamefont {Luzum},\ and\ \citenamefont
  {Ollitrault}}]{Gardim:2019xjs}%
  \BibitemOpen
  \bibfield  {author} {\bibinfo {author} {\bibfnamefont {F.~G.}\ \bibnamefont
  {Gardim}}, \bibinfo {author} {\bibfnamefont {G.}~\bibnamefont {Giacalone}},
  \bibinfo {author} {\bibfnamefont {M.}~\bibnamefont {Luzum}}, \ and\ \bibinfo
  {author} {\bibfnamefont {J.-Y.}\ \bibnamefont {Ollitrault}},\ }\href
  {\doibase 10.1038/s41567-020-0846-4} {\bibfield  {journal} {\bibinfo
  {journal} {Nature Phys.}\ }\textbf {\bibinfo {volume} {16}},\ \bibinfo
  {pages} {615} (\bibinfo {year} {2020})},\ \Eprint
  {http://arxiv.org/abs/1908.09728} {arXiv:1908.09728 [nucl-th]} \BibitemShut
  {NoStop}%
\bibitem [{\citenamefont {Schenke}\ \emph
  {et~al.}(2020{\natexlab{a}})\citenamefont {Schenke}, \citenamefont {Shen},\
  and\ \citenamefont {Tribedy}}]{Schenke:2020mbo}%
  \BibitemOpen
  \bibfield  {author} {\bibinfo {author} {\bibfnamefont {B.}~\bibnamefont
  {Schenke}}, \bibinfo {author} {\bibfnamefont {C.}~\bibnamefont {Shen}}, \
  and\ \bibinfo {author} {\bibfnamefont {P.}~\bibnamefont {Tribedy}},\ }\href
  {\doibase 10.1103/PhysRevC.102.044905} {\bibfield  {journal} {\bibinfo
  {journal} {Phys. Rev. C}\ }\textbf {\bibinfo {volume} {102}},\ \bibinfo
  {pages} {044905} (\bibinfo {year} {2020}{\natexlab{a}})},\ \Eprint
  {http://arxiv.org/abs/2005.14682} {arXiv:2005.14682 [nucl-th]} \BibitemShut
  {NoStop}%
\bibitem [{\citenamefont {Bozek}(2012)}]{Bozek:2011if}%
  \BibitemOpen
  \bibfield  {author} {\bibinfo {author} {\bibfnamefont {P.}~\bibnamefont
  {Bozek}},\ }\href {\doibase 10.1103/PhysRevC.85.014911} {\bibfield  {journal}
  {\bibinfo  {journal} {Phys. Rev. C}\ }\textbf {\bibinfo {volume} {85}},\
  \bibinfo {pages} {014911} (\bibinfo {year} {2012})},\ \Eprint
  {http://arxiv.org/abs/1112.0915} {arXiv:1112.0915 [hep-ph]} \BibitemShut
  {NoStop}%
\bibitem [{\citenamefont {Bozek}\ and\ \citenamefont
  {Broniowski}(2013{\natexlab{a}})}]{Bozek:2012gr}%
  \BibitemOpen
  \bibfield  {author} {\bibinfo {author} {\bibfnamefont {P.}~\bibnamefont
  {Bozek}}\ and\ \bibinfo {author} {\bibfnamefont {W.}~\bibnamefont
  {Broniowski}},\ }\href {\doibase 10.1016/j.physletb.2012.12.051} {\bibfield
  {journal} {\bibinfo  {journal} {Phys. Lett. B}\ }\textbf {\bibinfo {volume}
  {718}},\ \bibinfo {pages} {1557} (\bibinfo {year} {2013}{\natexlab{a}})},\
  \Eprint {http://arxiv.org/abs/1211.0845} {arXiv:1211.0845 [nucl-th]}
  \BibitemShut {NoStop}%
\bibitem [{\citenamefont {Bozek}\ and\ \citenamefont
  {Broniowski}(2013{\natexlab{b}})}]{Bozek:2013df}%
  \BibitemOpen
  \bibfield  {author} {\bibinfo {author} {\bibfnamefont {P.}~\bibnamefont
  {Bozek}}\ and\ \bibinfo {author} {\bibfnamefont {W.}~\bibnamefont
  {Broniowski}},\ }\href {\doibase 10.1016/j.physletb.2013.02.014} {\bibfield
  {journal} {\bibinfo  {journal} {Phys. Lett. B}\ }\textbf {\bibinfo {volume}
  {720}},\ \bibinfo {pages} {250} (\bibinfo {year} {2013}{\natexlab{b}})},\
  \Eprint {http://arxiv.org/abs/1301.3314} {arXiv:1301.3314 [nucl-th]}
  \BibitemShut {NoStop}%
\bibitem [{\citenamefont {Bozek}\ and\ \citenamefont
  {Broniowski}(2013{\natexlab{c}})}]{Bozek:2013uha}%
  \BibitemOpen
  \bibfield  {author} {\bibinfo {author} {\bibfnamefont {P.}~\bibnamefont
  {Bozek}}\ and\ \bibinfo {author} {\bibfnamefont {W.}~\bibnamefont
  {Broniowski}},\ }\href {\doibase 10.1103/PhysRevC.88.014903} {\bibfield
  {journal} {\bibinfo  {journal} {Phys. Rev. C}\ }\textbf {\bibinfo {volume}
  {88}},\ \bibinfo {pages} {014903} (\bibinfo {year} {2013}{\natexlab{c}})},\
  \Eprint {http://arxiv.org/abs/1304.3044} {arXiv:1304.3044 [nucl-th]}
  \BibitemShut {NoStop}%
\bibitem [{\citenamefont {Bozek}\ \emph {et~al.}(2013)\citenamefont {Bozek},
  \citenamefont {Broniowski},\ and\ \citenamefont {Torrieri}}]{Bozek:2013ska}%
  \BibitemOpen
  \bibfield  {author} {\bibinfo {author} {\bibfnamefont {P.}~\bibnamefont
  {Bozek}}, \bibinfo {author} {\bibfnamefont {W.}~\bibnamefont {Broniowski}}, \
  and\ \bibinfo {author} {\bibfnamefont {G.}~\bibnamefont {Torrieri}},\ }\href
  {\doibase 10.1103/PhysRevLett.111.172303} {\bibfield  {journal} {\bibinfo
  {journal} {Phys. Rev. Lett.}\ }\textbf {\bibinfo {volume} {111}},\ \bibinfo
  {pages} {172303} (\bibinfo {year} {2013})},\ \Eprint
  {http://arxiv.org/abs/1307.5060} {arXiv:1307.5060 [nucl-th]} \BibitemShut
  {NoStop}%
\bibitem [{\citenamefont {Bzdak}\ \emph {et~al.}(2013)\citenamefont {Bzdak},
  \citenamefont {Schenke}, \citenamefont {Tribedy},\ and\ \citenamefont
  {Venugopalan}}]{Bzdak:2013zma}%
  \BibitemOpen
  \bibfield  {author} {\bibinfo {author} {\bibfnamefont {A.}~\bibnamefont
  {Bzdak}}, \bibinfo {author} {\bibfnamefont {B.}~\bibnamefont {Schenke}},
  \bibinfo {author} {\bibfnamefont {P.}~\bibnamefont {Tribedy}}, \ and\
  \bibinfo {author} {\bibfnamefont {R.}~\bibnamefont {Venugopalan}},\ }\href
  {\doibase 10.1103/PhysRevC.87.064906} {\bibfield  {journal} {\bibinfo
  {journal} {Phys. Rev. C}\ }\textbf {\bibinfo {volume} {87}},\ \bibinfo
  {pages} {064906} (\bibinfo {year} {2013})},\ \Eprint
  {http://arxiv.org/abs/1304.3403} {arXiv:1304.3403 [nucl-th]} \BibitemShut
  {NoStop}%
\bibitem [{\citenamefont {Qin}\ and\ \citenamefont
  {M\"uller}(2014)}]{Qin:2013bha}%
  \BibitemOpen
  \bibfield  {author} {\bibinfo {author} {\bibfnamefont {G.-Y.}\ \bibnamefont
  {Qin}}\ and\ \bibinfo {author} {\bibfnamefont {B.}~\bibnamefont {M\"uller}},\
  }\href {\doibase 10.1103/PhysRevC.89.044902} {\bibfield  {journal} {\bibinfo
  {journal} {Phys. Rev. C}\ }\textbf {\bibinfo {volume} {89}},\ \bibinfo
  {pages} {044902} (\bibinfo {year} {2014})},\ \Eprint
  {http://arxiv.org/abs/1306.3439} {arXiv:1306.3439 [nucl-th]} \BibitemShut
  {NoStop}%
\bibitem [{\citenamefont {Werner}\ \emph {et~al.}(2014)\citenamefont {Werner},
  \citenamefont {Bleicher}, \citenamefont {Guiot}, \citenamefont {Karpenko},\
  and\ \citenamefont {Pierog}}]{Werner:2013ipa}%
  \BibitemOpen
  \bibfield  {author} {\bibinfo {author} {\bibfnamefont {K.}~\bibnamefont
  {Werner}}, \bibinfo {author} {\bibfnamefont {M.}~\bibnamefont {Bleicher}},
  \bibinfo {author} {\bibfnamefont {B.}~\bibnamefont {Guiot}}, \bibinfo
  {author} {\bibfnamefont {I.}~\bibnamefont {Karpenko}}, \ and\ \bibinfo
  {author} {\bibfnamefont {T.}~\bibnamefont {Pierog}},\ }\href {\doibase
  10.1103/PhysRevLett.112.232301} {\bibfield  {journal} {\bibinfo  {journal}
  {Phys. Rev. Lett.}\ }\textbf {\bibinfo {volume} {112}},\ \bibinfo {pages}
  {232301} (\bibinfo {year} {2014})},\ \Eprint {http://arxiv.org/abs/1307.4379}
  {arXiv:1307.4379 [nucl-th]} \BibitemShut {NoStop}%
\bibitem [{\citenamefont {Kozlov}\ \emph {et~al.}(2014)\citenamefont {Kozlov},
  \citenamefont {Luzum}, \citenamefont {Denicol}, \citenamefont {Jeon},\ and\
  \citenamefont {Gale}}]{Kozlov:2014fqa}%
  \BibitemOpen
  \bibfield  {author} {\bibinfo {author} {\bibfnamefont {I.}~\bibnamefont
  {Kozlov}}, \bibinfo {author} {\bibfnamefont {M.}~\bibnamefont {Luzum}},
  \bibinfo {author} {\bibfnamefont {G.}~\bibnamefont {Denicol}}, \bibinfo
  {author} {\bibfnamefont {S.}~\bibnamefont {Jeon}}, \ and\ \bibinfo {author}
  {\bibfnamefont {C.}~\bibnamefont {Gale}},\ }\href@noop {} {\  (\bibinfo
  {year} {2014})},\ \Eprint {http://arxiv.org/abs/1405.3976} {arXiv:1405.3976
  [nucl-th]} \BibitemShut {NoStop}%
\bibitem [{\citenamefont {Schenke}\ and\ \citenamefont
  {Venugopalan}(2014)}]{Schenke:2014zha}%
  \BibitemOpen
  \bibfield  {author} {\bibinfo {author} {\bibfnamefont {B.}~\bibnamefont
  {Schenke}}\ and\ \bibinfo {author} {\bibfnamefont {R.}~\bibnamefont
  {Venugopalan}},\ }\href {\doibase 10.1103/PhysRevLett.113.102301} {\bibfield
  {journal} {\bibinfo  {journal} {Phys. Rev. Lett.}\ }\textbf {\bibinfo
  {volume} {113}},\ \bibinfo {pages} {102301} (\bibinfo {year} {2014})},\
  \Eprint {http://arxiv.org/abs/1405.3605} {arXiv:1405.3605 [nucl-th]}
  \BibitemShut {NoStop}%
\bibitem [{\citenamefont {Romatschke}(2015)}]{Romatschke:2015gxa}%
  \BibitemOpen
  \bibfield  {author} {\bibinfo {author} {\bibfnamefont {P.}~\bibnamefont
  {Romatschke}},\ }\href {\doibase 10.1140/epjc/s10052-015-3509-3} {\bibfield
  {journal} {\bibinfo  {journal} {Eur. Phys. J. C}\ }\textbf {\bibinfo {volume}
  {75}},\ \bibinfo {pages} {305} (\bibinfo {year} {2015})},\ \Eprint
  {http://arxiv.org/abs/1502.04745} {arXiv:1502.04745 [nucl-th]} \BibitemShut
  {NoStop}%
\bibitem [{\citenamefont {Shen}\ \emph {et~al.}(2017)\citenamefont {Shen},
  \citenamefont {Paquet}, \citenamefont {Denicol}, \citenamefont {Jeon},\ and\
  \citenamefont {Gale}}]{Shen:2016zpp}%
  \BibitemOpen
  \bibfield  {author} {\bibinfo {author} {\bibfnamefont {C.}~\bibnamefont
  {Shen}}, \bibinfo {author} {\bibfnamefont {J.-F.}\ \bibnamefont {Paquet}},
  \bibinfo {author} {\bibfnamefont {G.~S.}\ \bibnamefont {Denicol}}, \bibinfo
  {author} {\bibfnamefont {S.}~\bibnamefont {Jeon}}, \ and\ \bibinfo {author}
  {\bibfnamefont {C.}~\bibnamefont {Gale}},\ }\href {\doibase
  10.1103/PhysRevC.95.014906} {\bibfield  {journal} {\bibinfo  {journal} {Phys.
  Rev. C}\ }\textbf {\bibinfo {volume} {95}},\ \bibinfo {pages} {014906}
  (\bibinfo {year} {2017})},\ \Eprint {http://arxiv.org/abs/1609.02590}
  {arXiv:1609.02590 [nucl-th]} \BibitemShut {NoStop}%
\bibitem [{\citenamefont {Weller}\ and\ \citenamefont
  {Romatschke}(2017)}]{Weller:2017tsr}%
  \BibitemOpen
  \bibfield  {author} {\bibinfo {author} {\bibfnamefont {R.~D.}\ \bibnamefont
  {Weller}}\ and\ \bibinfo {author} {\bibfnamefont {P.}~\bibnamefont
  {Romatschke}},\ }\href {\doibase 10.1016/j.physletb.2017.09.077} {\bibfield
  {journal} {\bibinfo  {journal} {Phys. Lett. B}\ }\textbf {\bibinfo {volume}
  {774}},\ \bibinfo {pages} {351} (\bibinfo {year} {2017})},\ \Eprint
  {http://arxiv.org/abs/1701.07145} {arXiv:1701.07145 [nucl-th]} \BibitemShut
  {NoStop}%
\bibitem [{\citenamefont {M\"antysaari}\ \emph {et~al.}(2017)\citenamefont
  {M\"antysaari}, \citenamefont {Schenke}, \citenamefont {Shen},\ and\
  \citenamefont {Tribedy}}]{Mantysaari:2017cni}%
  \BibitemOpen
  \bibfield  {author} {\bibinfo {author} {\bibfnamefont {H.}~\bibnamefont
  {M\"antysaari}}, \bibinfo {author} {\bibfnamefont {B.}~\bibnamefont
  {Schenke}}, \bibinfo {author} {\bibfnamefont {C.}~\bibnamefont {Shen}}, \
  and\ \bibinfo {author} {\bibfnamefont {P.}~\bibnamefont {Tribedy}},\ }\href
  {\doibase 10.1016/j.physletb.2017.07.038} {\bibfield  {journal} {\bibinfo
  {journal} {Phys. Lett. B}\ }\textbf {\bibinfo {volume} {772}},\ \bibinfo
  {pages} {681} (\bibinfo {year} {2017})},\ \Eprint
  {http://arxiv.org/abs/1705.03177} {arXiv:1705.03177 [nucl-th]} \BibitemShut
  {NoStop}%
\bibitem [{\citenamefont {Schenke}\ \emph
  {et~al.}(2020{\natexlab{b}})\citenamefont {Schenke}, \citenamefont {Shen},\
  and\ \citenamefont {Tribedy}}]{Schenke:2019pmk}%
  \BibitemOpen
  \bibfield  {author} {\bibinfo {author} {\bibfnamefont {B.}~\bibnamefont
  {Schenke}}, \bibinfo {author} {\bibfnamefont {C.}~\bibnamefont {Shen}}, \
  and\ \bibinfo {author} {\bibfnamefont {P.}~\bibnamefont {Tribedy}},\ }\href
  {\doibase 10.1016/j.physletb.2020.135322} {\bibfield  {journal} {\bibinfo
  {journal} {Phys. Lett. B}\ }\textbf {\bibinfo {volume} {803}},\ \bibinfo
  {pages} {135322} (\bibinfo {year} {2020}{\natexlab{b}})},\ \Eprint
  {http://arxiv.org/abs/1908.06212} {arXiv:1908.06212 [nucl-th]} \BibitemShut
  {NoStop}%
\bibitem [{\citenamefont {Dusling}\ \emph {et~al.}(2016)\citenamefont
  {Dusling}, \citenamefont {Li},\ and\ \citenamefont
  {Schenke}}]{Dusling:2015gta}%
  \BibitemOpen
  \bibfield  {author} {\bibinfo {author} {\bibfnamefont {K.}~\bibnamefont
  {Dusling}}, \bibinfo {author} {\bibfnamefont {W.}~\bibnamefont {Li}}, \ and\
  \bibinfo {author} {\bibfnamefont {B.}~\bibnamefont {Schenke}},\ }\href
  {\doibase 10.1142/S0218301316300022} {\bibfield  {journal} {\bibinfo
  {journal} {Int. J. Mod. Phys. E}\ }\textbf {\bibinfo {volume} {25}},\
  \bibinfo {pages} {1630002} (\bibinfo {year} {2016})},\ \Eprint
  {http://arxiv.org/abs/1509.07939} {arXiv:1509.07939 [nucl-ex]} \BibitemShut
  {NoStop}%
\bibitem [{\citenamefont {Loizides}(2016)}]{Loizides:2016tew}%
  \BibitemOpen
  \bibfield  {author} {\bibinfo {author} {\bibfnamefont {C.}~\bibnamefont
  {Loizides}},\ }\href {\doibase 10.1016/j.nuclphysa.2016.04.022} {\bibfield
  {journal} {\bibinfo  {journal} {Nucl. Phys. A}\ }\textbf {\bibinfo {volume}
  {956}},\ \bibinfo {pages} {200} (\bibinfo {year} {2016})},\ \Eprint
  {http://arxiv.org/abs/1602.09138} {arXiv:1602.09138 [nucl-ex]} \BibitemShut
  {NoStop}%
\bibitem [{\citenamefont {Nagle}\ and\ \citenamefont
  {Zajc}(2018)}]{Nagle:2018nvi}%
  \BibitemOpen
  \bibfield  {author} {\bibinfo {author} {\bibfnamefont {J.~L.}\ \bibnamefont
  {Nagle}}\ and\ \bibinfo {author} {\bibfnamefont {W.~A.}\ \bibnamefont
  {Zajc}},\ }\href {\doibase 10.1146/annurev-nucl-101916-123209} {\bibfield
  {journal} {\bibinfo  {journal} {Ann. Rev. Nucl. Part. Sci.}\ }\textbf
  {\bibinfo {volume} {68}},\ \bibinfo {pages} {211} (\bibinfo {year} {2018})},\
  \Eprint {http://arxiv.org/abs/1801.03477} {arXiv:1801.03477 [nucl-ex]}
  \BibitemShut {NoStop}%
\bibitem [{\citenamefont {Schenke}(2021)}]{Schenke:2021mxx}%
  \BibitemOpen
  \bibfield  {author} {\bibinfo {author} {\bibfnamefont {B.}~\bibnamefont
  {Schenke}},\ }\href@noop {} {\  (\bibinfo {year} {2021})},\ \Eprint
  {http://arxiv.org/abs/2102.11189} {arXiv:2102.11189 [nucl-th]} \BibitemShut
  {NoStop}%
\bibitem [{\citenamefont {Demirci}\ \emph {et~al.}(2021)\citenamefont
  {Demirci}, \citenamefont {Lappi},\ and\ \citenamefont
  {Schlichting}}]{Demirci:2021kya}%
  \BibitemOpen
  \bibfield  {author} {\bibinfo {author} {\bibfnamefont {S.}~\bibnamefont
  {Demirci}}, \bibinfo {author} {\bibfnamefont {T.}~\bibnamefont {Lappi}}, \
  and\ \bibinfo {author} {\bibfnamefont {S.}~\bibnamefont {Schlichting}},\
  }\href {\doibase 10.1103/PhysRevD.103.094025} {\bibfield  {journal} {\bibinfo
   {journal} {Phys. Rev. D}\ }\textbf {\bibinfo {volume} {103}},\ \bibinfo
  {pages} {094025} (\bibinfo {year} {2021})},\ \Eprint
  {http://arxiv.org/abs/2101.03791} {arXiv:2101.03791 [hep-ph]} \BibitemShut
  {NoStop}%
\bibitem [{\citenamefont {Schlichting}\ and\ \citenamefont
  {Teaney}(2019)}]{Schlichting:2019abc}%
  \BibitemOpen
  \bibfield  {author} {\bibinfo {author} {\bibfnamefont {S.}~\bibnamefont
  {Schlichting}}\ and\ \bibinfo {author} {\bibfnamefont {D.}~\bibnamefont
  {Teaney}},\ }\href {\doibase 10.1146/annurev-nucl-101918-023825} {\bibfield
  {journal} {\bibinfo  {journal} {Ann. Rev. Nucl. Part. Sci.}\ }\textbf
  {\bibinfo {volume} {69}},\ \bibinfo {pages} {447} (\bibinfo {year} {2019})},\
  \Eprint {http://arxiv.org/abs/1908.02113} {arXiv:1908.02113 [nucl-th]}
  \BibitemShut {NoStop}%
\bibitem [{\citenamefont {Berges}\ \emph {et~al.}(2021)\citenamefont {Berges},
  \citenamefont {Heller}, \citenamefont {Mazeliauskas},\ and\ \citenamefont
  {Venugopalan}}]{Berges:2020fwq}%
  \BibitemOpen
  \bibfield  {author} {\bibinfo {author} {\bibfnamefont {J.}~\bibnamefont
  {Berges}}, \bibinfo {author} {\bibfnamefont {M.~P.}\ \bibnamefont {Heller}},
  \bibinfo {author} {\bibfnamefont {A.}~\bibnamefont {Mazeliauskas}}, \ and\
  \bibinfo {author} {\bibfnamefont {R.}~\bibnamefont {Venugopalan}},\ }\href
  {\doibase 10.1103/RevModPhys.93.035003} {\bibfield  {journal} {\bibinfo
  {journal} {Rev. Mod. Phys.}\ }\textbf {\bibinfo {volume} {93}},\ \bibinfo
  {pages} {035003} (\bibinfo {year} {2021})},\ \Eprint
  {http://arxiv.org/abs/2005.12299} {arXiv:2005.12299 [hep-th]} \BibitemShut
  {NoStop}%
\bibitem [{\citenamefont {Schenke}\ \emph {et~al.}(2015)\citenamefont
  {Schenke}, \citenamefont {Schlichting},\ and\ \citenamefont
  {Venugopalan}}]{Schenke:2015aqa}%
  \BibitemOpen
  \bibfield  {author} {\bibinfo {author} {\bibfnamefont {B.}~\bibnamefont
  {Schenke}}, \bibinfo {author} {\bibfnamefont {S.}~\bibnamefont
  {Schlichting}}, \ and\ \bibinfo {author} {\bibfnamefont {R.}~\bibnamefont
  {Venugopalan}},\ }\href {\doibase 10.1016/j.physletb.2015.05.051} {\bibfield
  {journal} {\bibinfo  {journal} {Phys. Lett. B}\ }\textbf {\bibinfo {volume}
  {747}},\ \bibinfo {pages} {76} (\bibinfo {year} {2015})},\ \Eprint
  {http://arxiv.org/abs/1502.01331} {arXiv:1502.01331 [hep-ph]} \BibitemShut
  {NoStop}%
\bibitem [{\citenamefont {McLerran}\ and\ \citenamefont
  {Skokov}(2016)}]{McLerran:2015sva}%
  \BibitemOpen
  \bibfield  {author} {\bibinfo {author} {\bibfnamefont {L.}~\bibnamefont
  {McLerran}}\ and\ \bibinfo {author} {\bibfnamefont {V.}~\bibnamefont
  {Skokov}},\ }\href {\doibase 10.1016/j.nuclphysa.2015.12.005} {\bibfield
  {journal} {\bibinfo  {journal} {Nucl. Phys. A}\ }\textbf {\bibinfo {volume}
  {947}},\ \bibinfo {pages} {142} (\bibinfo {year} {2016})},\ \Eprint
  {http://arxiv.org/abs/1510.08072} {arXiv:1510.08072 [hep-ph]} \BibitemShut
  {NoStop}%
\bibitem [{\citenamefont {Schenke}\ \emph {et~al.}(2016)\citenamefont
  {Schenke}, \citenamefont {Schlichting}, \citenamefont {Tribedy},\ and\
  \citenamefont {Venugopalan}}]{Schenke:2016lrs}%
  \BibitemOpen
  \bibfield  {author} {\bibinfo {author} {\bibfnamefont {B.}~\bibnamefont
  {Schenke}}, \bibinfo {author} {\bibfnamefont {S.}~\bibnamefont
  {Schlichting}}, \bibinfo {author} {\bibfnamefont {P.}~\bibnamefont
  {Tribedy}}, \ and\ \bibinfo {author} {\bibfnamefont {R.}~\bibnamefont
  {Venugopalan}},\ }\href {\doibase 10.1103/PhysRevLett.117.162301} {\bibfield
  {journal} {\bibinfo  {journal} {Phys. Rev. Lett.}\ }\textbf {\bibinfo
  {volume} {117}},\ \bibinfo {pages} {162301} (\bibinfo {year} {2016})},\
  \Eprint {http://arxiv.org/abs/1607.02496} {arXiv:1607.02496 [hep-ph]}
  \BibitemShut {NoStop}%
\bibitem [{\citenamefont {Dusling}\ \emph
  {et~al.}(2018{\natexlab{a}})\citenamefont {Dusling}, \citenamefont {Mace},\
  and\ \citenamefont {Venugopalan}}]{Dusling:2017dqg}%
  \BibitemOpen
  \bibfield  {author} {\bibinfo {author} {\bibfnamefont {K.}~\bibnamefont
  {Dusling}}, \bibinfo {author} {\bibfnamefont {M.}~\bibnamefont {Mace}}, \
  and\ \bibinfo {author} {\bibfnamefont {R.}~\bibnamefont {Venugopalan}},\
  }\href {\doibase 10.1103/PhysRevLett.120.042002} {\bibfield  {journal}
  {\bibinfo  {journal} {Phys. Rev. Lett.}\ }\textbf {\bibinfo {volume} {120}},\
  \bibinfo {pages} {042002} (\bibinfo {year} {2018}{\natexlab{a}})},\ \Eprint
  {http://arxiv.org/abs/1705.00745} {arXiv:1705.00745 [hep-ph]} \BibitemShut
  {NoStop}%
\bibitem [{\citenamefont {Dusling}\ \emph
  {et~al.}(2018{\natexlab{b}})\citenamefont {Dusling}, \citenamefont {Mace},\
  and\ \citenamefont {Venugopalan}}]{Dusling:2017aot}%
  \BibitemOpen
  \bibfield  {author} {\bibinfo {author} {\bibfnamefont {K.}~\bibnamefont
  {Dusling}}, \bibinfo {author} {\bibfnamefont {M.}~\bibnamefont {Mace}}, \
  and\ \bibinfo {author} {\bibfnamefont {R.}~\bibnamefont {Venugopalan}},\
  }\href {\doibase 10.1103/PhysRevD.97.016014} {\bibfield  {journal} {\bibinfo
  {journal} {Phys. Rev. D}\ }\textbf {\bibinfo {volume} {97}},\ \bibinfo
  {pages} {016014} (\bibinfo {year} {2018}{\natexlab{b}})},\ \Eprint
  {http://arxiv.org/abs/1706.06260} {arXiv:1706.06260 [hep-ph]} \BibitemShut
  {NoStop}%
\bibitem [{\citenamefont {Greif}\ \emph {et~al.}(2017)\citenamefont {Greif},
  \citenamefont {Greiner}, \citenamefont {Schenke}, \citenamefont
  {Schlichting},\ and\ \citenamefont {Xu}}]{Greif:2017bnr}%
  \BibitemOpen
  \bibfield  {author} {\bibinfo {author} {\bibfnamefont {M.}~\bibnamefont
  {Greif}}, \bibinfo {author} {\bibfnamefont {C.}~\bibnamefont {Greiner}},
  \bibinfo {author} {\bibfnamefont {B.}~\bibnamefont {Schenke}}, \bibinfo
  {author} {\bibfnamefont {S.}~\bibnamefont {Schlichting}}, \ and\ \bibinfo
  {author} {\bibfnamefont {Z.}~\bibnamefont {Xu}},\ }\href {\doibase
  10.1103/PhysRevD.96.091504} {\bibfield  {journal} {\bibinfo  {journal} {Phys.
  Rev. D}\ }\textbf {\bibinfo {volume} {96}},\ \bibinfo {pages} {091504}
  (\bibinfo {year} {2017})},\ \Eprint {http://arxiv.org/abs/1708.02076}
  {arXiv:1708.02076 [hep-ph]} \BibitemShut {NoStop}%
\bibitem [{\citenamefont {Mace}\ \emph {et~al.}(2018)\citenamefont {Mace},
  \citenamefont {Skokov}, \citenamefont {Tribedy},\ and\ \citenamefont
  {Venugopalan}}]{Mace:2018vwq}%
  \BibitemOpen
  \bibfield  {author} {\bibinfo {author} {\bibfnamefont {M.}~\bibnamefont
  {Mace}}, \bibinfo {author} {\bibfnamefont {V.~V.}\ \bibnamefont {Skokov}},
  \bibinfo {author} {\bibfnamefont {P.}~\bibnamefont {Tribedy}}, \ and\
  \bibinfo {author} {\bibfnamefont {R.}~\bibnamefont {Venugopalan}},\ }\href
  {\doibase 10.1103/PhysRevLett.121.052301} {\bibfield  {journal} {\bibinfo
  {journal} {Phys. Rev. Lett.}\ }\textbf {\bibinfo {volume} {121}},\ \bibinfo
  {pages} {052301} (\bibinfo {year} {2018})},\ \bibinfo {note} {[Erratum:
  Phys.Rev.Lett. 123, 039901 (2019)]},\ \Eprint
  {http://arxiv.org/abs/1805.09342} {arXiv:1805.09342 [hep-ph]} \BibitemShut
  {NoStop}%
\bibitem [{\citenamefont {Mace}\ \emph {et~al.}(2019)\citenamefont {Mace},
  \citenamefont {Skokov}, \citenamefont {Tribedy},\ and\ \citenamefont
  {Venugopalan}}]{Mace:2018yvl}%
  \BibitemOpen
  \bibfield  {author} {\bibinfo {author} {\bibfnamefont {M.}~\bibnamefont
  {Mace}}, \bibinfo {author} {\bibfnamefont {V.~V.}\ \bibnamefont {Skokov}},
  \bibinfo {author} {\bibfnamefont {P.}~\bibnamefont {Tribedy}}, \ and\
  \bibinfo {author} {\bibfnamefont {R.}~\bibnamefont {Venugopalan}},\ }\href
  {\doibase 10.1016/j.physletb.2018.09.064} {\bibfield  {journal} {\bibinfo
  {journal} {Phys. Lett. B}\ }\textbf {\bibinfo {volume} {788}},\ \bibinfo
  {pages} {161} (\bibinfo {year} {2019})},\ \bibinfo {note} {[Erratum:
  Phys.Lett.B 799, 135006 (2019)]},\ \Eprint {http://arxiv.org/abs/1807.00825}
  {arXiv:1807.00825 [hep-ph]} \BibitemShut {NoStop}%
\bibitem [{\citenamefont {Kovner}\ and\ \citenamefont
  {Skokov}(2018)}]{Kovner:2018fxj}%
  \BibitemOpen
  \bibfield  {author} {\bibinfo {author} {\bibfnamefont {A.}~\bibnamefont
  {Kovner}}\ and\ \bibinfo {author} {\bibfnamefont {V.~V.}\ \bibnamefont
  {Skokov}},\ }\href {\doibase 10.1016/j.physletb.2018.09.001} {\bibfield
  {journal} {\bibinfo  {journal} {Phys. Lett. B}\ }\textbf {\bibinfo {volume}
  {785}},\ \bibinfo {pages} {372} (\bibinfo {year} {2018})},\ \Eprint
  {http://arxiv.org/abs/1805.09297} {arXiv:1805.09297 [hep-ph]} \BibitemShut
  {NoStop}%
\bibitem [{\citenamefont {Greif}\ \emph {et~al.}(2021)\citenamefont {Greif},
  \citenamefont {Greiner}, \citenamefont {Pl\"atzer}, \citenamefont {Schenke},\
  and\ \citenamefont {Schlichting}}]{Greif:2020rhi}%
  \BibitemOpen
  \bibfield  {author} {\bibinfo {author} {\bibfnamefont {M.}~\bibnamefont
  {Greif}}, \bibinfo {author} {\bibfnamefont {C.}~\bibnamefont {Greiner}},
  \bibinfo {author} {\bibfnamefont {S.}~\bibnamefont {Pl\"atzer}}, \bibinfo
  {author} {\bibfnamefont {B.}~\bibnamefont {Schenke}}, \ and\ \bibinfo
  {author} {\bibfnamefont {S.}~\bibnamefont {Schlichting}},\ }\href {\doibase
  10.1103/PhysRevD.103.054011} {\bibfield  {journal} {\bibinfo  {journal}
  {Phys. Rev. D}\ }\textbf {\bibinfo {volume} {103}},\ \bibinfo {pages}
  {054011} (\bibinfo {year} {2021})},\ \Eprint
  {http://arxiv.org/abs/2012.08493} {arXiv:2012.08493 [hep-ph]} \BibitemShut
  {NoStop}%
\bibitem [{\citenamefont {Agostini}\ \emph {et~al.}(2021)\citenamefont
  {Agostini}, \citenamefont {Altinoluk},\ and\ \citenamefont
  {Armesto}}]{Agostini:2021xca}%
  \BibitemOpen
  \bibfield  {author} {\bibinfo {author} {\bibfnamefont {P.}~\bibnamefont
  {Agostini}}, \bibinfo {author} {\bibfnamefont {T.}~\bibnamefont {Altinoluk}},
  \ and\ \bibinfo {author} {\bibfnamefont {N.}~\bibnamefont {Armesto}},\ }\href
  {\doibase 10.1140/epjc/s10052-021-09475-0} {\bibfield  {journal} {\bibinfo
  {journal} {Eur. Phys. J. C}\ }\textbf {\bibinfo {volume} {81}},\ \bibinfo
  {pages} {760} (\bibinfo {year} {2021})},\ \Eprint
  {http://arxiv.org/abs/2103.08485} {arXiv:2103.08485 [hep-ph]} \BibitemShut
  {NoStop}%
\bibitem [{\citenamefont {Abramovsky}\ \emph {et~al.}(1988)\citenamefont
  {Abramovsky}, \citenamefont {Gedalin}, \citenamefont {Gurvich},\ and\
  \citenamefont {Kancheli}}]{Abramovsky:1988zh}%
  \BibitemOpen
  \bibfield  {author} {\bibinfo {author} {\bibfnamefont {V.~A.}\ \bibnamefont
  {Abramovsky}}, \bibinfo {author} {\bibfnamefont {E.~V.}\ \bibnamefont
  {Gedalin}}, \bibinfo {author} {\bibfnamefont {E.~G.}\ \bibnamefont
  {Gurvich}}, \ and\ \bibinfo {author} {\bibfnamefont {O.~V.}\ \bibnamefont
  {Kancheli}},\ }\href@noop {} {\bibfield  {journal} {\bibinfo  {journal} {JETP
  Lett.}\ }\textbf {\bibinfo {volume} {47}},\ \bibinfo {pages} {337} (\bibinfo
  {year} {1988})}\BibitemShut {NoStop}%
\bibitem [{\citenamefont {Ortiz~Velasquez}\ \emph {et~al.}(2013)\citenamefont
  {Ortiz~Velasquez}, \citenamefont {Christiansen}, \citenamefont
  {Cuautle~Flores}, \citenamefont {Maldonado~Cervantes},\ and\ \citenamefont
  {Pai\'c}}]{OrtizVelasquez:2013ofg}%
  \BibitemOpen
  \bibfield  {author} {\bibinfo {author} {\bibfnamefont {A.}~\bibnamefont
  {Ortiz~Velasquez}}, \bibinfo {author} {\bibfnamefont {P.}~\bibnamefont
  {Christiansen}}, \bibinfo {author} {\bibfnamefont {E.}~\bibnamefont
  {Cuautle~Flores}}, \bibinfo {author} {\bibfnamefont {I.}~\bibnamefont
  {Maldonado~Cervantes}}, \ and\ \bibinfo {author} {\bibfnamefont
  {G.}~\bibnamefont {Pai\'c}},\ }\href {\doibase
  10.1103/PhysRevLett.111.042001} {\bibfield  {journal} {\bibinfo  {journal}
  {Phys. Rev. Lett.}\ }\textbf {\bibinfo {volume} {111}},\ \bibinfo {pages}
  {042001} (\bibinfo {year} {2013})},\ \Eprint {http://arxiv.org/abs/1303.6326}
  {arXiv:1303.6326 [hep-ph]} \BibitemShut {NoStop}%
\bibitem [{\citenamefont {Bierlich}\ \emph {et~al.}(2018)\citenamefont
  {Bierlich}, \citenamefont {Gustafson},\ and\ \citenamefont
  {L\"onnblad}}]{Bierlich:2017vhg}%
  \BibitemOpen
  \bibfield  {author} {\bibinfo {author} {\bibfnamefont {C.}~\bibnamefont
  {Bierlich}}, \bibinfo {author} {\bibfnamefont {G.}~\bibnamefont {Gustafson}},
  \ and\ \bibinfo {author} {\bibfnamefont {L.}~\bibnamefont {L\"onnblad}},\
  }\href {\doibase 10.1016/j.physletb.2018.01.069} {\bibfield  {journal}
  {\bibinfo  {journal} {Phys. Lett. B}\ }\textbf {\bibinfo {volume} {779}},\
  \bibinfo {pages} {58} (\bibinfo {year} {2018})},\ \Eprint
  {http://arxiv.org/abs/1710.09725} {arXiv:1710.09725 [hep-ph]} \BibitemShut
  {NoStop}%
\bibitem [{\citenamefont {Bierlich}\ \emph {et~al.}(2021)\citenamefont
  {Bierlich}, \citenamefont {Chakraborty}, \citenamefont {Gustafson},\ and\
  \citenamefont {L\"onnblad}}]{Bierlich:2020naj}%
  \BibitemOpen
  \bibfield  {author} {\bibinfo {author} {\bibfnamefont {C.}~\bibnamefont
  {Bierlich}}, \bibinfo {author} {\bibfnamefont {S.}~\bibnamefont
  {Chakraborty}}, \bibinfo {author} {\bibfnamefont {G.}~\bibnamefont
  {Gustafson}}, \ and\ \bibinfo {author} {\bibfnamefont {L.}~\bibnamefont
  {L\"onnblad}},\ }\href {\doibase 10.1007/JHEP03(2021)270} {\bibfield
  {journal} {\bibinfo  {journal} {JHEP}\ }\textbf {\bibinfo {volume} {03}},\
  \bibinfo {pages} {270} (\bibinfo {year} {2021})},\ \Eprint
  {http://arxiv.org/abs/2010.07595} {arXiv:2010.07595 [hep-ph]} \BibitemShut
  {NoStop}%
\bibitem [{\citenamefont {Wiedemann}(2021)}]{Wiedemann:2021bwz}%
  \BibitemOpen
  \bibfield  {author} {\bibinfo {author} {\bibfnamefont {U.~A.}\ \bibnamefont
  {Wiedemann}},\ }\href@noop {} {\  (\bibinfo {year} {2021})},\ \Eprint
  {http://arxiv.org/abs/2101.01971} {arXiv:2101.01971 [hep-ph]} \BibitemShut
  {NoStop}%
\bibitem [{\citenamefont {Heiselberg}\ and\ \citenamefont
  {Levy}(1999)}]{Heiselberg:1998es}%
  \BibitemOpen
  \bibfield  {author} {\bibinfo {author} {\bibfnamefont {H.}~\bibnamefont
  {Heiselberg}}\ and\ \bibinfo {author} {\bibfnamefont {A.-M.}\ \bibnamefont
  {Levy}},\ }\href {\doibase 10.1103/PhysRevC.59.2716} {\bibfield  {journal}
  {\bibinfo  {journal} {Phys. Rev. C}\ }\textbf {\bibinfo {volume} {59}},\
  \bibinfo {pages} {2716} (\bibinfo {year} {1999})},\ \Eprint
  {http://arxiv.org/abs/nucl-th/9812034} {arXiv:nucl-th/9812034} \BibitemShut
  {NoStop}%
\bibitem [{\citenamefont {Borghini}\ and\ \citenamefont
  {Gombeaud}(2011)}]{Borghini:2010hy}%
  \BibitemOpen
  \bibfield  {author} {\bibinfo {author} {\bibfnamefont {N.}~\bibnamefont
  {Borghini}}\ and\ \bibinfo {author} {\bibfnamefont {C.}~\bibnamefont
  {Gombeaud}},\ }\href {\doibase 10.1140/epjc/s10052-011-1612-7} {\bibfield
  {journal} {\bibinfo  {journal} {Eur. Phys. J. C}\ }\textbf {\bibinfo {volume}
  {71}},\ \bibinfo {pages} {1612} (\bibinfo {year} {2011})},\ \Eprint
  {http://arxiv.org/abs/1012.0899} {arXiv:1012.0899 [nucl-th]} \BibitemShut
  {NoStop}%
\bibitem [{\citenamefont {Romatschke}(2018)}]{Romatschke:2018wgi}%
  \BibitemOpen
  \bibfield  {author} {\bibinfo {author} {\bibfnamefont {P.}~\bibnamefont
  {Romatschke}},\ }\href {\doibase 10.1140/epjc/s10052-018-6112-6} {\bibfield
  {journal} {\bibinfo  {journal} {Eur. Phys. J. C}\ }\textbf {\bibinfo {volume}
  {78}},\ \bibinfo {pages} {636} (\bibinfo {year} {2018})},\ \Eprint
  {http://arxiv.org/abs/1802.06804} {arXiv:1802.06804 [nucl-th]} \BibitemShut
  {NoStop}%
\bibitem [{\citenamefont {Kersting}\ \emph {et~al.}(2019)\citenamefont
  {Kersting}, \citenamefont {Borghini},\ and\ \citenamefont
  {Feld}}]{Kersting:2018qvi}%
  \BibitemOpen
  \bibfield  {author} {\bibinfo {author} {\bibfnamefont {N.}~\bibnamefont
  {Kersting}}, \bibinfo {author} {\bibfnamefont {N.}~\bibnamefont {Borghini}},
  \ and\ \bibinfo {author} {\bibfnamefont {S.}~\bibnamefont {Feld}},\ }\href
  {\doibase 10.3390/proceedings2019010016} {\bibfield  {journal} {\bibinfo
  {journal} {MDPI Proc.}\ }\textbf {\bibinfo {volume} {10}},\ \bibinfo {pages}
  {16} (\bibinfo {year} {2019})},\ \Eprint {http://arxiv.org/abs/1811.06195}
  {arXiv:1811.06195 [nucl-th]} \BibitemShut {NoStop}%
\bibitem [{\citenamefont {Kurkela}\ \emph
  {et~al.}(2019{\natexlab{a}})\citenamefont {Kurkela}, \citenamefont
  {Wiedemann},\ and\ \citenamefont {Wu}}]{Kurkela:2019kip}%
  \BibitemOpen
  \bibfield  {author} {\bibinfo {author} {\bibfnamefont {A.}~\bibnamefont
  {Kurkela}}, \bibinfo {author} {\bibfnamefont {U.~A.}\ \bibnamefont
  {Wiedemann}}, \ and\ \bibinfo {author} {\bibfnamefont {B.}~\bibnamefont
  {Wu}},\ }\href {\doibase 10.1140/epjc/s10052-019-7428-6} {\bibfield
  {journal} {\bibinfo  {journal} {Eur. Phys. J. C}\ }\textbf {\bibinfo {volume}
  {79}},\ \bibinfo {pages} {965} (\bibinfo {year} {2019}{\natexlab{a}})},\
  \Eprint {http://arxiv.org/abs/1905.05139} {arXiv:1905.05139 [hep-ph]}
  \BibitemShut {NoStop}%
\bibitem [{\citenamefont {Kurkela}\ \emph {et~al.}(2018)\citenamefont
  {Kurkela}, \citenamefont {Wiedemann},\ and\ \citenamefont
  {Wu}}]{Kurkela:2018ygx}%
  \BibitemOpen
  \bibfield  {author} {\bibinfo {author} {\bibfnamefont {A.}~\bibnamefont
  {Kurkela}}, \bibinfo {author} {\bibfnamefont {U.~A.}\ \bibnamefont
  {Wiedemann}}, \ and\ \bibinfo {author} {\bibfnamefont {B.}~\bibnamefont
  {Wu}},\ }\href {\doibase 10.1016/j.physletb.2018.06.064} {\bibfield
  {journal} {\bibinfo  {journal} {Phys. Lett. B}\ }\textbf {\bibinfo {volume}
  {783}},\ \bibinfo {pages} {274} (\bibinfo {year} {2018})},\ \Eprint
  {http://arxiv.org/abs/1803.02072} {arXiv:1803.02072 [hep-ph]} \BibitemShut
  {NoStop}%
\bibitem [{\citenamefont {Borghini}\ \emph {et~al.}(2018)\citenamefont
  {Borghini}, \citenamefont {Feld},\ and\ \citenamefont
  {Kersting}}]{Borghini:2018xum}%
  \BibitemOpen
  \bibfield  {author} {\bibinfo {author} {\bibfnamefont {N.}~\bibnamefont
  {Borghini}}, \bibinfo {author} {\bibfnamefont {S.}~\bibnamefont {Feld}}, \
  and\ \bibinfo {author} {\bibfnamefont {N.}~\bibnamefont {Kersting}},\ }\href
  {\doibase 10.1140/epjc/s10052-018-6313-z} {\bibfield  {journal} {\bibinfo
  {journal} {Eur. Phys. J. C}\ }\textbf {\bibinfo {volume} {78}},\ \bibinfo
  {pages} {832} (\bibinfo {year} {2018})},\ \Eprint
  {http://arxiv.org/abs/1804.05729} {arXiv:1804.05729 [nucl-th]} \BibitemShut
  {NoStop}%
\bibitem [{\citenamefont {Kurkela}\ \emph {et~al.}(2021)\citenamefont
  {Kurkela}, \citenamefont {Mazeliauskas},\ and\ \citenamefont
  {T\"ornkvist}}]{Kurkela:2021ctp}%
  \BibitemOpen
  \bibfield  {author} {\bibinfo {author} {\bibfnamefont {A.}~\bibnamefont
  {Kurkela}}, \bibinfo {author} {\bibfnamefont {A.}~\bibnamefont
  {Mazeliauskas}}, \ and\ \bibinfo {author} {\bibfnamefont {R.}~\bibnamefont
  {T\"ornkvist}},\ }\href@noop {} {\  (\bibinfo {year} {2021})},\ \Eprint
  {http://arxiv.org/abs/2104.08179} {arXiv:2104.08179 [hep-ph]} \BibitemShut
  {NoStop}%
\bibitem [{\citenamefont {He}\ \emph {et~al.}(2016)\citenamefont {He},
  \citenamefont {Edmonds}, \citenamefont {Lin}, \citenamefont {Liu},
  \citenamefont {Molnar},\ and\ \citenamefont {Wang}}]{He:2015hfa}%
  \BibitemOpen
  \bibfield  {author} {\bibinfo {author} {\bibfnamefont {L.}~\bibnamefont
  {He}}, \bibinfo {author} {\bibfnamefont {T.}~\bibnamefont {Edmonds}},
  \bibinfo {author} {\bibfnamefont {Z.-W.}\ \bibnamefont {Lin}}, \bibinfo
  {author} {\bibfnamefont {F.}~\bibnamefont {Liu}}, \bibinfo {author}
  {\bibfnamefont {D.}~\bibnamefont {Molnar}}, \ and\ \bibinfo {author}
  {\bibfnamefont {F.}~\bibnamefont {Wang}},\ }\href {\doibase
  10.1016/j.physletb.2015.12.051} {\bibfield  {journal} {\bibinfo  {journal}
  {Phys. Lett. B}\ }\textbf {\bibinfo {volume} {753}},\ \bibinfo {pages} {506}
  (\bibinfo {year} {2016})},\ \Eprint {http://arxiv.org/abs/1502.05572}
  {arXiv:1502.05572 [nucl-th]} \BibitemShut {NoStop}%
\bibitem [{\citenamefont {Kurkela}\ \emph
  {et~al.}(2020{\natexlab{a}})\citenamefont {Kurkela}, \citenamefont {Taghavi},
  \citenamefont {Wiedemann},\ and\ \citenamefont {Wu}}]{Kurkela:2020wwb}%
  \BibitemOpen
  \bibfield  {author} {\bibinfo {author} {\bibfnamefont {A.}~\bibnamefont
  {Kurkela}}, \bibinfo {author} {\bibfnamefont {S.~F.}\ \bibnamefont
  {Taghavi}}, \bibinfo {author} {\bibfnamefont {U.~A.}\ \bibnamefont
  {Wiedemann}}, \ and\ \bibinfo {author} {\bibfnamefont {B.}~\bibnamefont
  {Wu}},\ }\href {\doibase 10.1016/j.physletb.2020.135901} {\bibfield
  {journal} {\bibinfo  {journal} {Phys. Lett. B}\ }\textbf {\bibinfo {volume}
  {811}},\ \bibinfo {pages} {135901} (\bibinfo {year} {2020}{\natexlab{a}})},\
  \Eprint {http://arxiv.org/abs/2007.06851} {arXiv:2007.06851 [hep-ph]}
  \BibitemShut {NoStop}%
\bibitem [{\citenamefont {Roch}\ and\ \citenamefont
  {Borghini}(2021)}]{Roch:2020zdl}%
  \BibitemOpen
  \bibfield  {author} {\bibinfo {author} {\bibfnamefont {H.}~\bibnamefont
  {Roch}}\ and\ \bibinfo {author} {\bibfnamefont {N.}~\bibnamefont
  {Borghini}},\ }\href {\doibase 10.1140/epjc/s10052-021-09147-z} {\bibfield
  {journal} {\bibinfo  {journal} {Eur. Phys. J. C}\ }\textbf {\bibinfo {volume}
  {81}},\ \bibinfo {pages} {380} (\bibinfo {year} {2021})},\ \Eprint
  {http://arxiv.org/abs/2012.02138} {arXiv:2012.02138 [nucl-th]} \BibitemShut
  {NoStop}%
\bibitem [{\citenamefont {Mueller}(2000)}]{Mueller:1999pi}%
  \BibitemOpen
  \bibfield  {author} {\bibinfo {author} {\bibfnamefont {A.~H.}\ \bibnamefont
  {Mueller}},\ }\href {\doibase 10.1016/S0370-2693(00)00084-8} {\bibfield
  {journal} {\bibinfo  {journal} {Phys. Lett. B}\ }\textbf {\bibinfo {volume}
  {475}},\ \bibinfo {pages} {220} (\bibinfo {year} {2000})},\ \Eprint
  {http://arxiv.org/abs/hep-ph/9909388} {arXiv:hep-ph/9909388} \BibitemShut
  {NoStop}%
\bibitem [{\citenamefont {Anderson}\ and\ \citenamefont
  {Witting}(1974{\natexlab{a}})}]{Anderson:1974}%
  \BibitemOpen
  \bibfield  {author} {\bibinfo {author} {\bibfnamefont {J.}~\bibnamefont
  {Anderson}}\ and\ \bibinfo {author} {\bibfnamefont {H.}~\bibnamefont
  {Witting}},\ }\href {\doibase 10.1016/0031-8914(74)90355-3} {\bibfield
  {journal} {\bibinfo  {journal} {Physica}\ }\textbf {\bibinfo {volume} {74}},\
  \bibinfo {pages} {466} (\bibinfo {year} {1974}{\natexlab{a}})}\BibitemShut
  {NoStop}%
\bibitem [{\citenamefont {Anderson}\ and\ \citenamefont
  {Witting}(1974{\natexlab{b}})}]{Anderson:1974b}%
  \BibitemOpen
  \bibfield  {author} {\bibinfo {author} {\bibfnamefont {J.}~\bibnamefont
  {Anderson}}\ and\ \bibinfo {author} {\bibfnamefont {H.}~\bibnamefont
  {Witting}},\ }\href {\doibase 10.1016/0031-8914(74)90356-5} {\bibfield
  {journal} {\bibinfo  {journal} {Physica}\ }\textbf {\bibinfo {volume} {74}},\
  \bibinfo {pages} {489} (\bibinfo {year} {1974}{\natexlab{b}})}\BibitemShut
  {NoStop}%
\bibitem [{\citenamefont {Cercignani}\ and\ \citenamefont
  {Kremer}(2002)}]{Cercignani:2002}%
  \BibitemOpen
  \bibfield  {author} {\bibinfo {author} {\bibfnamefont {C.}~\bibnamefont
  {Cercignani}}\ and\ \bibinfo {author} {\bibfnamefont {G.~M.}\ \bibnamefont
  {Kremer}},\ }\href@noop {} {\emph {\bibinfo {title} {The relativistic
  {B}oltzmann equation: theory and applications}}}\ (\bibinfo  {publisher}
  {Birkh\"{a}user Verlag},\ \bibinfo {address} {Basel, Switzerland},\ \bibinfo
  {year} {2002})\BibitemShut {NoStop}%
\bibitem [{\citenamefont {Rezzolla}\ and\ \citenamefont
  {Zanotti}(2013)}]{Rezzolla:2013}%
  \BibitemOpen
  \bibfield  {author} {\bibinfo {author} {\bibfnamefont {L.}~\bibnamefont
  {Rezzolla}}\ and\ \bibinfo {author} {\bibfnamefont {O.}~\bibnamefont
  {Zanotti}},\ }\href@noop {} {\emph {\bibinfo {title} {Relativistic
  hydrodynamics}}}\ (\bibinfo  {publisher} {Oxford University Press},\ \bibinfo
  {address} {Oxford, UK},\ \bibinfo {year} {2013})\BibitemShut {NoStop}%
\bibitem [{\citenamefont {Rocha}\ \emph {et~al.}(2021)\citenamefont {Rocha},
  \citenamefont {Denicol},\ and\ \citenamefont {Noronha}}]{Rocha:2021zcw}%
  \BibitemOpen
  \bibfield  {author} {\bibinfo {author} {\bibfnamefont {G.~S.}\ \bibnamefont
  {Rocha}}, \bibinfo {author} {\bibfnamefont {G.~S.}\ \bibnamefont {Denicol}},
  \ and\ \bibinfo {author} {\bibfnamefont {J.}~\bibnamefont {Noronha}},\ }\href
  {\doibase 10.1103/PhysRevLett.127.042301} {\bibfield  {journal} {\bibinfo
  {journal} {Phys. Rev. Lett.}\ }\textbf {\bibinfo {volume} {127}},\ \bibinfo
  {pages} {042301} (\bibinfo {year} {2021})},\ \Eprint
  {http://arxiv.org/abs/2103.07489} {arXiv:2103.07489 [nucl-th]} \BibitemShut
  {NoStop}%
\bibitem [{\citenamefont {Gelis}\ \emph {et~al.}(2010)\citenamefont {Gelis},
  \citenamefont {Iancu}, \citenamefont {Jalilian-Marian},\ and\ \citenamefont
  {Venugopalan}}]{Gelis:2010nm}%
  \BibitemOpen
  \bibfield  {author} {\bibinfo {author} {\bibfnamefont {F.}~\bibnamefont
  {Gelis}}, \bibinfo {author} {\bibfnamefont {E.}~\bibnamefont {Iancu}},
  \bibinfo {author} {\bibfnamefont {J.}~\bibnamefont {Jalilian-Marian}}, \ and\
  \bibinfo {author} {\bibfnamefont {R.}~\bibnamefont {Venugopalan}},\ }\href
  {\doibase 10.1146/annurev.nucl.010909.083629} {\bibfield  {journal} {\bibinfo
   {journal} {Ann. Rev. Nucl. Part. Sci.}\ }\textbf {\bibinfo {volume} {60}},\
  \bibinfo {pages} {463} (\bibinfo {year} {2010})},\ \Eprint
  {http://arxiv.org/abs/1002.0333} {arXiv:1002.0333 [hep-ph]} \BibitemShut
  {NoStop}%
\bibitem [{\citenamefont {Teaney}\ and\ \citenamefont
  {Yan}(2011)}]{Teaney:2010vd}%
  \BibitemOpen
  \bibfield  {author} {\bibinfo {author} {\bibfnamefont {D.}~\bibnamefont
  {Teaney}}\ and\ \bibinfo {author} {\bibfnamefont {L.}~\bibnamefont {Yan}},\
  }\href {\doibase 10.1103/PhysRevC.83.064904} {\bibfield  {journal} {\bibinfo
  {journal} {Phys. Rev. C}\ }\textbf {\bibinfo {volume} {83}},\ \bibinfo
  {pages} {064904} (\bibinfo {year} {2011})},\ \Eprint
  {http://arxiv.org/abs/1010.1876} {arXiv:1010.1876 [nucl-th]} \BibitemShut
  {NoStop}%
\bibitem [{\citenamefont {Bhalerao}\ \emph {et~al.}(2011)\citenamefont
  {Bhalerao}, \citenamefont {Luzum},\ and\ \citenamefont
  {Ollitrault}}]{Bhalerao:2011yg}%
  \BibitemOpen
  \bibfield  {author} {\bibinfo {author} {\bibfnamefont {R.~S.}\ \bibnamefont
  {Bhalerao}}, \bibinfo {author} {\bibfnamefont {M.}~\bibnamefont {Luzum}}, \
  and\ \bibinfo {author} {\bibfnamefont {J.-Y.}\ \bibnamefont {Ollitrault}},\
  }\href {\doibase 10.1103/PhysRevC.84.034910} {\bibfield  {journal} {\bibinfo
  {journal} {Phys. Rev. C}\ }\textbf {\bibinfo {volume} {84}},\ \bibinfo
  {pages} {034910} (\bibinfo {year} {2011})},\ \Eprint
  {http://arxiv.org/abs/1104.4740} {arXiv:1104.4740 [nucl-th]} \BibitemShut
  {NoStop}%
\bibitem [{\citenamefont {Voloshin}\ and\ \citenamefont
  {Zhang}(1996)}]{Voloshin:1994mz}%
  \BibitemOpen
  \bibfield  {author} {\bibinfo {author} {\bibfnamefont {S.}~\bibnamefont
  {Voloshin}}\ and\ \bibinfo {author} {\bibfnamefont {Y.}~\bibnamefont
  {Zhang}},\ }\href {\doibase 10.1007/s002880050141} {\bibfield  {journal}
  {\bibinfo  {journal} {Z. Phys. C}\ }\textbf {\bibinfo {volume} {70}},\
  \bibinfo {pages} {665} (\bibinfo {year} {1996})},\ \Eprint
  {http://arxiv.org/abs/hep-ph/9407282} {arXiv:hep-ph/9407282} \BibitemShut
  {NoStop}%
\bibitem [{\citenamefont {Borghini}\ \emph {et~al.}(2001)\citenamefont
  {Borghini}, \citenamefont {Dinh},\ and\ \citenamefont
  {Ollitrault}}]{Borghini:2000sa}%
  \BibitemOpen
  \bibfield  {author} {\bibinfo {author} {\bibfnamefont {N.}~\bibnamefont
  {Borghini}}, \bibinfo {author} {\bibfnamefont {P.~M.}\ \bibnamefont {Dinh}},
  \ and\ \bibinfo {author} {\bibfnamefont {J.-Y.}\ \bibnamefont {Ollitrault}},\
  }\href {\doibase 10.1103/PhysRevC.63.054906} {\bibfield  {journal} {\bibinfo
  {journal} {Phys. Rev. C}\ }\textbf {\bibinfo {volume} {63}},\ \bibinfo
  {pages} {054906} (\bibinfo {year} {2001})},\ \Eprint
  {http://arxiv.org/abs/nucl-th/0007063} {arXiv:nucl-th/0007063} \BibitemShut
  {NoStop}%
\bibitem [{\citenamefont {Ollitrault}(1992)}]{Ollitrault:1992bk}%
  \BibitemOpen
  \bibfield  {author} {\bibinfo {author} {\bibfnamefont {J.-Y.}\ \bibnamefont
  {Ollitrault}},\ }\href {\doibase 10.1103/PhysRevD.46.229} {\bibfield
  {journal} {\bibinfo  {journal} {Phys. Rev. D}\ }\textbf {\bibinfo {volume}
  {46}},\ \bibinfo {pages} {229} (\bibinfo {year} {1992})}\BibitemShut
  {NoStop}%
\bibitem [{\citenamefont {Song}\ and\ \citenamefont
  {Heinz}(2008)}]{Song:2007ux}%
  \BibitemOpen
  \bibfield  {author} {\bibinfo {author} {\bibfnamefont {H.}~\bibnamefont
  {Song}}\ and\ \bibinfo {author} {\bibfnamefont {U.~W.}\ \bibnamefont
  {Heinz}},\ }\href {\doibase 10.1103/PhysRevC.77.064901} {\bibfield  {journal}
  {\bibinfo  {journal} {Phys. Rev. C}\ }\textbf {\bibinfo {volume} {77}},\
  \bibinfo {pages} {064901} (\bibinfo {year} {2008})},\ \Eprint
  {http://arxiv.org/abs/0712.3715} {arXiv:0712.3715 [nucl-th]} \BibitemShut
  {NoStop}%
\bibitem [{\citenamefont {Karpenko}\ \emph {et~al.}(2014)\citenamefont
  {Karpenko}, \citenamefont {Huovinen},\ and\ \citenamefont
  {Bleicher}}]{Karpenko:2013wva}%
  \BibitemOpen
  \bibfield  {author} {\bibinfo {author} {\bibfnamefont {I.}~\bibnamefont
  {Karpenko}}, \bibinfo {author} {\bibfnamefont {P.}~\bibnamefont {Huovinen}},
  \ and\ \bibinfo {author} {\bibfnamefont {M.}~\bibnamefont {Bleicher}},\
  }\href {\doibase 10.1016/j.cpc.2014.07.010} {\bibfield  {journal} {\bibinfo
  {journal} {Comput. Phys. Commun.}\ }\textbf {\bibinfo {volume} {185}},\
  \bibinfo {pages} {3016} (\bibinfo {year} {2014})},\ \Eprint
  {http://arxiv.org/abs/1312.4160} {arXiv:1312.4160 [nucl-th]} \BibitemShut
  {NoStop}%
\bibitem [{\citenamefont {Kamata}\ \emph {et~al.}(2020)\citenamefont {Kamata},
  \citenamefont {Martinez}, \citenamefont {Plaschke}, \citenamefont
  {Ochsenfeld},\ and\ \citenamefont {Schlichting}}]{Kamata:2020mka}%
  \BibitemOpen
  \bibfield  {author} {\bibinfo {author} {\bibfnamefont {S.}~\bibnamefont
  {Kamata}}, \bibinfo {author} {\bibfnamefont {M.}~\bibnamefont {Martinez}},
  \bibinfo {author} {\bibfnamefont {P.}~\bibnamefont {Plaschke}}, \bibinfo
  {author} {\bibfnamefont {S.}~\bibnamefont {Ochsenfeld}}, \ and\ \bibinfo
  {author} {\bibfnamefont {S.}~\bibnamefont {Schlichting}},\ }\href {\doibase
  10.1103/PhysRevD.102.056003} {\bibfield  {journal} {\bibinfo  {journal}
  {Phys. Rev. D}\ }\textbf {\bibinfo {volume} {102}},\ \bibinfo {pages}
  {056003} (\bibinfo {year} {2020})},\ \Eprint
  {http://arxiv.org/abs/2004.06751} {arXiv:2004.06751 [hep-ph]} \BibitemShut
  {NoStop}%
\bibitem [{\citenamefont {Guennebaud}\ \emph {et~al.}(2010)\citenamefont
  {Guennebaud}, \citenamefont {Jacob} \emph {et~al.}}]{eigenweb}%
  \BibitemOpen
  \bibfield  {author} {\bibinfo {author} {\bibfnamefont {G.}~\bibnamefont
  {Guennebaud}}, \bibinfo {author} {\bibfnamefont {B.}~\bibnamefont {Jacob}},
  \emph {et~al.},\ }\href@noop {} {\enquote {\bibinfo {title} {Eigen v3},}\
  }\bibinfo {howpublished} {http://eigen.tuxfamily.org} (\bibinfo {year}
  {2010})\BibitemShut {NoStop}%
\bibitem [{\citenamefont {Romatschke}\ \emph {et~al.}(2011)\citenamefont
  {Romatschke}, \citenamefont {Mendoza},\ and\ \citenamefont
  {Succi}}]{Romatschke:2011hm}%
  \BibitemOpen
  \bibfield  {author} {\bibinfo {author} {\bibfnamefont {P.}~\bibnamefont
  {Romatschke}}, \bibinfo {author} {\bibfnamefont {M.}~\bibnamefont {Mendoza}},
  \ and\ \bibinfo {author} {\bibfnamefont {S.}~\bibnamefont {Succi}},\ }\href
  {\doibase 10.1103/PhysRevC.84.034903} {\bibfield  {journal} {\bibinfo
  {journal} {Phys. Rev. C}\ }\textbf {\bibinfo {volume} {84}},\ \bibinfo
  {pages} {034903} (\bibinfo {year} {2011})},\ \Eprint
  {http://arxiv.org/abs/1106.1093} {arXiv:1106.1093 [nucl-th]} \BibitemShut
  {NoStop}%
\bibitem [{\citenamefont {Ambru\cb{s}}\ and\ \citenamefont
  {Blaga}(2018)}]{Ambrus:2018kug}%
  \BibitemOpen
  \bibfield  {author} {\bibinfo {author} {\bibfnamefont {V.~E.}\ \bibnamefont
  {Ambru\cb{s}}}\ and\ \bibinfo {author} {\bibfnamefont {R.}~\bibnamefont
  {Blaga}},\ }\href {\doibase 10.1103/PhysRevC.98.035201} {\bibfield  {journal}
  {\bibinfo  {journal} {Phys. Rev. C}\ }\textbf {\bibinfo {volume} {98}},\
  \bibinfo {pages} {035201} (\bibinfo {year} {2018})},\ \Eprint
  {http://arxiv.org/abs/1612.01287} {arXiv:1612.01287 [physics.flu-dyn]}
  \BibitemShut {NoStop}%
\bibitem [{\citenamefont {Succi}(2018)}]{Succi:2018}%
  \BibitemOpen
  \bibfield  {author} {\bibinfo {author} {\bibfnamefont {S.}~\bibnamefont
  {Succi}},\ }\href {\doibase 10.1093/oso/9780199592357.001.0001} {\emph
  {\bibinfo {title} {{The Lattice Boltzmann Equation: For Complex States of
  Flowing Matter }}}}\ (\bibinfo  {publisher} {Oxford Univ. Press},\ \bibinfo
  {address} {Oxford, UK},\ \bibinfo {year} {2018})\BibitemShut {NoStop}%
\bibitem [{\citenamefont {Gabbana}\ \emph {et~al.}(2020)\citenamefont
  {Gabbana}, \citenamefont {Simeoni}, \citenamefont {Succi},\ and\
  \citenamefont {Tripiccione}}]{Gabbana:2019ydb}%
  \BibitemOpen
  \bibfield  {author} {\bibinfo {author} {\bibfnamefont {A.}~\bibnamefont
  {Gabbana}}, \bibinfo {author} {\bibfnamefont {D.}~\bibnamefont {Simeoni}},
  \bibinfo {author} {\bibfnamefont {S.}~\bibnamefont {Succi}}, \ and\ \bibinfo
  {author} {\bibfnamefont {R.}~\bibnamefont {Tripiccione}},\ }\href {\doibase
  10.1016/j.physrep.2020.03.004} {\bibfield  {journal} {\bibinfo  {journal}
  {Phys. Rept.}\ }\textbf {\bibinfo {volume} {863}},\ \bibinfo {pages} {1}
  (\bibinfo {year} {2020})},\ \Eprint {http://arxiv.org/abs/1909.04502}
  {arXiv:1909.04502 [hep-lat]} \BibitemShut {NoStop}%
\bibitem [{\citenamefont {Bazzanini}\ \emph {et~al.}(2020)\citenamefont
  {Bazzanini}, \citenamefont {Gabbana}, \citenamefont {Simeoni}, \citenamefont
  {Succi},\ and\ \citenamefont {Tripiccione}}]{Gabbana:2020oka}%
  \BibitemOpen
  \bibfield  {author} {\bibinfo {author} {\bibfnamefont {L.}~\bibnamefont
  {Bazzanini}}, \bibinfo {author} {\bibfnamefont {A.}~\bibnamefont {Gabbana}},
  \bibinfo {author} {\bibfnamefont {D.}~\bibnamefont {Simeoni}}, \bibinfo
  {author} {\bibfnamefont {S.}~\bibnamefont {Succi}}, \ and\ \bibinfo {author}
  {\bibfnamefont {R.}~\bibnamefont {Tripiccione}}\ }(\bibinfo {year} {2020})\
  \Eprint {http://arxiv.org/abs/2011.06856} {arXiv:2011.06856
  [physics.flu-dyn]} \BibitemShut {NoStop}%
\bibitem [{\citenamefont {Kurkela}\ \emph
  {et~al.}(2019{\natexlab{b}})\citenamefont {Kurkela}, \citenamefont
  {Wiedemann},\ and\ \citenamefont {Wu}}]{Kurkela:2018qeb}%
  \BibitemOpen
  \bibfield  {author} {\bibinfo {author} {\bibfnamefont {A.}~\bibnamefont
  {Kurkela}}, \bibinfo {author} {\bibfnamefont {U.~A.}\ \bibnamefont
  {Wiedemann}}, \ and\ \bibinfo {author} {\bibfnamefont {B.}~\bibnamefont
  {Wu}},\ }\href {\doibase 10.1140/epjc/s10052-019-7262-x} {\bibfield
  {journal} {\bibinfo  {journal} {Eur. Phys. J. C}\ }\textbf {\bibinfo {volume}
  {79}},\ \bibinfo {pages} {759} (\bibinfo {year} {2019}{\natexlab{b}})},\
  \Eprint {http://arxiv.org/abs/1805.04081} {arXiv:1805.04081 [hep-ph]}
  \BibitemShut {NoStop}%
\bibitem [{\citenamefont {Romatschke}\ and\ \citenamefont
  {Strickland}(2003)}]{Romatschke:2003ms}%
  \BibitemOpen
  \bibfield  {author} {\bibinfo {author} {\bibfnamefont {P.}~\bibnamefont
  {Romatschke}}\ and\ \bibinfo {author} {\bibfnamefont {M.}~\bibnamefont
  {Strickland}},\ }\href {\doibase 10.1103/PhysRevD.68.036004} {\bibfield
  {journal} {\bibinfo  {journal} {Phys. Rev. D}\ }\textbf {\bibinfo {volume}
  {68}},\ \bibinfo {pages} {036004} (\bibinfo {year} {2003})},\ \Eprint
  {http://arxiv.org/abs/hep-ph/0304092} {arXiv:hep-ph/0304092} \BibitemShut
  {NoStop}%
\bibitem [{\citenamefont {Moln\'ar}\ \emph {et~al.}(2016)\citenamefont
  {Moln\'ar}, \citenamefont {Niemi},\ and\ \citenamefont
  {Rischke}}]{Molnar:2016gwq}%
  \BibitemOpen
  \bibfield  {author} {\bibinfo {author} {\bibfnamefont {E.}~\bibnamefont
  {Moln\'ar}}, \bibinfo {author} {\bibfnamefont {H.}~\bibnamefont {Niemi}}, \
  and\ \bibinfo {author} {\bibfnamefont {D.~H.}\ \bibnamefont {Rischke}},\
  }\href {\doibase 10.1103/PhysRevD.94.125003} {\bibfield  {journal} {\bibinfo
  {journal} {Phys. Rev. D}\ }\textbf {\bibinfo {volume} {94}},\ \bibinfo
  {pages} {125003} (\bibinfo {year} {2016})},\ \Eprint
  {http://arxiv.org/abs/1606.09019} {arXiv:1606.09019 [nucl-th]} \BibitemShut
  {NoStop}%
\bibitem [{\citenamefont {Mysovskikh}(2003)}]{Mysovskikh:1987}%
  \BibitemOpen
  \bibfield  {author} {\bibinfo {author} {\bibfnamefont {I.~P.}\ \bibnamefont
  {Mysovskikh}},\ }\href {\doibase 10.1103/PhysRevD.68.023006} {\bibfield
  {journal} {\bibinfo  {journal} {Dokl. Akad. Nauk SSSR}\ }\textbf {\bibinfo
  {volume} {296}},\ \bibinfo {pages} {023006} (\bibinfo {year} {2003})},\
  \bibinfo {note} {sov. Math. Dokl. \textbf{36}, 229 (1988)}\BibitemShut
  {NoStop}%
\bibitem [{\citenamefont {Mieussens}(2000)}]{Mieussens:2000}%
  \BibitemOpen
  \bibfield  {author} {\bibinfo {author} {\bibfnamefont {L.}~\bibnamefont
  {Mieussens}},\ }\href {\doibase https://doi.org/10.1006/jcph.2000.6548}
  {\bibfield  {journal} {\bibinfo  {journal} {Journal of Computational
  Physics}\ }\textbf {\bibinfo {volume} {162}},\ \bibinfo {pages} {429}
  (\bibinfo {year} {2000})}\BibitemShut {NoStop}%
\bibitem [{\citenamefont {Weih}\ \emph {et~al.}(2020)\citenamefont {Weih},
  \citenamefont {Gabbana}, \citenamefont {Simeoni}, \citenamefont {Rezzolla},
  \citenamefont {Succi},\ and\ \citenamefont {Tripiccione}}]{Weih:2020qyh}%
  \BibitemOpen
  \bibfield  {author} {\bibinfo {author} {\bibfnamefont {L.~R.}\ \bibnamefont
  {Weih}}, \bibinfo {author} {\bibfnamefont {A.}~\bibnamefont {Gabbana}},
  \bibinfo {author} {\bibfnamefont {D.}~\bibnamefont {Simeoni}}, \bibinfo
  {author} {\bibfnamefont {L.}~\bibnamefont {Rezzolla}}, \bibinfo {author}
  {\bibfnamefont {S.}~\bibnamefont {Succi}}, \ and\ \bibinfo {author}
  {\bibfnamefont {R.}~\bibnamefont {Tripiccione}},\ }\href {\doibase
  10.1093/mnras/staa2575} {\bibfield  {journal} {\bibinfo  {journal} {Mon. Not.
  Roy. Astron. Soc.}\ }\textbf {\bibinfo {volume} {498}},\ \bibinfo {pages}
  {3374} (\bibinfo {year} {2020})},\ \Eprint {http://arxiv.org/abs/2007.05718}
  {arXiv:2007.05718 [physics.comp-ph]} \BibitemShut {NoStop}%
\bibitem [{\citenamefont {Shu}\ and\ \citenamefont {Osher}(1988)}]{Shu:1988}%
  \BibitemOpen
  \bibfield  {author} {\bibinfo {author} {\bibfnamefont {C.-W.}\ \bibnamefont
  {Shu}}\ and\ \bibinfo {author} {\bibfnamefont {S.}~\bibnamefont {Osher}},\
  }\href {\doibase 10.1016/0021-9991(88)90177-5} {\bibfield  {journal}
  {\bibinfo  {journal} {J. Comput. Phys.}\ }\textbf {\bibinfo {volume} {77}},\
  \bibinfo {pages} {439} (\bibinfo {year} {1988})}\BibitemShut {NoStop}%
\bibitem [{\citenamefont {Gottlieb}\ and\ \citenamefont
  {Shu}(1998)}]{Gottlieb:1998}%
  \BibitemOpen
  \bibfield  {author} {\bibinfo {author} {\bibfnamefont {S.}~\bibnamefont
  {Gottlieb}}\ and\ \bibinfo {author} {\bibfnamefont {C.-W.}\ \bibnamefont
  {Shu}},\ }\href {\doibase 10.1090/S0025-5718-98-00913-2} {\bibfield
  {journal} {\bibinfo  {journal} {Math. Comp.}\ }\textbf {\bibinfo {volume}
  {67}},\ \bibinfo {pages} {73} (\bibinfo {year} {1998})}\BibitemShut {NoStop}%
\bibitem [{\citenamefont {Jiang}\ and\ \citenamefont {Shu}(1996)}]{Jiang:1996}%
  \BibitemOpen
  \bibfield  {author} {\bibinfo {author} {\bibfnamefont {G.~S.}\ \bibnamefont
  {Jiang}}\ and\ \bibinfo {author} {\bibfnamefont {C.~W.}\ \bibnamefont
  {Shu}},\ }\href {\doibase 10.1006/jcph.1996.0130} {\bibfield  {journal}
  {\bibinfo  {journal} {J. Comput. Phys.}\ }\textbf {\bibinfo {volume} {126}},\
  \bibinfo {pages} {202} (\bibinfo {year} {1996})}\BibitemShut {NoStop}%
\bibitem [{\citenamefont {Busuioc}\ and\ \citenamefont
  {Ambru{\cb{s}}}(2019)}]{Busuioc:2019}%
  \BibitemOpen
  \bibfield  {author} {\bibinfo {author} {\bibfnamefont {S.}~\bibnamefont
  {Busuioc}}\ and\ \bibinfo {author} {\bibfnamefont {V.~E.}\ \bibnamefont
  {Ambru{\cb{s}}}},\ }\href {\doibase 10.1103/PhysRevE.99.033304} {\bibfield
  {journal} {\bibinfo  {journal} {Phys. Rev. E}\ }\textbf {\bibinfo {volume}
  {99}},\ \bibinfo {pages} {033304} (\bibinfo {year} {2019})}\BibitemShut
  {NoStop}%
\bibitem [{\citenamefont {Giacalone}\ \emph {et~al.}(2019)\citenamefont
  {Giacalone}, \citenamefont {Mazeliauskas},\ and\ \citenamefont
  {Schlichting}}]{Giacalone:2019ldn}%
  \BibitemOpen
  \bibfield  {author} {\bibinfo {author} {\bibfnamefont {G.}~\bibnamefont
  {Giacalone}}, \bibinfo {author} {\bibfnamefont {A.}~\bibnamefont
  {Mazeliauskas}}, \ and\ \bibinfo {author} {\bibfnamefont {S.}~\bibnamefont
  {Schlichting}},\ }\href {\doibase 10.1103/PhysRevLett.123.262301} {\bibfield
  {journal} {\bibinfo  {journal} {Phys. Rev. Lett.}\ }\textbf {\bibinfo
  {volume} {123}},\ \bibinfo {pages} {262301} (\bibinfo {year} {2019})},\
  \Eprint {http://arxiv.org/abs/1908.02866} {arXiv:1908.02866 [hep-ph]}
  \BibitemShut {NoStop}%
\bibitem [{\citenamefont {Noronha-Hostler}\ \emph {et~al.}(2016)\citenamefont
  {Noronha-Hostler}, \citenamefont {Yan}, \citenamefont {Gardim},\ and\
  \citenamefont {Ollitrault}}]{Noronha-Hostler:2015dbi}%
  \BibitemOpen
  \bibfield  {author} {\bibinfo {author} {\bibfnamefont {J.}~\bibnamefont
  {Noronha-Hostler}}, \bibinfo {author} {\bibfnamefont {L.}~\bibnamefont
  {Yan}}, \bibinfo {author} {\bibfnamefont {F.~G.}\ \bibnamefont {Gardim}}, \
  and\ \bibinfo {author} {\bibfnamefont {J.-Y.}\ \bibnamefont {Ollitrault}},\
  }\href {\doibase 10.1103/PhysRevC.93.014909} {\bibfield  {journal} {\bibinfo
  {journal} {Phys. Rev. C}\ }\textbf {\bibinfo {volume} {93}},\ \bibinfo
  {pages} {014909} (\bibinfo {year} {2016})},\ \Eprint
  {http://arxiv.org/abs/1511.03896} {arXiv:1511.03896 [nucl-th]} \BibitemShut
  {NoStop}%
\bibitem [{\citenamefont {Niemi}\ \emph {et~al.}(2016)\citenamefont {Niemi},
  \citenamefont {Eskola},\ and\ \citenamefont {Paatelainen}}]{Niemi:2015qia}%
  \BibitemOpen
  \bibfield  {author} {\bibinfo {author} {\bibfnamefont {H.}~\bibnamefont
  {Niemi}}, \bibinfo {author} {\bibfnamefont {K.~J.}\ \bibnamefont {Eskola}}, \
  and\ \bibinfo {author} {\bibfnamefont {R.}~\bibnamefont {Paatelainen}},\
  }\href {\doibase 10.1103/PhysRevC.93.024907} {\bibfield  {journal} {\bibinfo
  {journal} {Phys. Rev. C}\ }\textbf {\bibinfo {volume} {93}},\ \bibinfo
  {pages} {024907} (\bibinfo {year} {2016})},\ \Eprint
  {http://arxiv.org/abs/1505.02677} {arXiv:1505.02677 [hep-ph]} \BibitemShut
  {NoStop}%
\bibitem [{\citenamefont {Bernhard}\ \emph {et~al.}(2016)\citenamefont
  {Bernhard}, \citenamefont {Moreland}, \citenamefont {Bass}, \citenamefont
  {Liu},\ and\ \citenamefont {Heinz}}]{Bernhard:2016tnd}%
  \BibitemOpen
  \bibfield  {author} {\bibinfo {author} {\bibfnamefont {J.~E.}\ \bibnamefont
  {Bernhard}}, \bibinfo {author} {\bibfnamefont {J.~S.}\ \bibnamefont
  {Moreland}}, \bibinfo {author} {\bibfnamefont {S.~A.}\ \bibnamefont {Bass}},
  \bibinfo {author} {\bibfnamefont {J.}~\bibnamefont {Liu}}, \ and\ \bibinfo
  {author} {\bibfnamefont {U.}~\bibnamefont {Heinz}},\ }\href {\doibase
  10.1103/PhysRevC.94.024907} {\bibfield  {journal} {\bibinfo  {journal} {Phys.
  Rev. C}\ }\textbf {\bibinfo {volume} {94}},\ \bibinfo {pages} {024907}
  (\bibinfo {year} {2016})},\ \Eprint {http://arxiv.org/abs/1605.03954}
  {arXiv:1605.03954 [nucl-th]} \BibitemShut {NoStop}%
\bibitem [{\citenamefont {Denicol}\ \emph {et~al.}(2014)\citenamefont
  {Denicol}, \citenamefont {Jeon},\ and\ \citenamefont
  {Gale}}]{Denicol:2014vaa}%
  \BibitemOpen
  \bibfield  {author} {\bibinfo {author} {\bibfnamefont {G.~S.}\ \bibnamefont
  {Denicol}}, \bibinfo {author} {\bibfnamefont {S.}~\bibnamefont {Jeon}}, \
  and\ \bibinfo {author} {\bibfnamefont {C.}~\bibnamefont {Gale}},\ }\href
  {\doibase 10.1103/PhysRevC.90.024912} {\bibfield  {journal} {\bibinfo
  {journal} {Phys. Rev. C}\ }\textbf {\bibinfo {volume} {90}},\ \bibinfo
  {pages} {024912} (\bibinfo {year} {2014})},\ \Eprint
  {http://arxiv.org/abs/1403.0962} {arXiv:1403.0962 [nucl-th]} \BibitemShut
  {NoStop}%
\bibitem [{\citenamefont {Kurkela}\ \emph
  {et~al.}(2020{\natexlab{b}})\citenamefont {Kurkela}, \citenamefont {van~der
  Schee}, \citenamefont {Wiedemann},\ and\ \citenamefont
  {Wu}}]{Kurkela:2019set}%
  \BibitemOpen
  \bibfield  {author} {\bibinfo {author} {\bibfnamefont {A.}~\bibnamefont
  {Kurkela}}, \bibinfo {author} {\bibfnamefont {W.}~\bibnamefont {van~der
  Schee}}, \bibinfo {author} {\bibfnamefont {U.~A.}\ \bibnamefont {Wiedemann}},
  \ and\ \bibinfo {author} {\bibfnamefont {B.}~\bibnamefont {Wu}},\ }\href
  {\doibase 10.1103/PhysRevLett.124.102301} {\bibfield  {journal} {\bibinfo
  {journal} {Phys. Rev. Lett.}\ }\textbf {\bibinfo {volume} {124}},\ \bibinfo
  {pages} {102301} (\bibinfo {year} {2020}{\natexlab{b}})},\ \Eprint
  {http://arxiv.org/abs/1907.08101} {arXiv:1907.08101 [hep-ph]} \BibitemShut
  {NoStop}%
\bibitem [{\citenamefont {Kurkela}\ \emph
  {et~al.}(2019{\natexlab{c}})\citenamefont {Kurkela}, \citenamefont
  {Mazeliauskas}, \citenamefont {Paquet}, \citenamefont {Schlichting},\ and\
  \citenamefont {Teaney}}]{Kurkela:2018wud}%
  \BibitemOpen
  \bibfield  {author} {\bibinfo {author} {\bibfnamefont {A.}~\bibnamefont
  {Kurkela}}, \bibinfo {author} {\bibfnamefont {A.}~\bibnamefont
  {Mazeliauskas}}, \bibinfo {author} {\bibfnamefont {J.-F.}\ \bibnamefont
  {Paquet}}, \bibinfo {author} {\bibfnamefont {S.}~\bibnamefont {Schlichting}},
  \ and\ \bibinfo {author} {\bibfnamefont {D.}~\bibnamefont {Teaney}},\ }\href
  {\doibase 10.1103/PhysRevLett.122.122302} {\bibfield  {journal} {\bibinfo
  {journal} {Phys. Rev. Lett.}\ }\textbf {\bibinfo {volume} {122}},\ \bibinfo
  {pages} {122302} (\bibinfo {year} {2019}{\natexlab{c}})},\ \Eprint
  {http://arxiv.org/abs/1805.01604} {arXiv:1805.01604 [hep-ph]} \BibitemShut
  {NoStop}%
\bibitem [{\citenamefont {Kurkela}\ \emph
  {et~al.}(2019{\natexlab{d}})\citenamefont {Kurkela}, \citenamefont
  {Mazeliauskas}, \citenamefont {Paquet}, \citenamefont {Schlichting},\ and\
  \citenamefont {Teaney}}]{Kurkela:2018vqr}%
  \BibitemOpen
  \bibfield  {author} {\bibinfo {author} {\bibfnamefont {A.}~\bibnamefont
  {Kurkela}}, \bibinfo {author} {\bibfnamefont {A.}~\bibnamefont
  {Mazeliauskas}}, \bibinfo {author} {\bibfnamefont {J.-F.}\ \bibnamefont
  {Paquet}}, \bibinfo {author} {\bibfnamefont {S.}~\bibnamefont {Schlichting}},
  \ and\ \bibinfo {author} {\bibfnamefont {D.}~\bibnamefont {Teaney}},\ }\href
  {\doibase 10.1103/PhysRevC.99.034910} {\bibfield  {journal} {\bibinfo
  {journal} {Phys. Rev. C}\ }\textbf {\bibinfo {volume} {99}},\ \bibinfo
  {pages} {034910} (\bibinfo {year} {2019}{\natexlab{d}})},\ \Eprint
  {http://arxiv.org/abs/1805.00961} {arXiv:1805.00961 [hep-ph]} \BibitemShut
  {NoStop}%
\bibitem [{\citenamefont {Martinez}\ and\ \citenamefont
  {Strickland}(2010)}]{Martinez:2010sc}%
  \BibitemOpen
  \bibfield  {author} {\bibinfo {author} {\bibfnamefont {M.}~\bibnamefont
  {Martinez}}\ and\ \bibinfo {author} {\bibfnamefont {M.}~\bibnamefont
  {Strickland}},\ }\href {\doibase 10.1016/j.nuclphysa.2010.08.011} {\bibfield
  {journal} {\bibinfo  {journal} {Nucl. Phys. A}\ }\textbf {\bibinfo {volume}
  {848}},\ \bibinfo {pages} {183} (\bibinfo {year} {2010})},\ \Eprint
  {http://arxiv.org/abs/1007.0889} {arXiv:1007.0889 [nucl-th]} \BibitemShut
  {NoStop}%
\bibitem [{\citenamefont {Florkowski}\ and\ \citenamefont
  {Ryblewski}(2011)}]{Florkowski:2010cf}%
  \BibitemOpen
  \bibfield  {author} {\bibinfo {author} {\bibfnamefont {W.}~\bibnamefont
  {Florkowski}}\ and\ \bibinfo {author} {\bibfnamefont {R.}~\bibnamefont
  {Ryblewski}},\ }\href {\doibase 10.1103/PhysRevC.83.034907} {\bibfield
  {journal} {\bibinfo  {journal} {Phys. Rev. C}\ }\textbf {\bibinfo {volume}
  {83}},\ \bibinfo {pages} {034907} (\bibinfo {year} {2011})},\ \Eprint
  {http://arxiv.org/abs/1007.0130} {arXiv:1007.0130 [nucl-th]} \BibitemShut
  {NoStop}%
\bibitem [{\citenamefont {Florkowski}\ \emph {et~al.}(2013)\citenamefont
  {Florkowski}, \citenamefont {Ryblewski},\ and\ \citenamefont
  {Strickland}}]{Florkowski:2013lya}%
  \BibitemOpen
  \bibfield  {author} {\bibinfo {author} {\bibfnamefont {W.}~\bibnamefont
  {Florkowski}}, \bibinfo {author} {\bibfnamefont {R.}~\bibnamefont
  {Ryblewski}}, \ and\ \bibinfo {author} {\bibfnamefont {M.}~\bibnamefont
  {Strickland}},\ }\href {\doibase 10.1103/PhysRevC.88.024903} {\bibfield
  {journal} {\bibinfo  {journal} {Phys. Rev. C}\ }\textbf {\bibinfo {volume}
  {88}},\ \bibinfo {pages} {024903} (\bibinfo {year} {2013})},\ \Eprint
  {http://arxiv.org/abs/1305.7234} {arXiv:1305.7234 [nucl-th]} \BibitemShut
  {NoStop}%
\bibitem [{\citenamefont {Martinez}\ \emph {et~al.}(2012)\citenamefont
  {Martinez}, \citenamefont {Ryblewski},\ and\ \citenamefont
  {Strickland}}]{Martinez:2012tu}%
  \BibitemOpen
  \bibfield  {author} {\bibinfo {author} {\bibfnamefont {M.}~\bibnamefont
  {Martinez}}, \bibinfo {author} {\bibfnamefont {R.}~\bibnamefont {Ryblewski}},
  \ and\ \bibinfo {author} {\bibfnamefont {M.}~\bibnamefont {Strickland}},\
  }\href {\doibase 10.1103/PhysRevC.85.064913} {\bibfield  {journal} {\bibinfo
  {journal} {Phys. Rev. C}\ }\textbf {\bibinfo {volume} {85}},\ \bibinfo
  {pages} {064913} (\bibinfo {year} {2012})},\ \Eprint
  {http://arxiv.org/abs/1204.1473} {arXiv:1204.1473 [nucl-th]} \BibitemShut
  {NoStop}%
\bibitem [{\citenamefont {McNelis}\ \emph {et~al.}(2021)\citenamefont
  {McNelis}, \citenamefont {Bazow},\ and\ \citenamefont
  {Heinz}}]{McNelis:2021zji}%
  \BibitemOpen
  \bibfield  {author} {\bibinfo {author} {\bibfnamefont {M.}~\bibnamefont
  {McNelis}}, \bibinfo {author} {\bibfnamefont {D.}~\bibnamefont {Bazow}}, \
  and\ \bibinfo {author} {\bibfnamefont {U.}~\bibnamefont {Heinz}},\ }\href
  {\doibase 10.1016/j.cpc.2021.108077} {\bibfield  {journal} {\bibinfo
  {journal} {Comput. Phys. Commun.}\ }\textbf {\bibinfo {volume} {267}},\
  \bibinfo {pages} {108077} (\bibinfo {year} {2021})},\ \Eprint
  {http://arxiv.org/abs/2101.02827} {arXiv:2101.02827 [nucl-th]} \BibitemShut
  {NoStop}%
\bibitem [{{\relax DLMF}()}]{NIST:DLMF}%
  \BibitemOpen
  {\relax DLMF},\ \href {http://dlmf.nist.gov/} {\enquote {\bibinfo {title}
  {{\it NIST Digital Library of Mathematical Functions}},}\ }\bibinfo
  {howpublished} {http://dlmf.nist.gov/, Release 1.1.2 of 2021-06-15},\
  \bibinfo {note} {f.~W.~J. Olver, A.~B. {Olde Daalhuis}, D.~W. Lozier, B.~I.
  Schneider, R.~F. Boisvert, C.~W. Clark, B.~R. Miller, B.~V. Saunders, H.~S.
  Cohl, and M.~A. McClain, eds.}\BibitemShut {Stop}%
\end{thebibliography}%

\end{document}